\documentclass[10pt,a4paper,twoside]{phdthesis}

\usepackage[T2A]{fontenc}
\usepackage[utf8]{inputenc}

\usepackage{palatino}
\usepackage{calligra}

\usepackage{fancyhdr}
\usepackage{amsmath}
\usepackage{amssymb}
\usepackage{amsfonts}
\usepackage{epsfig}
\usepackage{url}
\usepackage[english,dutch,bulgarian,russian]{babel}
\usepackage{subfigure}
\usepackage{dsfont}
\usepackage{multirow}
\usepackage[sort&compress,numbers]{natbib}

\usepackage[bf]{caption}

\input cyracc.def

\usepackage[paperheight=24cm,paperwidth=17cm,vmarginratio=1:1,twoside,vmargin=3cm,hmargin=1.50cm,bindingoffset=0.50cm]{geometry}

\usepackage{ifpdf}
\ifpdf
\RequirePackage[pdftex]{hyperref}
\else
\RequirePackage[hypertex]{hyperref}
\fi
\hypersetup{hyperindex,pagebackref,pdfpagemode={UseNone},draft=false}

\ifpdf\AtBeginDocument{%
  \hypersetup{pdftitle={Title},pdfauthor={Kiril Hristov}}
}\fi

\newcommand{\superN}{{\mathcal N}}
\newcommand{\kahlerK}{{\mathcal K}}

\newcommand{\smallminus}{\!-\!}

\newcommand{\smallequiv}{\! \equiv \!}

\DeclareMathOperator{\rank}{rank}

\usepackage{chngcntr}
\counterwithout{footnote}{chapter}

\let\oldbibliography\thebibliography
\renewcommand{\thebibliography}[1]{%
  \oldbibliography{#1}%
  \setlength{\itemsep}{2pt}%
}

\begin{document}

\selectlanguage{english}
\pagenumbering{arabic}
\pagestyle{empty}

\renewcommand{\sectionmark}[1]{\markright{\it \thesection\ #1}}
\renewcommand{\chaptermark}[1]{\markboth{\it \thechapter.\ #1}{}}

\begin{flushright} \small
ITP--UU--12/26 \\ SPIN--12/24
\end{flushright}
\bigskip
\begin{center}
 {\Large\bfseries Lessons from the Vacuum Structure\\ of 4d $N=2$ Supergravity }\\[5mm]
{\Large Kiril Hristov} \\[3mm]
 {\large\slshape
 Institute for Theoretical Physics \emph{and} Spinoza Institute, \\
 Utrecht University, 3508 TD Utrecht, The Netherlands \\
\medskip
 {\upshape\ttfamily K.P.Hristov@uu.nl}\\[3mm]}
\end{center}
\vspace{5mm} \hrule\bigskip \centerline{\large \bfseries Abstract}
\bigskip
This PhD thesis is devoted to the study of supersymmetry preserving background solutions of $N=2$ supergravity in $4$ dimensions. The theories in consideration include arbitrary electric gaugings in the vector- and hypermultiplet sectors introduced in the beginning of the thesis. The main contents are divided into three major parts. Most of the chapters are based on previously published results with the exception of chapter \ref{chapter::no-go} in part \ref{part::3}, which is genuinely new. 
\bigskip

In part \ref{part::1} we consider vacua that can be fully analyzed by requiring preserved supersymmetry. We determine and analyze maximally supersymmetric configurations, preserving eight supercharges. We present several examples of such solutions and connect some of them to vacuum solutions of flux compactifications in string theory. We also provide a supersymmetry preserving consistent truncation of the gauged theory by integrating out massive supermultiplets. 

\bigskip

The second part focuses on the topic of supersymmetric black holes. These can be asymptotically flat or asymptotically anti-de Sitter (AdS), and we analyze both cases in detail. We construct BPS black hole solutions in Minkowski space with charged hypermultiplets. We find solutions with vanishing scalar hair that resemble already known black holes, while the genuinely new solutions with hair that we find contain ghost modes. We also elaborate on the static magnetic supersymmetric AdS black holes, investigating thoroughly the BPS constraints for spherical symmetry in gauged supergravity. We find Killing spinors that preserve two of the original eight supercharges and investigate the conditions for genuine black holes free of naked singularities. The existence of a horizon is intimately related with the requirement that the scalars are not constant, but given in terms of harmonic functions in analogy to the attractor flow in ungauged supergravity. We also briefly comment on the toroidal and hyperbolic BPS black holes in AdS.

\bigskip

The third major topic of this thesis is BPS bounds, and in this context we discuss asymptotically Minkowski and AdS solutions in full generality. Concerning asymptotically AdS spacetimes, we find that there exist two disconnected BPS ground states of the theory, depending on the presence of magnetic charge. Each of these ground states comes with a different superalgebra and a different BPS bound, which we derive. As a byproduct, we also demonstrate how the supersymmetry algebra has a built-in holographic renormalization method to define finite conserved charges. We derive the general form of the charges for all asymptotically flat, AdS, and magnetic AdS spacetimes. Some particular black hole examples from part \ref{part::2} are considered to explicitly demonstrate how AdS and mAdS masses differ when solutions with non-trivial scalar profiles are considered. Finally, chapter \ref{chapter::no-go} includes a comprehensive study of the superalgebras of the static black holes and their near-horizon geometries in $4d$ $N=2$ supergravity. We derive a no-go theorem for genuine BPS black holes in AdS in the absence of hypermultiplets and give the conditions for potential hypermultiplet gaugings that can evade it. We briefly comment on the analogous implications for supersymmetric non-rotating black holes and black rings in $5d$. 

\begin{center}
\fontsize{16}{16}\selectfont
Lessons from  \\[.5mm]
\fontsize{24}{24}\selectfont
the Vacuum Structure \\[.6mm]
\fontsize{20}{20}\selectfont
of 4d $N=2$ Supergravity

\vspace{1.5cm}

\large
Lessen uit \\[-1mm]
\huge
de Vacu\"{u}mstructuur \\[.5mm]
\Large
van 4d $N=2$ Superzwaartekracht

\vspace{0.5cm}

\vspace{2.0cm}

\Large Proefschrift

\vspace{1.0cm}

\parbox{0.8\textwidth}{
\normalsize
ter verkrijging van de graad van doctor aan de Universiteit Utrecht op
gezag van de rector magnificus, prof.\ dr.\ G.\ J.\ van der Zwaan, ingevolge het
besluit van het college voor promoties in het openbaar te verdedigen
op vrijdag 1 juni 2012 des middags te 2.30 uur \\
}

\vspace{1.0cm}

\large door

\vspace{1.0cm}

\LARGE Kiril Petrov Hristov

\vspace{0.4cm}

\large geboren op 30 maart 1985 te Sofia, Bulgarije

\end{center}

\newpage

\noindent
\begin{tabular}{ll}
Promotor:& Prof.\ dr.\, S.\ J.\ G.\ Vandoren \\
\end{tabular}

\vfill

This thesis was accomplished with financial support from the Huygens Scholarship Programme of the Netherlands Organization for International Cooperation in Higher Education (NUFFIC) and from the Netherlands Organization for Scientific Research (NWO) under the VICI grant 680-47-603.


\pagestyle{fancyplain}

\setcounter{tocdepth}{1}
\tableofcontents


\chapter*{Publications}
\addcontentsline{toc}{chapter}{Publications}

The main chapters of this thesis are based on the following (mostly published) works:

\begin{itemize}
\item
chapter 3:\\ \cite{Hristov:2009uj} K.~Hristov, H.~Looyestijn and S.~Vandoren,
    \emph{Maximally supersymmetric solutions of $D=4$ $N=2$ gauged
      supergravity},
\textit {JHEP} \textbf {11} (2009) 115.

\item
chapters 4, 5:\\ \cite{Hristov:2010eu} K.~Hristov, H.~Looyestijn and S.~Vandoren,
\emph{BPS black holes in $N=2$ $D=4$ gauged supergravities},
\textit {JHEP} \textbf {08} (2010) 103.

\item
chapter 6:\\ \cite{Hristov:2010ri} K.~Hristov and S.~Vandoren,
\emph{Static supersymmetric black holes in AdS$_4$ with spherical symmetry},
\textit {JHEP} \textbf {04} (2011) 047.

\item
chapters 7, 8, 9:\\ \cite{Hristov:2011ye} K.~Hristov, C.~Toldo and S.~Vandoren,
    \emph{On BPS bounds in $D=4$ $N=2$ gauged supergravity},
    \textit {JHEP} \textbf {12} (2011) 014. \\
\cite{Hristov:2011qr} K.~Hristov,
    \emph{On BPS Bounds in $D=4$ $N=2$ Gauged Supergravity II: General Matter couplings and Black Hole Masses},  \textit {JHEP} \textbf {03} (2012) 095.\\
\cite{Hristov:2012bd} K.~Hristov, C.~Toldo and S.~Vandoren,
    \emph{Black branes in AdS: BPS bounds and asymptotic charges},
    proceedings of the "XVII European Workshop on String Theory 2011", Padova, Italy, 5-9 September 2011

\item chapter 10:\\ K.~Hristov, new unpublished material.
\end{itemize}

Other publications, to which the author has contributed:
\begin{itemize}
\item \cite{Hristov:2008if} K.~Hristov, \emph{Axion Stabilization in Type IIB Flux Compactifications},
\textit {JHEP} \textbf {01} (2009) 046.
\end{itemize}

\newcommand{\publ}{}

\lhead[\thepage]{\fancyplain{}{}}
\chead[\fancyplain{}{}]{\fancyplain{}{}}
\rhead[\fancyplain{}{\leftmark}]{\fancyplain{}{\thepage}}
\lfoot[]{}
\cfoot[]{}
\rfoot[]{}


\chapter{Introduction and motivation}
\label{chapter::introduction}

The main subject of this work, the analysis of the vacuum structure of four-dimensional $N=2$ supergravity, might at first seem rather technical and disconnected from the fundamental questions of modern high energy physics. This thesis is after all just a case study of a particular theory that is ill-defined at high energies and does not seem to describe the physics of our universe even only at an effective level. The very meaning of ``vacuum structure'' is somewhat obscure and the question which background solutions are ``vacua'' will be further pursued in the main body of this work. Nevertheless, I will try to argue that classifying and understanding the various solutions in $D=4$ $N=2$ supergravity is in fact very much relevant for a number of important topics. This particular supergravity theory is connected in numerous direct or more subtle ways to different branches of theoretical physics, a more detailed account of which follows. This list encompasses a number of major topics in high energy physics at present. However, due to the vast amount of research topics, it is far from being comprehensive and some interesting implications have been omitted.

\subsubsection{General Relativity}
General Relativity (GR) has been one of the two hugely successful and groundbreaking ideas in 20th century physics. It is essentially a classical theory of gravity, developed by Einstein to reconcile his special theory of relativity with the notions of gravitational acceleration and space and time. GR introduces the revolutionary idea that the space and time are not just a background for physics to happen, but a dynamic part of it. Thus matter can influence and change the curvature of the spacetime, which in turn dictates the motion of matter. These ideas have been tested and verified in numerous experiments since they were first published in 1915, and GR is at present the best established and least controversial theory in physics\footnote{The superluminal neutrinos found in a recent experiment \cite{Adam:2011zb} seem to be potentially incompatible with the theories of special and general relativity, but further evidence and analysis are needed before more definitive statements can be made. At present, it seems likely that the final outcome of the experiment will be in accordance with Einstein's theory after some technical errors are corrected.}. Crucially, GR predicts its own failure at small scales due to the generic appearance of singularities in spacetimes. In this sense it is an incomplete theory and it is generally thought that GR should be viewed as the classic limit at low energies (large scales) of a more fundamental theory of quantum gravity.

Notice that, although GR is a relatively old and well-understood theory, we are very far from having a full classification of its possible solutions, i.e.\ of the spacetime and matter configurations permitted by the Einstein equations. Many attempts have been made but due to the infinite possibilities for matter couplings to gravity this is a hopeless task in its full generality. It is also important to understand which solutions are physically possible, starting from some realistic assumptions for the matter inside our universe. This is related with the well-known cosmic censorship conjectures (see e.g.\ \cite{Wald:1997wa} for more details). It is here that studies of $D=4$ $N=2$ supergravity solutions can be connected with the theory of GR. At the classical level, $N=2$ supergravity is just a particular type of relativistic theory coupled with matter. However, due to the presence of supersymmetry its symmetric solutions are governed by first instead of second order differential equations. This makes classification of solutions a much more feasible task, although it has also not been accomplished yet. Further details on the progress towards classification will be presented in the main body of this thesis.

\subsubsection{Quantum gravity}
The other paradigm shift of 20th century physics was the quantum theory, explaining how things at the very small scales behave. The foundations of the quantum theory were laid by Planck's work on black body radiation, giving birth to quantum mechanics. Subsequent work of Einstein on the
photo-electric effect and of Bohr on the atomic model led to further confirmation of the quantum theory of particle physics. The non-intuitive concepts of uncertainty and probabilistic interpretation, developed further by Heisenberg and Schr\"{o}dinger among others, remain some of the most puzzling physics facts. These quantum principles have been of major importance to humanity, giving rise to a number of technical applications. A further quantum treatment of relativistic electrodynamics led to the formulation of quantum field theory (QFT), the framework describing all particle interactions.

As successful as they have been until now, the principles of quantum mechanics have not yet been reconciled with the Einstein theory of General Relativity. As already mentioned, GR cannot be a complete theory and it is generally desired and believed that a theory of quantum gravity can be formulated. Any consistent quantum theory that in its classical (large scale) limit reduces to GR would in principle qualify for a theory of quantum gravity. However, such a theory is notoriously difficult to formulate. There have been numerous proposals in literature, but they are all a subject of current research and are heavily disputed. Therefore it is fair to say that a theory of quantum gravity has not been established yet, while mentioning as most notable candidates at present loop quantum gravity and string theory. From this point of view, $D=4$ $N=2$ supergravity is usually placed inside the broader framework of string theory, but it does not need to necessarily take a stand in this competition (see more later).

It is worth mentioning that we can explore some aspects of the quantum nature of gravity even without advocating any of the above mentioned theories. Although $N=2$ supergravity might not be a sensible quantum theory at very high energies, it is believed that the solutions preserving some of the fermionic symmetries of the theory remain stable at all energies. This can be used to probe some of the quantum properties of gravity. Black hole solutions are particularly interesting in this sense because they possess a classical entropy that remains the same in the quantum regime. A full theory of quantum gravity has to then explain the microscopic origin of this entropy in terms of fundamental degrees of freedom \cite{Strominger:1996sh,Maldacena:1997de}. This quantum aspect is less emphasized in the main text, but we will nevertheless be able to gain some intuition and comment on it in the concluding chapter.

\subsubsection{String theory}
One of the leading candidates for a theory of quantum gravity, string theory is based on a relatively new idea developed in 1970. It assumes that all elementary ingredients of nature are strings, instead of particles, propagating in a $10$-dimensional spacetime. The gravitational interaction in this context is just one of the infinitely many string excitations. All other particles that we observe in nature are also supposed to arise via the same mechanism. At low-energies string theory can still be described by effective particle theories, such as supergravities, and the hope is that eventually this will lead to the Standard Model of particle physics.

The Standard Model is at present the quantum field theory that describes best all physical phenomena, except gravity. It is being constantly checked and updated by the particle accelerators, presently by CERN's Large Hadron Collider (LHC). In the context of string theory, the Standard Model is thought of as a low-energy limit of the theories of supergravity, which in turn arise as low-energy limits of string theory. This topic within string theory is usually called flux compactification because one needs to split the original $10$ dimensions into $6$ small and compact directions and the $4$ dimensions we can all see. This way $4$-dimensional supergravities arise, and the less amount of supersymmetry there is (quantified by the value of $N$), the closer to real world physics we are. In other words, we live in an $N=0$ universe. However, this could still happen within an $N>0$ theory by the process of supersymmetry breaking. It is conceivable that the universe is a $N=0$ symmetric vacuum that can be found in $N=2$ supergravity, which in turn is a limiting case of string theory. Whether this is the correct approach to embedding our universe in string theory remains highly speculative and there is neither theoretical, nor experimental evidence to substantiate it. We will see in more detail in what follows that in fact de Sitter (dS) universes like ours are very rarely found and generally unstable solutions of supergravity.

\subsubsection{AdS/CFT correspondence}
A field that originated from string theory and now enjoying life of its own, the AdS/CFT correspondence is presently generating most of the interest in high energy physics. The correspondence was discovered by Maldacena in 1997 \cite{Maldacena:1997re} and has its roots in earlier ideas of 't Hooft \cite{tHooft:1973jz,tHooft:1993gx}. It relates quantum gravity on a $d$-dimensional anti-de Sitter (AdS) spacetime and a $(d-1)$-dimensional conformal field theory (CFT) in the absence of gravity. Although the precise correspondence involves the full quantum theories, it can also be partially verified in the corresponding classic or semi-classic limits. This is where supergravity enters the picture, being the low-energy effective action of string theory via flux compactifications. For some particular cases and dimensions the duality between the two sides can be made very precise and extremely non-trivial checks in the case of AdS$_5$/CFT$_4$ have been performed, suggesting that the general concept is correct. However, mathematically speaking, AdS/CFT remains a conjecture and one is even tempted to invert the statement and use it as a definition of quantum gravity given complete knowledge of the dual CFT.

Connection with the fundamental theory of quantum gravity aside, the real importance of AdS/CFT in very recent years has been in its various applications in experimentally reachable areas of physics. It turns out that even with simple classical solutions, e.g.\ black holes that asymptote to AdS, one can simulate to a good approximation a number of physically relevant field theories at strong coupling, such as the quark-gluon plasma \cite{Policastro:2001yc,Kovtun:2004de} tested at particle accelerators like the LHC and numerous condensed matter systems, e.g.\ \cite{Hartnoll:2009sz,Horowitz:2010gk,Hartnoll:2007ai}. It is therefore very important to understand well the possible asymptotic AdS solutions in supergravity, a subject that is directly related with the topic of this thesis.

\subsubsection{Supersymmetry and supergravity}
A fundamental particle in a quantum theory can be either a fermion or a boson. This defines the quantum statistics it obeys -
only a single fermion can occupy a particular quantum state at a given time, while no such restriction applies for bosons. Matter we know of is made entirely of fermions, e.g.\ the electrons, the protons and neutrons made out of quarks, atoms and molecules made out of electrons, protons and neutrons. Bosons on the other hand are the mediators of forces, such as the particle of light, the photon, that carries the electromagnetic force. Supersymmetry relates these two different types of particles. Each boson has a fermionic partner, and vice versa, forming a pair of superpartners. There can be situations where more than two particles are related this way, leading to a higher amount of supersymmetry (i.e.\ higher $N$, see the more precise meaning in the next chapter). As mentioned above, the Standard Model does not exhibit such a symmetry, whereas it is an essential component for supergravities and (super)string theory.

The other essential ingredient of supergravity is of course gravity, i.e.\ we need to describe the graviton, the boson carrying the force of gravity. Its superpartner fermion is standardly called the gravitino. There will be $N$ gravitini in any theory of supergravity, thus one can think of $N=2$ supergravity as the collection of all possible theories of gravity with supersymmetry and exactly two gravitini. There is no further restriction on the form of matter that can be included (as long as it exhibits the required symmetry), thus we will see that different possibilities for matter-coupled $N=2$ supergravities exist.

Originally, supergravities were discovered independently of any connection with string theory as separate candidates for theories of quantum gravity \cite{Freedman:1976xh}, but it was soon realized that they are non-renormalizable, i.e.\ not well behaved at high energies. However, recent progress suggests that in fact $D=4$ $N=8$ supergravity may be finite and thus well-behaved even if non-renormalizable by power counting. All supergravity theories are related via a complicated web of dualities, compactifications, and reductions. Their interconnection and intrinsic similarity mean that studying one particular theory can lead to a better understanding of all of them. In this sense the particular case of $N=2$ among the other $4$-dimensional theories is best-suited for making relations with others. It does not have so much symmetry to only allow for very restricted classes of solutions, and possesses just enough symmetry to be still mathematically tractable. This will be explained in more details in the following with explicit examples. Many of the tools used for classifying solutions immediately generalize to other supergravities. One is thus able to infer many implications about supergravity as a general concept from the case study of $D=4$ $N=2$ supergravity.

Having introduced supersymmetry and supergravity, we need to say that the notion of supersymmetric, or BPS, states is central for this thesis. These are background solutions that remain invariant under some of the supersymmetry transformations of the theory they exist in.  The BPS states, together with the corresponding BPS bounds they obey, are the building blocks of the vacuum structure of any supergravity theory. Leaving the technical terminology for later chapters, one of the important conceptual ideas in this thesis is that every supersymmetric solution in supergravity can be viewed as a vacuum in the sense that it defines a ground state. Another related notion, which has a prominent role in this thesis, is the supergroup of symmetries corresponding to a given BPS state and described by its superalgebra. BPS states and their superalgebras are a bridge between classical and quantum gravity since the existence of supersymmetry protects solutions from any high energy corrections that can destabilize them. This means that a BPS state in an effective theory such as 4d $N=2$ supergravity continues to exist and keeps its main properties, i.e.\ superalgebra, in the full quantum theory.

\subsubsection{Thesis contribution}
The relation between the main topic of this thesis and the above short list of major branches in theoretical physics might at this point seem rather obscure and sometimes contradictory. $D=4$ $N=2$ supergravity has been placed among both effective and fundamental theories, useful for its BPS solutions while real world physics does not exhibit any supersymmetry. This reflects well the state of the art in high energy physics at present: there is a general fuzzy pattern of what seem to be the relevant pieces of the puzzle and their internal connections. Still, there are many loose ends that need to be fitted in the full picture. In this sense the present work has the aim of understanding some particular ideas, representing just a single piece in the puzzle of physics. Although the results of this work might not be of great importance for solving the puzzle, many of the general concepts could turn out to be useful in the future.

In more precise terms, the study of the vacuum structure of $D=4$ $N=2$ supergravity will enable us to make concrete statements on some of the issues mentioned above. This unfortunately cannot be achieved with a pure physics intuition and without a certain degree of technicality in the discussions. I will therefore use exact definitions and try to present only fully rigorous results in the remaining chapters of the thesis, returning to more general and speculative discussions in the concluding chapter.

The direct contributions of this thesis to the major topics introduced above can be summarized briefly as follows. In the context of General Relativity, the vacuum structure of $N=2$ supergravity gives a new perspective on the cosmological censorship conjecture. We are going to see how BPS bounds provide a stability criterion on classical solutions in GR and project out some undesired solutions. We further use BPS bounds and superalgebras to learn more about the quantum gravity aspects of black holes and discuss possible outcomes for the faith of certain (classically) nakedly singular spacetimes in a full quantum gravity regime. Supersymmetric black hole solutions are also discussed from string theory perspective, since we show that certain novel classes of four-dimensional static black holes can be embedded in M-theory. These are interesting for the study of microscopic entropy from brane constructions. Since these new black holes have AdS asymptotics, their dual field theories can be of potential use for direct applications of the AdS/CFT dictionary. We give a very detailed analysis of the symmetries of AdS and the conserved charges on its boundary, showing that there are several distinct AdS-like vacua in $N=2$ supergravity. The techniques we use for determining supersymmetric solutions, BPS bounds and conserved charges, are directly applicable to other supersymmetry and supergravity theories in various dimensions. Several of the main chapters in this thesis can thus be used for applications in 10 and 11 dimensional supergravities that are of more fundamental nature from string/M-theory point of view.

These main lessons from the vacuum structure of $N=2$ supergravity are spread somewhat eclectically in the main body of this thesis. This is inevitably due to the different mathematical tools needed in the separate parts of the thesis, a more detailed account of which follows.

\subsubsection{Thesis content}
In chapter~\ref{chapter::supergravity} the theory at question, namely $D=4$ $N=2$ supergravity, is introduced in some detail. This is done separately for pure supergravity and for the possible matter couplings to vector and hypermultiplets. A clear distinction between the gauged and ungauged theories is made with explanation of the physical significance in each case. This chapter is the groundwork for all further discussions and results in this thesis and is thus
of central importance. The reader needs to posses a reasonable understanding of supergravity theory before reading further.

The remaining main chapters of this thesis are divided into three parts to ease the reader in choosing topics. Each of these parts can in principle be read independently of the others, although the understanding of certain technical points in part \ref{part::2} can be facilitated by first reading part \ref{part::1} while part \ref{part::2} provides some examples for the general concepts in part \ref{part::3}.

In part \ref{part::1} we start with the ``easiest'' solutions one can find in supergravity theories, the ones that follow entirely from supersymmetry. This means that the equations to be solved are only first order partial differential equations instead of second order. As shown in chapter \ref{chapter::n2vac}, this is enough to completely determine the solutions that preserve all supersymmetries of the action. This is equivalent with saying that only fully-supersymmetric (or fully-BPS) solutions are discussed. We then solve the differential equations and determine certain algebraic conditions on the fields of the theory that will ensure one can find such solutions. As far as the allowed spacetimes are concerned, there are only four possible configurations: three flat vacua, amongst which four-dimensional Minkowski space, and one vacuum with constant non-zero curvature, AdS$_4$. These solutions are also important in the later chapters when we try to find asymptotically flat (Minkowski) or asymptotically AdS black hole solutions (see later in part \ref{part::2}). The next chapter of this part, \ref{chapter::solution_generator}, is very similar in spirit. It uses some of the identities discovered in the study of fully-supersymmetric vacua in order to propose a method of generating new solutions with extra matter from already known ones. Alternatively, one can view the results as describing a consistent supersymmetric truncation for the matter-coupled $D=4$ $N=2$ theories. This leads to a reduced number of degrees of freedom and thus to a simplification of the theory whenever needed.

Part \ref{part::2} deals primarily with the study of black hole solutions in $D=4$ $N=2$ ungauged and gauged supergravity. It contains a general summary of all known black hole solutions that asymptote either to Minkowski or to AdS space, but focuses on the supersymmetric solutions that can be to a large extent fully classified. In chapter \ref{chapter::blackholes_flat} the focus is entirely on asymptotically flat black holes. The supersymmetric solutions in this case are fully classified in the ungauged theory, and some generic statements can be made in the case of gaugings that leave the asymptotics flat. An attempt to find a qualitatively new class of solutions with charged hypers is made, but it turns out that some of the fields in this case need to be ghosts, which generally renders such solutions unphysical. In the next chapter only asymptotically AdS black holes are considered. These solutions can only be found in the theories with gauging. They are relevant for some applications of the AdS/CFT correspondence and exhibit qualitatively very different relation with supersymmetry compared to their flat analogs. We discuss in details particularly the case of static supersymmetric black holes that are shown to exist only when nontrivial scalar profiles are considered \cite{Cacciatori:2009iz,Dall'Agata:2010gj}. This is an example where the addition of extra matter can crucially change the properties of the spacetime and teaches us an important lesson for the supersymmetric vacuum structure. This topic comes into focus also in the next part of this thesis.

The last main part of this work is devoted to superalgebras and BPS bounds in $D=4$ $N=2$ supergravity. It provides a different viewpoint towards the classification of solutions and some other issues discussed in parts \ref{part::1} and \ref{part::2}. It zooms out of the particular details of a given solution and concentrates on large classes that have the same asymptotics, e.g.\ asympotically flat or asymptotically AdS solutions of any type. In chapter \ref{chapter::BPS-general}, the general procedure of finding the superalgebra of a given field configuration is outlined. Although it is focused on the specific supergravity theory at hand, the principles are in fact completely general and thus applicable for any field theory possessing supersymmetry (i.e.\ all supersymmetric field theories and supergravities). It turns out that this approach leads to well-defined asymptotic charges for all solutions that asymptote to a supersymmetric vacuum. Thus the technique of holographic renormalization \cite{Henningson:1998gx,Henningson:1998ey,deHaro:2000xn,Skenderis:2000in,Bianchi:2001de,Papadimitriou:2004rz,Papadimitriou:2005ii,Cheng:2005wk} at the boundary of AdS is no longer explicitly needed. In chapter \ref{chapter::BPS-minimal} we then concentrate on the simplest type of gauged $D=4$ $N=2$ supergravity. The focus of this chapter is on the surprising fact that there are two different vacua with the same AdS asymptotics, differentiated by the value of the magnetic charge. The two different superalgebras and BPS bounds are discussed in details and some previous confusion in literature is clarified. The next chapter then discusses $D=4$ $N=2$ supergravity with matter couplings. The main fermionic anticommutator of the superalgebra is derived, allowing one to find the relevant superalgebra practically in every possible case allowed by the theory. Some particular examples are discussed, showing how the black holes from chapter \ref{chapter::blackholes_AdS} fit in the story. The final chapter of this part of the thesis, \ref{chapter::no-go}, presents a very explicit application of the superalgebra analysis to the purpose of classifying solutions. The different static BPS solutions of part \ref{part::2} are discussed and their corresponding superalgebras are derived. This allows us to prove a no-go theorem for the existence of spherically symmetric BPS black holes in AdS$_4$ in absence of hypermultiplets, while giving a clear prediction on the attractor mechanism for such black holes in theories with suitable hypermultiplet gaugings. We also show that the near-horizon geometries for black holes in Minkowski and magnetic AdS are unique and distinct from each other. We close this chapter with some remarks on the possible black hole superalgebras in $5$ dimensions, which turn out to be strikingly similar to their four-dimensional counterparts. We then give analogous conditions for the potential existence of non-rotating BPS black holes and black rings in AdS$_5$.

In the end of the main body of this work, chapter \ref{chapter::conclusions} summarizes and emphasizes the important lessons learned from the analysis of the vacuum structure of $D=4$ $N=2$ supergravity. Connections with the issues raised in the present chapter are made and some more general statements and conjectures are discussed. The outlook section includes a proposal for the microscopic description of black holes in AdS$_4$ via a version of the AdS/CFT correspondence. This chapter concludes the relevant physical discussion in this thesis. In order to make the main text less technical, some more mathematical aspects are left to the appendices. They are referred to inside the main text and are included to ensure that this work is self-contained. Additionally, there are plenty of references to original work used in this thesis and therefore the interested reader should be able to explicitly check and understand all equations and discussions.


\chapter{$D=4$ $N=2$ Supergravity}
\label{chapter::supergravity}

This chapter gives an introduction to $D=4$ $N=2$ supergravity. It is a central chapter in this work as it explains the lagrangian and field content that will be used throughout the thesis, along with the main notation and conventions. However, it is not a comprehensive review and concentrates on the features of supergravity that are important for the purposes of the later chapters. A more thorough introduction to supersymmetry can be found
in~\cite{Bilal:2001nv, FigueroaO'Farrill:2001tr,wessbagger}, whereas supergravity
is covered in~\cite{deWit:1984hw,Kleijnthesis, vProeyen, Andrianopoli:1996cm}. The connection between
the four- and ten-dimensional supergravities is reviewed in~\cite{Grana:2005jc,Douglas:2006es,Blumenhagen:2006ci}.

We will start directly with the most general bosonic lagrangian of $D=4$ $N=2$ supergravity with electric gauging, explaining the main field content and supersymmetry variations. The remaining parts of the chapter discuss more carefully the main supersymmetry multiplets: the gravity multiplet that is always present in a theory of supergravity, and the vector and hypermultiplets that provide additional matter couplings. Particular emphasis will be put on the difference between ungauged and gauged theories and the respective scalar potentials. We finish with a qualitative discussion about the other local and global symmetries of the theory.

\section{The full bosonic lagrangian}
The standard (two derivative) bosonic part of the on-shell lagrangian of $D=4$ $N=2$ electrically gauged supergravity with $n_V$ vector and $n_H$ hypermultiplets is given by\footnote{We always use the action $S = \int d^4x \sqrt{-g} \mathcal L$.}:
\begin{align}\label{lagr}
  {\cal L}=& \frac{1}{2}R(g)+g_{i\bar \jmath}\nabla^\mu z^i
\nabla_\mu
{\bar z}^{\bar \jmath} + h_{uv}\nabla^\mu q^u \nabla_\mu q^v + I_{\Lambda\Sigma}F_{\mu\nu}^{\Lambda}F^{\Sigma\,\mu\nu}
+\frac 12 R_{\Lambda\Sigma}\epsilon^{\mu\nu\rho\sigma}
F_{\mu\nu}^{\Lambda}F^{\Sigma}_{\rho\sigma}\nonumber\\&-\frac 13 g\,c_{\Lambda,\Sigma\Pi}\,\epsilon^{\mu\nu\rho\sigma}A_\mu^\Lambda
A_\nu^\Sigma \left( \partial_\rho
A_\sigma^\Pi-\frac{3}{8}f_{\Omega\Gamma}{}^\Pi A_\rho^\Omega
A_\sigma^\Gamma \right) - g^2 V(z,\bar z,q)\,,
\end{align}
with potential
\begin{align}\label{eq:scalar-potential}
V(z,\bar z,q)=\Big[ (g_{i\bar \jmath}k^i_\Lambda k^{\bar \jmath}_\Sigma + 4
h_{uv}k^u_\Lambda k^v_\Sigma){\bar L}^\Lambda L^\Sigma + (g^{i\bar
\jmath}f_i^\Lambda {\bar f}_{\bar \jmath}^\Sigma -3{\bar
L}^\Lambda L^\Sigma)P^x_\Lambda P^x_{\Sigma}\Big]\,.
\end{align}
In what follows we always put fermions to zero on classical solutions, therefore we concentrate on the bosonic part of the lagrangian. The fundamental bosonic fields are the metric $g_{\mu \nu}$, $n_V+1$ vector fields $A_{\mu}^{\Lambda}$ ($\Lambda = 0, 1,..., n_V$), $n_V$ complex scalar fields $z^i$ ($i = 1,...,n_V$), and $4 n_H$ real scalar fields $q^u$ ($u=1,...,4 n_H$) called hypers. The fermionic degrees of freedom match exactly the bosonic ones because of supersymmetry. There are two gravitinos $\psi_{\mu A}$ (spin $3/2$), $2 n_V$ gauginos $\lambda^{iA}$ (spin $1/2$), and $2 n_H$ hyperinos $\zeta_\alpha$ (spin $1/2$), where $A = 1, 2$. The full lagrangian is supersymmetric, i.e.\ invariant under transforming the bosons and fermions into each other in a particular way. Since we eventually put all fermions to zero, all bosonic variations vanish, while the fermionic variations read:
\begin{align}
\nonumber \delta_\varepsilon \psi_{\mu A}&= \big(\partial_\mu - \frac 14 \omega_\mu^{ab}
  \gamma_{ab}\big)\varepsilon_A + \frac i 2 A_\mu \varepsilon_A +
  \omega_{\mu A}{}^B \varepsilon_B + \\ &+
T^-_{\mu\nu}\gamma^\nu \epsilon_{AB}\varepsilon^B + ig S_{A B}
\gamma_\mu\varepsilon^B \equiv \tilde{\mathcal{D}}_\mu\varepsilon_A\,,\label{susy-gravi}
\\
  \delta_\varepsilon\lambda^{iA}&=i\nabla_\mu z^i
\gamma^\mu\varepsilon^A + G_{\mu\nu}^{-i}
\gamma^{\mu\nu}\epsilon^{AB}\varepsilon_B+g W^{iAB} \varepsilon_B\,,\label{susygluino}
\\
\delta_\varepsilon \zeta_\alpha &= i\,
\mathcal{U}^{B\beta}_u\nabla_\mu q^u \gamma^\mu
\epsilon_{AB}\mathbb{C}_{\alpha\beta} \varepsilon^A + g N_\alpha^A \varepsilon_A \,,\label{susyhyperino}
\end{align}
upto terms of higher order in fermions. The two Weyl spinors $\varepsilon_A$ are the arbitrary supersymmetry transformation parameters that carry a total of $8$ free parameters (in $4$ dimensions). We therefore say that this is an $N=2$ theory with $8$ supercharges (the number of supercharges depends on the spacetime dimension).

The matrices $W^{iAB}$, $N^A_\alpha$ and $S_{AB}$ are called the
gaugino, hyperino and gravitino mass matrices respectively, and are given
by
\begin{align}
  W^{iAB} &= k^i_\Lambda \bar L^\Lambda \epsilon^{AB} + i g^{i \bar
    \jmath} f_{\bar \jmath}^\Lambda P^x_\Lambda \sigma^{AB}_x\,,\\
  N^A_\alpha &= 2 \mathcal U^A_{\alpha u} \tilde k^u_\Lambda \bar L^\Lambda\,,\\
  S_{AB} &= \frac i 2 P^x_\Lambda L^\Lambda \sigma^x_{AB}\,.\label{mass-gravitino}
\end{align}
The scalar potential \eqref{eq:scalar-potential} can be written in terms of the mass
matrices as
\begin{align}
V=-6S^{AB}S_{AB}+\frac{1}{2}g_{i\bar \jmath}W^{iAB} W^{\bar \jmath}_{AB}+N_\alpha^A N^\alpha_A\ .
\end{align}
The quantities $g_{i\bar \jmath}, I_{\Lambda\Sigma}, R_{\Lambda\Sigma}, L^\Lambda, f_{\bar \jmath}^\Lambda, k^i_\Lambda, T_{\mu \nu}, G^i_{\mu \nu}, A_{\mu}$ are related with the special geometry of the vector multiplet moduli space and are explained in detail a bit later, just like $h_{uv}, U^A_{\alpha u}, \tilde k^u_\Lambda, \omega_{\mu A}{}^B$ that come from the quaternionic geometry of the hypermultiplet moduli space. The structure constants $f_{\Omega\Gamma}{}^\Pi, c_{\Lambda,\Sigma\Pi}$ are also considered in more detail further in the text, while the conventions for the $\gamma, \sigma, \epsilon, \mathbb{C}$ matrices, as well as some other notational issues, are left for App. \ref{appendixA}.

\section{The graviton multiplet}
The $N=2$ graviton multiplet~\cite{deWit:1979ug, Fradkin:1979cw} consists of the graviton $g_{\mu\nu}$, a doublet of
gravitinos $\psi_{\mu A}$ with positive chirality\footnote{Chiral spinors are the eigenspinors of $\gamma^5$. Spinors of positive (negative) chirality therefore obey:\begin{align} \gamma^5 \chi = \pm \chi\ . \end{align}} and a gauge field $A^g_\mu$, called
the graviphoton (the reason for the index $0$ will become clear when we introduce the vector multiplets). As everywhere in this thesis, the negative chirality fermions are given by $\psi^A_\mu \equiv (\psi_{\mu A})^*$. The supersymmetric theories that only consist of the graviton multiplet are called minimal supergragravities.
\subsection{Ungauged}
The bosonic part of the supersymmetric action for the ungauged $D=4$ $N=2$ minimal theory is:
\begin{align}\label{lagr_minimal_ungauged}
  {\cal L}=& \frac{1}{2}R(g) - \frac{1}{2} F_{\mu\nu}^{g}F^{g\,\mu\nu}\ ,
\end{align}
with
\begin{align}
    F_{\mu\nu}^{g} = \frac 12 \left(\partial_\mu A_\nu^g - \partial_\nu A_\mu^g \right)\ .
\end{align}

The gravitino variation in this case is just
\begin{align}
\nonumber \delta_\varepsilon \psi_{\mu A}&= \big(\partial_\mu - \frac 14 \omega_\mu^{ab}
  \gamma_{ab}\big)\varepsilon_A +
F^g_{\mu\nu}\gamma^\nu \epsilon_{AB}\varepsilon^B\ .
\end{align}
The theory remains invariant under global rotations of the gravitino doublet $(\psi_{\mu 1},\psi_{\mu 2})$, i.e. there is a a $U(1)_R \times SU(2)_R$ symmetry group. This symmetry of the action is called R-symmetry and is generally present in all $N > 1$ ungauged supergravities, corresponding to the rotations between the $N$ gravitinos. There is also a local $U(1)$ gauge symmetry acting on the vector field.

\subsection{Gauged}
We can use the gauge symmetry to gauge some of the global isometries of the theory, i.e. the R-symmetry group. One then picks a $U(1)$ subgroup of the $U(1)_R\times SU(2)_R$ and promotes it to a local symmetry (in the same time explicitly breaking the remaining global symmetries). This results in the so called minimal gauged supergravity, where the gravitinos become charged under the vector field (via a covariant derivative) with a charge (coupling constant) $g$. Due to supersymmetry, the bosonic action also gets modified by a negative constant proportional to $g^2$, which effectively plays the role of a cosmological constant term:
\begin{align}\label{lagr_minimal_gauged}
  {\cal L}=& \frac{1}{2}R(g) - \frac{1}{2} F_{\mu\nu}^{g}F^{g\,\mu\nu}\ + 3 g^2\ .
\end{align}
The gravitino variation becomes
\begin{align}
\nonumber \delta_\varepsilon \psi_{\mu A}&= \big(\partial_\mu - \frac 14 \omega_\mu^{ab}
  \gamma_{ab}\big)\varepsilon_A +
F^g_{\mu\nu}\gamma^\nu \epsilon_{AB}\varepsilon^B - \frac 12 g \sigma^a_{AB} \varepsilon^B\ ,
\end{align}
with $a = 1, 2,$ or $3$ corresponding to the $3$ different ways of embedding the $U(1)$ gauge group in the original $U(1)_R\times SU(2)_R$ symmetry.

Notice that although only slightly different, the lagrangians of the gauged and the ungauged theories \eqref{lagr_minimal_gauged} and \eqref{lagr_minimal_ungauged} will have qualitatively different solutions. In the ungauged case one can only have spacetimes with vanishing Ricci scalar as solutions, while in gauged case the scalar curvature is forced to be negative: $R = - 3 g^2$. Therefore already at this stage we can predict that the ungauged theory will give rise to Minkowski and other more general asymptotically flat solutions, while the gauged theory has an AdS$_4$ vacuum and asymptotically AdS solutions of different types. On the other hand, it is certain that de Sitter vacua are not allowed in these simple minimal supergravity theories. We in fact need to include more general matter couplings to the gravity multiplet in order to make the vacuum structure richer and more interesting.

\section{Vector multiplets}
The $N=2$ vector multiplet contains a
complex scalar $z$, a doublet of chiral fermions
$\lambda^A$, called the gauginos, and a
vector field $A_\mu$.

We then couple a number $n_V$ of vector multiplets, with an index
$i=1,\ldots,n_V$, to the graviton multiplet. We use a common index $\Lambda
= 0,\ldots, n_V$ to group the gauge fields $A^g_\mu$ and $A^i_\mu$
together\footnote{In the lagrangian \eqref{lagr} the graviphoton $A_{\mu}^g$ and vector fields $A_{\mu}^i$ mix between each other and appear as vector fields $A_{\mu}^{\Lambda}$, $\Lambda = 0,...,n_V$, with corresponding field strengths $F_{\mu \nu}^{\Lambda}$.}.

\subsubsection{Special geometry}
Supersymmetry requires that the moduli space, described by the metric $g_{i \bar
  \jmath} (z,\bar{z})$, is a special K\"ahler manifold~\cite{deWit:1984pk}. The K\"ahler property implies
that $g_{i \bar \jmath}$ is locally the derivative of a real function
$\mathcal K$
\begin{align}\label{eq:definitie-kahler}
  g_{i \bar \jmath} = \partial_i \partial_{\bar \jmath} \mathcal K\,,
\end{align}
where $\partial_i = \frac{\partial}{\partial z^i}$. The function $\mathcal K$ is called the K\"ahler potential. It is
not unique as the metric $g_{i \bar \jmath}$ remains the same under
the K\"ahler transformation
\begin{align}\label{eq:kahlertransformation}
  \mathcal K \to \mathcal K + f(z) + \bar f(\bar z)\,.
\end{align}
The metric $g_{i \bar \jmath}$ typically depends on the fields $z^i$,
and could contain regions in field space where it is no longer
positive definite. Therefore, one has to restrict the fields to the
so-called positivity domain, where the metric is positive definite.

The extra property of special K\"ahler manifolds is the existence of sections $X^\Lambda$ and $F_\Lambda$, which are
holomorphic functions of $z^i$. The K\"ahler potential $\mathcal K$ for the metric
$g_{i \bar \jmath}$, as in~\eqref{eq:definitie-kahler}, is then given by
\begin{align}
  \label{eq:def-kahler}
 \kahlerK(z, \bar z) = - \ln\left[i \bar X^\Lambda F_\Lambda - i X^\Lambda
    \bar F_\Lambda\right]\,.
\end{align}
From the sections $X^\Lambda$ and $F_\Lambda$, we can construct
\begin{align}
 L^\Lambda &\equiv {\rm e}^{\kahlerK/2} X^\Lambda\,,& M_\Lambda &\equiv e^{\kahlerK/2}
 F_\Lambda\,,\\
  f^\Lambda_i &\equiv {\rm e}^{\kahlerK/2} (\partial_i + \partial_i \kahlerK) X^\Lambda\,,&
  h_{\Lambda|i} &\equiv {\rm e}^{\kahlerK/2} (\partial_i + \partial_i \kahlerK) F_\Lambda\,,
\end{align}
where $\partial_i \kahlerK \equiv \frac {\partial}{\partial
  z^i} \kahlerK$. A further requirement of special
K\"ahler geometry is
\begin{align}
  X^\Lambda h_{\Lambda|i} - F_\Lambda f^\Lambda_i = 0\,.
\end{align}

The terms proportional to $\partial_i \kahlerK$ make $f^\Lambda_i$
and $h_{\Lambda|i}$ transform covariantly
under K\"ahler transformations~\eqref{eq:kahlertransformation}. These terms define the $U(1)$ K\"ahler connection
\begin{align}\label{eq:def-u1-kahler}
  A_\mu \equiv -\frac i 2 (\partial_i \mathcal K \partial_\mu z^i
  - \partial_{\bar \imath} \mathcal K \partial_\mu z^{\bar \imath})\,,
\end{align}
that appears in the supercovariant derivative \eqref{susy-gravi}.

The period matrix $\mathcal
N_{\Lambda\Sigma}$ is defined by the properties
\begin{align}
  F_\Lambda = \mathcal N_{\Lambda\Sigma} X^\Sigma,\qquad
  h_{\Lambda|\bar \imath}
  = \mathcal N_{\Lambda\Sigma} f_{\bar \imath}^\Sigma.
\end{align}
It can be shown~\cite{Craps:1997gp} that the matrix $(L^\Lambda\  f_{\bar \imath}^\Lambda)$ is invertible, which gives the expression
\begin{align}\label{eq:definitie-period2}
  \mathcal N_{\Lambda\Sigma} = \begin{pmatrix}
    h_{\Lambda | \bar \imath} \\  M_\Lambda  \end{pmatrix} \cdot \begin{pmatrix}
f^\Sigma_{\bar \imath}\\    L^\Sigma  \end{pmatrix}^{-1}\,,
\end{align}
and it additionally follows that $\mathcal N_{\Lambda\Sigma}$ is symmetric. We define
\begin{align}
  I_{\Lambda\Sigma} \equiv {\rm Im} \mathcal N_{\Lambda\Sigma}\,,\qquad
  R_{\Lambda\Sigma} \equiv {\rm Re} \mathcal N_{\Lambda\Sigma}\,,
\end{align}
and these matrices appear as couplings in the
Lagrangian. $I_{\Lambda\Sigma}$ is invertible and negative definite~\cite{Craps:1997gp}, and therefore each gauge field
has a kinetic term with a positive sign.

Some further identities one can derive are
\begin{align}\label{eq:period-identities}
  L^\Lambda I_{\Lambda \Sigma} \bar L^\Sigma &= -\frac 12\,,&  L^\Lambda I_{\Lambda\Sigma} f^\Sigma_{i} &= 0\,,\\
  f^\Lambda_i I_{\Lambda\Sigma}  L^\Sigma &= 0\,,&  f^\Lambda_i I_{\Lambda\Sigma} f^\Sigma_{\bar \jmath} &=-\frac 12 g_{i \bar \jmath}\,,
\end{align}
and
\begin{align}
  f_i^\Lambda g^{i \bar \jmath} f_{\bar \jmath}^\Sigma  = -\frac 12 I^{\Lambda\Sigma} - \bar
  L^\Lambda L^\Sigma\,.
\end{align}

It is sometimes possible to specify the sections $X^\Lambda$ and
$F_\Lambda$ in terms of a single holomorphic function $F(X^\Lambda)$, called the
prepotential. In applications to supersymmetry, $F$ is then given as
$F = \frac 12 X^\Lambda F_\Lambda$, and is homogeneous of second
degree. We then have $F_\Lambda = \partial_{X^\Lambda} F$ and $z^i =
X^i / X^0$.

Using the period matrix, we define the linear combinations
\begin{align}
  \label{eq:def-tg}
\begin{split}
  T^-_{\mu\nu} &= 2 i L^\Lambda I_{\Sigma\Lambda}
   F_{\mu\nu}^{\Sigma-} \,,\\
  G^{i-}_{\mu\nu} &= - g^{i \bar \jmath} f_{\bar \jmath}^\Lambda
  I_{\Lambda\Sigma} F^{\Sigma-}_{\mu\nu}\,,
\end{split}
\end{align}
which are the anti-selfdual parts of the graviphoton and matter field strengths (earlier denoted by $F^g_{\mu\nu}, F^i_{\mu\nu}$),
respectively. These relations can be inverted to yield
\begin{align}\label{eq:def-tg-inverse}
F_{\mu\nu}^{\Lambda-}=i{\bar L}^\Lambda
T^-_{\mu\nu}+2f_i^\Lambda G^{i-}_{\mu\nu}\,.
\end{align}

\subsubsection{Examples}\label{sect:sg_examples}
A simple example of special geometry is given by the prepotential
\begin{align}
  F = -\frac i 2 (X^0 X^0 - X^1 X^1)\,.
\end{align}
Using a coordinate $z = X^1 / X^0$, and choosing the gauge $X^0 = 1$, the sections $X^\Lambda$ and $F_\Lambda$ read
\begin{align}
X^0 &= 1\,, & X^1 &= z\,,  & F_0 &= -i\,,  & F_1 &= i z \,,
\end{align}
which are clearly holomorphic functions of $z$. The metric $g_{i \bar
  \jmath}$ can easily be computed from~\eqref{eq:def-kahler} and reads
$ g = (1 - z \bar z)^{-2}\, {\rm d}z \,{\rm d}\bar z\,.$ We see we have to restrict ourselves to $|z| < 1$ and recognize this as the metric on the Poincar\'e disk.

Another important class of examples, which arise in Calabi-Yau
compactifications in string theory, is given by
\begin{align}\label{calabi-yau-prepot}
  F = -\frac 1 6 \frac{\mathcal K_{ijk} X^iX^jX^k}{X^0}\,,
\end{align}
where the $\mathcal K_{ijk}$ are constant, real numbers, determined by the
topology of the Calabi-Yau manifold.

\subsection{Ungauged}
The bosonic terms in the lagrangian are given by
\begin{align}\label{eq:lagr-vectors-ungauged}
\begin{split}
\mathcal L &=\frac{1}{2}R(g)+g_{i\bar
\jmath}(z) \partial^\mu z^i
\partial_\mu
{\bar z}^{\bar \jmath} +
I_{\Lambda\Sigma}(z)F_{\mu\nu}^{\Lambda}F^{\Sigma\,\mu\nu}
+\frac{1}{2}R_{\Lambda\Sigma}(z)\epsilon^{\mu\nu\rho\sigma}
F_{\mu\nu}^{\Lambda}F^{\Sigma}_{\rho\sigma} \,.
\end{split}
\end{align}
As before, we have the Einstein-Hilbert term that now comes together with
a non--linear sigma model for the complex scalars $z^i$. The third
term is the kinetic term for the gauge fields, and the last term is a
generalization of the $\theta$-angle term of Maxwell theory. The fermionic
supersymmetry transformations now include both gravitinos and gauginos:
\begin{align}
\delta_\varepsilon \psi_{\mu A}&= \big(\partial_\mu - \frac 14 \omega_\mu^{ab}
  \gamma_{ab}\big)\varepsilon_A + \frac i 2 A_\mu \varepsilon_A +
T^-_{\mu\nu}\gamma^\nu \epsilon_{AB}\varepsilon^B\,,
\\
  \delta_\varepsilon\lambda^{iA}&=i\partial_\mu z^i
\gamma^\mu\varepsilon^A + G_{\mu\nu}^{-i}
\gamma^{\mu\nu}\epsilon^{AB}\varepsilon_B\,.
\end{align}

\subsubsection{Electro-magnetic duality}
The introduction of vectors allows for one important symmetry of the equations of motion in the ungauged theory called electro-magnetic, or e/m, duality. If we define\footnote{We are using the standard notation in literature, which unfortunately uses the very similar notations $G_{\mu \nu \Lambda}$ and $G^i_{\mu \nu}$ for different field strengths.}
\begin{align}
G_{\mu \nu \Lambda} \equiv i \epsilon_{\mu \nu \rho \sigma} \frac{\delta \mathcal{L}}{\delta F_{\rho \sigma}^{\Lambda}}\ ,
\end{align}
it turns out that the Bianchi identities for the ``electric'' field strengths $F_{\mu \nu}^{\Lambda}$ are the same as the equations of motion for the ``magnetic'' field strengths $G_{\mu \nu \Lambda}$,
\begin{align}\label{e/m dual equations}
D_{\mu} \begin{pmatrix} F^{+ \mu \nu \Lambda} - F^{- \mu \nu \Lambda} \\ G^{+ \mu \nu}_{\Lambda} - G^{- \mu \nu}_{\Lambda} \end{pmatrix} = 0\ .
\end{align}
One can therefore rotate $F_{\mu \nu}^{\Lambda}$ and $G_{\mu \nu \Lambda}$ between each other while keeping the equations of motion invariant. It turns out that the relevant symmetry group is the symplectic group $Sp(2 (n_V+1), \mathbb{Z})$. The electric and magnetic field strengths therefore make up a single symplectic vector,
\begin{align}
    \begin{pmatrix} F^{\mu \nu \Lambda}\\ G^{\mu \nu}_{\Lambda}\end{pmatrix} \rightarrow \begin{pmatrix} \tilde{F}^{\mu \nu \Lambda}\\ \tilde{G}^{\mu \nu}_{\Lambda}\end{pmatrix} = \begin{pmatrix} U^{\Lambda}{}_{\Sigma} & Z^{\Lambda \Sigma}\\ W_{\Lambda \Sigma} & V_{\Lambda}{}^{\Sigma}\end{pmatrix} \begin{pmatrix} F^{\mu \nu \Sigma}\\ G^{\mu \nu}_{\Sigma}\end{pmatrix}\ .
\end{align}
The e/m duality also rotates the sections $X^{\Lambda}$ and $F_{\Lambda}$ as a symplectic vector in the above way in order to keep the equations of motion of the scalar fields invariant. It is a symmetry of the equations of motion, but not of the lagrangian. Nevertheless, it only changes the lagrangian in a very restricted sense, i.e. it leaves it in the general form \eqref{eq:lagr-vectors-ungauged} with different $g_{i \bar{\jmath}}, I_{\Lambda \Sigma}, R_{\Lambda \Sigma}$ due to the change of the sections and field strengths. This means that e/m duality transforms between different $N=2$ supergravities with the same dynamics.

\subsubsection{Example}
E/m duality essentially relates theories with different prepotentials/sections. A simple and often used example is the (electric) STU model that is a special case of \eqref{calabi-yau-prepot} with sections $X^0...X^3$:
\begin{align}\label{el_STU}
  F_{el} = \frac{X^1X^2X^3}{X^0}\,.
\end{align}
If one now rotates the $X^{\Lambda}$'s and the corresponding $F_{\Lambda}$'s with all possible symplectic matrices one will generate the full duality orbit of theories that have the same solutions. One particular choice of symplectic rotation (we will give the explicit details in chapter \ref{6:Klemm} when we really need them) leads to a different prepotential in the same orbit, the magnetic STU model:
\begin{align}\label{magn_STU}
  F_{magn} = -2 i \sqrt{X^0X^1X^2X^3}\,.
\end{align}
Note that not all lagrangians in this or any other given duality orbit come with a prepotential. However, in any given duality orbit there is at least one lagrangian with a prepotential. Due to the more compact notation and often simpler calculations one usually prefers to deal only with theories with prepotentials, knowing that this is enough to capture the dynamics of all possible duality orbits in ungauged $D=4$ $N=2$ supergravity.

\subsection{Gauging internal isometries}
We first consider the $N=2$ vector multiplets and assume the
scalar sector to be invariant under the isometries
\begin{align}\label{eq:Gtrasnfs-scalars}
\delta_G z^i = -g k^i_\Lambda
\alpha^\Lambda,
\end{align}
where $\alpha^\Lambda$ are the parameters of the
transformations, and we have included a coupling constant $g$. To preserve supersymmetry
when we gauge these isometries, the Killing vector fields $k^i_\Lambda$ must
be holomorphic.

To close the gauge algebra on the scalars, the Killing vector
fields must span a Lie-algebra with commutation relations
\begin{equation}\label{Lie-alg}
[k_\Lambda,k_\Sigma]=f_{\Lambda\Sigma}{}^{\Gamma} k_{\Gamma}\,,
\end{equation}
and structure constants $f_{\Lambda\Sigma}{}^{\Gamma}$ of some
Lie-group $G$ that one wishes to gauge. Not all holomorphic
isometries can be gauged within $N=2$ supergravity. The induced
change on the sections needs to be consistent with the symplectic
structure of the theory, and this requires the holomorphic
sections to transform as
\begin{equation}\label{Gtransf-sections}
\delta_G \begin{pmatrix} X^\Lambda \\ F_\Lambda
\end{pmatrix}=-g\alpha^\Sigma\bigg[T_\Sigma \cdot\begin{pmatrix}
X^\Lambda \\ F_\Lambda \end{pmatrix}+r_\Sigma(z)\begin{pmatrix}
X^\Lambda \\ F_\Lambda \end{pmatrix}\bigg]\, .
\end{equation}
The first term on the right-hand-side of
\eqref{Gtransf-sections} contains a constant matrix $T_\Sigma$
that acts on the sections as infinitesimal symplectic
transformations.  For electric gaugings, which we mostly consider in this
thesis, we mean, by definition, that the representation is of the
form
\begin{equation}\label{el-gauge}
T_\Lambda =\begin{pmatrix} -f_\Lambda & 0 \\ c_\Lambda &
f_\Lambda^t \end{pmatrix},
\end{equation}
where $f_\Lambda$ denotes the matrix
$(f_\Lambda)_{\Sigma}{}^\Pi=f_{\Lambda\Sigma}{}^\Pi$ and
$f^t_\Lambda$ is its transpose. The tensor
$c_{\Lambda,\Sigma\Pi}\equiv (c_\Lambda)_{\Sigma\Pi}$ is required
to be symmetric for $T_\Lambda$ to be a symplectic generator.
Moreover, there are some additional constraints on the $c_\Lambda$
in order for the $T_\Lambda$ to be symplectically embedded within
the same Lie-algebra as in
\eqref{Lie-alg}. The second term in \eqref{Gtransf-sections} induces a K\"ahler transformation on the K\"ahler
potential
\begin{equation}\label{Gtransf-K}
\delta_G {\cal K}(z,\bar z)=g\alpha^\Lambda\big(r_\Lambda(z)+{\bar
r}_\Lambda(\bar z)\big)\,,
\end{equation}
for some holomorphic functions $r_\Lambda(z)$.  Finally, closure of the gauge transformations on the
K\"ahler potential requires that
\begin{equation}\label{equiv2}
k^i_\Lambda \partial_ir_\Sigma - k^i_\Sigma \partial_ir_\Lambda =
f_{\Lambda\Sigma}{}^\Gamma r_\Gamma\,.
\end{equation}
We summarize some other identities on vector multiplet
gauging in appendix~\ref{app:specialkahler}.

Magnetic gaugings allow also non-zero entries in the upper--right
corner of $T_\Lambda$, but we will not consider them here. The
gauged action, in particular the scalar potential, that we
consider below is not invariant under magnetic gauge
transformation. To restore this invariance, one needs to introduce
massive tensor multiplets, but the most general lagrangian with
both electric and magnetic gauging is not fully understood yet
(for some partial results see
\cite{Theis:2003jj,D'Auria:2004yi,deWit:2005ub,deVroome:2007zd,Andrianopoli:2007ep}). The theory with mutually compatible electric and magnetic gaugings was recently derived in \cite{deWit:2011gk}.

Given a choice for the gauge group \eqref{el-gauge}, one can
reverse the order of logic and determine the form of the Killing
vectors, and therefore the gauge transformations of the scalar
fields $z^i$. This analysis was done in \cite{vProeyen}, and the
result is written in the appendix, see \eqref{Kill-vect}.

In the lagrangian, one replaces the partial derivatives with the
covariant derivatives
\begin{align}
  \nabla_\mu z^i = \partial_\mu z^i + g k^i_\Lambda A^\Lambda_\mu\,,
\end{align}
where the gauge fields $A^\Lambda_\mu$ transform as $\delta_G
A_\mu^\Lambda = \partial_\mu \alpha^\Lambda$. Notice that the additional terms effectively introduce additional couplings for the ``electric'' gauge fields $A_{\mu}^{\Lambda}$ but not for their magnetic counterparts (one can define them to be the gauge fields $B_{\mu \Lambda}$ from $G_{\mu \nu \Lambda}$). This is why such gaugings are called electric and they break electromagnetic duality. Therefore the gauged supergravity lagrangians in this work can no longer be related by e/m duality\footnote{In the context of the example presented above with the electric and magnetic STU models, they are equivalent only for ungauged supergravity. When gauged, these two models lead to different physics as we will show explicitly in chapters \ref{6:Klemm}.}.
 Furthermore, the lagrangian contains the full nonabelian field strengths
\begin{align}
  F_{\mu\nu}^\Lambda = \frac 12 \left(\partial_\mu A_\nu
    - \partial_\nu  A_\mu\right) + \frac 12 f_{\Sigma\Gamma}{}^\Lambda
    A_\mu^\Sigma A_\nu^\Gamma\,.
\end{align}
Finally, to preserve supersymmetry, we have to modify the
supersymmetry transformations and add additional terms to
the lagrangian~\cite{Andrianopoli:1996cm}, such as mass terms for the fermions (c.f.\ \eqref{mass-gravitino}). Also a scalar potential has to be added, given by
\begin{align}
  V = g_{i \bar \jmath} k^i_\Lambda k^{\bar \jmath}_\Sigma \bar L^\Lambda  L^\Sigma\,.
\end{align}

The lagrangian and susy variations in this case take essentially the most general form as written in the beginning of the chapter, without the terms that concern the hypermultiplet sector which we are yet to discuss. However, we first need to give some more details on the gauging, concerning the notion of moment maps and the tensor $c_{\Lambda, \Sigma \Pi}$.

\subsubsection{Moment maps}
The Killing vectors $k^i_\Lambda$ are holomorphic. Using the Killing
equation, one finds that they can be written as
\begin{equation}\label{eq:def-mom-maps}
k^i_\Lambda=-ig^{i\bar \jmath}\partial_{\bar \jmath}P_\Lambda\,,
\end{equation}
where the real, scalar functions $P_\Lambda$ are called moment maps. For special
K\"ahler spaces, it is convenient to use instead a definition
\begin{equation}\label{mom-maps}
P_\Lambda \equiv i(k^i_\Lambda \partial_i{\cal K}+r_\Lambda)\,,
\end{equation}
where $r_\Lambda$ was defined in equation~\eqref{Gtransf-sections}. Since the K\"ahler potential satisfies \eqref{Gtransf-K}, it is
easy to show that $P_\Lambda$ is real. From this definition, it is
easy to verify~\eqref{eq:def-mom-maps}. Hence the $P_\Lambda$ can be called moment maps, but they are not subject to arbitrary additive constants. Using \eqref{equiv2}
and \eqref{mom-maps}, it is now easy to prove the relation
\begin{equation}\label{equivar}
k^i_\Lambda g_{i\bar\jmath}k^{\bar\jmath}_\Sigma-k^i_\Sigma
g_{i\bar\jmath}k^{\bar\jmath}_\Lambda=if_{\Lambda\Sigma}{}^\Gamma
P_\Gamma\,,
\end{equation}
also called the equivariance condition. The $U(1)$ K\"ahler connection
also gets additional terms due to the gauging and reads
\begin{align}\label{eq:kahler-connection-gauged}
  A_\mu\equiv -\frac{i}{2}\Big(\partial_i{\cal K}\nabla_\mu
z^i-\partial_{\bar\iota}{\cal K}\nabla_\mu{\bar z}^{\bar
\iota}\Big)-\frac{i}{2}gA_\mu^\Lambda (r_\Lambda -{\bar
r}_\Lambda)\,.
\end{align}

\subsubsection{Gauge invariance}\label{sec:gauge-invariance}
Under the gauge transformations~\eqref{eq:Gtrasnfs-scalars}, the
period matrix $\mathcal N_{\Lambda\Sigma}$ transforms. From~\eqref{eq:definitie-period2} one finds
\begin{equation}\label{eq:transf-mathn}
\delta_G {\cal N}_{\Lambda\Sigma}=-g\alpha^\Pi \left( f_{\Pi
\Lambda}{}^{\Gamma} {\cal N}_{\Gamma \Sigma} + f_{\Pi
\Sigma}{}^{\Gamma} {\cal N}_{\Gamma \Lambda} + c_{\Pi,\Lambda
\Sigma} \right).
\end{equation}
To compensate for this transformation, we need to add an additional
term to the lagrangian~\cite{deWit:1984px}, which involves the $c_\Lambda$ tensor. There
are some additional constraints on this tensor. In the abelian case, the only
constraint is that the totally symmetrized $c$-tensor vanishes,
i.e.\
\begin{equation}\label{cyclicity}
c_{\Lambda,\Sigma\Pi}+c_{\Pi,\Lambda\Sigma}+c_{\Sigma,\Pi\Lambda}=0\,.
\end{equation}
This implies that for a single vector field, the $c_\Lambda$ tensor
term vanishes. The additional constraints for nonabelian gaugings
involve the structure constants \cite{deWit:1984px}:
\begin{equation}\label{struct const condition}
  f_{\Lambda \Sigma}{}^{\Gamma} c_{\Gamma, \Pi \Omega} + f_{\Omega
  \Sigma}{}^{\Gamma} c_{\Lambda, \Gamma \Pi} + f_{\Pi
  \Sigma}{}^{\Gamma} c_{\Lambda, \Gamma \Omega} + f_{\Lambda
  \Omega}{}^{\Gamma} c_{\Sigma, \Gamma \Pi} + f_{\Lambda
  \Pi}{}^{\Gamma}c_{\Sigma, \Gamma \Omega} = 0\,.
\end{equation}

\subsection{Fayet-Iliopoulos (FI) gauging}
Instead of gauging an internal isometry, one can also proceed in analogy to the mininal gauged case, i.e. promote a subgroup of the $U(1)_R\times SU(2)_R$ to a local symmetry. This is standardly called Fayet-Iliopoulos or FI gauging and involves a number of constants $\xi^x_{\Lambda}$ that appear in the scalar potential and specify the charge of the gravitinos. These will become more clear once we discuss hypermultiplet moment maps in the next section. Here we will just give the lagrangian and susy variations for the $U(1)$ FI gauged supergravity that will be used in the later chapters of the thesis,
\begin{align}
\begin{split}
  {\cal L} &=\frac{1}{2}R(g)+g_{i\bar \jmath}\partial^\mu z^i
\partial_\mu
{\bar z}^{\bar \jmath} + I_{\Lambda\Sigma}F_{\mu\nu}^{\Lambda}F^{\Sigma\,\mu\nu}
+\frac 12 R_{\Lambda\Sigma}\epsilon^{\mu\nu\rho\sigma}
F_{\mu\nu}^{\Lambda}F^{\Sigma}_{\rho\sigma}\\ &- g^2 (g^{i\bar \jmath}f_i^\Lambda {\bar f}_{\bar \jmath}^\Sigma
-3{\bar L}^\Lambda L^\Sigma)\xi^a_\Lambda \xi^a_{\Sigma}\,,
\end{split}
\end{align}
with FI parameters $\xi^a_{\Lambda}, a = 1, 2,$ or $3$ and susy rules
\begin{align}
\nonumber \delta_\varepsilon \psi_{\mu A}&= \big(\partial_\mu - \frac 14 \omega_\mu^{ab}
  \gamma_{ab}\big)\varepsilon_A + \frac i 2 A_\mu \varepsilon_A +
  g \xi^a_{\Lambda} A_{\mu}^{\Lambda} \sigma^a_{A}{}^B \varepsilon_B +
T^-_{\mu\nu}\gamma^\nu \epsilon_{AB}\varepsilon^B - \frac 12 g \xi^a_{\Lambda} L^{\Lambda} \sigma^a_{A B}
\gamma_\mu\varepsilon^B\,,
\\
  \delta_\varepsilon\lambda^{iA}&=i\partial_\mu z^i
\gamma^\mu\varepsilon^A + G_{\mu\nu}^{-i}
\gamma^{\mu\nu}\epsilon^{AB}\varepsilon_B+i g g^{i \bar
    \jmath} f_{\bar \jmath}^\Lambda \xi^a_\Lambda \sigma^{a, AB} \varepsilon_B\,.
\end{align}

If there are enough (at least three) vector fields, one can alternatively choose to gauge the $SU(2)_R$ symmetry. This results in a nonabelian gauged supergravity with FI terms, the theory in this case is discussed in \cite{Bergshoeff:2002qk,Bergshoeff:2004kh}. The abelian and nonabelian gauged supergravities with FI terms can be more easily understood once we discuss hypermultiplet moduli spaces, to which topic we turn our attention now.

\section{Hypermultiplets}
An on-shell $N=2$ hypermultiplet consists of four real scalars $q^u$
and two chiral fermions $\zeta_\alpha$. For $n_H$ hypermultiplets,
we have bosonic fields $q^u$, with $u=1,\ldots,4 n_H$ and fermions
$\zeta_\alpha$ with $\alpha=1,\ldots, 2n_H$. The bosonic lagrangian
for the hypermultiplets is a non-linear sigma model
\begin{align}\label{eq:lagr-hyper-ungauged}
  \mathcal L = h_{uv} \partial_\mu q^u \partial^\mu q^v.
\end{align}
Supersymmetry~\cite{Bagger:1983tt} now requires the $4 n_H$-dimensional
metric $h_{uv}$ to be a quaternionic-K\"ahler space, of negative
scalar curvature\footnote{A
  quaternionic-K\"ahler space need not be K\"ahler, and by a slight abuse of nomenclature, we will
  refer to them as quaternionic spaces.}. This requires
that there are three almost complex structures $J^x$, $x=1,2,3$, that
satisfy a quaternionic algebra
\begin{align}
  J^x J^y = -\delta^{xy} + \epsilon^{xyz} J^z.
\end{align}
The metric $h_{uv}$ is hermitean with respect to each $J^x$, and we can
define three quaternionic two-forms $K^x_{uv} = h_{uw} (J^x){}^w{}_v$. They are not closed, but they are covariantly constant with respect to
an $SU(2)$ connection $\omega^x$:
\begin{align}\label{eq:def-u2-quaternionic}
  D K^x \equiv d K^x - \epsilon^{xyz} \omega^y \wedge K^z = 0\,.
\end{align}
The $SU(2)$ connection $\omega^x$ defines the $SU(2)$ curvature
$\Omega^x \smallequiv d \omega^x \smallminus \frac 12 \epsilon^{xyz}
  \omega^y \wedge \omega^z$, and then we have the relation
\begin{align}
  \Omega^x = \lambda K^x\,,
\end{align}
where $\lambda$ is a non-zero\footnote{When $\lambda=0$, we have a hyper-K\"ahler manifold, which features in the rigid
$N=2$ hypermultiplet.}  constant. Supersymmetry requires this
constant to be related to Newton's
constant as $\lambda=-\kappa^2$, and therefore we have $\lambda =
-1$ (we work in natural units where all constants are set to $1$). With these units, the Ricci scalar curvature of the quaternionic
manifold is given as $R = - 8 n_H (n_H+2)$, and is therefore always negative.

We can decompose the metric $h_{uv}$ in quaternionic vielbein $ \mathcal U^{A\alpha}_u $ as
\begin{align}
  h_{uv} = \mathcal U^{A\alpha}_u \mathcal U^{B \beta}_v \mathbb
  C_{\alpha\beta} \epsilon_{AB}\,,
\end{align}
where $\mathbb C_{\alpha\beta}$ and $\epsilon_{AB}$ are the antisymmetric
symplectic and $SU(2)$ tensors.

\subsubsection{The universal hypermultiplet}\label{ssect:uhm}
As an example of a quaternionic space, we discuss the so-called
``universal hypermultiplet''. In compactifications of type
II string theory one typically finds a number of hypermultiplets. One of them is always
present, and is therefore called the universal hypermultiplet.

It is possible to construct consistent truncations, such that it is
the only hypermultiplet. The metric is then the coset space
$SU(2,1)/U(2)$, which can be written in
terms of the real coordinates $\{r, \chi, \varphi, \sigma\}$ as
\begin{align}\label{eq:def_uhm}
  {\rm ds}^2 = \frac 1 {r^2} \Bigl( {\rm d}r^2 + r \, ({\rm d}\chi^2 +
  {\rm d}\varphi^2) + \big({\rm d}\sigma + \chi {\rm d}\varphi\big)^2 \Bigr)\ .
\end{align}
The field $r$ is called the dilaton and the metric is restricted to
the region where $r>0$. Its expectation value $\langle r \rangle$ determines
the string coupling constant $g_s$ via $g_s = \mathcal V
\langle r \rangle^{-1/2}$, where $\mathcal V$ is the volume of the Calabi-Yau
space. For more details about the universal hypermultiplet and its isometries see appendix \ref{app:UHM}.

\subsection{Ungauged hypermultiplet sector}
In the case when the hypermultiplet sector is not gauged, one can just directly add the bosonic piece \eqref{eq:lagr-hyper-ungauged} to the full supergravity lagrangian, i.e.\ the ungauged hypermultiplets couple minimally with the gravity multiplet and do not directly couple to the vectormultiplet fields. One is therefore still free to gauge an internal isometry in the vector multiplet sector and keep the hypermultiplets ungauged. Gaugings of the R-symmetry are however no longer possible if we want to keep the hypermultiplet scalars ungauged. We still need to mention the hyperino supersymmetry variation, which is simply
\begin{align}
\delta_\varepsilon \zeta_\alpha = i\,
\mathcal{U}^{B\beta}_u\partial_\mu q^u \gamma^\mu
\epsilon_{AB}\mathbb{C}_{\alpha\beta} \varepsilon^A\,.\label{susyhyperino_ungauged}
\end{align}

\subsection{Gauged quaternionic isometries}
Similar to the vector multiplet scalars, the hypermultiplet
lagrangian~\eqref{eq:lagr-hyper-ungauged} has its isometries as global
symmetries, and we can gauge them. The transformation of the scalars
is denoted as
\begin{align}
  \delta_G q^u = -g \tilde k^u_\Lambda \alpha^\Lambda\,,
\end{align}
and these Killing vectors $\tilde k^u_\Lambda$ form a representation
of the same gauge algebra as in~\eqref{Lie-alg}:
\begin{align}
  [\tilde k_\Lambda, \tilde k_\Sigma] = f_{\Lambda\Sigma}{}^\Gamma
  \tilde k_\Gamma\,.
\end{align}
These gaugings again introduce additional fermionic terms and a scalar potential, as
given in~\eqref{eq:scalar-potential}. The full bosonic lagrangian and supersymmetry transformations for general hypermultiplet and vector multiplet gaugings were already given in the beginning of this chapter.

\subsubsection{Moment maps}
Although the quaternionic spaces are not complex, we can still define
moment maps for the quaternionic Killing vectors. The moment maps
$P^x_\Lambda$ are defined by
\begin{align}
  K^x_{uv} k^v_\Lambda = D_u P^x_\Lambda \equiv \partial_u P^x_\Lambda -
  \epsilon^{xyz} \omega^y_u P^x_\Lambda\,.
\end{align}
Using these, we find the equivariance condition
\begin{align}
  K^x_{uv} k^u_\Lambda k^v_\Sigma + \frac 12 \epsilon^{xyz}
  P^y_\Lambda P^z_\Sigma = \frac 12 f_{\Lambda\Sigma}{}^\Gamma P^x_\Gamma\,.
\end{align}
The $SU(2)$ connection gets modified to
\begin{align}\label{gauged-Sp1-supergravity}
 \omega_{\mu \,A}{}^B\equiv\partial_\mu q^u \omega_{u\,A}{}^B+gA_\mu^\Lambda P_{\Lambda\,A}{}^B\,,
 \end{align}
where $P_{\Lambda\,A}{}^B = \frac i 2 \sigma^x{}_A{}^B P^x_\Lambda$.
In absence of hypermultiplets, $n_H=0$, and
for suitable structure constants
$f_{\Lambda\Sigma}{}^\Gamma$, it is possible to keep the $P^x_\Lambda$'s as non-zero constants, standardly denoted $\xi^x_{\Lambda}$. This is only possible if the gauge group
contains $SU(2)_R$ or $U(1)_R$ factors. Such constants are called Fayet-Iliopoulos (FI) terms, as already explained.

\section{Symmetries}
We have already mentioned several types of symmetries of the lagrangian, but here we go over the symmetries of $D=4$ $N=2$ supergravity a bit more comprehensively, skipping technical details. A detailed technical account can be found in \cite{Ortin}. We will separate the symmetries in three categories: local symmetries, global symmetries, and dualities, or equation of motion symmetries. It is important to stress that the following symmetries concern only the supergravity action, not its solutions. Any given solution could preserve a part of the original symmetries, but this is strictly case dependent and will be discussed when we turn our attention to solutions of the theory in the remainder of this thesis.

\subsubsection{Local symmetries}
The local symmetries are: general coordinate transformations, local Lorentz transformations, gauge invariance, and of course local $N=2$ transformations. We have given details about the supersymmetry transformations, while the others are very standard physical symmetries so we will not spell them out more explicitly here. Note that the general coordinate, local Lorentz, and gauge transformations are all parametrized by a bosonic parameter, i.e.\ they are bosonic symmetries. On the other hand, supersymmetry is a fermionic symmetry (parametrized by the spinors $\varepsilon_A$). The commutator of two supersymmetry variations is then also a bosonic quantity, and is in fact always a linear combination of the bosonic symmetries present. In this sense, the supersymmetry transformations of a given theory are always explicitly dependent on the bosonic symmetries. Therefore $D=4$ $N=2$ supergravity always comes with all the above listed local symmetries, while the global symmetries and dualities can depend on the details of the multiplets included.

\subsubsection{Global symmetries}
The different global symmetries were already mentioned above. Apart from the R-symmetry group that is fixed by $N$ being $2$, the other global symmetries need not always be the same. These are the internal symmetries of the special and quaternionic K\"{a}hler manifolds of the vector and hypermultiplet scalars. All the global symmetries can be promoted to local ones using the gauge fields. This is the process of gauging that was explained for the various parts of the lagrangian above. Therefore it is not always easy to distinguish between local and global symmetries of the action. In some cases promoting a part of a global symmetry to a local one can break the remaining global symmetry, e.g.\ the FI gaugings breaking the $SU(2)_R\times U(1)_R$ to a gauged $U(1)_R$ or $SU(2)_R$. Note that although local and global symmetries might seem hard to distinguish and are related to each other mathematically, their physical meaning is strikingly different. Local symmetries in the theory signify redundant degrees of freedom, i.e.\ they are not real physical symmetries. On the contrary, the global symmetries are truly characterizing the physical system and thus provide means of understading any given vacuum at consideration, as will be discussed more explicitly in part \ref{part::3} of this thesis.

\subsubsection{Dualities}
We already discussed the meaning of dualities when we introduced electro-magnetic duality above. They are important for this work because they are symmetries of the equations of motion, i.e.\ solutions of the theory are forced to respect them by construction. This often turns out to be a good guiding principle for constructing general classes of solutions. E/m duality can even be used for the construction of the supergravity lagrangian itself. Restoring it in the case of gauged supergravity leads to the formulation of magnetically gauged theories \cite{deWit:2005ub,deVroome:2007zd,deWit:2011gk} that will be discussed in more details when needed.


\part{Maximal Supersymmetry}\label{part::1}
\chapter{Fully supersymmetric vacua}\label{chapter::n2vac}

\section{Introduction}
In this chapter, we study the configurations that preserve maximal
supersymmetry, i.e.\ the classical solutions of the general lagrangian \eqref{lagr} that preserve all eight supercharges. These models include arbitrary electric gaugings in the vector and hypermultiplet sectors. We present several examples of such solutions and connect some of them to vacuum
solutions of flux compactifications in string theory. It is of general interest to study four-dimensional supersymmetric string vacua and
their low-energy effective supergravity descriptions. Firstly, in the context of flux compactifications and
gauged supergravities, one is motivated by the problem of moduli
stabilization and the properties of string vacua in which these
moduli are stabilized. Often, one
focuses on supersymmetric vacua since there is better control over
the dynamics of the theory, though for more realistic situations,
e.g. in accelerating cosmologies, the vacuum must break all
supersymmetry. Secondly, we are motivated to look for new versions
of the AdS$_4$/CFT$_3$ correspondence. The recently proposed
dualities studied in \cite{Aharony:2008ug} are based on AdS$_4$ string vacua
preserving 32 or 24 supersymmetries. Versions of the
AdS/CFT correspondence with less amount of supersymmetry
are not yet well established, but are
important for studying aspects of four-dimensional quantum
gravity, and potentially also for certain condensed matter systems
at criticality described by three-dimensional conformal field
theories.

In string theory, ungauged $N = 2$ models arise e.g.\ from Calabi-Yau
compactifications of type II string theories or $K3\times T^2$
compactifications of the heterotic string. Both models are known
to have a rich dynamical structure with controllable quantum
effects in both vector and hypermultiplet sectors that are
relatively well understood. Gaugings in $N=2$ supergravity are
well studied and have a long history
\cite{deWit:1984pk,deWit:1983rz,Derendinger:1983rc,deWit:1984px,D'Auria:1990fj,Andrianopoli:1996cm,D'Auria:2001kv,deWit:2001bk}. Their analysis in terms of
string compactifications with fluxes started in~\cite{Polchinski:1995sm,Michelson:1996pn,Dall'Agata:2001zh},
and is an ongoing research topic. For a (partial) list of
references, see~\cite{Andrianopoli:2003jf, Kachru:2004jr,
  Grana:2005ny, Grana:2006hr, Aharony:2008rx, Gauntlett:2009zw,
  Cassani:2009ck, Louis:2009dq}.

In the ungauged case, a complete classification of all the
supersymmetric solutions already existed~\cite{Behrndt:1997ny,
  LopesCardoso:2000qm,Meessen:2006tu,Huebscher:2006mr}, and there were some partial results
in the gauged case for (abelian) vector multiplets~\cite{Cacciatori:2008ek,Hubscher:2008yz,Klemm:2009uw}.
We extend this by taking completely general vector and
hypermultiplet sectors. Since we concentrate only on the maximally
supersymmetric solutions, we use different methods than the ones
in the above references. The spacetime conditions we
obtain for our solutions closely resemble other maximally
supersymmetric solutions in other dimensions such as~\cite{FigueroaO'Farrill:2002ft}.

\subsubsection*{Plan of this chapter}
In section \ref{3.2} we analyze the supersymmetry rules and derive the conditions for maximally
supersymmetric vacua. The possible solutions divide in two classes
of space-times, with zero scalar curvature and with negative scalar
curvature, and we explicitly list all the possible outcomes. We
discuss further the scalar potential for the obtained
vacua in section \ref{3.3} and show that they automatically satisfy all equations of motion. In section \ref{3.4}, we
consider explicit cases from string theory compactifications and
general supergravity considerations that exemplify the use of our
maximal supersymmetry conditions. Some helpful identities and notation are left for the appendices, where we also
present some intermediate formulas that are important
for our results.

\section{Supersymmetry transformations}\label{3.2}
We consider in this section vector multiplets, hypermultiplets and
the gravitational multiplet, with arbitrary electric gaugings as given by the general lagrangian \eqref{lagr}. It can be seen by inspection that the maximally supersymmetric configurations\footnote{We remind the reader that in this thesis we use interchangeably the terms supersymmetric configurations and BPS
configurations, meaning the field expectation values that are invariant under some
supercharges in the theory.} are purely bosonic, and the fermions
need to be zero. This follows from the supersymmetry variations of
the bosonic fields, which can be read off from \cite{Andrianopoli:1996cm}.
Therefore, we can restrict ourselves to the supersymmetry
variations of the fermions only.

\subsection{Gauginos}

As seen in the previous chapter, the transformation of the gauginos is given by
\begin{equation}\label{eq:n2:susygluino}
\delta_\varepsilon\lambda^{iA}=i\nabla_\mu z^i
\gamma^\mu\varepsilon^A + G_{\mu\nu}^{i-}
\gamma^{\mu\nu}\epsilon^{AB}\varepsilon_B+gW^{iAB}\varepsilon_B\,,
\end{equation}
up to terms that are higher order in the fermions and which vanish
for purely bosonic configurations.

A maximally supersymmetric configuration preserves all eight
supercharges, hence the variation of the fermions should vanish
for all choices of the supersymmetry parameters. Since at each
point in spacetime they are linearly independent, the first term
on the right hand side of~\eqref{eq:n2:susygluino} must vanish
separately from the others,
\begin{equation}\label{z=covconst}
\nabla_\mu z^i\equiv \partial_\mu z^i + g A_\mu^\Lambda
k^i_\Lambda =0\,.
\end{equation}
It implies the integrability condition\footnote{We will assume in
the remainder of this chapter that the gauge coupling constant $g\neq
0$. The case of $g=0$ is treated in the literature in e.g.
\cite{Meessen:2006tu,Huebscher:2006mr}.}
\begin{equation}\label{int-cond-vectors}
F_{\mu\nu}^\Lambda\, k_\Lambda^i =0\,,
\end{equation}
and complex conjugate. Here, $F^{\Lambda}_{\mu\nu}$ is the full
non-abelian field strength, given by
\begin{align}\label{field-str-def}
  F^\Lambda_{\mu\nu} = \frac 12 (\partial_\mu A_\nu - \partial_\nu
  A_\mu) + \frac 12 f_{\Sigma\Gamma}{}^\Lambda A^\Sigma_\mu
  A^\Gamma_\nu\,.
\end{align}

The second and third term in the supersymmetry variation of the
gauginos, equation \eqref{eq:n2:susygluino}, need also to vanish
separately, since they multiply independent spinors of the same
chirality. For the second term, this leads to
\begin{equation}\label{eq:Giszero}
G_{\mu\nu}^{i\,-} = 0\,,
\end{equation}
where $g^{i\bar \jmath}$ is the inverse K\"ahler metric, with
K\"ahler potential ${\cal K}$ from \eqref{eq:def-kahler}.

Finally, setting the third term in the supersymmetry variation to
zero leads to
\begin{equation}\label{defW}
W^{iAB}\equiv k^i_\Lambda {\bar L}^\Lambda \epsilon^{AB} + i
g^{i\bar \jmath}{f}^\Lambda_{\bar
\jmath}P^x_\Lambda\sigma_x^{AB}=0\,.
\end{equation}
Close inspection of \eqref{defW} shows that both terms are
linearly independent in $SU(2)_R$ space, hence they must vanish
separately,
\begin{equation}\label{Pf}
k^i_\Lambda {\bar L}^\Lambda =0 \ ,\qquad P^x_\Lambda
f^\Lambda_i=0\,,
\end{equation}
and their complex conjugates.

\subsection{Hyperinos}\label{hyperinos}

The hyperinos transform as
\begin{equation}\label{eq:n2:susy-hyperino}
\delta_\varepsilon \zeta_\alpha = i\,
\mathcal{U}^{B\beta}_u\nabla_\mu q^u \gamma^\mu \varepsilon^A
\epsilon_{AB}\mathbb{C}_{\alpha\beta} + g N_\alpha^A\varepsilon_A
\,,
\end{equation}
again, up to terms that are of higher order in the fermions. The
hyperino mass matrix $N^A_\alpha$ is defined by
\begin{equation}
N_\alpha^A\equiv 2\,\mathcal{U}^A_{\alpha\,u}{\tilde k}^u_\Lambda
{\bar L}^\Lambda\,.
\end{equation}

Similarly as for the gauginos, $N=2$ supersymmetric configurations
require the two terms in \eqref{eq:n2:susy-hyperino} to vanish separately.
Since the quaternionic vielbeine are invertible and nowhere
vanishing, the scalars need to be covariantly constant,
\begin{equation}
\nabla_\mu q^u\equiv \partial_\mu q^u + g A_\mu^\Lambda {\tilde
k}_\Lambda^u =0\,,
\end{equation}
implying the integrability conditions
\begin{equation}\label{int-cond-hypers}
F_{\mu\nu}^\Lambda\, {\tilde k}_\Lambda^u=0\,.
\end{equation}
Furthermore, there is a second condition from
\eqref{eq:n2:susy-hyperino} coming from the vanishing of the hyperino
mass matrix $N_\alpha^A$. This leads to
\begin{equation}\label{kHL=0}
{\tilde k}^u_\Lambda L^\Lambda =0\,,
\end{equation}
and complex conjugate.

In the absence of hypermultiplets, i.e. when $n_H=0$,
the $N=2$ conditions from the variations of the hyperinos
disappear. However, the second condition in \eqref{Pf} remains, with
the moment maps replaced by FI parameters. Our formalism therefore
automatically includes the case $n_H=0$.

\subsection{Gravitinos}\label{3.2.3}

The supersymmetry transformations of the
gravitinos are (up to irrelevant higher
order terms in the fermions)
\begin{equation}\label{eq:n2:susy-gravi}
\delta_\varepsilon \psi_{\mu A}=\nabla_\mu\varepsilon_A +
T^-_{\mu\nu}\gamma^\nu
\epsilon_{AB}\varepsilon^B+igS_{AB}\gamma_\mu\varepsilon^B\,.
\end{equation}
Here, $\nabla_\mu\varepsilon_A$ is the gauged supercovariant
derivative (specified in equation~\eqref{susy-gravi}).

Notice again that for $n_H=0$, in fact even also in the absence of vector
multiplets when $n_V=0$, the gravitino mass-matrix $S_{AB}$ can be
non-vanishing and constant. In the lagrangian, which we discuss in
the next section, this leads to a (negative) cosmological constant
term. The anti-selfdual part of the graviphoton field strength
$T_{\mu\nu}$ satisfies the identity~\eqref{eq:def-tg-inverse}
\begin{equation} \label{graviphoton-identity}
F_{\mu\nu}^{\Lambda\,-}=i{\bar L}^\Lambda
T^-_{\mu\nu}+2f_i^\Lambda G^{i\,-}_{\mu\nu}\,,
\end{equation}
with $G_{\mu\nu}^{i\,-}$ defined in \eqref{eq:def-tg}. From the
vanishing of the gaugino variation, we have that
$G_{\mu\nu}^{i\,-}=0$, so a maximally supersymmetric
configuration must satisfy $F_{\mu\nu}^{\Lambda\,-}=i{\bar
L}^\Lambda T^-_{\mu\nu}$, or
\begin{equation}\label{F-LT}
  F_{\mu \nu}^\Lambda = i \overline L^\Lambda T^-_{\mu \nu} - i
  L^\Lambda T^+_{\mu \nu}\,.
\end{equation}
Using this, we then see that equation~\eqref{kHL=0} implies
the integrability conditions~\eqref{int-cond-hypers} in the
hypermultiplet sector. For the integrability equations in the
vector multiplet sector, the situation is more subtle, as the
Killing vectors are complex and holomorphic. Now, the BPS
condition \eqref{Pf} only implies that
\begin{align}\label{int-relation}
  k^i_\Lambda F^\Lambda_{\mu \nu} = -i k^i_\Lambda L^\Lambda T^+_{\mu
    \nu}\,.
\end{align}
In appendix~\ref{app:specialkahler} we show that $k^i_\Lambda L^\Lambda = 0$ is an identity of the theory,
and hence the integrability condition is always satisfied. The integrability condition
might only locally be sufficient, but this is fine for our purposes. One might however check in
addition whether the covariant constancy of the vector multiplet scalars imposes further (global) restrictions.

To solve the constraints from the gravitino variation, we must
first look at the gauged supercovariant derivative on the
supersymmetry parameter, c.f.\ \eqref{susy-gravi}
\begin{equation}\label{scov-eps}
\nabla_\mu\varepsilon_A=(\partial_\mu -\frac{1}{4}{\cal
\omega}_\mu^{ab}\gamma_{ab})\varepsilon_A+\frac{i}{2} A_\mu
\varepsilon_A+\omega_{\mu\,A}{}^B\varepsilon_B\,.
\end{equation}
Besides the spin connection $\omega_\mu^{ab}$, there appear
two other connections associated to the special K\"ahler and
quaternion-K\"ahler manifolds. We need to compute their curvatures
since they enter the integrability conditions that
follow from the Killing spinor equations. The first one is called
the gauged $U(1)$ K\"ahler-connection, defined by
\cite{Andrianopoli:1996cm,vProeyen}
\begin{equation}\label{gaugedU1}
A_\mu\equiv -\frac{i}{2}\Big(\partial_i{\cal K}\nabla_\mu
z^i-\partial_{\bar\iota}{\cal K}\nabla_\mu{\bar z}^{\bar
\iota}\Big)-\frac{i}{2}gA_\mu^\Lambda (r_\Lambda -{\bar
r}_\Lambda)\,.
\end{equation}
Under a gauge transformation, one finds that
\begin{equation}
\delta_G A_\mu =
\frac{i}{2}g\,\partial_\mu\Big[\alpha^\Lambda(r_\Lambda -{\bar
r}_\Lambda)\Big]\,.
\end{equation}
The curvature of this connection can be computed to be
\begin{equation}
F_{\mu\nu} =ig_{i\bar\jmath}\nabla_{[\mu}z^i\nabla_{\nu ]}{\bar
z}^{\bar\jmath}-g F_{\mu\nu}^\Lambda P_\Lambda\,,
\end{equation}
where $P_\Lambda$ is the moment map, defined in \eqref{mom-maps},
and we have used the equivariance condition \eqref{equivar}. For
maximally supersymmetric configurations, the scalars are
covariantly constant and hence the curvature of the K\"ahler
connections satisfies $F_{\mu\nu}=-g F_{\mu\nu}^\Lambda P_\Lambda$.

The second connection appearing in the gravitino supersymmetry
variation is the gauged $Sp(1)$ connection of the
quaternion-K\"ahler manifold~\eqref{gauged-Sp1-supergravity}. It reads
\begin{equation}\label{gauged-Sp1}
 \omega_{\mu \,A}{}^B\equiv\partial_\mu q^u \omega_{u\,A}{}^B+gA_\mu^\Lambda P_{\Lambda\,A}{}^B\,,
 \end{equation}
 where  $\omega_{u\,A}{}^B$ is the (ungauged) $Sp(1)$ connection of the quaternion-K\"ahler manifold, whose curvatures are related to the three quaternionic two-forms. The effect of the gauging is to add the second term on the right hand side of \eqref{gauged-Sp1}, proportional to the triplet of moment maps of the quaternionic isometries,
 with $P_{\Lambda\,A}{}^B= \frac i 2 P^x_\Lambda (\sigma^x)_A{}^B$. The curvature of \eqref{gauged-Sp1} can then be computed to be
 \begin{align}
 \Omega_{\mu\nu\,A}{}^B =2\Omega_{uv\,A}{}^B\nabla_{[\mu}q^u\nabla_{\nu]}q^v+gF_{\mu\nu}^\Lambda P_{\Lambda\,A}{}^B\,,
 \end{align}
where $\Omega_{uv\,A}{}^B$ is the quaternionic curvature.
 For
fully BPS solutions, we therefore have
$\Omega_{\mu\nu\,A}{}^B=gF_{\mu\nu}^\Lambda P_{\Lambda\,A}{}^B$.

We can now investigate the integrability  conditions that follow
from the vanishing of the gravitino transformation rules
\eqref{eq:n2:susy-gravi}. From the definition of the supercovariant
derivative \eqref{scov-eps}, we find\footnote{Strictly speaking,
we get the supercovariant curvatures appearing in
\eqref{comm-eps}, which also contain fermion bilinears. Since the
fermions are zero on maximally supersymmetric configurations, only
the bosonic part of the curvatures remains.}
 \begin{equation}\label{comm-eps}
 [\nabla_\mu,\nabla_\nu]\varepsilon_A=-\frac{1}{4}R_{\mu\nu}{}^{ab}\gamma_{ab}\,\varepsilon_A-
 ig F_{\mu\nu}^\Lambda P_\Lambda\varepsilon_A + 2gF_{\mu\nu}^\Lambda P_{\Lambda\,A}{}^B\varepsilon_B\,,
 \end{equation}
 where we have used the covariant constancy of the scalars. We recall that $P_\Lambda$ are the moment maps on the special K\"ahler geometry, whereas $P_{\Lambda\,A}{}^B$ are the quaternion-K\"ahler moment maps. Alternatively, we can compute the commutator from
 the vanishing of the gravitino variations spelled out in \eqref{susy-gravi}. By equating this to the result of \eqref{comm-eps}, we get a set of constraints.
 Details of the calculation are given in appendix \ref{app:specialkahler}, and the results can be summarized as follows. First of all, we find the covariant constancy of the graviphoton field
 strength\footnote{Recall that $T^+$ and $T^-$ are related by complex conjugation, and hence the vanishing of $DT^+$ implies $DT=0$.}
 \begin{equation}\label{cov-cons-graviphoton}
D_\rho T_{\mu\nu}^+=0\,.
\end{equation}
Secondly, we get that the quaternionic moment maps must satisfy
\begin{equation}
\epsilon^{xyz}P^y \overline {P^z}=0\,,\qquad P^x\equiv L^\Lambda
P_\Lambda^x\,.
\end{equation}
Moreover, there are cross terms between the graviphoton and the
moment maps, which enforce the conditions
\begin{equation}
T^+_{\mu\nu}\,P^x=0\,.
\end{equation}
This equation separates the classification of BPS configurations
in two sectors, those with a solution of $P^x=0$ at a particular
point (or locus) in field space, and those with non-vanishing
$P^x$ (for at least one index $x$) but $T_{\mu\nu}=0$. We will see
later on that this distinction corresponds to zero or non-zero
(and negative) cosmological constant in the spacetime.

Another requirement that follows from the gravitino integrability
conditions is
\begin{equation}\label{int-field-moment}
F_{\mu\nu}^\Lambda P_\Lambda =0\,,
\end{equation}
where $P_\Lambda$ is defined in \eqref{mom-maps}, and is real. Using
\eqref{F-LT}, this is equivalent to the condition
\begin{equation}
{\bar L}^\Lambda P_\Lambda T^-_{\mu\nu}=L^\Lambda P_\Lambda T^+_{\mu\nu}\,,
\end{equation}
which is satisfied as $P_\Lambda L^\Lambda = 0$,  so \eqref{int-field-moment} does not lead to any new constraint.

Finally, there is the condition on the spacetime Riemann
curvature. It reads
\begin{equation}\label{spacetime-Riemann}
R_{\mu\nu\rho\sigma}=4T^+_{\mu [\sigma}T^-_{\rho
]\nu}+g^2P^x{\overline
{P^x}}g_{\mu\sigma}g_{\nu\rho}-(\mu\leftrightarrow \nu)\,.
\end{equation}
It can be checked that this leads to a vanishing Weyl tensor, implying
conformal flatness. From the curvature, we can compute the value of the Ricci-scalar to be
\begin{equation}
R=-12g^2 P^x{\overline {P^x}}\,.
\end{equation}
Hence, the classification of fully supersymmetric configurations
separates into negative scalar curvature with $P^x \overline
{P^x}\neq 0$, and zero curvature with $P^x=0$ at the
supersymmetric point. In both of these cases there are important
simplifications.

\subsubsection{Negative scalar curvature}
The case of negative scalar curvature is characterized by
$T_{\mu\nu}=0$ and $P^x \overline {P^x} \neq 0$ at the supersymmetric point. Since the BPS
conditions imply that then both $T_{\mu\nu}$ and
$G^{i-}_{\mu\nu}=0$ (see equation~\eqref{eq:Giszero}), we find that all
field strengths should be zero: $F^\Lambda_{\mu \nu} = 0$. The
gauge fields then are required to be pure gauge, but can still be
topologically non-trivial. Furthermore, because of the vanishing field strengths,
the integrability conditions on the scalar fields are satisfied,
and a solution for the sections  $X^\Lambda(z)$ is obtained by a
gauge transformation on the constant (in spacetime) sections.
Finally, the Riemann tensor is  given by
\begin{align}
R_{\mu\nu\rho\sigma}&=g^2P^x{\overline
{P^x}}\left(g_{\mu\sigma}g_{\nu\rho}-g_{\nu\sigma}g_{\mu\rho}\right)\
,
\end{align}
which shows that the space is maximally symmetric, and therefore locally AdS$_4$. The scalar
curvature is $R=-12g^2 P^x \overline {P^x}$.

\subsubsection{Zero scalar curvature}
The class of zero curvature is characterized by configurations for
which $P^x=0$ at the supersymmetric point.  In this case, we can combine the conditions
$P^x_\Lambda f_i^\Lambda = 0$ and $P^x \equiv P^x_\Lambda
L^\Lambda =0$ into
\begin{align}
  P^x_\Lambda \begin{pmatrix} \bar L^\Lambda \\
    f_i^\Lambda \end{pmatrix} =0\,.
\end{align}
The matrix appearing here is the invertible matrix of special geometry (as used in~\eqref{eq:definitie-period2}), hence
we conclude that $P^x_\Lambda = 0$. The Riemann tensor is then
\begin{align}
R_{\mu\nu\rho\sigma}&=4T^+_{\mu [\sigma}T^-_{\rho
  ]\nu}-(\mu\leftrightarrow \nu)\,.
\end{align}
From the covariant constancy of the graviphoton,
condition~\eqref{cov-cons-graviphoton}, we find $D_\rho R_{\mu \nu
\sigma\tau} = 0$. Spaces with covariantly constant Riemann tensor
are called locally symmetric, and they are classified, see e.g.
\cite{CahenWallach,Ortin,FigueroaO'Farrill:2002ft}. In our case we also have zero
scalar curvature, and then only three spaces are possible:
\begin{enumerate}
\item Minkowski space M$_4$ ($T_{\mu\nu}=0$)
\item AdS$_2 \times $S$^2$
\item The pp-wave solution
\end{enumerate}
The explicit metrics and field strengths for the latter two cases are listed in appendix~\ref{app:metrics}, while M$_4$ and AdS$_4$ are further discussed in great detail in the coming chapters of this thesis.

\subsection{Summary}\label{summary}
Let us now summarize the results. There are two different classes:
negative scalar curvature (leading to AdS$_4$) and zero scalar
curvature solutions (leading to $M_4$, AdS$_2 \times $S$^2$ or the pp--wave).

The result of our analysis is that all the conditions on the
spacetime dependent part are explicitly solved\footnote{This is apart from
the scalar fields and Killing spinors, which are spacetime
dependent. The integrability conditions that we have imposed
guarantee locally the existence of a solution, although we did not
explicitly construct it. Its construction cannot be done in closed
form in full generality, but can be worked out in any given
example \cite{Ortin}.}, and the remaining conditions are
purely algebraic, and depend only on the geometry of the special
K\"ahler and quaternionic  manifolds. The solutions to these
algebraic equations define the configuration space of maximally
supersymmetric configurations. There are two separate cases:

\subsubsection{ Negative scalar curvature (AdS$_4$)}\label{2.4.1}
This case is characterized by configurations for which $P^x
\overline{P^x} \neq 0$ at the supersymmetric point. The BPS conditions are
\begin{align}\label{fullyBPSnegative}
\begin{split}
 k^i_\Lambda \overline L^\Lambda &=0\,, \qquad
\tilde k^u_\Lambda L^\Lambda =0\,, \\
P^x_\Lambda f_i^\Lambda &=0\,, \qquad  \epsilon^{xyz} P^y \overline {P^z} = 0\,,
\end{split}
\end{align}
which should be satisfied at a point (or a locus) in field
space. The field strengths are zero, $F^\Lambda_{\mu \nu}=0$, and
the space--time is AdS$_4$ with scalar curvature $R = -12 g^2 P^x
\overline{P^x}$.

\subsubsection{Zero scalar curvature ($M_4$, AdS$_2 \times $S$^2$ or pp--wave)}\label{2.4.2}
In this case, the BPS conditions are \begin{align}\label{fullyBPSzero}
k^i_\Lambda \overline L^\Lambda =0\,, \quad  \tilde k^u_\Lambda L^\Lambda =0\,, \quad P^x_\Lambda =0\,.
\end{align} We remind that, when $T_{\mu\nu}=0$ (Minkowski space), all field
strengths are vanishing ($F_{\mu\nu}^\Lambda=0$).

\section{Scalar potential and equations of motion}\label{3.3}

As shown in the previous chapter, the scalar potential can be written in terms of the mass-matrices,
\begin{equation}
V=-6S^{AB}S_{AB}+\frac{1}{2}g_{i\bar \jmath}W^{iAB}{\overline
W}^{\bar \jmath}_{AB}+N_\alpha^A N^\alpha_A\ .
\end{equation}
Since the gaugino and hyperino mass-matrices, $W^{iAB}$ and
$N^A_\alpha$ respectively, vanish on $N=2$ supersymmetric
configurations, one sees that the scalar potential is
semi-negative definite, and determined by the gravitino
mass-matrix $S_{AB}$. Even in the absence of vector and
hypermultiplets, the gravitino mass-matrix can be non-vanishing,
leading to a negative cosmological constant in the lagrangian.
We thus find for $N=2$ preserving configurations
\begin{equation}
V=-3g^2{\bar L}^\Lambda L^\Sigma P^x_\Lambda P^x_{\Sigma}\ .
\end{equation}
In the absence of hypermultiplets, $N=2$ preserving AdS$_4$ vacua
can therefore only be generated by non-trivial Fayet-Illiopoulos terms.

It can be verified that maximally supersymmetric configurations also solve the equations of motion.
To show this, one varies
the lagrangian~\eqref{lagr} and uses the
identities~\eqref{cyclicity}, \eqref{struct const condition} and
the formulas in section~\ref{summary}. After a somewhat tedious but
straightforward computation one sees that all equations of motion
are indeed satisfied by the maximally supersymmetric
configurations.

\section{Examples}\label{3.4}
In this section we list some (string theory motivated) examples of
$N=2, D=4$ theories, leading to $N=2$ supersymmetric configurations.
We first mention briefly
some already known and relatively well-understood $N=2$ vacua from
string theory and then concentrate on our two main examples in
subsections \ref{su(3) compactification} and \ref{SE_7} that
exhibit best the different features discussed above. In the last
subsection we include some supergravity models, not necessarily obtained from
string compactifications, leading
to AdS$_4$ vacua that can be of interest.

Obtaining gauged $N=2, D=4$ supergravity seems to be important for
string theory compactifications since it is an intermediate step
between the more realistic $N=1$ models and the mathematically controllable
theories. Thus in the last decade there has been much literature
on the subject. An incomplete list of examples consists
of \cite{Andrianopoli:2003jf,Aharony:2008rx,Gauntlett:2009zw,Cassani:2009ck,Louis:2009dq} and it is
straightforward to impose and solve the maximal supersymmetry
constraints in each case. In some cases the vacua have been
already discussed or must exist from general string
theory/M-theory considerations.

For example, it was found that the coset compactifications studied
in \cite{Cassani:2009ck} do not lead to $N=2$ supersymmetric configurations. This can also
be seen from imposing the constraints in section
\ref{summary}. In contrast, the compactification on $K3 \times
T^2/\mathbb{Z}_2$ presented in~\cite{Andrianopoli:2003jf} does exhibit $N=2$
solutions with non-trivial hypermultiplet gaugings. The authors of~\cite{Andrianopoli:2003jf} explicitly found $N=2$ Minkowski vacua by satisfying the
same susy conditions as in section \ref{summary}. From our analysis, it trivially follows that
also the pp-wave and the AdS$_2 \times $S$^2$ backgrounds can also be found in these theories.

A similar example is provided by the  (twisted) $K3 \times T^2$ compactification of the heterotic string,
recently analyzed in~\cite{Louis:2009dq}. For abelian gaugings, one can verify that the three zero scalar curvature vacua are present in these models.

We now turn to discuss the remaining models in more detail.

\subsection[M-theory compactification on $SU(3)$ structure
manifolds]{M-theory compactification on $\mathbf{SU(3)}$ structure
manifolds}\label{su(3) compactification}

There is a very interesting model for $N=2, D=4$ supergravity with
non-abelian gauging of the vector multiplet sector, arising from compactifications of M-theory on
seven-manifolds with $SU(3)$ structure \cite{Aharony:2008rx}. More
precisely, they consider Calabi-Yau (CY) threefolds fibered over a
circle. The $c$-tensor, introduced
in~\eqref{eq:transf-mathn}-\eqref{cyclicity}, is non-trivial in these
models. For the precise
M-theory set-up, we refer the reader to the original
paper~\cite{Aharony:2008rx}; here we only discuss the relevant data for
analyzing the maximal supersymmetry conditions:

\begin{itemize}
\item The vector multiplet space can be parametrized by special
  coordinates, $X^\Lambda \!=\! (1, t^i)$, $t^i = b^i + i v^i$, and prepotential
\begin{equation}
F(X) = - \frac{1}{6} \mathcal K_{i j k} \frac{X^i X^j X^k}{X^0}\ ,
\end{equation}

  with the triple intersection numbers $\mathcal K_{ijk}$ that depend on the particular choice of the CY-manifold.
  This gives the K\"{a}hler potential
  \begin{equation}\label{aharony example kaehler}
    {\cal K} = - \log \Big[ \frac{i}{6} \kappa_{i j k} (t^i-\bar{t}^i) (t^j-\bar{t}^j) (t^k-\bar{t}^k)\Big] \equiv - \log \mathcal{V}\ ,
  \end{equation}
  where $\mathcal{V}$ denotes the volume of the compact manifold. The gauge group is non-abelian with  structure
  constants
  \begin{equation}
  f_{\Lambda\Sigma}{}^0 = 0 = f_{i j}{}^k, \qquad f_{i 0}{}^j = - M_i^j\ ,
  \end{equation}
  and a  $c$-tensor whose only non-vanishing components are
  \begin{equation}
  c_{i, j k} = \frac{1}{2} M_i^l \mathcal K_{l j k}\ .
  \end{equation}
The constant matrix $M_i^j$ specifies the Killing vectors and moment-maps of the special K\"ahler manifold:
  \begin{equation}\label{aharony killing vectors}
    k^j_0 = - M^j_k t^k\ , \qquad \qquad k^j_i = M_i^j\ ,
  \end{equation}
  and
    \begin{equation}\label{aharony moment maps}
      P_0 = - M_i^j t^i \partial_j {\cal K}\ , \qquad \qquad P_i = M_i^j \partial_j {\cal K}\ .
    \end{equation}
    Not for any choice of $M_i^j$ is the Killing equation satisfied. As explained in \cite{Aharony:2008rx}, this is only the case when the relation \eqref{cyclicity} holds. This also ensures that \eqref{struct const condition} is satisfied, as one can easily check.

\item Generally in this class of compactifications there always appear hypermultiplet scalars, but there is no gauging of this sector,
so the Killing vectors and the moment maps $P^x_{\Lambda}$ are
vanishing.
\end{itemize}

The scalar potential in this case reduces to the simple formula
  \begin{equation}\label{aharony scalar potential}
    V = - \frac{8}{\mathcal{V}^2} M_i^k M_j^l \mathcal K_{k l m} v^i v^j v^m\ ,
  \end{equation}
  which is positive semi-definite.

Analyzing the susy conditions is rather straightforward. Since $P^x = 0$, the
only allowed $N=2$ vacua are the ones with zero-scalar curvature. What is left for us to check are the conditions
$k^i_{\Lambda} \bar{L}^{\Lambda} = 0$ and $P_{\Lambda} L^{\Lambda}
= 0$. The latter is very easy to check and holds as an identity at every point in the special K\"ahler manifold. Also, it is equivalent to the relation $k^i_{\Lambda} L^{\Lambda} = 0$ which is  satisfied
whenever there exists a prepotential \cite{D'Auria:1990fj}.
The condition $k^i_{\Lambda} \bar{L}^{\Lambda} = 0$ eventually
leads to
\begin{equation}\label{aharony susy condition}
     \frac{M^i_j (t^j - \bar{t}^j)}{\mathcal{V}} = 2 i \frac{M^i_j v^j}{\mathcal{V}} = 0\ , \qquad \forall i\ .
    \end{equation}
The solution to the above equation that always exists is the
decompactification limit when $\mathcal{V} \rightarrow \infty$.
The other more interesting solutions depend on the explicit form
of the matrix $M$. In case $M_i^j$ is invertible there are no
further solutions to \eqref{aharony susy condition}. On the other
hand, when $M$ has zero eigenvalues we can have $N=2$ M-theory
vacua, given by (a linear combination of) the corresponding zero
eigenvectors of $M$. For the supergravity approximation to hold,
one might require that this solution leads to a non-vanishing (and
large) volume of the CY three-fold. Each eigenvector will correspond to a
flat direction of the scalar potential, and with $V=0$ along these
directions. The case where the full matrix $M$ is zero corresponds
to a completely flat potential, the one of a standard M-theory
compactification on $CY \times S^1$ without gauging.

Thus it is clear that $M_i^j$ is an important object for this type
of M-theory compactifications and we now give a few more details
on its geometrical meaning \cite{Aharony:2008rx}. In the above class of
M-theory compactifications we have a very specific fibration of
the Calabi-Yau manifold over the circle. It is chosen such that
only the second cohomology $H^{(1,1)} (CY)$ is twisted with
respect to the circle, while the third cohomology $H^3 (CY)$ is
unaffected. Thus the hypermultiplet sector remains ungauged as in
regular $CY \times S^1$ compactification, while the vector
multiplets feel the twisting and are gauged. This twisting is
parametrized exactly by the matrix $M$, as it determines the
differential relations of the harmonic (on the CY manifold) two-forms:
\begin{equation}\label{aharony twisting matrix}
  {\rm d} \omega_i = M^j_i \omega_j \wedge {\rm d}z\ ,
\end{equation}
where $z$ is the circle coordinate.

Let us now zoom in on the interesting case when we have nontrivial
zero eigenvectors of $M$, corresponding to non-vanishing volume of
the CY manifold. For a vanishing volume, or a vanishing two-cycle, the effective supergravity description
might break down due
to additional massless modes appearing in string
theory\footnote{For a detailed analysis of the possibilities in a
completely analogous case in five dimensions see \cite{Mohaupt:2001be}.}.
Therefore the really consistent and relevant examples for $N=2$
vacua are only those when the matrix $M$ is non-invertible with
corresponding zero eigenvectors that give nonzero value for every
$v^i$.

To illustrate this better, we consider a particular example, given
in section 2.5 of \cite{Aharony:2008rx}, of a compactification where the
CY three-fold is a $K3$-fibration. In this setting one can explicitly
construct an $M$-matrix, compatible with the intersection numbers
$\mathcal K_{i j k}$. Here one can find many explicit cases where all
of the above described scenarios happen. As a very simple and
suggestive example we consider the 5-scalar case with $\mathcal K_{1 4 4}
= \mathcal K_{1 5 5} = 2$, $\mathcal K_{1 2 3}
= -1$, and twist-matrix
\begin{equation}\label{aharony M-matrix}
M =\begin{pmatrix} 0 & 0 & 0 & 0 & 0 \\
0 & 4 & 0 & -2 & -2 \\
0 & 0 & -4 & 2 & 2 \\
0 & 1 & -1 & 0 & 0 \\
0 & 1 & -1 & 0 & 0
\end{pmatrix}\ .
\end{equation}
The general solution of $M \cdot \vec{v} = 0$ is
\begin{equation}\label{aharony solution}
\vec{v} =\lambda \begin{pmatrix} 1 \\ 0 \\ 0 \\ 0 \\ 0
\end{pmatrix}\ + \mu \begin{pmatrix} 0 \\ 1 \\ 1 \\ 2 \\ 0
\end{pmatrix}\ + \nu \begin{pmatrix} 0 \\ 1 \\ 1 \\ 0 \\ 2
\end{pmatrix}\ ,
\end{equation}
and the resulting volume is
\begin{equation}
\mathcal {V} = 8 \lambda \left( 2
\mu^2 + 2 \nu^2 + (\mu - \nu)^2 \right)\ ,
\end{equation}
which is clearly
positive semi-definite. In the case when either $\mu$ or $\nu$
vanishes we have a singular manifold that is still a solution to
the maximal supersymmetry conditions. When all three coefficients
(that are essentially the remaining unstabilized moduli fields)
are non-zero, we have a completely proper solution both from
supergravity and string theory point of view, thus providing an
example of $SU(3)$ structure compactifications with zero-curvature
$N=2$ vacua. This example can be straightforwardly generalized to
a higher number of vector multiplets, as well as to the lower
number of 4 scalars (there cannot be less than 4 vector multiplets
in this particular case).

Finally we note that a special case of the
general setup described above was already known for more than
twenty years in \cite{deWit:1984px} (3.21), where $M_1^1 = -2$, $M_2^2 =
1,$ and $\mathcal K_{1 2 2} = 2$. It was derived purely from 4d
supergravity considerations, but it now seems that one can embed
it in string theory.

\subsection{Reduction of M-theory on Sasaki-Einstein$_7$}\label{SE_7}

There has been much advance in the last years in understanding
Sasaki-Einstein manifolds and their relevance for M-theory
compactifications, both from mathematical and physical
perspective. A metric ${\rm ds^2}$ is Sasaki-Einstein if and only if the cone
metric, defined as
$
  {\rm ds}_{\rm cone}^2 \equiv {\rm d}r^2 + r^2 {\rm ds^2}
$,
is K\"ahler and Ricci-flat. These spaces are good candidates for examples of the
AdS$_4$/CFT$_3$ correspondence and an explicit reduction to $D=4$
has been recently obtained in \cite{Gauntlett:2009zw}. Originally the
effective lagrangian includes magnetic gauging and a scalar-tensor
multiplet, but after a symplectic rotation it can be formulated in
the standard $N=2$ formalism discussed here. After the dualization
of the original tensor to a scalar we have the following data for
the multiplets, needed for finding maximally supersymmetric vacua:
\begin{itemize}
  \item There is one vector multiplet, given by $X^{\Lambda} = (1, \tau^2)$ and $F(X) =
  \sqrt{X^0 (X^1)^3}$, leading to $F_{\Lambda} = (\frac{1}{2} \tau^3,\frac{3}{2} \tau^2)$ and K\"{a}hler potential
  \begin{equation}\label{SE_7 tensor kaehler}
    {\cal K} = - \log \frac{i}{2} (\tau - \bar{\tau})^3\ .
  \end{equation}
  There is no gauging in this sector, i.e.\ $k^i_{\Lambda} = 0$ and $P_{\Lambda} = 0$ for all $i,
  \Lambda$. This also means that both $f_{\Lambda \Sigma}{}^{\Pi}$ and $c_{\Lambda, \Sigma
  \Pi}$ vanish.
  \item The hypermultiplet scalars are $\{ r, \chi, \varphi, \sigma
    \}$ with the universal hypermultiplet metric\footnote{The
      relation with the coordinates $\{\rho, \sigma, \xi, \bar \xi\}$
      used in~\cite{Gauntlett:2009zw} is given by
      $\rho = r, \sigma_{\rm there} = \sigma_{\rm here} + \frac 12 u
      v$ and $\xi = \frac 12 (\chi + i \varphi)$. Furthermore, there
      is an overall factor $\frac 14$ in their definition of the
      universal hypermultiplet. Finally, they use a different $SU(2)$
      frame to calculate the moment maps $P^x_\Lambda$, which is why
      they are rotated with respect to the ones displayed here.}, introduced in~\eqref{eq:def_uhm}:
  \begin{equation}\label{quaternion metric SE_7 tensors}
    {\rm d}s^2 = \frac{1}{r^2} \left( {\rm d} r^2 +  r ({\rm d} \chi^2
      + {\rm d} \varphi^2) + ({\rm d}\sigma +
      \chi {\rm d}\varphi)^2 \right)\ .
  \end{equation}

  We have an abelian gauging (see app.\ \ref{app:UHM} for all isometries of the universal hypermultiplet), given by:
  \begin{align}\label{killing vector SE_7 tensors}
\begin{split}
    \tilde{k}_0 &= 24 \partial_{\sigma} +4\bigl(\chi \partial_\varphi -
    \varphi \partial_\chi  + \frac 12 (\phi^2 -
    \chi^2) \partial_\sigma \bigr)\,,\\
 \tilde{k}_1 &= 24 \partial_{\sigma}\ ,
\end{split}
  \end{align}
and the moment maps, calculated in appendix~\ref{app:UHM}, are
  \begin{align}\label{moment map SE_7 tensors}
   \nonumber P_0^1 &= \frac{4 \chi}{\sqrt r}\ ,& P_0^2 &=  \frac
   {4\varphi} {\sqrt r}\ , & P_0^3 &= -\frac{12}{r} + 4 - \frac{\chi^2 + \varphi^2}{r}\ ,\\
   P_1^1& = 0\ , & P_1^2 &= 0\ ,&  P_1^3& = -\frac{12}{r}\ .
  \end{align}
\end{itemize}
We can now proceed to solving the maximal supersymmetry
constraints. The conditions involving vector multiplet gauging are
satisfied trivially, while from $\tilde{k}^u_{\Lambda} L^{\Lambda}
= 0$ we obtain the conditions $\chi = \varphi = 0$ and $1 +
\tau^2 = 0$. Therefore $\tau = i$ (the solution $\tau = -i$ makes
the K\"{a}hler potential ill-defined) and ${\cal K} = - \log 4$.
However, not all the moment maps at this vacuum can be zero
simultaneously, leaving AdS$_4$ as the only possibility for a
$N=2$ vacuum solution. One can then see that $\epsilon_{x y z} P^y
\overline{P^z} = 0$ is satisfied, so the only remaining condition
is $P^3_{\Lambda} f^{\Lambda}_{\tau} = 0$. This fixes $r = 4$.
Therefore we have stabilized all (ungauged) directions in moduli
space: $\chi = \varphi = 0, \tau = i, r = 4$. The potential is
nonzero in this vacuum since $P^3 = 2$, which means the only
possibility for the space-time is to be AdS$_4$ with vanishing
field strengths. This is indeed expected since $SE_7$
compactifications of M-theory lead to an $N=2$ AdS$_4$ vacuum, the
one just described by us in the dimensionally reduced theory.

One can verify that this vacuum is stable under deformations in
the hypermultiplet sector of the type discussed in
\cite{Antoniadis:2003sw,Anguelova:2004sj}. To show this, first observe that the
condition $\tilde{k}^u_{\Lambda} L^{\Lambda} = 0$ for $u = \chi$ and
$u = \varphi$
always ensures vanishing $\chi$ and $\varphi$. Secondly, one may verify that the
deformations to the quaternionic moment maps are proportional to
$\chi$ or $\varphi$, and hence the remaining $N = 2$ conditions from section
\ref{2.4.1} are satisfied. It would be interesting to understand
if this deformation corresponds to a perturbative one-loop
correction in this particular type of M-theory compactification.

\subsection[Other gaugings exhibiting AdS$_4$ vacua]{Other gaugings exhibiting $\mathbf{AdS_4}$ vacua}

Another example of an AdS$_4$ supersymmetric vacuum can be
obtained from the universal hypermultiplet. In the same
coordinates $\{r, \chi, \varphi, \sigma\}$ as used in the
previous example, the metric is again given by \eqref{quaternion
metric SE_7 tensors}. This space has a rotational isometry acting
on $\chi$ and $\varphi$, given by ${\tilde k}_1-{\tilde k}_0$ in
the notation of \eqref{killing vector SE_7 tensors}. We leave the
vector multiplet sector unspecified for the moment, and gauge the
rotation isometry by a linear combination of the gauge fields
$A_\mu^\Lambda$. This can be done by writing the Killing vector as
\begin{align}
\tilde k^u_\Lambda = \alpha_\Lambda \Bigl(0,-\varphi,\chi,\frac 12 (\varphi^2
-\chi^2)\Bigr)\ ,
\end{align}
for some real constant parameters $\alpha_\Lambda$. The
quaternionic moment maps are given by (see appendix~\ref{app:UHM})
\begin{align}
  P^x_\Lambda = \alpha_\Lambda \left(\frac {\varphi}{\sqrt r}, \frac
    {\chi}{\sqrt r}, 1 - \frac{\varphi^2+\chi^2}{4 r}
  \right)\ .
\end{align}
It can be seen that there are no points for which
${P}^x_\Lambda=0, \forall x$, so this means that only AdS$_4$
$N=2$ vacua are possible. To complete the example, we have to
specify the vector multiplet space, and solve the conditions
${P}^x_\Lambda f^\Lambda_i=0$ and $\tilde k^u_\Lambda L^\Lambda
=0$. The latter can be solved as $\chi = \varphi = 0$, and then
also $\epsilon^{xyz}P^y\overline{P^z}=0$. The first one then
reduces to $\alpha_\Lambda f^\Lambda_i = 0$. This condition is
trivially satisfied when e.g. $n_V=0$. A more complicated example
is to take the special K\"ahler space of the previous subsection
with no gauging in the vector multiplet sector. There is one
complex scalar $\tau$, a section $X^\Lambda = (1,\tau^2)$ and a
prepotential $F = \sqrt{X^0 (X^1)^3}$. We then find a solution for
$\tau = i \sqrt{\frac{- 3 \alpha_0}{\alpha_1}}$, under the
condition that $\alpha_0$ and $\alpha_1$ are non-vanishing real
constants of opposite sign. More complicated examples with more
vector multiplets may be constructed as well. It would be
interesting to study if such examples can be embedded into string
theory.

A similar situation arises in the absence of hypermultiplets. As
mentioned in the end of section \ref{hyperinos}, we can have
non-vanishing moment maps that can be chosen as $P^x_{\Lambda} =
\alpha_{\Lambda} \delta^{x 3}$. Then we again need to satisfy the
same condition $\alpha_\Lambda f^\Lambda_i = 0$ as above, and we
already discussed the possible solutions.

\chapter{Supersymmetry preserving Higgs mechanism}\label{chapter::solution_generator}
\section{Introduction}
This chapter is a bridge between the fully supersymmetric solutions that were described in the previous discussion and the solutions that break some of the supersymmetries, which will be a subject of the coming chapters. We use the results derived in chapter \ref{chapter::n2vac} to define a consistent truncation of gauged $D=4$ $N=2$ supergravity into another (un)gauged $D=4$ $N=2$ lagrangian with reduced number of multiplets. The main idea is simple - in gauged supergravity, one can give expectation values to some of the
scalars (from both the vector and hypermultiplets) such that one
breaks the gauge symmetry spontaneously in a maximally
supersymmetric $N=2$ vacua, specified by the conditions
\eqref{fullyBPSnegative} or \eqref{fullyBPSzero}. Let us suppose for simplicity
that the vacuum has zero cosmological constant, the argument can
be repeated for $N=2$ preserving anti-de Sitter vacua. Due to the
Higgs mechanism some of the fields become massive, and as a
consequence of the $N=2$ preserving vacua, the gravitinos remain
massless and the heavy modes form massive $N=2$ vector multiplets.
As a second step, we can set the heavy fields to zero, and the
theory gets truncated to an ungauged $N=2$ supergravity. These
truncations are consistent due to the fact that supersymmetry is
unbroken. Black holes and other types of solutions can then be found by taking any
solution of the ungauged theory and augmenting it with the massive
fields that were set to zero. In fact, it is clear from this
procedure that one can even implement a non-BPS
solution of the ungauged theory into the gauged lagrangian. This procedure works for non-abelian gaugings, as
long as it is broken spontaneously to an abelian subgroup with
residual $N=2$ supersymmetry.

Let us now illustrate the above mechanism in some more detail. We
restrict ourselves first to spontaneous symmetry breaking in
Minkowski vacua, where one has $\langle P^x_\Lambda \rangle = 0$
and $\langle {\tilde k}^u_\Lambda L^\Lambda\rangle = \langle k^i_\Lambda \overline{L}^\Lambda\rangle = 0$ according
to \eqref{fullyBPSzero}. At such a point in the moduli space, the resulting scalar potential is
zero as required by a Minkowski vacuum. After
the hypermultiplet scalar fields take their vacuum expectation
values, the lagrangian~\eqref{lagr} contains a mass term for some
of the gauge fields, given by
\begin{equation}\label{eq:vectormass}
  \mathcal L^V_{\text {mass}} = M_{\Lambda\Sigma}
  A_\mu^\Lambda A^{\mu\Sigma}\ ,\qquad M_{\Lambda\Sigma} \equiv g^2 \langle h_{uv}  \tilde k^u_\Lambda \tilde k^v_\Sigma\rangle \ .
 \end{equation}
There is no contribution to the mass matrix for the vector fields
coming from expectation values of the vector multiplet scalars,
since the gauging was chosen to be abelian. The number of massive
vectors is then given by the rank of $M_{\Lambda\Sigma}$, and as
$h_{uv}$ is positive definite, one has $\rank(M_{\Lambda\Sigma}) =
\rank(\tilde k^u_\Lambda)$. Hence, the massive vector fields are encoded by the linear combinations
${\tilde k}^u_\Lambda A_\mu^\Lambda$.
Similarly, some of the vector and
hypermultiplet scalars acquire a mass, determined by expanding the
scalar  potential to quadratic order in the fields. Then one reads off the mass
matrix, and in general there can be off-diagonal mass terms
between vector and hypermultiplet scalars. Massive vector
multiplets can then be formed out of a massive vector, a massive
complex scalar from the vector multiplet, and 3 hypermultiplet
scalars. The fourth hypermultiplet scalar is the Goldstone mode
that is eaten by the vector field. We will illustrate this more
explicitly in some concrete examples below.

Upon setting the massive fields to zero (or integrating them out),
one obtains a supergravity theory with only massless fields.
Because of $\langle P^x_\Lambda \rangle =0$, the mass matrix for
the gravitinos is zero as follows from \eqref{mass-gravitino}.
Therefore, the resulting theory is an ungauged supergravity theory. Black hole
solutions and other solutions of interest can then be simply copied from already existing literature. By going through the Higgs mechanism in
reverse order, one can uplift this solution easily to the gauged
theory by augmenting it with the necessary expectation values of
the scalars. The original solution is not
charged with respect to the gauge fields that acquired a mass.

The situation for spontaneous symmetry breaking in an AdS vacuum is
similar. To generate a negative cosmological constant from the
potential, we must have a $\langle P^x_\Lambda\rangle
\neq 0$ in the vacuum. The conditions for unbroken $N=2$
supersymmetry  are given in \eqref{fullyBPSnegative}. After expanding the
fields around this vacuum, one can truncate the theory further to
a lagrangian with a bare cosmological constant, in which one can
use already known solutions of minimal gauged supergravity. We will discuss explicit black hole examples both in Minkowski and AdS$_4$ in the next part of this thesis.

\section{Solution generating technique}\label{sect:solutiontechnique}
We now elaborate on constructing solutions more
explicitly. As explained above, the general technique is to embed
a (BPS) solution in ungauged supergravity into a gauged
supergravity. The considerations in this section also apply for
the more general case of non-abelian gaugings, although explicit use of this technique will only be made in the abelian case in following chapters. First, to illustrate the systematics of our procedure, we analyze a simpler setup in
which we embed solutions from pure supergravity into a model with
vector multiplets only. Then we extend the models to include both
hypermultiplets and vector multiplets, i.e.\ the most general
(electrically) gauged supergravities. We always consider solutions
with vanishing fermions, i.e. the discussion concerns only the
bosonic fields.

\subsection{Vector multiplets}
We start from pure $N = 2$ supergravity, i.e.\ only the gravity
multiplet normalized as $\mathcal L=\frac{1}{2}R(g) - \frac{1}{2}
F_{\mu\nu} F^{\mu\nu} - \Lambda$. Let us assume we have found a
solution of this lagrangian, which we denote by
$\mathring{g}_{\mu \nu}, \mathring{F}_{\mu \nu}$. We can embed
this into a supergravity theory with only vector multiplets as
follows. If we have a theory with (gauged) vector multiplets we can
find a corresponding solution to it by satisfying
\begin{equation}\label{vectors}
    \nabla_{\mu} z^i = 0\ , \qquad G_{\mu \nu}^{i} = 0\ , \qquad k^i_{\Lambda} \bar{L}^{\Lambda} = 0\ .
\end{equation}
Note that the integrability condition following from $\nabla_{\mu}
z^i = 0$ is always satisfied given the other constraints. We further have the relations
\begin{equation}\label{eq:trivial_vectors}g_{\mu \nu} =
\mathring{g}_{\mu \nu}\ , \quad \sqrt{2 I_{\Lambda\Sigma} \bar{L}^{\Lambda} \bar{L}^{\Sigma}}\, T^-_{\mu
\nu} = \mathring{F}^-_{\mu \nu}\ .
\end{equation}
The last equality
is to be used for determining $T^-_{\mu \nu}$. Then we can find
the solution for our new set of gauge field strengths by
$F^{\Lambda -}_{\mu \nu} = i \bar{L}^{\Lambda} T^-_{\mu \nu}$
since we already know that $G_{\mu \nu}^{i} = 0$.

The new configuration will, by construction,  satisfy all equations of
motion of the theory and will preserve the same amount of
supersymmetry (if any) as the original one. This can be checked explicitly from the supersymmetry
transformation rules \eqref{susy-gravi} and \eqref{susygluino}, combined with the
results from the previous chapter. Indeed, \eqref{vectors} comes from imposing
the vanishing of \eqref{susygluino}, while \eqref{eq:trivial_vectors} is required by the Einstein
equations.

\subsection{Hypermultiplets} Given any solution of $N = 2$
supergravity with no hypermultiplets, we can obtain a new solution
with (gauged) hypermultiplets preserving the same amount of
supersymmetry as the original one. We require the theory to remain
the same in the other sectors (vector and gravity multiplets with
solution $\mathring{g}_{\mu \nu}, \mathring{F}^{\Lambda}_{\mu
\nu}, \mathring{z}^i$) and impose some additional constraints that
have to be satisfied in addition to the already given solution. We
then simply require the fields of our new theory to be
\begin{equation}\label{eq:trivialhypers}g_{\mu \nu} =
\mathring{g}_{\mu \nu}\ , \qquad F^{\Lambda}_{\mu \nu} =
\mathring{F}^{\Lambda}_{\mu \nu}\ , \qquad z^i =
\mathring{z}^i\ ,
\end{equation} under the following restriction that
has to be solved for the hypers. Here we are left with two cases:
the original theory was either with or without Fayet-Iliopoulos (FI) terms
(cosmological constant). In absence of FI terms, a new solution
after adding hypers is given by imposing the constraints:
\begin{equation}\label{hypers no FI terms}
    \nabla_{\mu} q^u = 0\  \Rightarrow \ \tilde{k}^u_{\Lambda} F^{\Lambda}_{\mu \nu} = 0\ , \qquad P^x_{\Lambda} = 0 \ ,\qquad \tilde{k}^u_{\Lambda}
    L^{\Lambda}
    = 0\ ,
\end{equation}
while in the case of original solution with FI terms we have a
solution after adding hypers (thus no longer allowing for FI terms
but keeping $P^x_{\Lambda} L^{\Lambda}$ the same) with:
\begin{equation}\label{hypers with FI terms}
    \nabla_{\mu} q^u = 0 \Rightarrow \tilde{k}^u_{\Lambda}
    F^{\Lambda}_{\mu \nu} = 0\ , \quad P^x_{\Lambda} f^{\Lambda}_i =
    0\ ,
    \quad \epsilon^{x y z} P^y_{\Lambda} P^z_{\Sigma} L^{\Lambda}
    \bar{L}^{\Sigma} = 0\ , \quad \tilde{k}^u_{\Lambda}
    L^{\Lambda} = 0\ .
\end{equation}
The new field configuration (given it can be found from the original data)
again satisfies all equations of motion and
preserves the same amount of supersymmetry as the original
one. This is true because the susy variations of gluinos and gravitinos remain the same
as in the original solution, and also the variations for the newly
introduced hyperinos are zero.

\subsection{Vector and hypermultiplets} This case is just
combining the two cases above. If we start with no FI terms the
new solution will be generated by imposing equations (\ref{hypers
no FI terms}) and (\ref{vectors}). If we have a solution with a
cosmological constant we need to impose (\ref{hypers with FI
terms}) and (\ref{vectors}). Then the integrability condition
following from $\nabla_{\mu} q^u = 0$ is automatically satisfied
in both cases, using relations \eqref{eq:trivial_vectors}.

\part{Black Hole Solutions}\label{part::2}

\chapter{Asymptotically flat black holes}\label{chapter::blackholes_flat}

\section{Introduction}\label{5:intro}
One of the interesting predictions of general relativity is the
existence of black hole solutions. It was followed by various experimental indications for the
presence of such objects in the universe and even in our own galaxy. Black holes are places in spacetime where
gravity becomes strong enough to stop even light from escaping. There is
an event horizon, and particles that pass the
horizon can (classically) never return to the external spacetime.

In 1974, Stephen Hawking discovered that black holes do emit
radiation~\cite{Hawking:1974rv}, due to quantum effects. It was shown that black holes follow the rules of thermodynamics, and one can
derive their temperature and entropy~\cite{Bekenstein:1973ur,Hawking:1974rv}. It is
then a challenge for a theory of quantum gravity to provide the microscopic description of black hole thermodynamics.

From the point of view of this thesis, black holes are interesting for a variety of reasons. They are solutions one
finds in supergravity and can often be embedded in the full-fledged
string theory. If string theory is a viable theory of quantum gravity,
it should provide a complete, microscopic description of a black
hole. One can compare this with the macroscopic description in
supergravity, which is its low energy effective action. For instance,
in the macroscopic picture, one can compute the entropy. In the
microscopic picture, this corresponds to counting the number of
microstates, which was shown to exactly reproduce the same entropy in a number of examples~\cite{Strominger:1996sh}.

Let us now briefly consider the basic black hole solutions that asymptote to Minkowski spacetime.

\subsubsection{The Schwarzschild solution}
The simplest and best studied black hole solution is the Schwarzschild spacetime. It is a solution of pure gravity without any matter, i.e.\ all fields of $N=2$ supergravity except for the graviton are frozen to zero. The Schwarzschild metric is typically written in spherical coordinates due to its manifest spherical symmetry:
\begin{align}\label{eq5:sch-metric}
  {\rm ds}^2 = \left(1-\frac{2 M}{r}\right) {\rm d}t^2 - \frac{{\rm d}r^2}{\left(1-\frac{2 M}{r}\right)} - r^2 {\rm d} \Omega_2^2\ ,
\end{align}
with ${\rm d} \Omega_2^2 = \left({\rm d} \theta^2 + \sin^2 \theta {\rm d} \varphi^2\right)$ and $M$ an arbitrary parameter that corresponds to the mass of black hole with respect to the ground state ($M=0$ clearly corresponds to Minkowski). In the limit $r\rightarrow\infty$ one recovers flat spacetime, while $r=0$ is a genuine point singularity of the solution. Additionally, there is a coordinate singularity at $r_h = 2 M$, which is a spherical shell and corresponds to the event horizon. One then typically says that the black hole has a radius $r_h$ and area $A = 4 \pi r_h^2$. Note that in case $M < 0$, $r_h$ is negative and thus no horizon exists to shield the singularity. The spacetime is then said to contain a naked singularity and is typically no longer of physical interest. Such a configuration will not form under gravitational collapse of a spherical mass shell, see e.g.~\cite{Townsend:1997ku}.

From the rules of black hole thermodynamics it can be shown that the Schwarzschild black hole has a temperature that is inversely proportional to the mass, $T \sim M^{-1}$ and an entropy proportional to the area\footnote{The entropy-area law holds for all black holes and other black objects in any spacetime dimension, while the temperature is proportional to the surface gravity and thus depends explicitly on the metric that is considered.}, $S \sim A \sim M^2$.

The last thing we need to mention in this very brief review are the symmetries of the Schwarzschild solution. The metric \eqref{eq5:sch-metric} has 4 Killing vectors, corresponding to time translations ($\mathbb{R}$) and space rotations ($SO(3)$). The full Lorentz symmetry $SO(1,3)$ is restored only asymptotically as $r \rightarrow \infty$. The metric is therefore static, spherically symmetric and asymptotically flat.

\subsubsection{The Reissner-Nordstr\"{o}m solution}
The Reissner-Nordstr\"om solution describes a charged black
hole solution in the theory of general relativity with a Maxwell
field. It can be then directly embedded in minimal ungauged supergravity (with vanishing gravitino expectation values), or more general lagrangians of the sort \eqref{lagr} where the additional fields are frozen to zero or constant and the moment maps are vanishing. The gauge field has non-vanishing components
\begin{align}\label{eq:rn-gauge}
  A_t = \frac {2Q} r\,,\qquad A_\varphi = -2P \cos \theta\,,
\end{align}
where $Q$ is the electric and $P$ the magnetic charge of the black hole. The metric can be written as
\begin{align}\label{eq:rn-metric}
  {\rm ds}^2 = U^2(r) {\rm d}t^2 - \frac{{\rm d}r^2}{U^2(r)} - r^2 {\rm d} \Omega_2^2\ ,
\end{align}
with the metric function
\begin{align}
  U^2(r) = 1 - \frac{2M}r + \frac{Z^2}{r^2}\,,\qquad Z^2 \equiv Q^2 +P^2\,.
\end{align}
In the limit as $r$ goes to infinity the metric approaches flat
Minkowski space and the electric and magnetic fields computed
from~\eqref{eq:rn-gauge} vanish. Besides the point $r=0$,
which is a true singularity, there are other possible values of $r$ where
$V(r)=0$, corresponding to event horizons. They follow from the equations
\begin{align}
  r_\pm = M \pm \sqrt{M^2 - Z^2}\ ,
\end{align}
which could have two, one or zero real solutions for $r_\pm$ depending on:
\begin{description}
\item[$M^2 > Z^2$]\hfill\\
In this case, there are two different roots of $V$,
  given by $r = r_\pm$. There is an inner and an outer horizon.

\item[$M^2 = Z^2$]\hfill\\
  If the charge balances the mass, we call the black hole an extremal
  black hole. The real singularity is shielded by a single event horizon at
  $r = r_+ = r_-$.

\item[$M^2 < Z^2$]\hfill\\
 If the charge exceeds the mass, there are no roots
  of $V$, and any observer can travel to the real singularity at
  $r=0$, which is not shielded by an event horizon. This is again a naked
  singularity and is deemed unphysical.
\end{description}

Note that the Reissner-Nordstr\"{o}m solution is also static and spherically symmetric (global symmetry group $\mathbb{R} \times SO(3)$), just as the Schwarzschild one. However, it enjoys very different thermodynamic properties. The temperature in this case is proportional to the difference between the two horizons, $T \sim r_+ - r_-$, while the entropy is proportional to the area of the outer horizon, $S \sim A \sim r_+^2$. We see now that something very special happens for extremal black holes when $r_+ = r_-$: the temperature vanishes while keeping the entropy and event horizon non-zero. This was not possible for Schwarzschild solutions and is generally a very rare example of a physical system with vanishing temperature and macroscopic disorder.

The extremal black hole is also very particular from the point of view of supersymmetry. Supersymmetry is incompatible with temperature, therefore only extremal black holes can potentially preserve some amounts of supersymmetry in a theory of supergravity. This is also the case with the example at hand - the extremal Reissner-Nordstr\"{o}m solution preserves half of the supercharges of $N=2$ supergravity, thus it is 1/2 BPS. Let us examine more closely the metric in this extremal case. If we redefine the radial coordinate as $r \to r+M$, the metric
is given by
\begin{align}
{\rm ds}^2 =  \frac {r^2}{(r+M)^2} {\rm d}t^2 - \frac{(r+M)^2}{r^2}
\Bigl( {\rm d}r^2 + r^2 {\rm d} \Omega_2^2 \Bigr)\ .
\end{align}
The  true singularity is now at $r=-M$, and the horizon is at
$r=0$. Close to the horizon, we can approximate $r + M \simeq M$, and
we find
\begin{align}
  {\rm ds}^2 =  \frac {r^2}{M^2} {\rm d}t^2 - \frac{M^2}{r^2} {\rm
    d}r^2 + M^2 {\rm d} \Omega_2^2\ .
\end{align}
The metric becomes a product metric: we have an AdS$_2$ space,
parametrized by $t$ and $r$, and a sphere S$^2$, parametrized by
$\theta$ and $\varphi$. The mass $M$ determines the curvature of both
spaces, which are equal in magnitude; as AdS$_2$ has negative curvature and S$^2$ positive, the
total curvature at the horizon is zero. The spacetime outside of the black hole therefore interpolates between the fully supersymmetric
configurations AdS$_2 \times $$S^2$ and flat Minkowski space, discussed in chapter \ref{chapter::supergravity}.

\subsubsection{The Kerr-Newman solution}
This is the most general asymptotically flat black hole solution of the Einstein-Maxwell lagrangian. In addition to mass and charges, the Kerr-Newman spacetime also possesses a constant angular momentum, i.e.\ the black hole rotates around a given axis. The spacetime is therefore stationary but not static, with a symmetry group $\mathbb{R} \times U(1)$ (time translations and axial symmetry). The metric of the Kerr-Newman spacetime is typically written in the so called Boyer-Lindquist coordinates and reads
\begin{align}\label{eq5:Kerr-Newman}
   \begin{split}
    {\rm d} s^2 &= - \left(1- \frac{2 M r - Z^2}{\rho^2} \right) {\rm d} t^2 - \frac{(2 M r-Z^2) 2 a \sin^2 \theta}{\rho^2} {\rm d} t {\rm d} \varphi\\
    &+\frac{\rho^2}{\Delta} {\rm d} r^2 + \rho^2 {\rm d} \theta^2 + \left(r^2+a^2+\frac{(2 M r-Z^2) 2 a \sin^2 \theta}{\rho^2} \right) \sin^2 \theta {\rm d} \varphi^2\ ,
\end{split}
\end{align}
where \begin{equation}\label{ineq}a^2+Z^2 \leq M^2\end{equation}for a genuine black hole with an event horizon, and
$$M \equiv {\rm mass}, \quad Z^2 = Q^2+P^2 \equiv {\rm charge}, \quad a = \frac{J}{M} \equiv {\rm angular}\ {\rm momentum}\ {\rm per}\ {\rm unit}\ {\rm mass},$$
$$\Delta \equiv r^2-2 M r +a^2+Z^2\ ,\qquad \rho^2 \equiv r^2+a^2 \cos^2 \theta\ .$$ One can easily see that when the angular momentum parameter vanishes we recover the Reissner-Nordstr\"{o}m solution.

We postpone the discussion of the thermodynamic and supersymmetric properties of the Kerr-Newman solution here since they are somewhat more involved in comparison with the static cases. For the purposes of this introductory section it is enough to mention that Kerr-Newman black holes that satisfy \eqref{ineq} can never be supersymmetric, even in the extremal limit. Additionally, the Kerr-Newman spacetime is somewhat special in another aspect - it contains a closed timelike curve inside its inner horizon, which is causality violating and often considered very unphysical. Furthermore, due to the fact that the spacetime has a constant rotation, it follows that the parts of the spacetime very far from the black hole will need to move faster than the speed of light in order to keep up with the rest. The solution therefore cannot be taken seriously in its full domain, but can nevertheless be used as an important example in this thesis. It often serves also as a good first approximation to the physics of cosmological black holes.

\subsubsection{Plan of this chapter}

The plan of this chapter is as follows. We split the discussion of asymptotically black holes in ungauged and gauged supergravity. There are many general results in ungauged supergravity where the BPS black hole solutions are fully understood and there exist various advances towards full classification of extremal and non-extremal black holes. We then move to the less well-explored topic of Minkowski black holes in gauged supergravity, where we provide the first attempts of generalizing the ungauged solutions.

First, in section \ref{5:BPS}, we give a summary of the known supersymmetric black hole
solutions in ungauged $N=2$ supergravity with neutral hypermultiplets. We then also briefly comment on the results concerning non-BPS extremal and non-extremal black holes with non-trivial scalar profiles\footnote{The classification of solutions without scalars was already presented above, therefore from the spacetime point of view one does not expect qualitatively new solutions to appear. We will however see that this point of view is too naive when discussing asymptotically AdS black holes in the next chapter.} in section \ref{5:nonBPS}.

Moving on to the black holes in gauged supergravity, in section \ref{5:trivial} we make use of the
Higgs mechanism for spontaneous gauge symmetry breaking explained in the previous chapter, in order
to obtain effective $N = 2$ ungauged theories from a general
gauged $N = 2$ supergravity. We show how this method can be used to embed
already known black hole solutions into gauged supergravities
and explain the physical meaning of the new solutions. We illustrate this with
an explicit example of a static, asymptotically flat black hole with the
well-known STU model and one gauged hypermultiplet (the universal
hypermultiplet).

In section \ref{5:BLS} we discuss in more general terms asymptotically flat,
stationary spacetimes preserving half of the supersymmetries. We analyze
the fermion susy variations in gauged supergravity after choosing a particular ansatz for
the Killing spinor. One finds two separate cases,
defined by $T^-_{\mu \nu} = 0$ and $P^x_\Lambda = 0$,
respectively. Whereas the former case contains only Minkowski and
AdS$_4$ solutions, the latter leads to a class of solutions
that generalize the standard black hole solutions of ungauged
supergravity. We analyze this in full detail and give the complete set of equations that
guarantees a half-BPS solution. We then explain how this fits to
the solutions obtained in section \ref{5:trivial}.

Finally, in section \ref{5:hair}, we study
asymptotically flat black holes with scalar hair\footnote{By scalar hair, in this thesis, we mean a scalar field that is zero at the horizon of the black hole, but non-zero outside of the horizon. According to this definition, the vector multiplet scalars subject to the attractor mechanism in $N=2$ ungauged supergravity, do not form black holes with scalar hair. The solutions that we discuss in section \ref{5:hair}, however, will have hair.} . We find two separate classes of such solutions. One is a purely bosonic solution with
scalar hair, but with the shortcoming of having ghost modes in the
spacetime. The other class of solutions has no ghosts but along
with scalar hair we also find fermionic hair, i.e. the fermions
are not vanishing in such a vacuum.

Some of the more technical aspects of this chapter, including
explicit hypermultiplet gaugings leading to Minkowski asymptotics, are presented in the appendices.

\section{Ungauged theory}\label{5:section2}
\subsection{Supersymmetric stationary solutions}\label{5:BPS}

Asymptotically flat and stationary BPS black hole solutions of
ungauged supergravity have been a very fruitful field of research
in the last decades. In absence of vector multiplets $(n_V=0)$,
with only the graviphoton present, the supersymmetric solution is
just the well-known extremal Reissner-Nordstr\"om (RN) black hole\footnote{The Kerr-Newman spacetime with $M = |Z|$ and an arbitrary angular momentum is also a supersymmetric solution, but it does not have an event horizon and is therefore a naked singularity rather than black hole.}.
This solution was later generalized to include a number of vector
multiplets \cite{Ferrara:1995ih}. The most general classification of
the BPS solutions, including multicentered black holes, was given
by Behrndt, L\"ust and Sabra \cite{Behrndt:1997ny} and we will refer to those
as BLS solutions. The hypermultiplet scalars
$q^u$ do not mix with the other fields (apart from the graviton)
at the level of the equations of motion, and it is therefore
consistent to set them to a constant value. We will briefly list the main points of the
solutions, as they will play an important role in what follows. For a comprehensive review of these black holes solutions and their important implications we refer the interested reader to \cite{Mohaupt:2000mj}.

To characterize the black hole solutions, we first denote the
imaginary parts of the holomorphic sections by
\begin{equation}\label{eq:sections}
    H^{\Lambda}\equiv i (X^{\Lambda} - \bar{X}^{\Lambda})
        \ ,  \qquad  H_{\Lambda}\equiv i (F_{\Lambda} - \bar{F}_{\Lambda}) \ .
\end{equation}
We assume stationary solutions with axial symmetry\footnote{Thus we do not allow for multicentered solutions with more than two centers. It this thesis we are mostly concerned with the general classification of solutions and will not elaborate much on the multicentered case.} parametrized by an angular
coordinate $\varphi$. The result of the BPS analysis is that the metric takes the form\footnote{Note
that all the results are in spherical coordinates, see
\cite{Behrndt:1997ny,Meessen:2006tu} for the coordinate independent results.}
\begin{equation}\label{eq:BLSmetric}
    {\rm ds}^2 = {\rm e}^{{\cal K}} ({\rm d} t + \omega_{\varphi} {\rm d}
    \varphi)^2 - {\rm e}^{- {\cal K}} \left( {\rm d} r^2 + r^2 {\rm d}
    \Omega_2^2 \right)\ ,
\end{equation}
where ${\cal K}$ is the K\"{a}hler potential~\eqref{eq:def-kahler} of special geometry.
The metric components and the symplectic vector $\big( H^{\Lambda}, H_{\Lambda} \big)$ only depend
on the radial variable $r$ and the second angular coordinate $\theta$, and the BPS conditions imply the differential equations on $\omega_{\varphi}$
\begin{align}
\begin{split}
  \frac{1}{r^2 \sin \theta} \partial_{\theta} \omega_\varphi =
  H_{\Lambda}\partial_r H^{\Lambda} - H^{\Lambda}
  \partial_r H_{\Lambda}\ ,\qquad
    -\frac{1}{\sin \theta} \partial_{r} \omega_\varphi =
  H_{\Lambda}\partial_{\theta} H^{\Lambda} - H^{\Lambda}
  \partial_{\theta} H_{\Lambda}\ .
  \end{split}
\end{align}
From this follows the integrability condition  $H_{\Lambda} \square
H^{\Lambda} - H^{\Lambda}
  \square   H_{\Lambda} = 0$, where $\square$ is the
  3-dimensional Laplacian.

What is left to specify are the gauge field strengths
$F^{\Lambda}_{\mu \nu}$. First we define (c.f.\ the discussion about electro-magnetic duality in chapter \ref{chapter::supergravity}) the magnetic field strengths
\begin{align}\label{eq:defg}
G_{\Lambda}{}_{\mu \nu} \equiv R_{\Lambda
    \Sigma} F^{\Sigma}_{\mu \nu} - \frac 12 I_{\Lambda \Sigma}\,
  \epsilon_{\mu \nu \gamma \delta} F^{\Sigma \gamma \delta}\ ,
\end{align}
such that the Maxwell equations and Bianchi identities take the
simple form
\begin{equation}\label{eq:blsmaxwell}
    \epsilon^{\mu \nu \rho \sigma} \partial_{\nu} G_{\Lambda}{}_{\rho \sigma} =
    0, \quad \epsilon^{\mu \nu \rho \sigma} \partial_{\nu} F^{\Lambda}_{\rho \sigma} =
    0\ ,
\end{equation}
such that $(F^\Lambda, G_\Lambda)$ transforms as a vector under electric-magnetic duality transformations.

For the full solution it is enough to specify half of the
components of $F^\Lambda_{\mu \nu}$ and $G_{\Lambda \mu \nu}$, since the other half can be
found from~\eqref{eq:defg}. In spherical coordinates, the BPS equations imply the
non-vanishing components\footnote{The BPS conditions also imply $F^\Lambda_{r\theta}=G_{\Lambda r\theta}=0$
due to axial symmetry.}
\begin{equation}\label{eq:FasHtilde} F^{\Lambda}_{r \varphi} =
\frac{-r^2 \sin \theta}{2}
\partial_{\theta} H^{\Lambda}, \qquad F^{\Lambda}_{\theta \varphi}
= \frac{r^2 \sin \theta}{2} \partial_r H^{\Lambda},
\end{equation} and
\begin{equation}\label{eq:GasH}
G_{\Lambda}{}_{r \varphi} = \frac{-r^2 \sin \theta}{2}
\partial_{\theta} H_{\Lambda}, \qquad G_{\Lambda}{}_{\theta \varphi} =
\frac{r^2 \sin \theta}{2} \partial_r H_{\Lambda}.
\end{equation}
From \eqref{eq:blsmaxwell} it now follows that $H_{\Lambda}$ and $
H^{\Lambda}$ are harmonic functions. With the above
identities we can always find the vector multiplet scalars $z^i$,
given that we know explicitly how they are defined in terms of the
sections $X^{\Lambda}$ and $F_{\Lambda}$.
The integration constants of the harmonic functions specify the asymptotic behavior of the fields at
the black hole horizon(s) (the constants can be seen to be the black hole electric and
magnetic charges) and at spatial infinity.

The complete proof that these are indeed all the supersymmetric
black hole solutions with abelian vector multiplets and no
cosmological constant was given in \cite{Meessen:2006tu}. Note that the
BLS solutions describe half-BPS stationary spacetimes with (only
in the multicentered cases) or without angular momentum. The
near-horizon geometry around each center is always AdS$_2 \times $S$^2$ with equal
radii of the two spaces, determined by the charges of the black
hole (see \cite{Kunduri:2007vf} for a more general discussion of possible near-horizon geometries). All solutions exhibit the so-called attractor mechanism
\cite{Ferrara:1995ih,Strominger:1996kf,Ferrara:1996dd}. This means that the (vector multiplet) scalar
fields get attracted to constant values at the horizon that only depend on the black hole charges.
As the scalars can be arbitrary constants at
infinity we also find the so-called attractor flow, i.e.\ the
scalars flow from their asymptotic value to the fixed constant at
the horizon.

\subsection{Non-supersymmetric black holes}\label{5:nonBPS}
The full classification of non-BPS black hole solutions and attractors as opposed to the supersymmetric ones is much more involved and is still in
progress. The presence or absence of unbroken supercharges at technical level results in a difference between solving first and second order partial differential equations, respectively. Therefore it is not a surprise to find that much less is known about the general structure of black hole solutions and the microscopic theories behind them.

The simplest generalization from BPS to non-BPS black holes is of course to still consider zero temperature solutions. They share some similar features to the BPS solutions, notably the attractor phenomenon \cite{Ferrara:1997tw,Sen:2005wa,Goldstein:2005hq} and the fact that in certain cases they can also be found from first order differential equations after a suitable rewriting of the action \cite{Denef:2000nb}. In contract to the BPS case, non-BPS attractors exhibit flat directions, i.e.\ the scalars are not completely fixed at the horizon by the values of the charges. A few examples have been found where changing some of the signs in the BPS solutions result in non-BPS extremal solutions \cite{Tripathy:2005qp,Kallosh:2006ib}, but such examples are not generic enough. The role of electro-magnetic duality is very prominent in the attempts to find general classes of solutions, since full duality orbits can be found just by having a single (generic enough) solution.

When discussing non-BPS solutions, we need to remember that there is another symmetry that is allowed to be broken - the spherical symmetry. There can be various extremal and non-extremal generalizations of the Kerr-Newman black hole, usually referred to as overrotating black holes (since the limit of vanishing angular momentum for these solutions leads to a naked singularity). Their underrotating counterparts, on the other hand, are rotating generalizations of the Reissner-Nordstr\"{o}m solutions. The organizing principle for the general structure of either classes is unfortunately still unclear.

Recently, there has been some advance in simplifying the equations of motion also for non-extremal black holes \cite{Galli:2011fq}, but results on this topic are even more scarce and disorganized. Virtually all solutions of asymptotically flat black holes can be embedded in supergravity theories. We will therefore not go into details on the topic since it extends very far from the techniques one can employ from supersymmetry considerations. The situation at present is even more hopeless for theories with gauging, where the equations of motion allow for even broader classes of solutions than in the ungauged case. In what follows, however, we only concentrate on solutions that can be analyzed purely from a supersymmetric point of view.

\section{Gauged theory}\label{5:section3}
\subsection{Solutions via spontaneous symmetry breaking}\label{5:trivial}
The simplest solutions one can have in gauged supergravity are the ones that exist also in ungauged supergravity. As we saw above, there are many results in the literature on the subject of black holes in ungauged supergravity and via the method of spontaneous symmetry breaking they can all be used in general classes of gauged supergravities. These, in a way trivial, solutions follow directly from our discussion in chapter \ref{chapter::solution_generator} and we find them easier to understand in the context of a canonical example from BPS black holes in ungauged supergravity: the STU model.

\subsubsection{The STU model with gauged universal hypermultiplet}\label{sect:stu}
Let us consider an $N = 2$ theory with the universal hypermultiplet (UHM). Its quaternionic
metric and isometries are given in appendix \ref{app:UHM}, and isometry 5 is chosen to be gauged. This allows for asymptotically flat black holes, since we can find solutions of \eqref{hypers no FI terms}, as we shall see below\footnote{A
suitable combination of isometries 1 and 4 would also do the job.
Note that typically in string theory isometry 5 gets broken
perturbatively while 1 and 4 remain also at quantum level. For the
present discussion it is irrelevant which one we choose since we
are not trying to directly obtain the model from string theory.}.
The quaternionic Killing vector and moment maps are given by
\begin{align}
{\tilde k}_{\Lambda} &= a_{\Lambda} \left(2 r \partial_R + \chi \partial_\chi + \varphi \partial_\varphi + 2 \sigma \partial_\sigma \right),\\
{\vec P}_{\Lambda} &= a_{\Lambda} \left\{ -\frac \chi {\sqrt R}, \frac \varphi {\sqrt R},
-\frac {\sigma + \frac 12 \varphi \chi} R \right\}\ ,
\end{align}
with $a_\Lambda$ arbitrary constants. In this chapter we will use $R$ to
denote the coordinate on the UHM, to avoid confusion with the radial,
spacetime coordinate $r$.

In the vector multiplet sector we take the so-called STU model, based on the prepotential
\begin{equation}
F = \frac{X^1 X^2 X^3}{X^0}\ ,
\end{equation}
together with $z^i = \frac{X^i}{X^0}; i=1,2,3$.  The gauge group is $U(1)^3$, but it will be broken
to $U(1)^2$ in the supersymmetric Minkowski vacua, in which we construct the black hole solution.
The conditions for a fully BPS Minkowski vacuum require
$F_{\mu \nu}^{\rm vev} = 0$, $z^{i}{}^{\rm vev} = \langle z^i \rangle = \langle
b^i\rangle+i \langle v^i \rangle$, $\chi^{\rm vev} = \varphi^{\rm vev} =
\sigma^{\rm vev} = 0$, $
R^{\rm vev} = \langle R \rangle $, with arbitrary constants $\langle z^i \rangle$ and $\langle R \rangle$.
Moreover, from \eqref{hypers no FI terms}, the
vector multiplets scalar vevs must obey $(a_{\Lambda} L^{\Lambda})^{\rm vev} =
0$ (which is an equation for the $\langle z^i \rangle$'s). Then, after expanding around this vacuum, the mass terms for the scalar fields are given by the quadratic terms in the scalar potential
\eqref{eq:scalar-potential}. Now, if we make the definition $z \equiv
a_{\Lambda} L^{\Lambda}$, we have $z^{\rm vev} = 0$. Expanding
the first term in \eqref{eq:scalar-potential} gives the mass term for
$z$,
$$\left(4
h_{uv}\tilde{k}^u_\Lambda \tilde{k}^v_\Sigma{\bar L}^\Lambda
L^\Sigma\right)^{\rm quadratic} = 16 z \bar{z}.$$ Expanding the second
term to quadratic order gives the mass for three of the hypers:
\begin{align}
  \left(g^{i\bar \jmath}f_i^\Lambda {\bar f}_{\bar
\jmath}^\Sigma P^x_\Lambda P^x_{\Sigma}\right)^{\rm quadratic} =
\frac{a_i^2 \langle v^i \rangle^2}{\langle v^1 v^2 v^3 \rangle
\langle R\rangle} \biggl(\chi^2+\varphi^2+\frac{(\sigma + \frac 12
\chi \varphi)^2}{\langle R\rangle}\ \biggr)\ ,
\end{align}
while the third term vanishes at quadratic order and does not contribute
to the mass matrix of the scalars.

Therefore two of the six vector multiplet scalars become massive
(i.e.\ the linear combination given by our definition for $z$),
together with three of the hypers. The fourth hyper $R$ remains
massless and is eaten up by the massive gauge field $a_{\Lambda}
A^{\Lambda}_{\mu}$ (with mass $4$ given by~\eqref{eq:vectormass}).
Thus we are left with an effective $N = 2$ supergravity theory of
one massive and two massless vector multiplets and no
hypermultiplets, which can be further consistently truncated to
only include the massless modes. One can then search for BPS
solutions in the remaining theory and the prescription for finding
black holes is again the one given by Behrndt, L\"{u}st and Sabra
and explained in section \ref{5:BPS}.

We now construct the black hole solution more explicitly, following the solution generating technique
of chapter \ref{chapter::solution_generator}. For this, we need to satisfy \eqref{eq:trivialhypers}
and \eqref{hypers no FI terms}. The condition $P^x_{\Lambda} = 0$
fixes $\chi = \varphi = \sigma =0$ and the remaining non-zero Killing vectors
are $k^R_{\Lambda} = 2 R a_{\Lambda}$. Now we have to satisfy the remaining conditions
$\tilde{k}^u_{\Lambda} X^{\Lambda} = 0$ and $\tilde{k}^u_{\Lambda}
F^{\Lambda}_{\mu \nu} = 0$. To do so, we use the BLS solution of the STU model. For simplicity we
take the static limit $\omega_m = 0$, discussed in detail in
section 4.6 of \cite{Behrndt:1997ny}. The solution is fully expressed in
terms of the harmonic functions
\begin{align}
H _0 = h_0 + \frac{q_0}{r}, \qquad H^i &= h^i +
\frac{p^i}{r}\ ,\qquad \  i=1,2,3\ ,
\end{align}
under the condition that one of them is negative definite. The
sections then read
\begin{align}
X^0 = \sqrt{- \frac{H^1 H^2 H^3}{4 H_0}}\ ,
\qquad X^i = - i \frac{H^i}{2}\ ,
\end{align}
with metric function
\begin{align}
{\rm e}^{-\mathcal K} = \sqrt{-4 H_0 H^1 H^2 H^3}.
\end{align}
In this case $F^0_{m n} = 0$ and the $F^i_{m n}$ components (here
$m,n$ are the spatial indices) are expressed solely in terms of derivatives
of $H^i$. After evaluating the period matrix we
obtain $F^i_{m t} = 0$ and $F^0_{m t}$ are given in terms of
derivatives of $H_0, H^i$. Thus the equations
$\tilde{k}^R_{\Lambda} X^{\Lambda} = 0$ and $\tilde{k}^R_{\Lambda}
F^{\Lambda}_{\mu \nu} = 0$ lead to
\begin{align}\label{STU-cond}
a_0 = 0\ , \qquad a_i h^i = 0\ , \qquad a_i p^i = 0\ .
\end{align}
The solution is qualitatively the same
as the original one, but  the charges $p^i$ and the
asymptotic constants $h^i$ are now related by \eqref{STU-cond}. So, effectively,
the number of independent scalars and vectors is decreased by one, consistent with
the results from spontaneous symmetry breaking.
The usual attractor mechanism for the remaining massless vector multiplet scalars holds while for the
hypermultiplet scalars we know that $\chi = \varphi = \sigma = 0$ and $R$ is
fixed to an arbitrary constant everywhere in spacetime with no
boundary conditions at the horizon. In other words the hypers are not
`attracted'.

Our construction can be generalized for non-BPS solutions as well.
In the particular case of the STU model, we can obtain a
completely analogous, non-BPS,  solution by following the
procedure described in \cite{Kallosh:2006ib}. We flip the sign of one
of the harmonic functions in \eqref{eq:sections} such that
\begin{align}
{\rm e}^{-\mathcal K} = \sqrt{4 H_0 H^1 H^2 H^3}\ .
\end{align}
This solution preserves no supersymmetry, but it is
extremal. By following our procedure above, we can embed this solution into
the gauged theory.

\subsection{1/2 BPS solutions}\label{5:BLS}
In this section we will take a more systematic approach to
studying the supersymmetric solutions of the general gauged theory~\eqref{lagr}. We search
for a solution where the expectation values of the fermions are
zero. This implies that the supersymmetry variations of the bosons
should be zero. The vanishing of the supersymmetry variations
\eqref{susy-gravi}-\eqref{susygluino} then guarantees some amount
of conserved supersymmetry. Depending on the number of independent
components of the variation parameters $\varepsilon_A$ we will
have different amount of conserved supersymmetry. Here we will
focus on particular solutions preserving (at least) 4
supercharges, i.e. half-BPS configurations. A BPS configuration
has to further satisfy the equations of motion in order to be a
real solution of the theory, so we also impose those. The
fermionic equations of motion vanish automatically, so we are left
with the equations of motion for the graviton $g_{\mu \nu}$, the
vector fields $A^{\Lambda}_{\mu}$, and the scalars $z^i$ and
$q^u$. We will come to the relation between the BPS constraints
and the field equations in due course, but we first introduce some
more relations for the Killing spinors $\varepsilon_A$.

\subsubsection{Killing spinor identities}
We will make use of the approach~\cite{Gauntlett:2003fk} where one
first assumes the existence of a Killing spinor. From this spinor,
various bilinears are defined, whose properties constrain the form
of the solution to a degree where a full classification is
possible. We use this method in $D=4, N=2$, which is generalizing
the main results of~\cite{Meessen:2006tu,Huebscher:2006mr} to include hypermultiplets in
the description. As it later turns out, we cannot completely use
this method to classify all the supersymmetric configurations, but
the method nevertheless gives useful information.

We define $\varepsilon_A$ to be a Killing spinor if it solves the
gravitino variation $\delta_\varepsilon \psi_{\mu A} = 0$, defined in
\eqref{susy-gravi}, and assume $\varepsilon_A$ to be a Killing spinor
in the remainder of this article. Such spinors anti-commute, but we
can expand them on a basis of Grassmann variables and only work with
the expansion coefficients. This leads to a commuting spinor,
which we also denote with $\varepsilon_A$, and we define\footnote{We will be brief
  on some technical points of the discussion, and refer
  to~\cite{Meessen:2006tu,Huebscher:2006mr} for more information.}
\begin{align}
\begin{split}
\overline {\varepsilon_A} &\equiv i (\varepsilon^{A})^{\dagger} \gamma_0\ ,\\
    X &\equiv \frac 12 \epsilon^{AB} \overline {\varepsilon_A} \varepsilon_B\ ,\\
 V_{\mu}{}^A{}_B &\equiv i \overline {\varepsilon^A} \gamma_\mu
 \varepsilon_B\ ,\\
\Phi_{AB\mu\nu} &\equiv \overline {\varepsilon_A} \gamma_{\mu\nu}
\varepsilon_B\ .
\end{split}
\end{align}
We now show that this implies that $V^\mu \equiv V^\mu{}^A{}_A$ is a Killing
vector. For its derivatives we find
\begin{align}
\begin{split}
  \nabla_\mu V_{\nu}{}^A{}_B &= i \delta^A{}_B (T^+_{\mu\nu} X -
  T^-_{\mu\nu} \bar X) - g_{\mu\nu} (S^{AC} \epsilon_{CB} X - S_{BC}
  \epsilon^{AC} \bar X)\\
&- i (\epsilon^{AC} T^+_\mu{}^\rho \Phi_{CB\rho\nu} +
\epsilon_{BC} T^-_{\mu}{}^\rho \Phi^{AC}{}_{\nu\rho}) - (S^{AC}
\Phi_{CB\mu\nu} + S_{BC} \Phi^{AC}{}_{\mu\nu})\ .
\end{split}
\end{align}
The second and third term are traceless, so they vanish when we
compute $\nabla_\mu V_\nu$. The other terms are antisymmetric in
$\mu\nu$, so this proves
\begin{align}
  \nabla_\mu V_\nu + \nabla_\nu V_\mu = 0\ ,
\end{align}
thus $V_\mu$ is a Killing vector. We make the decomposition $
  V^A{}_{B\mu} = \frac 12 V_\mu \delta^A{}_C + \frac 1 {\sqrt 2}
  \sigma^{xA}{}_B V^x_\mu$
and using Fierz identities one finds
\begin{align}\label{eq:decomposition}
  V_\mu{}^A{}_B V_\nu{}^B{}_A = V_\mu V_\nu - \frac 12 g_{\mu\nu} V^2\ .
\end{align}
One can show that $V_\mu V^\mu = 4|X|^2$, which shows that the
Killing vector $V_\mu$ is timelike or null. For the remainder of
this chapter we restrict ourselves to a timelike Killing spinor
ansatz, defined as one that leads to a timelike Killing vector. We
make this choice, as our goal is to find stationary black hole solutions,
which always have a timelike isometry.  In this case, by
definition, $V_\mu V^\mu = 4 |X|^2 \neq 0$, so we can
solve~\eqref{eq:decomposition} for the metric as
\begin{align}
  g_{\mu\nu}
&=\frac 1 {4|X|^2} \left(V_\mu V_\nu - 2 V_\mu^x V_\nu^x\right)\ .
\end{align}
It follows that
\begin{align}
  V_\mu = g_{\mu\nu} V^\nu = V_\mu - \frac 1{2|X|^2} V_\mu^x (V_\nu^x
  V^\nu)\ ,
\end{align}
so $V_\mu^x V^\mu = 0$. We define a time coordinate by $V^\mu
\partial_\mu = \sqrt 2 \partial_t$, which implies $V^x_t = 0$. We
decompose $V_\mu {\rm d}x^\mu = 2 \sqrt 2 X \bar X({\rm d}t + \omega)$, where
the factor in front of ${\rm d}t$ follows from $V^2 = 4 X \bar X$ and
$\omega$ has no ${\rm d}t$ component. The metric is then given by
\begin{align}\label{eq:metric}
  {\rm ds}^2 = 2 |X|^2 ({\rm d}t + \omega)^2 - \frac 1{2|X|^2}
  \gamma_{mn}{\rm d}x^m {\rm d}x^n\ ,
\end{align}
where $|X|, \omega$ and $\gamma_{mn}$ are independent of time.

Now we are ready to make a relation between the susy variations
\eqref{susy-gravi}-\eqref{susygluino} and the equations of
motion, using an elegant and simple argument of Kallosh and
Ortin \cite{Kallosh:1993wx} that was later generalized in
\cite{Meessen:2006tu,Huebscher:2006mr}. Assuming the existence of (any amount of)
unbroken supersymmetry, one can derive a set of equations relating
the equations of motion for the bosonic fields with derivatives of
the bosonic susy variations. For the theory at hand these read:
\begin{align}
\begin{split}
\mathcal E_\Lambda^\mu i f_i^\Lambda \gamma_{\mu}
\varepsilon^A
  \epsilon_{AB} + \mathcal E_i \varepsilon_B &= 0\ ,\\
  \mathcal E_a^\mu (-i \gamma^a \varepsilon^A) + \mathcal
  E_\Lambda^\mu \left(2 \bar L^\Lambda \varepsilon_B
    \epsilon^{AB}\right) &= 0\ ,\\
\mathcal E_u \mathcal U^{u}_{\alpha A} \varepsilon^A &= 0\ ,
\end{split}
\end{align}
where $\mathcal E$ is the equation of motion for the corresponding
field in subscript. More precisely, $\mathcal{E}_a^\mu$ is the
equation for the vielbein $e_\mu^a$ (the Einstein equations),
$\mathcal{E}_\Lambda^\mu$ corresponds to $A^{\Lambda}_\mu$ (the
Maxwell equations), $\mathcal{E}_u$ corresponds to $q^u$ and
$\mathcal{E}_i$ to $z^i$. Now, let us assume that the Maxwell
equations are satisfied, $\mathcal{E}_\Lambda^\mu = 0$. If we
multiply each of the remaining terms in the three equations by
$\overline{\varepsilon^B}$ and $\gamma^\nu \overline
{\varepsilon^B}$ and use the fact that the Killing spinor is
timelike such that $X \neq 0$ we directly obtain that the
remaining field equations are satisfied. So, apart from the BPS
conditions, only the Maxwell equations
\begin{equation}\label{maxwell}
\epsilon^{\mu \nu \rho \sigma} \partial_{\nu} G_{\Lambda\rho\sigma} =
- gh_{u v} {\tilde k}^u_{\Lambda} \nabla^{\mu}
  q^v\ ,
\end{equation}
need to be satisfied.

\subsubsection{Killing spinor ansatz}
Contracting the gaugino variation
(\ref{susygluino}) with $\varepsilon_A$ we find the condition
\begin{align}
  0 = - 2 i \bar X \nabla_\mu z^i + 4 i G^{-i}_{\rho\mu} V^\rho - i g
  k^i_\Lambda \bar L^\Lambda V_\mu - \sqrt 2 g g^{i\bar\jmath} \bar
  f_{\bar \jmath}^\Lambda P_\Lambda^x V_\mu^x\ .
\end{align}
Using this to eliminate $\nabla_\mu z^i$ and plugging back into
$\delta \lambda^{iA} = 0$ we find\footnote{One could, as done in
  e.g.~\cite{Meessen:2006tu,Huebscher:2006mr}, eliminate the gauge fields $G^{i-}_{\rho\mu}$ to
  obtain an equivalent relation.}
\begin{align}\label{eq:susy-gaugino-contracted}
  G^{i-}_{\rho\mu} \gamma^\mu \left(2 i V^\rho \varepsilon^A - \bar
    X \gamma^\rho \epsilon^{AB} \varepsilon_B\right) + g g^{i \bar
    \jmath} \bar f_{\bar \jmath}^\Lambda P^x_\Lambda \left(- \frac 1 {\sqrt
    2} V^x_\mu \gamma^\mu \varepsilon^A + i \bar X \sigma^{xAB}
  \varepsilon_B \right)=0\ .
\end{align}
It is here that we find an important difference with the ungauged
theories. In the latter case, $g = 0$, and the second term is
absent. Then, assuming that the gauge fields $G^{i-}_{\rho\mu}$
are non-zero, one can rewrite
equation~\eqref{eq:susy-gaugino-contracted} as
\begin{align}\label{eq:ansatz}
  \varepsilon^A + i {\rm e}^{-i\alpha} \gamma_0 \epsilon^{AB} \varepsilon_B
  = 0\ ,
\end{align}
where ${\rm e}^{i \alpha} \equiv \frac X {|X|}$. We have thus
derived the form of the Killing spinor, which is not an
ansatz anymore.

In gauged supergravity, $g \neq 0$, so there are
various ways to solve equation~\eqref{eq:susy-gaugino-contracted}.
One could, for instance, generalize~\eqref{eq:ansatz} to
\begin{align}\label{eq:gen-ansatz}
  \varepsilon^A = b \gamma^0 \epsilon^{AB} \varepsilon_B + a^x_m
  \gamma^m \sigma^{xAB} \varepsilon_B\ .
\end{align}
Plugging this back into \eqref{eq:susy-gaugino-contracted}, one obtains BPS conditions on the fields which one can then
try to solve. While this is hard  in general, it has been done in a specific case. Namely, the ansatz used for the AdS-RN black holes in minimally gauged supergravity (with a bare cosmological constant), as analyzed by
Romans~\cite{Romans:1991nq}, fits into~\eqref{eq:gen-ansatz}, but
not in~\eqref{eq:ansatz}. In fact, we will see later that
with~\eqref{eq:ansatz} one cannot find AdS black holes.

In the remainder of this section and chapter, we will use \eqref{eq:ansatz} as a
particular ansatz, hoping to find new BPS black hole solutions that
are asymptotically flat. The reader should keep in mind that more
general Killing spinors are possible. The search for BPS black holes that asymptote to AdS$_4$, and their Killing spinors, is postponed for the next chapter.

\subsubsection{Metric and gauge field ansatz}
We will further make the extra assumption that the solution for the spacetime metric, field strengths and
scalars, is axisymmetric, i.e. there is a well-defined axis of rotation, such that $\omega = \omega_{\varphi}
{\rm d} \varphi$ lies along the angle of rotation (we choose to call it $\varphi$) in \eqref{eq:metric}. For a
stationary axisymmetric black hole solution the symmetries constrain the metric not to depend on $t$ and
$\varphi$. These symmetries also constrain the scalars and gauge field strengths to depend only on the remaining
coordinates, which we choose to call $r$ and $\theta$. We further assume $F^{\Lambda}_{r \theta} = 0$, such that
(after also using the gauge freedom) we can set $A^{\Lambda}_r =
A^{\Lambda}_{\theta} = 0$ for all $\Lambda$.

\subsubsection{Gaugino variation}
Plugging the ansatz~\eqref{eq:ansatz} into the gaugino variation
$\delta \lambda^{iA} = 0$ gives
\begin{equation}\label{eq:Pfiszero}
   P_\Lambda^x f_i^\Lambda = 0\ ,
\end{equation}
and
\begin{equation}\label{eq:vector-conditions}
  \left({\rm e}^{-i \alpha} \partial_\mu z^i \gamma^{\mu} \gamma^0 + G^{- i}_{\mu \nu} \gamma^{\mu \nu} \right) \varepsilon_A  =
  0\ .
\end{equation}
The latter condition can be simplified further, but we will see in
what follows that it automatically becomes simpler or gets
satisfied in certain cases, so we will come back to
\eqref{eq:vector-conditions} later. We will make use of condition
\eqref{eq:Pfiszero} when solving the gravitino integrability
conditions.

\subsubsection{Hyperino variation}\label{sect:hyperino-variation}
With the ansatz~\eqref{eq:ansatz}, setting the hyperino variation to zero gives the condition
\begin{align}
{\rm  e}^{-i\alpha} \nabla_\mu q^u \gamma^\mu \gamma_0 + 2 g \tilde k^u_\Lambda
  \bar L^\Lambda = 0\ .
\end{align}
Using the independence of the gamma matrices, one finds
\begin{align}\label{eq:hyper-conditions}
\begin{split}
  \nabla_r q^u &= \nabla_{\theta} q^u = 0\,,\\
 \nabla_{\varphi} q^u &= \omega_\varphi \nabla_t q^u\,,\\
 \nabla_t q^u  &= -  \sqrt
2 g {\tilde k}^u_\Lambda \left(X \bar L^\Lambda  + \bar X
L^\Lambda
\right) , \\
0 &= \tilde k^u_\Lambda  \left( \bar X L^\Lambda - X \bar
L^\Lambda\right) \,.
\end{split}
\end{align}
Using axial symmetry and the gauge choice for the vector fields, $A^{\Lambda}_r =
A^{\Lambda}_{\theta} = 0$, it follows
that $\nabla_r q^u =
\partial_r q^u$ and $\nabla_{\theta} q^u =
\partial_{\theta} q^u$, and these both vanish from the BPS
conditions. Furthermore, the hypers cannot depend on
$t$ and $\varphi$, because this would induce such dependence also on
the vector fields and complex scalars via the Maxwell equations
\eqref{maxwell}. Thus the hypers cannot depend on any of the
spacetime coordinates, so they are constant. This will be
important when we analyze the gravitino variation.

\subsubsection{Gravitino variation}

The gravitino equation reads
\begin{align}
  \nabla_\mu \varepsilon_A = - {\rm e}^{-i\alpha} \left(T^-_{\mu\rho}
    \gamma^\rho \delta_A{}^C + g S_{AB} \epsilon^{BC}
    \gamma_\mu\right) \gamma_0 \varepsilon_C\ .
\end{align}
We study the integrability condition which follows from this
equation. The explicit computation is presented in appendix \ref{sect:integrability}.
The main result that we will first focus on is equation \eqref{eq:pxl=0},
\begin{align}
 T^-_{\mu\nu} P^x_\Lambda L^\Lambda &= 0\ ,
\end{align}
so that there are two separate cases: $T_{\mu\nu}^- = 0$ or
$P^x_\Lambda L^\Lambda = 0$. We will study these two cases in
different subsections.

\subsubsection{Case 1: T$^-_{\mu\nu} = 0$}\label{other solutions}
In this case the integrability conditions imply that the
spacetime is maximally symmetric with constant scalar curvature
$P^x_\Lambda L^\Lambda$, as further explained in appendix~\ref{sect:tzero} .
This corresponds either to Minkowski
space when $P^x_\Lambda L^\Lambda = 0$, or AdS$_4$ when the scalar
curvature is non-zero. Although there might be interesting half
BPS solutions here, they will certainly not describe black holes.

\subsubsection{Case 2: $P^x_\Lambda = 0$}\label{subsect:BLS}

The second case is $P^x_\Lambda L^\Lambda = 0$.
We combine this identity with $P^x_\Lambda
f^\Lambda_i = 0$ from~\eqref{eq:Pfiszero}. We now obtain
\begin{align}
P^x_\Lambda  \begin{pmatrix} \bar L^\Lambda \\ f^\Lambda_i \end{pmatrix} = 0\,.
\end{align}
The matrix between brackets on the left hand side is invertible. This follows from the properties of
special geometry, and we used it also in the characterization of the maximally supersymmetric vacua in
\cite{Hristov:2009uj}. We therefore conclude that
$P^x_\Lambda = 0$. Next, we show that in this case we have enough information
to solve the gravitino variation and give the metric
functions.

From the definition for $\nabla_\mu
\varepsilon_A$, the quaternionic $Sp(1)$ connection $\omega_{\mu A}{}^B$ vanishes, as the hypers
are constant by the arguments in
section~\ref{sect:hyperino-variation}. Combining this with
$P^x_\Lambda = 0$, we see that the gravitino
variation~\eqref{susy-gravi} is precisely the same as in a theory
without hypermultiplets and vanishing FI-terms. Thus our problem
reduces to finding the most general solution of the gravitino
variation in the ungauged theory. The answer, as proven by \cite{Meessen:2006tu,Huebscher:2006mr}, is that this is
the well-known BLS solution \cite{Behrndt:1997ny} for stationary black holes
(or naked singularities and monopoles in certain cases). Thus we
can use the BLS solution, which in fact also solves the
gaugino variation \eqref{eq:vector-conditions}. We now only have to
impose the Maxwell equations, which are not the same as in the BLS
setup, due to the gauging of the hypermultiplets.

The sections are again described by functions $H_\Lambda$ and
$H^\Lambda$, as in~\eqref{eq:sections}, although not all of
them are harmonic. The metric and
field strengths are given
by~\eqref{eq:BLSmetric},~\eqref{eq:FasHtilde} and~\eqref{eq:GasH}. In
terms of our original description~\eqref{eq:metric}, we have that $\gamma_{mn}$ is
three dimensional flat space and
\begin{align}
  {\rm e}^{\mathcal K} = 2 |X|^2\ .
\end{align}
In the ungauged case the Maxwell equations have no source term and the field strengths are thus
described by harmonic functions, while now in our case they will
be more complicated. We can then directly compare to the original
BLS solution described in section \ref{5:BLS} and see how
the new equations of motion change it. At this point we have
chosen the phase $\alpha$ in \eqref{eq:ansatz} to vanish, just
as it does in the BLS solution. We can do this without any loss of
generality since an arbitrary phase just appears in the
intermediate results for the symplectic sections
\eqref{eq:sections}, but drops out of the physical quantities such
as the metric and the field strengths.

We repeat that the Maxwell equations are given by~\eqref{maxwell},
\begin{equation}\label{eq:newMaxwell}
    \epsilon^{\mu \nu \rho \sigma} \partial_{\nu} G_{\Lambda}{}_{\rho \sigma} = - g h_{u v} {\tilde k}^u_{\Lambda} \nabla^{\mu}
    q^v\ ,
\end{equation}
with $G_{\mu\nu}$ defined as in ~\eqref{eq:defg}. Since our Bianchi identities are unmodified, and the same as in BLS, we again solve them by taking the $H^{\Lambda}$'s to be harmonic
functions. The difference is in the Maxwell equations.

We plug in the identities from~\eqref{eq:hyper-conditions},~\eqref{eq:BLSmetric}
and~\eqref{eq:GasH}. The components of \eqref{eq:newMaxwell} with $\mu\neq
t$ are then automatically satisfied. The only non-trivial equation
follows from $\mu=t$, and reads
\begin{equation}
\label{eq:maxwellforH}
 \bigg.  \qquad \square H_{\Lambda} = - 2 g^2 {\rm e}^{-\mathcal K} h_{u v}
\tilde{k}^u_{\Lambda} \tilde{k}^v_{\Sigma} X^{\Sigma}\ .
\end{equation}
Here, $\square$ is again the three-dimensional Laplacian in flat space.
The left hand side is real, and so is the right hand side, as a consequence of the last equation in \eqref{eq:hyper-conditions} and the fact that we have chosen the phase in $X/|X|$ (see \eqref{eq:ansatz} to vanish.
In other words, $X$ is real, and therefore also ${\tilde k}^u_\Lambda X^\Lambda$ is real.

We furthermore have a consistency condition for the field
strengths. The gauge potentials appear in~\eqref{eq:hyper-conditions},
but also in~\eqref{eq:GasH}, and these should lead to the same
solution. These consistency conditions were not present in the
ungauged case, since in that case there are no restrictions on $F^{\Lambda}$ from the hyperino
variation. The constraints can be easily derived from the integrability conditions of ~\eqref{eq:hyper-conditions}, and are given by
\begin{align}\label{constraints}
\begin{split}
  \tilde{k}^u_{\Lambda} H^{\Lambda} &= 0\ ,\\
    \tilde{k}^u_{\Lambda} F^{\Lambda}_{r \varphi} &= -
\tilde{k}^u_{\Lambda}
    \partial_{r} \left(\omega_\varphi {\rm e}^{\mathcal K} X^{\Lambda} \right),\\
\tilde{k}^u_{\Lambda} F^{\Lambda}_{\theta \varphi} &= -
\tilde{k}^u_{\Lambda}
    \partial_{\theta} \left(\omega_\varphi {\rm e}^{\mathcal K} X^{\Lambda} \right),\\
    \tilde{k}^u_{\Lambda} F^{\Lambda}_{r t} &= - \tilde{k}^u_{\Lambda}
    \partial_{r} \left( {\rm e}^{\mathcal K} X^{\Lambda} \right),\\
    \tilde{k}^u_{\Lambda} F^{\Lambda}_{\theta t} &= - \tilde{k}^u_{\Lambda}
    \partial_{\theta} \left( {\rm e}^{\mathcal K} X^{\Lambda} \right).
\end{split}
\end{align}

The first condition can always be satisfied as it merely implies
that some of the harmonic functions $H^\Lambda$ depend on the others (remember that the hypermultiplet scalars are constant, and therefore also the Killing vectors ${\tilde k}^u_\Lambda$). In more physical terms, this constraint
decreases the number of magnetic charges by the rank of $\tilde k^u_\Lambda$. The other
constraints have to be checked against the explicit form of the
field strengths \eqref{eq:FasHtilde} and \eqref{eq:GasH}. This
cannot be done generically and has to be checked once an explicit
model is taken.

In chapter \ref{chapter::solution_generator}, we explained how the vanishing of $\tilde{k}^u_{\Lambda}
L^{\Lambda} $ and $\tilde k^u_\Lambda A_{\mu}$ led to a BPS solution using spontaneous symmetry
breaking. We can see that also from the equations of this
section. When $\tilde{k}^u_{\Lambda}
L^{\Lambda} = 0$, the right hand side of~\eqref{eq:maxwellforH} is
zero. This equation is then solved by harmonic functions
$H_{\Lambda}$. Furthermore, as $\tilde k^u_\Lambda$ is
constant, we can move it inside the derivatives in~\eqref{constraints},
so the right hand sides are zero. The left hand
sides are zero as well, as $\tilde{k}^u_{\Lambda}
F^{\Lambda}_{\mu\nu} = 0$. Finally, the condition $\tilde{k}^u_{\Lambda}
H^{\Lambda} = 0$ is  satisfied as
$\tilde{k}^u_{\Lambda} L^{\Lambda}$ is already real.

\subsection{Solutions with scalar hair}\label{5:hair}
In this section, we search for solutions of the above BPS conditions that
do not fall in the class described in chapter \ref{chapter::solution_generator}.
They describe asymptotically flat black holes and would have non-trivial profiles for the massive vector and
scalar fields, i.e. they would be distinguishable by the scalar
hair degrees of freedom outside the black hole horizon.
Remarkably, we could not find models with pure scalar hair
solutions without the need to introduce some extra features, such
as ghost modes or non-vanishing fermions. Below, we describe two examples of solutions that lead to
at least one negative eigenvalue of the K\"{a}hler metric. We show
that if we require strictly positive definite kinetic terms in the
considered models, one cannot find scalar hair solutions, but only
the ones described in chapter \ref{chapter::solution_generator}. It is of course
hard to justify these ghost solutions physically. However, there
have been cases in literature where this is not necessarily a problem, e.g.\ in Seiberg-Witten theory
\cite{Seiberg:1994rs,Seiberg:1994aj} one has to perform duality transformation
such that the kinetic terms remain positive definite. Whether a
similar story holds in our case remains to be seen. If such
duality transformations exist they will have to map the ghost
black hole solutions of our abelian electrically gauged
supergravity to proper black hole solutions, possibly of
magnetically gauged supergravity. However, we cannot present any
direct evidence for such a possibility.

\subsubsection{Ghost solutions}

Before we present our examples, we start with a general comment.
We can obtain some more information from the Einstein equations.
The trace of the Einstein equations reads
\begin{align}
  R = T^q + T^z + 4 V\ ,
\end{align}
where $R$ is the Ricci scalar, and we have defined
\begin{align}
  T^q &= - 2 h_{uv} \nabla_\mu q^u \nabla^\mu q^v\ , & T^z = -2 g_{i \bar
    \jmath} \partial_\mu z^i \partial^\mu \bar z^{\bar \jmath}\ .
\end{align}
Using the BPS conditions in~\eqref{eq:hyper-conditions}, one
quickly finds $T^q = -2V$. Furthermore, as $\partial_t z^i = 0$,
we find\footnote{Recall that our spacetime signature convention is
$(+,-,-,-)$.} $T^z \geq 0$, and $V \geq 0$ by
equations~\eqref{eq:scalar-potential} and the condition $P^x_\Lambda = 0$. We
therefore find
\begin{align}\label{eq:traced-einstein}
  R = T^z + 2 V \geq 0\ ,
\end{align}
as long as the metric $g_{i \bar \jmath}$ is positive definite. So
the BPS conditions forbid the Ricci scalar $R$ to become negative.
In our examples below, the metric components will show some oscillatory behavior, as a consequence
of the non-linear differential equation \eqref{eq:maxwellforH}. Therefore, their derivatives, and hence the
Ricci scalar, will oscillate between positive and negative values. This would contradict the positivity bound
\eqref{eq:traced-einstein}, unless the K\"ahler metric $g_{i \bar \jmath}$ contains regions in which it is not positive definite. We now discuss this in detail with two examples.

\subsubsection{Quadratic prepotential} We start with two simple
models, which have
only one vector multiplet. They are described by the two prepotentials
\begin{align}
F = -\frac{i}{2} \left( X^0 X^0 \pm X^1 X^1 \right).
\end{align}
These lead to the special K\"ahler metrics
\begin{align}
  g_{z \bar z} = \frac {\mp 1}  {(1 \pm z \bar z)^2}\ ,
\end{align}
where $z = X^1/X^0$. With the upper sign, we therefore get a negative definite
K\"{a}hler metric and the vector multiplet scalar is a
ghost field. With the lower sign, we obtain a positive definite metric. We
couple this to the universal hypermultiplet, and gauge isometry 5 from appendix
\ref{app:UHM}, using $A^1_\mu$ as the gauge field. The condition
$P^x_{\Lambda} = 0$ fixes $\chi = \varphi = \sigma
=0$ and the only non-vanishing component of the Killing vectors is then
$\tilde{k}^R_1 = 2 R a_1$, where $a_1$ is a constant.

From the relations~\eqref{eq:sections} follows that $X^0 =
\frac{1}{2} (H_0 \!-\! i H^0)$ and $X^1 \!= \frac{1}{2} (\pm H_1\! -\!
i H^1)$. The K\"{a}hler potential~\eqref{eq:def-kahler} is then
\begin{align}
{\rm e}^{-\mathcal K} = 2 \left( X^0 \bar{X}^0 \pm X^1 \bar{X}^1 \right).
\end{align}
As we do not use $A^0_\mu$ for the gauging, $X^0$ remains
harmonic, such that even if the solution for $X^1$ is considerably
different, we still have hope of producing a black hole by having
$X^1$ as a small perturbation of the leading term $X^0$ in the
metric function ${\rm e}^{-\mathcal K}$. For simplicity, we
restrict ourself to the spherically symmetric single-centered
case, so now our constraints \eqref{constraints} lead to
$H^1 = 0$ and $\tilde{k}^u_{\Lambda} F^{\Lambda}_{r t}
= - \tilde{k}^u_{\Lambda}\partial_{r}\! \left( {\rm e}^{\mathcal
K} X^{\Lambda} \right)$. The latter eventually implies that
$H^0$ is constant. Since we can absorb this constant
by rescaling $H_0$, we will set $H^0 = 0$. Thus we are
left with $2 X^0 = H_0 =\sqrt{2} + \frac{q_0}{r}$ ($q_0 > 0$),
where we set the constant of the harmonic function to $\sqrt{2}$
to obtain canonically normalized Minkowski space as $r \rightarrow
\infty$.

The metric is given by~\eqref{eq:BLSmetric}, where
\begin{align}
{\rm e}^{-\mathcal K} = \frac{1}{2} \left( \left(\sqrt{2} +
\frac{q_0}{r}\right)^2 \pm H_1^2 \right).
\end{align}
The only undetermined function is $H_1$, which is subject
to the only equation left to be satisfied, \eqref{eq:maxwellforH},
which in this case is given by
\begin{align}\label{eq:diff-h1}
\square H_1 = \mp {\rm e}^{-\mathcal K} H_1 = \mp \frac{1}{2} \left(
\left(\sqrt{2} + \frac{q_0}{r}\right)^2 \pm H_1^2 \right) H_1,
\end{align}
after setting $g |{\tilde k}| = 1$. Besides the trivial solution $H_1 = 0$
(belonging to the class solutions from chapter \ref{chapter::solution_generator}), we could not find an analytic solution to
these equations. We can analyze the differential equation as $r
\to 0$ and $r \to \infty$. As $r \to \infty$, we require ${\rm
e}^{-\mathcal
  K} \to 1$, to obtain flat space at infinity. Likewise, we require,
as $r \to 0$, that ${\rm e}^{-\mathcal K} \to q^2 r^{-2}$, to obtain $AdS_2
\times S^2$ at the horizon. The constant $q$ (which is not necessarily equal to $q_0$) determines the (equal)
radii of $AdS_2$ and $S^2$. If we solve~\eqref{eq:diff-h1} for large
values of $r$, we have to solve $\square H_1 = \mp H_1$; for small
values of $r$ we have to solve $\square H_1 = \mp \frac 12 q^2 r^{-2} H_1$.
\begin{itemize}
\item With the upper sign (the ghost model), we find the general solution
  \begin{align}
    H_1 &= A \frac{\cos(r)}{r} + B \frac{\sin(r)}{r}\ ,  & r &\to
    \infty\ ,\\
    H_1 &= C r^{-\frac 12 - \frac 12 \sqrt{1-4 q^2}} + D r^{-\frac 12
      + \frac 12 \sqrt{1-4q^2}}\ ,& r &\to 0\ .
  \end{align}
As long as $4 q^2 < 1$, all the asymptotics are fine.

\item With the lower sign (the non-ghost model), we find the general solution
  \begin{align}\label{eq:h1-asympt}
    H_1 &= A \frac{{\rm e}^{-r}}{r} + B \frac{{\rm e}^r}{r}\ , & r &\to
    \infty\ ,\\
\label{eq:h1-asympt2}
    H_1 &= C r^{-\frac 12 - \frac 12 \sqrt{1+4 q^2}} + D r^{-\frac 12
      + \frac 12 \sqrt{1+4q^2}}\ ,& r &\to 0 \ .
  \end{align}
When $B$ is nonzero, this violates the boundary condition that  ${\rm e}^{-\mathcal
  K} \to 1$ as $r \to \infty$, so we have to set $B=0$. Likewise, we
have to set $C=0$. We will now prove that imposing such boundary conditions implies $H_1
= 0$. To do this, we use the identity
\begin{align}
  \int_0^\infty (r H_1) \partial_r^2 (r H_1) \,{\rm d}r = -\int_0^\infty \partial_r
  (r H_1) \partial_r  (r H_1) \,{\rm d}r + (rH_1) \partial_r (r H_1)
  \Big|^{r=\infty}_{r=0}\ .
\end{align}
Using~\eqref{eq:h1-asympt} and \eqref{eq:h1-asympt2} one finds that, for $B=C=0$, the boundary term
vanishes. On the left-hand side, we use~\eqref{eq:diff-h1}, and we
obtain (using $\square H_1 = r^{-1} \partial_r^2 (r H_1)$)
\begin{align}
  \int_0^\infty H_1 {\rm e}^{-\mathcal K} H_1 \,{\rm d}r = -\int_0^\infty \partial_r
  (r H_1) \partial_r  (r H_1) \,{\rm d}r\ .
\end{align}
The left-hand side is non-negative, whereas the right-hand side is
non-positive, so this proves $H_1 = 0$. This argument can easily be repeated for solutions
with only axial symmetry.

\end{itemize}

We could plot the solution with the upper sign numerically with
generic starting conditions, and the result is that the scalar field $z = H_1/H_0$ oscillates around zero in the region of interest. The metric function also gets oscillatory perturbations, while having its endpoints fixed to the
desired values. This is due to the fact that the function $H_1$ approaches zero as $r \rightarrow \infty$ in an
oscillatory fashion. To investigate the behavior near the horizon
at $r=0$, we also checked that $r H_1$ approaches zero, and hence
$H_1$ diverges slower than $1/r$. Both are in agreement with the asymptotic analysis above.

The numerics further show that the metric function for negative
values of $r$ yields the expected singularity at $r = -
\frac{q_0}{\sqrt{2}}$. We conclude that this is indeed a black
hole spacetime, having one electric charge $q_0$, and the
fluctuations around the usual form of the metric are due to the
effect of the abelian gauging of the hypermultiplet.

Let us now try to give a bit more physical interpretation of this
new black hole spacetime. After more careful inspection of the
solution, we see that at the horizon and asymptotically at infinity
we again have supersymmetry enhancement, since the
vector multiplet scalars are fixed to a constant value. It is
interesting that the electric charge, associated to the broken
gauge symmetry vanishes at the horizon, i.e. the black hole itself
is not charged with $q_1$ exactly as in the normal case without
ghosts. Yet there is a non-zero charge density for this charge
everywhere in the spacetime outside the black hole, which is the
qualitatively new feature of the ghost solutions. Clearly the fact
that there is non-vanishing charge density everywhere in
spacetime does not change the asymptotic behavior, but it seems
that it is physically responsible for the ripples that can be
observed in the metric function (of
course this is all related to the fact that we have propagating
ghost fields). We should note that these are not the first rippled
black hole solutions, similar behavior is found in the higher
derivative ungauged solutions, e.g. in \cite{Hubeny:2004ji}, where
one again finds ghost modes in the resulting theory. The detailed
analysis in section 4 of \cite{Hubeny:2004ji} holds in our case, i.e.\ the
main physical feature of the ripples is that gravitational force
changes from attractive to repulsive in some spacetime points.

\subsubsection{Cubic prepotential}
The example above shows already the general qualitatively new
features of this class of black holes with ghost fields, but is
still not interesting from a string theory point of view, since
Calabi-Yau compactifications lead to cubic prepotentials of the
form
\begin{align}
F = -\frac{ \mathcal K_{i j k} X^i X^j X^k}{6 X^0}\ .
\end{align}
The simplest case one can consider is the STU model of section
\ref{sect:stu}. We coupled it to the universal hypermultiplet with
a single gauged isometry and found it impossible to produce any
new solutions. However, other choices of $\mathcal K_{i j k}$ allow
for interesting numerical solutions of \eqref{eq:maxwellforH}. For
this purpose we consider a relatively simple model with three
vector multiplets:
\begin{align}
F = \frac{(X^1)^3 - (X^1)^2 X^2 - X^1 (X^3)^2}{2 X^0}\ .
\end{align}
We again use the universal hypermultiplet and gauge the same
isometry as before, but we now use only $A^3_\mu$ for our gauging.
Again, the condition $P^x_\Lambda = 0$ fixes $\chi=\varphi=\sigma=0$, and the
only non-vanishing component of the Killing vector is $\tilde
k^R_3 =  2 R a_3$. In parts of moduli space this model exhibits
proper Calabi-Yau behavior, i.e. the K\"{a}hler metric is positive
definite, but there are regions where $g_{i\bar \jmath}$ has
negative eigenvalues (or ${\rm e}^{-\mathcal
  K}$ becomes negative). There is no general expression for this
so-called positivity domain; one has to analyze an explicit model to
find the conditions.

For simplicity, we set $H^i = H_0 =
0$, so the non-vanishing functions are $H_i$ and $H^0$. Inverting \eqref{eq:sections} we obtain for the K\"{a}hler
potential
\begin{align}\label{eq:kahler-cy}
{\rm e}^{-\mathcal K} =  \sqrt {2 H_2} \sqrt{H^0} \left(H_1 + H_2 + \frac
  {H_3^2}{4H_2}\right)\ .
\end{align}
We see that, as is commonly encountered in these models, one has
to choose the signs of the functions $H_i$ and $H^0$ such
that this gives a real and positive quantity. With these we satisfy all conditions in \eqref{constraints} and
are left to solve \eqref{eq:maxwellforH} that explicitly reads:
\begin{align}\label{eq:diff-eq-cy}
\square H_3 = -a_3^2 H^0 \left(H_1 + H_2 + \frac {H_3^2}{4
  H_2}\right)  H_3\ ,
\end{align}
where $H^0, H_1$ and $H_2$ are harmonic functions, and we have
set $g |{\tilde k}| = 1$ for convenience.

We impose the same boundary conditions, so as $r \to \infty$, we require ${\rm
e}^{-\mathcal
  K} \to 1$, to obtain flat space at infinity. Likewise, we require,
as $r \to 0$, that ${\rm e}^{-\mathcal K} \to q^2 r^{-2}$, to obtain AdS$_2
\times$ S$^2$ at the horizon. Using~\eqref{eq:kahler-cy}, we then find
that we have to solve
\begin{align}
  \square H_3 &= -a_3^2 q^2 r^{-2} H_3\ ,&\text{ as }& r \to 0\ ,\\
  \square H_3 &= -a_3^2 c^2 H_3\ ,&\text{ as }& r \to \infty\ ,
\end{align}
where $c^2$ is also a constant, specified by the asymptotics of
$H^0, H_1$ and $H_2$. We therefore again find
\begin{align}
  H_3 &= A \frac{\cos (a_3 c r)}{r} + B \frac{\sin (a_3 c
    r)}{r}\ ,&\text{ as }& r \to \infty\ .
\end{align}
These functions are oscillating; therefore the K\"ahler
potential~\eqref{eq:kahler-cy} will also oscillate. This
causes the Ricci scalar to become negative, which is in violation of
the bound~\eqref{eq:traced-einstein}. Therefore, there is always a negative eigenvalue
of the metric, corresponding to a ghost mode.

We could only find a numerical solution to this equation, and the results
are qualitatively the same as the ones with quadratic prepotential, so we will not elaborate the details for this
model.

It is therefore possible to find black hole solutions in these
Calabi-Yau models, but they do contain regions in which some of the scalars become ghost-like.

\subsubsection{Fermionic hair}\label{ferm-hair}

There is a different way of generating scalar hair with properly
normalized positive-definite kinetic terms. As such, we can
thereby avoid the ghost-like behavior of the previously discussed
examples. The idea is simple and works for any solution that
breaks some supersymmetry. By acting with the broken susy
generators on a bosonic solution, we will turn on the fermionic
fields to yield the fermionic zero modes. These fermionic zero
modes solve the linearized equations of motion and produce
fermionic hair. In turn, the fermionic hair sources the equations
of motion for the bosonic field, and  in particular, the scalar
field equations will have a source term which is bilinear in the
fermions. The solution of this equation produces scalar hair and
can be found explicitly by iterating again with the broken
supersymmetries. This iteration procedure stops after a finite number of
steps and produces a new solution to the full non-linear equations
of motion. By starting with a BPS black hole solution of the type
discussed in chapter \ref{chapter::solution_generator}, one therefore produces
new solutions with both fermionic and scalar hair. For a
discussion on this for black holes in ungauged supergravity, see
\cite{Aichelburg:1986wv}.

The explicit realization of this idea is fairly complicated since
it requires  to explicitly find the Killing spinors preserving
supersymmetry. This can sometimes be done also just by considering
the possible bosonic and fermionic deformations of the theory, as
done in e.g. \cite{Banerjee:2009uk,Jatkar:2009yd} for black holes
in ungauged supergravity. The extension of this fermionic-hair analysis to
gauged supergravities would certainly be an interesting extension
of our work.

\chapter{Asymptotically AdS black holes}\label{chapter::blackholes_AdS}

\section{Introduction}\label{6:intro}
Compared to their Minkowski relatives, black holes in AdS$_4$ have been somewhat neglected in the literature. This is understandable, since such solutions seem not to be relevant for describing observable objects in our universe. However, due to the AdS/CFT correspondence, solutions in AdS presently enjoy a great deal of attention. The thermal black objects in AdS turn out to be relevant for describing the effective physics in quark-gluon plasmas and various condensed matter phenomena at strong coupling, such as high temperature superconductivity and quantum Hall effect. The problem of constructing and classifying solutions in AdS is therefore more pressing than ever.

Before going into a more technical discussion, it is important to explain that the usage of the term ``black hole'' in AdS has a much broader meaning than in Minkowski. Unlike the case for asymptotically flat static black holes, the topology of the horizon of AdS$_4$ black holes is not unique. The
horizon can be a Riemann surface of any genus as explained in \cite{Caldarelli:1998hg}. The black holes can thus be divided into three classes - spherical, or ordinary, black holes, toroidal black holes or black branes in the infinite volume limit, and higher genus black holes. All these objects share the same thermodynamics properties, like the entropy-area law.

In this zoo of black objects it is fairly easy to lose track of the discussion. In order to ease the presentation, we relegate the analysis of the three major types of black holes to separate sections. The focus of this work falls on spherical black holes, which are very thoroughly analyzed, but the interested reader can find a comprehensive (although much shorter) discussion of the other two cases.

\subsubsection{Plan of this chapter}

Section \ref{6:section2} is devoted to a literature review on spherical black holes in AdS with an emphasis on supersymmetric results. After a general introduction to the basic types of black hole solutions in minimal gauges supergravity, section \ref{6:sect2.1} and \ref{6:sect2.2} aim at generalizing these results in order to allow for additional matter multiplets. Section \ref{6:sect2.1} exemplifies the solution generating technique of chapter \ref{chapter::solution_generator}, while section \ref{6:sect2.2} is devoted to the static electric Reissner-Nordstr\"{o}m-AdS solutions with nontrivial scalars \cite{Sabra:1999ux}.

Section \ref{6:Klemm} is entirely focused on a new type of static spherically symmetric BPS black holes with magnetic charges. These solutions require non-constant scalar profiles in order to develop event horizons. As we will see in the following chapters, these black holes define a separate vacuum solution that is topologically disconnected from AdS$_4$. Still, their spacetime asymptotics provide a good evidence that a version of the gauge/gravity correspondence should be applicable.

Sections \ref{6:branes} and \ref{6:higher} discuss the cases of black branes and higher genus black holes, respectively. Many analogies are drawn with the spherically symmetric solutions, while the differences are clearly emphasized and pursued.

\section{Spherical solutions}\label{6:section2}
The main types of spherically symmetric black holes in AdS$_4$ closely resemble their flat analogs, with one additional parameter corresponding to the cosmological constant, or radius of AdS. As a start, we consider the solutions embeddable in minimal gauged supergravity, i.e.\ Einstein-Maxwell theory with a bare cosmological constant. As a consequence of the gauging, there is an additional subtlety in this theory. Recall that in minimal gauged supergravity the $U(1)_R$ symmetry of the gravitinos is promoted to a local one via the graviphoton. Although not explicitly given as a formula, this naturally means that the kinetic term of the gravitinos in the lagrangian acquires a gauge covariant derivative, with the term $g A_{\mu}^g$ in it, i.e.\ the gravitinos become charged under the graviphoton with an electric charge $g$, which is also giving the bare cosmological constant, $\Lambda = - 3 g^2$. Due to this electric charge (the gauging is electric, therefore the gravitinos remain with a vanishing magnetic charge), a Dirac quantization condition is imposed on every solution of the theory: each object needs to have a magnetic charge $P$ that obeys $2 g P = n, n \in \mathbb{Z}$. Since the magnetic charge of the gravitinos is zero, the electric charge of the solutions remains a continuous parameter, at least a priori. This subtlety of the quantum theory might on first sight be of little relevance for the topic of classical solutions, but we will soon find that supersymmetry is not oblivious to these issues. Without further delay, we can now discuss the different types of static and stationary solutions in minimal gauged supergravity.

\subsubsection{Static solutions}
First we focus on static spherically symmetric spacetimes with metrics of the form
\begin{equation}
{\rm d} s^2 = U^2(r)\, {\rm d}t^2 - U^{-2}(r)\, {\rm d}r^2 -
h^2(r)\, ({\rm d} \theta^2 + \sin^2 \theta {\rm d} \varphi^2)\ ,
\end{equation}
for some functions $U(r)$ and $h(r)$ to be determined from the BPS
conditions and/or the equations of motion.

For Minkowski spacetime, we had $U=1$ and $h=r$, and for
four-dimensional anti-de Sitter spacetime, one has
\begin{equation}
AdS_4: \qquad U^2(r)=1+g^2 r^2\ ,\qquad h(r)= r\ ,
\end{equation}
where $g$ is related to the
cosmological constant of AdS$_4$ through the scalar curvature relation
$R=-12 g^2$. So, in the standard conventions the cosmological
constant is $\Lambda = -3 g^2$. The simplest generalization of AdS is to Schwarzschild-AdS, with mass $M$ and vanishing charges:
\begin{equation}\label{Sch-AdS}
Schwarzschild-AdS_4: \qquad U^2(r)=1-\frac{2M}{r}+g^2 r^2\ ,\qquad h(r)= r\ .
\end{equation}
Just as in the asymptotically flat case, there is no extremal Schwarzschild-AdS black hole and only positive masses result in an event horizon. From the point of view of the AdS/CFT correspondence, a Schwarzschild-AdS with positive mass corresponds to a thermal field theory on the boundary.

We then move to the more interesting (from the supersymmetry point of view) case when we allow for non-vanishing graviphoton field strength. For the Reisnner-Nordstr\"om black hole solution in AdS$_4$ (RN-AdS), with mass $M$ and electric and magnetic charges $Q$ and $P$, we have
\begin{equation}\label{RN-AdS}
RN-AdS_4: \qquad U^2(r)=1-\frac{2M}{r}+\frac{Q^2+P^2}{r^2}+g^2
r^2\ ,\qquad h(r)= r\ .
\end{equation}
Imposing BPS conditions leads to exactly two different possibilities in pure supergravity without vector
multiplets, as analyzed long ago in \cite{Romans:1991nq}. One solution is usually referred to as "extreme RN-AdS
electric solution", it is half-BPS and it requires $M=Q, P=0$, hence
\begin{equation}
extreme \quad electric \quad RN-AdS_4: \qquad
U^2(r)=(1-\frac{Q}{r})^2+g^2 r^2\, \qquad h(r) = r\ .
\end{equation}
The function $U(r)$ has no zeroes and therefore
no event horizon exist. The point $r=0$ is then a naked singularity. The other
solution is referred to as an "exotic AdS solution" and is only
quarter-BPS, imposing $M = 0, P = 1/(2 g)$,
\begin{equation} exotic \quad AdS_4: \qquad
U^2(r)=(g r+\frac{1}{2 g r})^2+\frac{Q^2}{r^2}\ , \qquad h(r) = r\
.\end{equation} This case has no flat space limit for $g
\rightarrow 0$ and is therefore very different in behavior from
the first solution. Still, the solution has a naked singularity.

The aim of the following sections is to find generalizations of these BPS solutions that include vector and hypermultiplets. The extension of the extreme RN-AdS solutions for many vector multiplets and non-trivial scalars has been investigated in \cite{Sabra:1999ux} with the outcome of nakedly singular spacetimes once again, which will be discussed in detail in section \ref{6:sect2.2}. Some generalizations of the exotic solution also exist in the literature, e.g.\ in \cite{Chamseddine:2000bk}, but these set the scalars to constants and are thus not general enough to resolve the naked singularity. Our strategy will be to replace the cosmological constant with a nontrivial potential for the vector multiplet scalars that contains Fayet-Iliopoulos terms, see more in section \ref{6:Klemm}.

\subsubsection{Kerr-Newman-AdS}
The Kerr-Newman-AdS (KN-AdS) spacetime is the most general stationary solution in AdS$_4$. Just as the KN black hole in Minkowski, KN-AdS has a mass parameter $m$, electric and magnetic charges $q$ and $p$ (with $z^2 \equiv q^2 + p^2$), and angular momentum parameter $a$. The metric is axisymmetric and is written in Boyer-Lindquist-type coordinates,
\begin{align}
    {\rm d} s^2 = \frac{\Delta_r}{\Xi^2 \rho^2} \left({\rm d} t - a \sin^2 \theta {\rm d} \varphi \right)^2 - \frac{\rho^2}{\Delta_r} {\rm d}r^2 - \frac{\rho^2}{\Delta_{\theta}} {\rm d}\theta^2 - \frac{\Delta_{\theta} \sin^2 \theta}{\Xi^2 \rho^2} \left( a {\rm d}t - (r^2+a^2) {\rm d} \varphi \right)^2\ ,
\end{align}
with
$$\rho^2 \equiv r^2+a^2 \sin^2 \theta\ , \qquad \Xi = 1 - a g\ ,$$
and $$\Delta_r \equiv (r^2+a^2) (1+ g^2 r^2)-2 m r +z^2 \ , \qquad \Delta_{\theta}\equiv 1 - a^2 g^2 \cos^2 \theta \ .$$
From the metric we can see that the spacetime is well-defined only when $a g < 1$, i.e.\ the angular momentum is bounded from above.. This comes from the requirement that the angular velocity at the conformal AdS boundary ($r \rightarrow \infty$) does not exceed the speed of light, see \cite{Hawking:1998kw}. The analysis of KN-AdS black hole thermodynamics and extremality conditions is slightly more involved and can be found in \cite{Caldarelli:1998hg,Hawking:1998kw}.The conditions imposed from supercharge conservation on the parameters $m,p,q,a$ will be discussed in details in the next part of this thesis. Here we just note that the BPS KN-AdS solution is the only supersymmetric configuration in minimal gauged supergravity with a genuine horizon, corresponding to a black hole with a finite entropy.

\subsection{Solutions via spontaneous symmetry breaking}\label{6:sect2.1}
The simplest solutions in gauged supergravity are the of course the ones where the scalars are just constants. Because of the non-trivial scalar potential of the gauged lagrangian \eqref{lagr}, however, the scalars cannot be just frozen to arbitrary constants. In principle one can try to find all minima of the scalar potential in a given solution, but this is often an impossible task if the number of scalars is too big. A way out is provided by our analysis in chapter \ref{chapter::solution_generator} which essentially provides a way of finding the supersymmetric minima of the scalar potential.

Here we give a simple but yet qualitatively very general example
of how to apply the procedure outlined above to find
asymptotically anti-de Sitter black hole solutions with gauged hypers,
starting from already known black hole solutions without hypers.
In this case we start from a solution of pure
supergravity and add abelian gauged vector multiplets and
hypermultiplets. Alternatively, one can think of it as breaking the
gauge symmetry such that all hyper- and vector multiplets become
massive, and one is left with a gravity multiplet with cosmological
constant. Here we already know the full classification of black
hole solutions, as described above.

An already worked out example in section~\ref{SE_7} is the
case of the gauged supergravity, arising from a consistent reduction
to four dimensions of M-theory on a Sasaki-Einstein$_7$ manifold
\cite{Gauntlett:2009zw}. The resulting low-energy effective action has a
single vector multiplet and a single hypermultiplet (the universal
hypermultiplet). The special
geometry prepotential is given by
$$F = \sqrt{X^0 (X^1)^3},$$
with $X^{\Lambda} = \{ 1, \tau^2\}$, where $\tau$ is the vector
multiplet scalar, and the isometries on the UHM are given by
  \begin{align}
\begin{split}
    \tilde{k}_0 &= 24 \partial_{\sigma} +4\bigl(\chi \partial_\varphi -
    \varphi \partial_\chi  + \frac 12 (\varphi^2 -
    \chi^2) \partial_\sigma \bigr)\ ,\\
 \tilde{k}_1 &= 24 \partial_{\sigma}\ ,
\end{split}
  \end{align}
which is combination of isometries 1 and 4 from
appendix~\ref{app:UHM}. The corresponding moment maps, see
appendix~\ref{app:UHM}, are given by
  \begin{align}
   \nonumber P_0^1 &= \frac{4 \chi}{\sqrt R}\ ,& P_0^2 &=  \frac
   {4\varphi} {\sqrt R}\ , & P_0^3 &= -\frac{12}{R} + 4 - \frac{\chi^2 + \varphi^2}{R}\ ,\\
   P_1^1& = 0\ , & P_1^2 &= 0\ ,&  P_1^3& = -\frac{12}{R}\ .
  \end{align}
Maximally supersymmetric AdS$_4$ vacua were found in section~\ref{SE_7}.
The condition~\eqref{hypers with FI terms}
fixes the values of the vector multiplet scalar $\tau^{\rm vev} \equiv
(\tau_1 + i \tau_2)^{\rm vev} = i$ and two of the four hypers
$\chi^{\rm vev}
= \varphi^{\rm vev} = 0$. The third ungauged
hyper, which is the dilaton, is fixed to the constant non-zero value
$R^{\rm vev}
= 4$. The remaining hypermultiplet scalar is an arbitrary constant
$\sigma^{\rm vev}
= \langle \sigma \rangle$. All the gauge fields have vanishing
expectation values at this fully supersymmetric AdS$_4$ vacuum. If
we now expand the scalar field potential \eqref{eq:scalar-potential} up
to second order in fields we obtain the following mass terms
\begin{align}
V^{\rm quadratic} = -12 + 138(\tau_1^2 + \tau_2^2) + \frac{3}{4} R^2 + 6 R
\tau_2 + 10 (\chi^2 + \varphi^2)\,.
\end{align}
We can see that three of the hyperscalars and the (complex) vector
multiplet scalar acquire mass. There is also a mass term $m^2 =
36$ for the gauge field $A_0+A_1$, this field thus eats up the
remaining massless hyperscalar $\sigma$. So we observe the formation of
a massive $N=2$ vector multiplet consisting of one massive vector
and five massive scalars, and we can consistently set all these
fields to zero. The resulting Lagrangian is that of pure $N=2$ supergravity
with a cosmological constant $\Lambda = -12$. Using the static class of
black hole solutions in minimal gauged supergravity, it is straightforward to provide a solution of the gauged
supergravity theory. All the solutions described in the beginning of this section will
also be solutions in our considered model as they obey the
Einstein-Maxwell equations  of pure supergravity.

\subsection{1/2 BPS solutions}\label{6:sect2.2}
Here we briefly summarize the solutions of Sabra, \cite{Sabra:1999ux}, which generalize the extreme-electric RN-AdS configurations of minimal gauged supergravity. In short, the solutions of are in gauged supergravity with an arbitrary number of vector multiplets and $U(1)$ FI parameters $\xi_{\Lambda}$. The solutions are purely electric with arbitrary charges $q_{\Lambda}$. The metric and symplectic sections are
 \begin{align}\label{sabra_solution}
 \begin{split}
 {\rm d} s^2 = e^{\mathcal{K}} \left(1 + g^2 r^2 e^{-2 \mathcal{K}}\right) {\rm d} t^2 - \frac{e^{-\mathcal{K}} {\rm d} r^2}{\left(1 + g^2 r^2 e^{-2 \mathcal{K}}\right)} -e^{-\mathcal{K}} r^2 {\rm d} \Omega_2^2\ ,\\
 {\rm Im} X^{\Lambda} = 0, \qquad  2\ {\rm Im} F_{\Lambda} = H_{\Lambda} = \xi_{\Lambda} + \frac{q_{\Lambda}}{r}\ .
 \end{split}
 \end{align}
 These solutions represent nakedly singular spacetimes that preserve 4 of the original 8 supercharges. They seem to be the most general static supersymmetric solutions with spherical symmetry that strictly asymptote\footnote{The solutions presented in the following section have non-vanishing magnetic charge coming from the graviphoton and will not be counted as asymptotically AdS. This is explained carefully in the coming chapters.} to AdS$_4$. It is natural to expect that supersymmetry preserving generalizations of the KN-AdS spacetimes with non-constant scalars should exist, but these have not been constructed yet.

\section{Static magnetic spherical BPS black holes}\label{6:Klemm}
This section deals with a BPS extension of the exotic AdS$_4$ nakedly singular solution discussed briefly in the previous section. Recall that
 \begin{equation} exotic \quad AdS_4: \qquad
U^2(r)=(g r+\frac{1}{2 g r})^2+\frac{Q^2}{r^2}\ , \qquad h(r) = r\ .\end{equation}
We will see that there exist generalizations of this within $N=2$ gauged supergravity such that this naked singularity is resolved due to non-trivial scalar behavior.

Anticipating our results, we now briefly explain how the exotic solution is modified to make a proper black hole
in AdS$_4$. We set the electric charges to zero but allow for non-trivial scalars, which will in the end result
in changing the metric function $U$ to be\footnote{Here, the discussion is only schematic in order to underline
the main point, the actual solution is more involved as we explain in section \ref{6:general_solution}. There we also comment further on the other function in the metric, $h(r)$.}:
\begin{equation}U^2(r)=(g r+\frac{c}{2 g r})^2\ ,\end{equation}
with a constant $c \neq 1$ that depends on the explicit running of
the scalars. The important outcome from this is that in certain
cases we will have $c < 0$, and then a horizon will appear at $r_h
= \sqrt{\frac{-c}{2 g^2}}$ to shield the singularity. In this way,
one can find a static quarter-BPS asymptotically AdS$_4$ black
hole with nontrivial scalar fields and magnetic charges.

\subsection{Gauged supergravity with Fayet-Iliopoulos
parameters}\label{6:sect:2}

In this section we focus on abelian gauged $N=2$ supergravity in four dimensions in the absence of hypermultiplets. We consider $n_V$ vector multiplets and briefly recall some already discussed facts about $U(1)$ gaugings with FI parameters. This will be needed to make a distinction later when discussing lagrangians with magnetic FI parameters. Recall that as the gauge group is abelian, the vector multiplet scalars are neutral, and the only charged fields in the theory are the two gravitinos. The gauge fields that couple to the gravitinos appear in a linear combination of the graviphoton and the $n_V$ vectors from the vectormultiplets, $\xi_{\Lambda}
A^{\Lambda}_{\mu}$, with $\Lambda = 0, 1,..., n_V$. The bosonic part of the
Lagrangian for such a system is
\begin{align}\label{6:lagr}
\mathcal L=\frac{1}{2}R(g)+g_{i\bar \jmath}\partial^\mu z^i
\partial_\mu
{\bar z}^{\bar \jmath} +
I_{\Lambda\Sigma}F_{\mu\nu}^{\Lambda}F^{\Sigma\,\mu\nu}
+\frac{1}{2}R_{\Lambda\Sigma}\epsilon^{\mu\nu\rho\sigma}
F_{\mu\nu}^{\Lambda}F^{\Sigma}_{\rho\sigma} - g^2 V (z, \bar{z})\ ,
\end{align}
where
\begin{equation}\label{6:pot}
V= (g^{i\bar \jmath}f_i^\Lambda {\bar f}_{\bar \jmath}^\Sigma
-3{\bar L}^\Lambda L^\Sigma)\xi_\Lambda \xi_{\Sigma}
\end{equation}
is the scalar potential. The
supercovariant derivative of the spinor reads:
\begin{equation}\label{6:supercovar_der}
    \nabla_\mu\varepsilon_A = (\partial_\mu - \frac{1}{4} \omega^{a b}_{\mu} \gamma_{a
    b}) \varepsilon_A + \frac{1}{4} \left({\cal K}_i\partial_\mu
z^i-{\cal K}_{\bar\iota}\partial_\mu{\bar z}^{\bar \iota} \right)
\varepsilon_A + \frac{i}{2} g \xi_{\Lambda} A^{\Lambda}_{\mu}
\sigma^3_A{}^B \varepsilon_B\ ,
\end{equation}
and similarly for the gravitino's
\begin{equation}\label{6:supercovar_der-gravitino}
    \nabla_\mu\psi_{\nu\,A} = \partial_\mu \psi_{\nu\,A}
   +...+ \frac{i}{2} g \xi_{\Lambda} A^{\Lambda}_{\mu}
\sigma^3_A{}^B \psi_{\mu\,B}\ .
\end{equation}
The fact that only $\sigma^3$ appears in the supersymmetry transformation rules and covariant derivatives
reflects the fact that the $SU(2)_R$ symmetry is broken to $U(1)$.

We have to stress that the above theory is gauged only electrically,
since we have used only electric fields $A_{\mu}^{\Lambda}$ for the gauging of the
gravitino. Thus the FI parameters can be thought of as the
electric charges $\pm e_\Lambda$ of the gravitino fields, with
\begin{equation}\label{6:gravitino-charges}
e_\Lambda =  g\xi_\Lambda\ ,
\end{equation}
The fact that the gravitinos have opposite electric charge finds its origin from the eigenvalues of $\sigma^3$.
Generically in such a theory one encounters a Dirac-like quantization condition in the presence of magnetic
charges $p^{\Lambda}$,
\begin{equation}\label{6:dir_quantization}
  e_{\Lambda} p^{\Lambda} = n\ , \qquad n \in \mathbb{Z}\ ,
\end{equation}
as explained already in the previous section. Clearly,
\eqref{6:dir_quantization} is not a symplectic invariant, due to the choice of the gauging.
Later, in section \ref{6:electr-magn general solution}, we generalize this to include also magnetic gaugings.

\subsection{Black hole ansatz and Killing spinors}\label{6:sect:3}

As already stated above, we look for a supersymmetric solution
similar to the ``exotic AdS solution'' of \cite{Romans:1991nq}, but
with nonconstant scalar fields. We start with the general static
metric ansatz
\begin{equation}\label{6:metric-ansatz} {\rm d} s^2 = U^2(r)\, {\rm
d}t^2 - U^{-2}(r)\, {\rm d}r^2 - h^2(r)\, ({\rm d} \theta^2 +
\sin^2 \theta {\rm d} \varphi^2)\ ,
\end{equation}
and corresponding vielbein
\begin{equation}
e_{\mu}^a = {\rm diag}\Big(U(r), U^{-1}(r), h(r), h(r) \sin
\theta\Big)\ .
\end{equation}
The non-vanishing components of the spin connection turn out to
be:
\begin{equation}
\omega_t^{0 1} = U \partial_r U, \quad
\omega_{\theta}^{1 2} = - U \partial_r h, \quad
\omega_{\varphi}^{13} = - U
\partial_r h \sin \theta, \quad \omega_{\varphi}^{23} = - \cos
\theta\ .
\end{equation}
We further assume that the gauge field
strengths are given by
\begin{equation}\label{6:el_field_strengths}
F^{\Lambda}_{t r} = 0, \qquad F^{\Lambda}_{\theta \varphi}
=\frac{p^{\Lambda}}{2} \sin \theta,
\end{equation}
or alternatively
\begin{equation}\label{6:gauge_fields}
A^{\Lambda}_t =  A^{\Lambda}_r =
A^{\Lambda}_{\theta} = 0, \qquad A^{\Lambda}_{\varphi} = -
p^{\Lambda} \cos \theta,
\end{equation}
which are needed in the BPS equations below. If we allow also electric charges, we then
should use an electromagnetic basis $F_{\mu \nu}^{\Lambda},
G_{\mu \nu, \Lambda}$\footnote{Recall that the magnetic field strengths can be
defined from the lagrangian to be $$G_{\Lambda}{}_{\mu \nu} \equiv
R_{\Lambda
    \Sigma} F^{\Sigma}_{\mu \nu} - \frac 12 I_{\Lambda \Sigma}\,
  \epsilon_{\mu \nu \gamma \delta} F^{\Sigma \gamma \delta}\ .$$}, and require
\begin{equation}\label{6:solve_field_strengths}
G_{\Lambda, \theta \varphi} = \frac{q_{\Lambda}}{2} \sin \theta,
\qquad F^{\Lambda}_{\theta \varphi} = \frac{p^{\Lambda}}{2} \sin
\theta\ .
\end{equation}
These automatically solve the Maxwell equations and Bianchi
identities in full analogy to the case of ungauged
supergravity\footnote{Notice that the vector field part of the
lagrangian \eqref{6:lagr} is the same as in the ungauged
theory, so they have the same equations of motion.The only difference appears in the coupling to the gravitinos.} \cite{Behrndt:1997ny}. However, we start with a purely electric
gauging \eqref{6:lagr} and we set the electric charges of the black hole to zero since
otherwise we cannot directly solve for the gauge fields
$A_t^{\Lambda}$ that are needed for the BPS equations. This is a
particular choice we make at this point in view of the BPS
conditions we derive below. In section \ref{6:electr-magn general
solution} we will explain how to explicitly find a solution also
with electric charges in a more general electromagnetic gauging
frame.

\subsubsection{Killing spinor ansatz}

With the gamma matrix conventions spelled out in appendix \ref{appendixA}
we make the following ansatz for the (chiral) Killing spinors:
\begin{equation}\label{6:KS-ansatz}
\varepsilon_A =  e^{i \alpha}\, \epsilon_{A B} \gamma^0
\varepsilon^B, \qquad \varepsilon_A =  \pm e^{i
\alpha}\,{\sigma}^3_{A B}\,\, {\gamma}^1\, \varepsilon^B\ ,
\end{equation}
where $\alpha$ is an arbitrary constant phase, and the choice of
sign in the second condition will lead to two distinguishable
Killing spinor solutions with corresponding BPS equations. This Killing spinor ansatz corresponds
(in our conventions for chiral spinors) to the Killing spinor
projections derived in \cite{Romans:1991nq} for the exotic
solutions. Note that the choice of phase $\alpha$ is irrelevant
due to $U(1)_R$ symmetry, i.e. any value of $\alpha$
leads to the exact same physical solution. It will nevertheless
amount to putting the symplectic sections of the vector multiplet
moduli space in a particular frame, as we explain in more
detail in the next subsection. Furthermore, from the above
equations one can deduce that the Killing spinor can be
parametrized as follows. Using our convention from App. \ref{appendixA},
one finds that, $\forall a \in \mathbb{C}$, for the upper sign
(which we call type $I$) in \eqref{6:KS-ansatz}:
\begin{equation}
\varepsilon_1^I = a(x) \left(
\begin{array}{c}
1\\
i \\
-i \\
-1
 \end{array} \right), \qquad \varepsilon_2^I =
{\bar a(x)} e^{i \alpha}\,\left(
\begin{array}{c}
-i \\
1\\
1\\
-i
 \end{array} \right).
\end{equation}
For the negative sign (type II) one finds,
\begin{equation}
\varepsilon_1^{II} = a(x) \left(
\begin{array}{c}
1\\
i \\
i \\
1
 \end{array} \right), \qquad \varepsilon_2^{II} =
{\bar a(x)} e^{i \alpha}\,\left(
\begin{array}{c}
i \\
-1\\
1\\
-i
 \end{array} \right).
\end{equation}
This type of Killing spinors explicitly break $3/4$ of the
supersymmetry. The two degrees of freedom of the complex function
$a$ give the remaining two supercharges.

We look for spacetimes that are static and spherically symmetric,
so in particular invariant under the rotation group. This rotation
group acts on spinors, and can in general leave or not leave our
Killing spinor ansatz invariant. It will be a check on our
explicit solution for the Killing spinors that they should be also
rotationally invariant, just as in the original case for exotic
solutions \cite{Romans:1991nq}.

Note that our choice of Killing spinors makes them timelike, i.e.\ they give rise to a timelike Killing vector
(see \cite{Kallosh:1993wx,Meessen:2006tu,Huebscher:2006mr} for more details about Killing spinor identities). It was already shown in section \ref{5:BLS} that, to obtain a supersymmetric solution, one needs to check only the Maxwell equations and Bianchi
identities in addition to the BPS conditions. The equations of motion for the other fields then follow, due to
the timelike Killing spinor.

\subsubsection{BPS conditions and attractor flow}

With the above ans\"{a}tze for the spacetime and the Killing spinors one can show that the gaugino and
gravitino variations \eqref{susygluino}, \eqref{susy-gravi} simplify substantially but do not yet vanish
identically.

From the gaugino variation we obtain the following radial flow equations for the scalar fields:
\begin{equation}\label{6:gaugino-var}
e^{-i \alpha} U \partial_r z^i = g^{i \bar{j}}
\bar{f}^{\Lambda}_{\bar{j}} \left( \frac{I_{\Lambda \Sigma}
p^{\Sigma}}{h^2} \mp g \xi_{\Lambda}
 \right),
\end{equation}
where the two different signs correspond to the two types of
Killing spinors in the given order.

If we require the gravitino variation \eqref{susy-gravi} to vanish, we derive four extra equations that need to
be satisfied (one for each spacetime index). The equations for $t$ and $\theta$ determine the radial dependence
of the metric components,
\begin{equation}\label{6:gravitino-var1}
e^{i \alpha} \partial_r U = - \frac{L^{\Lambda} I_{\Lambda
\Sigma} p^{\Sigma}}{h^2} \pm g \xi_{\Lambda} L^{\Lambda},
\end{equation}
\begin{equation}\label{6:gravitino-var2}
e^{i \alpha} \frac{U}{h} \partial_r h  = \frac{L^{\Lambda}
I_{\Lambda \Sigma} p^{\Sigma}}{h^2} \pm g \xi_{\Lambda}
L^{\Lambda}\ .
\end{equation}
The $\varphi$ component of the gravitino variation further constrains
\begin{equation}\label{6:gravitino-var3}
g \xi_{\Lambda} p^{\Lambda} = \mp 1\ ,
\end{equation}
and the radial part gives a differential equation for the Killing spinor, solved by
\begin{equation}\label{6:gravitino-var4}
a(r) = a_0\ \sqrt{U(r)}\ e^{-\frac{i}{2}\int A_r(r)\, {\rm d} r}\
,
\end{equation}
with
\begin{equation}\label{6:K-connect}
A_r(r) =-\frac{i}{2} \left(\mathcal{K}_i \partial_r z^i -
\mathcal{K}_{\bar{j}} \partial_r z^{\bar{j}} \right)
\end{equation}
the $U(1)$ K\"{a}hler connection. These results are in agreement
with rotational symmetry since the Killing spinor is only a
function of $r$. The solution is 1/4 BPS and has two conserved
supercharges, corresponding to the two free numbers of the complex
constant $a_0$. We further see that \eqref{6:gravitino-var4} does
not give an extra constraint on the fields, but can be used to
determine the explicit radial dependence of the Killing spinor
parameter $a(r)$. One can always evaluate the integral of $A(r)$
for a given solution and thus the Killing spinor can be explicitly
found once the BPS equations
\eqref{6:gaugino-var}-\eqref{6:gravitino-var3} are satisfied.

Notice also that \eqref{6:gravitino-var3} is in accordance with the generalized Dirac quantization condition
\eqref{6:dir_quantization} with the smallest non-zero integer $n=\pm 1$. It will be interesting to understand how
one can generate other solutions with higher values of $n$ or whether supersymmetry always strictly constrains
$n$ as in the present case. Furthermore, it is easy to see that in the limit $g \rightarrow 0$ where the gauging
vanishes one recovers the well-known first order attractor flow equations of black holes in ungauged $N=2$
supergravity \cite{Ferrara:1995ih,Strominger:1996kf,Ferrara:1996dd}. The presence of the extra terms due to the
gauging is precisely where the difference between ungauged and gauged black holes lies. Thus we believe the BPS
equations are now written in a simpler and more suggestive form compared to \cite{Cacciatori:2009iz} (see also \cite{Dall'Agata:2010gj} where the above BPS equations are derived from a different perspective by a smart rewriting of the original lagrangian).

A short comment on the phase $\alpha$ is in order. One can see in
eqs. \eqref{6:gravitino-var1}, \eqref{6:gravitino-var2} that the
quantities $e^{-i \alpha} L^{\Lambda}$ must always be real. Thus,
if e.g. $\alpha = 0$ then $L^{\Lambda}$ will need to be real,
while if $\alpha = \frac{\pi}{2}$, $L^{\Lambda}$ have to be
imaginary. This $U(1)_R$ symmetry of the BPS conditions is of
course well understood in the ungauged case and there are
generally two ways of proceeding. One can just fix the phase to a
particular value and go on to write down the solutions, as
originally done in \cite{Behrndt:1997ny}, or one can also put
explicitly the phase factor in the definition of the sections as
done in \cite{Denef:2000nb}. Here we choose to fix $\alpha = 0$
for the rest of the section as it will minimize the factors of $i$
in what follows (note that \cite{Behrndt:1997ny} makes the
opposite choice and thus the solutions are given for the imaginary
instead of the real parts of the sections). It should be clear how
one can always plug back the factor of $e^{-i \alpha}$ and choose
a different phase if needed in different conventions. In
particular this choice implies that (after adding
\eqref{6:gravitino-var1} and \eqref{6:gravitino-var2})
\begin{equation}\label{6:imag_sections}
\xi_{\Lambda} {\rm Im}(X^{\Lambda}) = 0\ .
\end{equation}

\subsection{Black hole solutions}\label{6:general_solution}

Now we would like to find explicit solutions to eqs. \eqref{6:gaugino-var}-\eqref{6:gravitino-var2}. We already know
(by assumption) the solution for the vector field strengths \eqref{6:el_field_strengths}, so we search for
solutions of the metric functions $U(r), h(r)$ and the symplectic sections $X^{\Lambda} (r), F_{\Lambda} (r)$
that determine the scalars. We propose the following form for the solution of the BPS equations in the electric
frame (for the choice of phase $\alpha = 0$):
\begin{equation}\label{6:scalar_sol}
 \frac{1}{2} \left(X^{\Lambda} + \bar{X}^{\Lambda} \right) = H^{\Lambda}\ , \qquad \frac{1}{2} \left(F_{\Lambda} + \bar{F}_{\Lambda} \right) = 0\ ,
\end{equation}
$$H^{\Lambda} = \alpha^{\Lambda} + \frac{\beta^{\Lambda}}{r},$$
and
\begin{equation}\label{6:metric_sol}
  U(r) = e^{\mathcal{K}/2} \left(g r + \frac{c}{2 g r} \right)\ , \qquad h(r) = r e^{-
  \mathcal{K}/2}\ ,
\end{equation}
where $\mathcal{K}$ is the K\"{a}hler potential
\begin{equation}\label{6:Kaehler}
    e^{- \mathcal{K}} = i \left(
  \bar{X}^{\Lambda} F_{\Lambda} - X^{\Lambda} \bar{F}_{\Lambda}
  \right)\ ,
\end{equation}
and $c$ some constant. The line element of the spacetime is then
\begin{equation}\label{6:line-element_sol}
  {\rm d} s^2 = e^{\mathcal{K}} \left(g r+\frac{c}{2 g r} \right)^2 {\rm d}t^2 - \frac{e^{-\mathcal{K}} {\rm d}r^2}{\left(g r+\frac{c}{2 g r} \right)^2} - e^{-\mathcal{K}} r^2 {\rm d} \Omega_2^2\ .
\end{equation}
The constant $c$ above is not specified yet and depends explicitly
on the chosen model. This is also the case for the constants
$\alpha^{\Lambda}, \beta^{\Lambda}$ that may eventually be
expressed in terms of the FI parameters $\xi_{\Lambda}$ and the
magnetic charges $p^{\Lambda}$. We give some explicit examples in
section \ref{6sect:explcitexamples}. Here we just use the above
results to show how the BPS equations simplify to a form where
they can be explicitly solved given a particular model with a
prepotential (we further assume that \eqref{6:scalar_sol} implies
${\rm Im}(X^{\Lambda}) = 0$ in accordance with
\eqref{6:imag_sections}). Eqs.
\eqref{6:gravitino-var1}-\eqref{6:gravitino-var2}, together with
\eqref{6:scalar_sol}-\eqref{6:metric_sol}, lead to:
\begin{equation}\label{6:final_eq1}
  \xi_{\Lambda} \alpha^{\Lambda} = \pm 1, \qquad \xi_{\Lambda}
  \beta^{\Lambda} = 0\ ,
\end{equation}
\begin{equation}\label{6:final_eq2}
  F_{\Lambda} \left( -2 g^2 r \beta^{\Lambda} + c
  \alpha^{\Lambda}+g p^{\Lambda}\right)= 0\ .
\end{equation}

Multiplying \eqref{6:gaugino-var} with $f_i^{\Lambda}$ we eventually
obtain
\begin{align}\label{6:final_eq3}
\begin{split}
\left( g r + \frac{c}{2 g r} \right) \left( F_{\Sigma} X^{\Sigma} \partial_r X^{\Lambda} - X^{\Lambda} F_{\Sigma} \partial_r  X^{\Sigma} \right) &= -\frac{1}{2 r^2} F_{\Sigma} \left(X^{\Sigma} p^{\Lambda} - X^{\Lambda}  p^{\Sigma} \right) \\
& + g F_{\Sigma} X^{\Sigma} \left( X^{\Lambda} \pm i F_{\Pi}
X^{\Pi} (I^{-1})^{\Lambda \Gamma} \xi_{\Gamma} \right)\ .
\end{split}
\end{align}
We chose to rewrite it in this form in order to have equations
only for the symplectic sections, as standardly done also in
ungauged black holes literature \cite{Behrndt:1997ny}. In
principle however $f_i^{\Lambda}$ is non-invertible and thus
\eqref{6:final_eq3} does not strictly speaking imply
\eqref{6:gaugino-var}. Practically this never seems to be an issue since in
fact \eqref{6:final_eq3}  gives one extra equation. In all cases we solved explicitly the
equations, we found that the condition coming from the gaugino
variation is already automatically satisfied after solving
\eqref{6:final_eq1} and \eqref{6:final_eq2}. Unfortunately, we were
not able to prove that it must vanish identically with the above
ansatz.

Using \eqref{6:scalar_sol} it is straightforward to prove that the
K\"{a}hler connection \eqref{6:K-connect} vanishes identically (c.f.
Eq.(29) of \cite{Behrndt:1997ny}). Thus the functional dependence
of the Killing spinors becomes
\begin{equation}\label{6:wrong}
a(r) = \sqrt{U(r)}\ a_0\ ,
\end{equation}
just as in the original solution without scalars
\cite{Romans:1991nq}.

Note that with \eqref{6:scalar_sol} one can now also show that the
field strengths \eqref{6:el_field_strengths} identically solve the
Bianchi identities and the Maxwell equation as they fall in the
form \eqref{6:solve_field_strengths} with $q_{\Lambda} = 0$. Thus
any solution of \eqref{6:final_eq1}-\eqref{6:final_eq3} will be a
supersymmetric solution of the theory with no further constraints.

One particular solution (the only one in absence of vector multiplets) of the above equations that is always
present, is when $\alpha^{\Lambda} = -g p^{\Lambda}$, $\beta^{\Lambda} = 0$, for all $\Lambda$, and $c = 1$.
This solution is in fact the one discovered in \cite{Chamseddine:2000bk} with constant scalars ($X^{\Lambda}$ is
constant when $\beta^{\Lambda} = 0$). However, this solution has a naked singularity, since $c>0$. A horizon is
not present in this case,  since generally it will appear at $r_h^2 = -\frac{c}{2 g^2}$ and thus only for $c <
0$. We will see in section \ref{6sect:explcitexamples} that indeed there exist solutions of the above equations
in which $c<0$, such that a proper horizon shields the singularity. These solutions however necessarily have
nonzero $\beta^{\Lambda}$'s. Thus a proper black hole can only form in the presence of some sort of attractor
mechanism for the scalar fields.

\subsection{Black holes with electric and magnetic
charges}\label{6:electr-magn general solution}

We now explain how one can restore the broken electromagnetic duality invariance of the theory \eqref{6:lagr}. As
discussed in section \ref{6:sect:2}, the electric gaugings break electromagnetic invariance, i.e. performing
symplectic rotations leads us to a new lagrangian that will be of different form from \eqref{6:lagr}. One then
needs to allow for both electric and magnetic gaugings and change the form of the scalar potential in order to
recover the electromagnetic invariance of the ungauged theory. There have been various proposals in literature
for extending it to gauged supergravity \cite{Michelson:1996pn,Dall'Agata:2003yr,D'Auria:2004yi}. It turns out that the correct approach to
introducing real magnetic gaugings is the embedding tensor formalism, and we closely follow the analysis of
\cite{deWit:2005ub,deWit:2011gk}. It restores full electromagnetic duality invariance of the gauge theory (when the electric and
magnetic charges are mutually local) by introducing additional tensor fields in the Lagrangian. Unfortunately
the theory is not yet fully developed in general for supergravity (for rigid $N=2$ supersymmetry, see \cite{deVroome:2007zd}), but we will nevertheless be able to write down particular solutions due to the fact that we can
do duality transformations on the solutions of the electrically gauged theory.

Even though we cannot give the most general lagrangian and susy variations for the theory with electric and
magnetic gaugings, we know how the bosonic part of the Lagrangian should look like in this very special case of
FI gaugings. It is most instructive to integrate out the additional tensor field that has to be introduced,
following the procedure of section 5.1 of \cite{deWit:2005ub}. Exactly half of the gauge fields (we will originally
have both electric and magnetic gauge fields, $(A_{\mu}^{\Lambda}, A_{\mu, \Lambda})$) will also be integrated
out in this process. One first splits the index $\Lambda$ in two parts, $\{ \Lambda \} = \{ \Lambda',\Lambda''
\}$, for the non-vanishing electric and magnetic gauge fields respectively. The lagrangian will then consist
only of $A_{\mu}^{\Lambda'}, A_{\Lambda'', \mu}$, while $A_{\mu}^{\Lambda''}, A_{\Lambda', \mu}$ are integrated
out together with the additional tensor field. Thus the linear combination of fields used for the $U(1)$ FI
gauging is $\xi_{\Lambda'} A_{\mu}^{\Lambda'} - \xi^{\Lambda''} A_{\Lambda'', \mu}$. The $\xi^{\Lambda''}$'s are
the magnetic charges of the gravitinos, and the new generalized Dirac quantization condition for electric and
magnetic charges $(q_{\Lambda},p^{\Lambda})$ of any solution is
\begin{equation}\label{6:dir_quantization_em}
   e_{\Lambda'} p^{\Lambda'} - m^{\Lambda''} q_{\Lambda''}  = n, \qquad n \in \mathbb{Z}\ ,
\end{equation}
with electric and magnetic gravitino charges $e_{\Lambda'} \equiv g \xi_{\Lambda'}$ and $m^{\Lambda''} \equiv g
\xi^{\Lambda''}$. The scalar potential is then of the form
\begin{equation}\label{6:pot_em}
V= (g^{i\bar \jmath}f_i^{\Lambda'} {\bar f}_{\bar \jmath}^{\Sigma'} -3{\bar L}^{\Lambda'}
L^{\Sigma'})\xi_{\Lambda'} \xi_{\Sigma'} - (g^{i\bar \jmath} h_{i, \Lambda''} {\bar h}_{\bar{\jmath}, \Sigma''}
-3{\bar M}_{\Lambda''} M_{\Sigma''})\xi^{\Lambda''} \xi^{\Sigma''}\ .
\end{equation}
The main point about electromagnetic invariance is that the equations of motion are now invariant under the
group $Sp(2 (n_V+1), \mathbb{R})$, which at the same time rotates the lagrangian from a purely electric gauging
frame to a more general electromagnetic gauging. The symplectic vectors transforming under the symmetry group
are the sections $(F_{\Lambda}, X^{\Lambda})$ and the FI parameters $(\xi_{\Lambda}, \xi^{\Lambda})$, as well as
the vector field strengths $F_{\mu \nu}^{\Lambda}, G_{\mu \nu, \Lambda}$ (which come from the respective
electric and magnetic gauge potentials $(A_{\mu}^{\Lambda}, A_{\mu, \Lambda})$). One can then see how natural
equations \eqref{6:dir_quantization_em},\eqref{6:pot_em} are if we start from a purely electric frame with only
$\xi_{\Lambda}, F_{\mu \nu}^{\Lambda}$ nonzero and then perform an arbitrary symplectic transformation. The
important message is that once we have found a solution to the purely electric theory we can always perform any
symplectic transformation of the theory to see how the solution looks like in a more general electromagnetic
setting.

It is in fact easy to guess how the solution looks like in a more general theory with electric and magnetic
gaugings. We have not proven the existence of such a BPS solution due to the lack of a properly defined
lagrangian and supersymmetry variations, but we can nevertheless indirectly find it by symplectic rotations.
This procedure leads to  a solution, where the metric is again given by  \eqref{6:line-element_sol}, together with
\begin{align}
\begin{split}
F^{\Lambda'}_{t r} = 0\ , \qquad F^{\Lambda'}_{\theta \varphi} =
\frac{p^{\Lambda'}}{2} \sin \theta\ ,\\
G_{\Lambda'', t r} = 0\ , \qquad G_{\Lambda'',\theta \varphi} =
\frac{q_{\Lambda''}}{2} \sin \theta\ ,
\end{split}
\end{align}
and harmonic functions that determine the sections
\begin{align}
\begin{split}
\frac{1}{2} \left(X^{\Lambda'} + \bar{X}^{\Lambda'} \right) =
H^{\Lambda'}\ , \qquad \frac{1}{2} \left(F_{\Lambda'} +
\bar{F}_{\Lambda'} \right) = 0\ ,
\\
\frac{1}{2} \left(X^{\Lambda''} + \bar{X}^{\Lambda''} \right) = 0\
, \qquad \frac{1}{2} \left(F_{\Lambda''} + \bar{F}_{\Lambda''}
\right) = H_{\Lambda''}\ ,
\end{split}
\end{align}
$$H^{\Lambda'} = \alpha^{\Lambda'} + \frac{\beta^{\Lambda'}}{r}\ , \qquad H_{\Lambda''} = \alpha_{\Lambda''} + \frac{\beta_{\Lambda''}}{r}\ .$$
The above should give solutions provided that the following
identities (coming from the BPS conditions) are satisfied,
\begin{equation}\label{6:eq0_em}
  2 g ( \xi_{\Lambda'} p^{\Lambda'} - \xi^{\Lambda''} q_{\Lambda''}) = \mp
  1\ ,
\end{equation}
\begin{equation}\label{6:eq1_em}
  \xi_{\Lambda'} \alpha^{\Lambda'} - \xi^{\Lambda''} \alpha_{\Lambda''} = \pm 1, \qquad \xi_{\Lambda'}
  \beta^{\Lambda'} - \xi^{\Lambda''} \beta_{\Lambda''} = 0\ ,
\end{equation}
\begin{equation}\label{6:eq2_em}
  F_{\Lambda'} \left( -2 g^2 r \beta^{\Lambda'} + c
  \alpha^{\Lambda'}+g p^{\Lambda'}\right) - X^{\Lambda''} \left( -2
  g^2 r \beta_{\Lambda''} + c \alpha_{\Lambda''}+g q_{\Lambda''}
\right)= 0\ ,
\end{equation}
together with the symplectic invariant version of
\eqref{6:final_eq3} coming from contraction with $f_i^{\Lambda}$.
This expression becomes lengthy and cumbersome to check
and we will not write it down explicitly. In this case it will be
easier to explicitly check the symplectic invariant version of
\eqref{6:gaugino-var} by first defining the complex vector multiplet
scalars from the sections. Of course in case of confusion one can
always take a model and rotate it to the electric frame where the
susy variations are clearly spelled out (\eqref{susy-gravi}-\eqref{susygluino}).

\subsection{Explicit black hole solutions}\label{6sect:explcitexamples}

\subsubsection{$n_V = 1$ with $F =-2 i \sqrt{X^0 (X^1)^3}$}\label{6:sect:5.1}
This is the simplest prepotential in the ordinary electrically
gauged theory that leads to a black hole solution. We have one
vector multiplet with the prepotential
\begin{equation}F =-2 i \sqrt{X^0 (X^1)^3}\ ,\end{equation}
thus one finds $X^0 = \alpha^0+\frac{\beta^0}{r}, X^1 =
\alpha^1+\frac{\beta^1}{r}$ from \eqref{6:scalar_sol}. This theory
exhibits an AdS$_4$ vacuum at the minimum of the scalar potential
(corresponding to the cosmological constant)
\begin{equation}
V^* = \Lambda = - \frac{2 g^2}{\sqrt{3}} \sqrt{\xi_0 \xi_1^3}
\end{equation}
at $z^* = \sqrt{\frac{3 \xi_0}{\xi_1}}$ (defining $z \equiv
\frac{X^1}{X^0}$). This can be easily deduced using the results of
the previous chapters. Going through the BPS equations
\eqref{6:final_eq1}-\eqref{6:final_eq2}, we can fix all the constants
of the solution in terms of the FI parameters $\xi_0, \xi_1$ apart
from one free parameter (here we leave $\beta^1$ to be free for
convenience, but it can be traded for one of the magnetic charges
or for $\beta^0$). We obtain that the magnetic charges are given
by:
\begin{equation}\label{6:magn-charges}
p^0 = \mp \frac{2}{g \xi_0} \left(\frac{1}{8}+\frac{8 (g \xi_1 \beta^1)^2}{3} \right), \quad p^1 = \mp \frac{2}{g \xi_1} \left(\frac{3}{8}-\frac{8 (g \xi_1 \beta^1)^2}{3} \right)\ ,
\end{equation}
for spinor I and II respectively. The other constants in the
solution are
\begin{equation}\label{6:constants}
\beta^0 = -\frac{\xi_1 \beta^1}{\xi_0}, \qquad \alpha^0 = \frac{\pm 1}{4 \xi_0}, \qquad \alpha^1 = \frac{\pm 3}{4 \xi_1}, \qquad c = 1 - \frac{32}{3} (g \xi_1 \beta^1)^2\ .
\end{equation}
Using the definition of the gravitino charges \eqref{6:gravitino-charges}, $e_\Lambda = g \xi_\Lambda$, these relations imply
\begin{equation}
e_\Lambda \alpha^\Lambda = \pm g\ , \qquad e_\Lambda \beta^\Lambda = 0\ , \qquad e_\Lambda p^\Lambda = \mp 1\ ,
\end{equation}
and one can check that the complete solution is a function of the
variables $e_\Lambda, p^\Lambda$ and $g$. Note that in fact the
dependence on $g$ is artificial since it can always be absorbed in
the definition of the coordinates. In particular, the rescaling $g
r \rightarrow r, t \rightarrow g t$ makes the metric and the
scalar flow dependent only on $e_\Lambda, p^\Lambda$ as is also
the form of the solution presented in \cite{Cacciatori:2009iz}.

Interestingly, one can verify that the condition coming from the
gaugino variation, \eqref{6:final_eq3}, is automatically satisfied
with no further constraints. One can see that the two spinor types
in the end amount to having opposite magnetic charges and to
flipping some signs for the solution of the sections.

We now
analyze the physical properties of the solution. In this case it
is important to give explicitly the metric function in front of
the ${\rm d} t^2$ term. Using the form of the line element in \eqref{6:line-element_sol}, the specific form of the sections with constants given in \eqref{6:constants}, one can explicitly compute:
\begin{equation}
g_{tt} = \frac{2 \sqrt{\xi_0 \xi_1^3} r^2 \left(g r+\frac{1}{2 g
r}-\frac{16 g}{3 r} (\xi_1 \beta^1)^2 \right)^2}{\sqrt{(r \mp 4
\xi_1 \beta^1)(3 r \pm 4 \xi_1 \beta^1)^3}}\ .
\end{equation}
The leading terms of the (infinite) asymptotic expansion of the
metric for $r\rightarrow\infty$ are then
\begin{equation}
g_{tt} (r \rightarrow \infty) = -\frac{\Lambda r^2}{3} \left( 1 +
\frac{1}{2 g^2} (1+c) \frac{1}{r^2} - \frac{256 (\xi_1
\beta^1)^3}{27} \frac{1}{r^3} + \mathcal{O} \left(\frac{1}{r^4}
\right) \right)\ .
\end{equation}
Clearly, the metric has the correct AdS$_4$ asymptotics. Although
the constant term of the asymptotic expansion is not exactly $1$
when we compare to the RN-AdS metric, we
are still tempted to think that the coefficient in front of the
$1/r$ term determines the physical mass of the black hole,
\begin{equation}\label{6:proposedmass}
M = - \frac{128}{81} \Lambda (\xi_1 \beta^1)^3\ .
\end{equation}
The issue of defining the mass is a bit more subtle in
asymptotically AdS spacetimes and we address it more carefully in
the following chapters, where we show that this naive expectation is in fact wrong.

One can also notice that there are some subtleties for the radial
coordinate that usually do not appear for black hole spacetimes.
In particular, $r = 0$ is neither a horizon (where $g_{tt} = 0$),
nor a singularity (where $g_{tt} \rightarrow \infty$). In fact the
point $r = 0$ is never part of the spacetime, since the
singularity is always at a positive $r$, where the space should be
cut off. Thus the $r$ coordinate does not directly correspond to
the radial coordinate from the singularity. The horizon for both
signs is at
\begin{equation}
r_h = \sqrt{\frac{16}{3} (\xi_1
\beta^1)^2 - \frac{1}{2 g^2}}\ ,
\end{equation}
while genuine singularities will appear at $r_s = \pm 4 \xi_1 \beta^1, \mp \frac{4}{3} \xi_1 \beta^1$. The
spacetime will then continue only until the first singularity is encountered. If we want to have an actual black
hole spacetime we must insist that the horizon shields the singularity, i.e. $r_h
> r_s$, otherwise we again have a naked singularity and the sphere at $r_h$ will not be part of the spacetime.
This requirement further sets the constraints $|g \xi_1 \beta^1| >
\frac{3}{8}$, with $\xi_1 \beta^1 < 0$ for solution I (upper sign) and $\xi_1
\beta^1 > 0$ for solution II (lower sign). Since the parameter $\beta^1$ is at
our disposal, it can always be chosen to be within the required
range, thus the singularity can be shielded by a horizon in a
particular parameter range for $\beta^1$. So, putting together
both solutions, we know that a proper black hole with a horizon
will form in case $g \xi_1 \beta^1 \in (-\infty,
-\frac{3}{8})\bigcup (\frac{3}{8}, \infty)$, with the
corresponding relations given above between the magnetic charges
and $\xi_1 \beta^1$ for the two intervals. In between, we are dealing with naked singularities, which are of no
interest for us at present. The constant $c$ is always negative, and satisfies
\begin{equation}
c < -\frac{1}{2}\ ,
\end{equation}
which reflects again the existence of a horizon, as announced in section 2.

Let us now investigate further the properties of
these new black holes. Their entropy is proportional to the area
of the black hole at the horizon,
\begin{equation}
    S = \frac{A}{4} = \frac{3}{4 \Lambda} \sqrt{(r_h-r_{s,1}) (r_h-r_{s,2})^3} = \frac{\sqrt{(r_h \mp 4 \xi_1 \beta^1)(3 r_h \pm 4 \xi_1 \beta^1)^3}}{8 \sqrt{\xi_0 \xi_1^3}}\ ,
\end{equation}
so the entropy is effectively a function of $\xi_0, \xi_1,
\beta^1$, which can be rewritten in terms of the FI-terms and
magnetic charges. Thus the entropy is a function of the black hole
charges $p^{\Lambda}$ and the gravitino charges $e_{\Lambda}$. One
can further observe that in case of fixed gravitino charges
$e_{\Lambda}$, the entropy scales quadratically with the parameter
$\beta^1$ and thus linearly with the charges $p^0$ or $p^1$ in the
limit of large charges. The opposite limit of fixed magnetic
charges shows that the entropy remains constant for large
gravitino charge.

It is interesting to note that the fact that the scalars at the
horizon are fixed in terms of the gravitino and black hole charges
is not directly obvious from the general form of the solution. The
scalars depend on the constants $\alpha^{\Lambda}, \beta^{\Lambda}
$ that might not always be fully determined by $\xi_{\Lambda},
p^{\Lambda}$. One example of this is for the prepotential $F = -i
X^0 X^1$ where the magnetic black hole charges are fully fixed in
terms of FI parameters and either $\beta^0$ or $\beta^1$ can be
freely chosen. However, one can show that in this case there is no
parameter range for the $\beta^{\Lambda}$'s where the singularity
is shielded by the horizon, thus black holes do not exist. In all
the cases for which we checked that a black hole is possible we
could verify that indeed the scalar values at the horizon can be
expressed in terms of the charges and FI parameters, but we have
no general proof of this\footnote{The BPS equations
\eqref{6:final_eq1}-\eqref{6:final_eq3} can be relatively easily
solved in full generality for a prepotential of the form $F =
(X^0)^s (X^1)^{2-s}$. The outcome is that black holes exist for $s
\in (0,1)$. The solution for general $n$ is in full analogy to the
one presented here. There is only certain $n$ dependence in the
way the various constants depend on each other, which does not
lead to any qualitative differences. Here we chose to explicitly
describe the case with $n=1/2$ since it is the most relevant case
from a string theory point of view as we will see in the next
section.}.

Another interesting question is what the near-horizon geometry of
this black hole is. It is natural to expect that a static four
dimensional BPS black hole has a near-horizon geometry of AdS$_2
\times $S$^2$ and this is indeed the case. The radii of the two
spaces are
\begin{equation}
R_{S^2} = r_h
e^{-\mathcal{K}/2}|_{r=r_h}, \qquad R_{AdS_2} =
\frac{e^{-\mathcal{K}/2}|_{r=r_h}}{2 \sqrt{2} g}\ ,
\end{equation}
and it can be shown that $R_{S^2} > \sqrt{2} R_{AdS_2}$ from the
constraints on having a horizon. As the radii are inversely
proportional to the scalar curvature of these spaces, it follows
that the overall AdS$_2 \times $S$^2$ space has a negative
curvature, as expected for asymptotically AdS$_4$ black holes.
Thus it is clear that near the horizon we do not observe a
supersymmetry enhancement to a fully BPS vacuum as is the case for
the asymptotically flat static BPS black holes\footnote{We already saw that AdS$_2
\times $S$^2$ is maximally supersymmetric only for $R_{S^2} =
R_{AdS_2}$}. Nevertheless, there
is still a supersymmetry enhancement from a 1/4 BPS overall
solution to a 1/2 BPS vacuum near the horizon, as shown in \cite{deWit:2011gk}.

\subsubsection{$F =\frac{(X^1)^3}{X^0}$ in a mixed electromagnetic frame}
In order to give an example of black hole solutions in a more
general electromagnetic frame, one can rotate the sections and FI
parameters of the previous example by the symplectic matrix
\begin{equation}
\mathcal{S}=\begin{pmatrix} 1 & 0 & 0 & 0 \\ 0 & 0 & 0 & 1/3 \\
0 & 0 & 1 & 0\\ 0 & -3 & 0 & 0 \end{pmatrix}\ ,
\end{equation} such
that the prepotential after rotation corresponds to the
well-studied in ungauged supergravity $T^3$ model with
prepotential
\begin{equation}
F =\frac{(X^1)^3}{X^0}\ ,
\end{equation}
and the non-vanishing FI
parameters are $\xi_0, \xi^1$. The theory will then be
electrically gauged with $A_{\mu}^0$ and magneticaly gauged with
$A_{1, \mu}$. This prepotential cannot lead to an AdS BPS black
hole in the purely electric gauging, because it does not exhibit a
supersymmetric AdS$_4$ vacuum. However, in this mixed electromagnetic
gauging, the $T^3$ model does have a proper fully supersymmetric
AdS vacuum.

Now we can follow the more general procedure outlined in section
\ref{6:electr-magn general solution}. In this case it turns out that
$X^0 = \alpha^0+\frac{\beta^0}{r}, F_1 =
\alpha_1+\frac{\beta_1}{r}$. The black hole solution will then
have one magnetic charge $p^0$ and one electric charge $q_1$.
Going through the BPS equations \eqref{6:eq0_em}-\eqref{6:eq2_em}, we
can fix all the constants of the solution in terms of the FI
parameters $\xi_0, \xi^1$, apart from one free parameter which we
choose to be $\beta_1$. The charges are given by:
\begin{equation}
p^0 = \mp \frac{2}{g \xi_0} \left(\frac{1}{8}+\frac{8 (g \xi^1
\beta_1)^2}{3} \right), \quad q_1 = \pm \frac{2}{g \xi^1}
\left(\frac{3}{8}-\frac{8 (g \xi^1 \beta_1)^2}{3} \right)\ ,
\end{equation}
for spinor I and II respectively. The other constants in the
solution are \begin{equation}\beta^0 = \frac{\xi^1
\beta_1}{\xi_0}, \qquad \alpha^0 =\pm  \frac{1}{4 \xi_0}, \qquad
\alpha_1 = \mp \frac{ 3}{4 \xi^1}, \qquad c = 1 - \frac{32}{3} (g
\xi^1 \beta_1)^2\ .\end{equation} and one can see that the metric
and scalar profile in this case are analogous to the example in
the previous subsection, as expected. This confirms the
consistency of the results in section \ref{6:electr-magn general
solution}. The entropy of the black hole is now a function of the
electric and magnetic gravitino charges, $e_0 = g \xi_0$ and $m^1
= g \xi^1$, and the black hole charges $p^0$ and $q_1$.

Note that we could have for instance rotated the frame from a
fully electric to a fully magnetic frame, by the symplectic matrix
\begin{equation}
\mathcal{S}=\begin{pmatrix} 0 & 0 & -1 & 0 \\ 0 & 0 & 0 & -1/3 \\
1 & 0 & 0 & 0\\ 0 & 3 & 0 & 0 \end{pmatrix}\ ,
\end{equation} and
it turns out that the prepotential $F =-2 i \sqrt{X^0 (X^1)^3}$ is
in fact invariant under this transformation. The resulting
solution will be the same, but there will be two electric instead
of two magnetic charges.

\subsection{M-theory lift}\label{6:sect:M-theory reduction}

An explicit string theory example of abelian gauged $N=2$, $D=4$
supergravity with FI terms was found by a consistent truncation of
M-theory on $S^7$ in \cite{Duff:1999gh,Cvetic:1999xp}. A standard Kaluza-Klein
compactification on $S^7$ leads initially to an $SO(8)$ gauged
$N=8$ supergravity in four dimensions. To avoid some of the
complications of non-abelian gauge fields, the authors of
\cite{Duff:1999gh,Cvetic:1999xp} further defined a consistent truncation of this theory
to an $U(1)^4$ gauged $N=2$ supergravity. The 11-dimensional
metric ansatz is given by:
\begin{equation}\label{6:11dim metric}
    ds^2_{11} = \Delta^{2/3} ds_4^2 + 2 g^{-2} \Delta^{-1/3}
    \sum_{\Lambda = 0}^3 a_{\Lambda}^{-1} \left(d\mu_{\Lambda}^2 + \mu_{\Lambda}^2 (d\phi_{\Lambda}+\frac{g}{\sqrt{2}} A^{\Lambda})^2
    \right),
\end{equation}
where $\Delta = \sum_{\Lambda} a_{\Lambda} \mu_{\Lambda}^2$ with
the $\mu_{\Lambda}$'s satisfying $\sum_{\Lambda} \mu_{\Lambda}^2 =
1$. They can be parameterized by the angles on the 3-sphere as
explained in more detail in \cite{Duff:1999gh,Cvetic:1999xp}. The remaining 4 angles
$\phi_{\Lambda}$ together with the $\mu_{\Lambda}$ describe the
internal space, while $x^{\mu}$ are coordinates of the
four-dimensional spacetime on which the resulting $N=2, D=4$
gauged supergravity is defined. The factors $a_{\Lambda}$ depend
on the four-dimensional axio-dilaton scalars $\tau_i =
e^{-\varphi_i} + i \chi_i$ (defined below) and the gauge fields
$A^{\Lambda} = A^{\Lambda}_{\mu} dx^{\mu}$ are exactly the ones
appearing in the four-dimensional theory. Note that if all the
gauge fields are vanishing and the scalars are at the minimum of
the potential, the internal space becomes exactly $S^7$. Apart
from the metric, the field strength of the 11-dimensional three
form field is given by:
\begin{align}\label{6:11dim_gauge}
\begin{split}
  F_{4} &= \sqrt{2} g \sum_{\Lambda} (a_{\Lambda}^2 \mu_{\Lambda}^2-\Delta
  a_{\Lambda})\epsilon_4+\frac{1}{\sqrt{2} g} \sum_{\Lambda}
  a_{\Lambda}^{-1} \bar{*} d a_{\Lambda} \wedge
  d(\mu_{\Lambda}^2)\\ &- \frac{1}{g^2} \sum_{\Lambda} a_{\Lambda}^{-2}
  d(\mu_{\Lambda}^2) \wedge (d\phi_{\Lambda}+\frac{g}{\sqrt{2}} A^{\Lambda}) \wedge \bar{*} d A^{\Lambda},
\end{split}
\end{align}
with $\bar{*}$ the Hodge dual with respect to the four-dimensional
metric $ds_4$, and $\epsilon_4$ the corresponding volume form.

With these identifications, the four-dimensional $N=2$ bosonic
lagrangian, written in our conventions, reads
\begin{align}\label{6:duff_lagr}
\begin{split}
\mathcal L&=\frac{1}{2}R(g)+\frac{1}{4} \sum_{i = 1}^3
\left((\partial \varphi_i)^2+e^{2 \varphi_i} (\partial \chi_i)^2
\right) +
Im(\mathcal{M})_{\Lambda\Sigma}F_{\mu\nu}^{\Lambda}F^{\Sigma\,\mu\nu}
\\&+\frac{1}{2}Re(\mathcal{M})_{\Lambda\Sigma}\epsilon^{\mu\nu\rho\sigma}
F_{\mu\nu}^{\Lambda}F^{\Sigma}_{\rho\sigma} + 2 g^2 \sum_{i=1}^3
\left( \cosh \varphi_i + \frac{1}{2} \chi_i^2 e^{\varphi_i}
\right)\ .
\end{split}
\end{align}
One can then check explicitly (using also the particular result
for the matrix $\mathcal{M}$ given in \cite{Duff:1999gh,Cvetic:1999xp}) that the above
Lagrangian is indeed of the form of \eqref{6:lagr} with prepotential
\begin{equation}\label{6:duff_prepot}
  F = - 2 i \sqrt{X^0 X^1 X^2 X^3}\ ,
\end{equation}
where the sections $X^{\Lambda}$ define the three scalars $\tau_i$
by $\frac{X^1}{X^0} \equiv \tau_2 \tau_3, \frac{X^2}{X^0} \equiv
\tau_1 \tau_3, \frac{X^3}{X^0} \equiv \tau_1 \tau_2$. The FI
parameters take the particularly simple form \begin{equation}
\xi_0 = \xi_1 = \xi_2 = \xi_3 = 1\ .
\end{equation} In this theory
one can find a black hole solution in analogy to the example in
section \ref{6sect:explcitexamples}. Following the general results in section
\ref{6:general_solution}, $X^{\Lambda} =
\alpha^{\Lambda}+\frac{\beta^{\Lambda}}{r}$, and from
\eqref{6:final_eq1}-\eqref{6:final_eq2} one can find the full solution
with $\alpha^0 = \alpha^1 = \alpha^2 = \alpha^3 = \pm\frac{1}{4}$
and three arbitrary parameters $\beta^1,\beta^2,\beta^3$ (or
equivalently $p^1,p^2,p^3$). We will not write down the full
solution as the expressions for the constant $c$ and the magnetic
charges in terms of the $\beta^{\Lambda}$'s are very long and do
not lead to further insight. It is clear that the particular
solution when we choose $\beta^1=\beta^2=\beta^3$ in fact
coincides precisely with the solution in section \ref{6:sect:5.1}
and this means that in any case a genuine black hole of the
M-theory reduction exists particularly when the three complex
scalars are equal. In the full solution of course there is a wider
range of values for $\beta^1,\beta^2,\beta^3$ that will lead to a
black hole, but this will suffice for our purposes here.

We now comment on the meaning of these four-dimensional black
holes from the point of view of M-theory as a first step towards
constructing the corresponding microscopic theory. It is notable
that the particular M-theory reduction we have leads to an
electrically gauged $N=2$ supergravity and thus the resulting
solution has only magnetic charges. This in fact makes the higher
dimensional interpretation a bit more involved. There are two main
points one can notice about the full 11-dimensional geometry from
the form in \eqref{6:11dim metric}. First, due to the nonconstant
scalars $\tau_i$, the full space is a warped product of the
internal seven-dimensional space with the AdS$_4$ black hole
spacetime. Second, due to the non-vanishing gauge fields
$A^{\Lambda}_{\varphi} = -p^{\Lambda} \cos \theta$, there is an
explicit mixing between the four angles $\phi_{\Lambda}$ of the
internal space and the four-dimensional angle $\varphi$. This
leads to four topological charges of the 11-dimensional spacetime,
in analogy to NUT charges. Note that in case the charges were only
electric, i.e.~$A^{\Lambda}_t = \frac{q^{\Lambda}}{r}$, the time
coordinate would mix with the internal angles and we would obtain
four angular momenta, leading to the interpretation of the
spacetime as arising from the decoupling limit of rotating
M2-branes as explained in detail in \cite{Duff:1999gh,Cvetic:1999xp}. In the present
case however the interpretation of the four-dimensional black
holes from M-theory is more involved because apart from M2-branes
we need to have some Kaluza-Klein monopoles in the M-theory
solution, in order to account for the topological charge coming
from the magnetic charges in four dimensions. Unfortunately we
were not able to find an explicit example for this type of
solutions in the literature, which probably is also related to the
fact that they would break almost all supersymmetry\footnote{The
black hole solutions in four dimensions preserve only two
supercharges, i.e. they are 1/4 BPS in $N=2$. In $N=8$, they are
1/16 BPS. This means that at least 30 of the original 32
supercharges in the original 11-dimensional supergravity will have
to be broken for the conjectured bound state of M2-branes and
Kaluza-Klein monopoles.}.

\section{Black branes and toroidal black holes}\label{6:branes}
Many of the above results considering spherically symmetric black holes in fact can easily be extended to objects with a flat horizon, namely toroidal black hole and black brane solutions. A very good reference for such solutions in minimal gauged supergravity is \cite{Caldarelli:1998hg}. Here we will concentrate on the static solutions with charges, which already exhibit the similarity between toroidal and spherical solutions, and allow us to also point out the major differences. We will then explain how the BPS solutions can be extended to include non-constant scalars in matter coupled supergravity.

\subsection{Static solutions in minimal gauged supergravity}
We start from the minimal $D=4$ $N=2$ gauged supergravity and consider static metrics of the form
\begin{equation}\label{formbrane}
{\rm d}s^2= U^2(r) \, {\rm d}t^2 -\frac{1}{U^2(r)} {\rm d}r^2 - r^2 {\rm d}\sigma^2\ , \qquad \qquad
U^2(r)= g^2 r^2 - \frac{2 \eta}{r} + \frac{q^2 + p^2}{r^2}\,,
\end{equation}
with a toroidal area element (with $\mathcal{V}$ the volume of the torus and $\tau$ the complex structure)
\begin{equation}\label{area-element}
{\rm d}\sigma^2 = \frac{\mathcal{V}}{{\rm Im}\tau} ({\rm d}x^2 + 2 {\rm Re}\tau \, {\rm d}x {\rm d}y + |\tau|^2 {\rm d}y^2) \,.
\end{equation}
The electromagnetic 1-form (the graviphoton) and its corresponding field strength are:
\begin{equation}\label{formbrane2}
A = \frac{q}{r} {\rm d}t + p \mathcal{V} x {\rm d}y \,, \qquad F = \frac{q}{r^2}\ {\rm d}t \wedge {\rm d}r + p \mathcal{V}\ {\rm d}x \wedge {\rm d}y\ .
\end{equation}
The range of the coordinates is restricted to $x \in [0,1]$, $y \in [0, 1]$ with periodic boundary conditions. The case of black branes can be obtained by decompactifying the torus, e.g. by considering a rectangular torus with ${\cal V}=R_1R_2, \tau=iR_2/R_1$ and sending the radii $R_1$ and $R_2$ to infinity. Doing so, we can use the volume as a regulator for black branes, and mass and charge densities will therefore be finite and well-defined.

The above metric describes a class of static charged toroidal black hole solutions with compact horizons. They asymptote at $r\rightarrow \infty$ to the vacuum configuration with $\eta=q=p=0$, which is a quotient of AdS$_4$, due to the identifications on $x$ and $y$. This spacetime is sometimes referred to in the literature as Riemann-anti-de Sitter (RiAdS) \cite{Vanzo:1997gw}. Supersymmetric toroidal solutions with magnetic charge do not exist, as shown in \cite{Cacciatori:2009iz,Caldarelli:1998hg} from the analysis of the integrability condition. The conditions to have a supersymmetric solution of the form \eqref{formbrane}-\eqref{formbrane2} that asymptotes to RiAdS are:
\begin{equation}
\eta=p=0\
\end{equation}
 for arbitrary electric charge. The resulting spacetime has a naked singularity whenever $q \neq 0$, which is often considered unphysical. In minimal gauged supergravity there is therefore no genuine BPS black brane solution. To make the situation more appealing, we now turn to general gauged supergravities. We will see that turning on matter couplings allows us to generate non-zero central charge and mass for the BPS configurations, which also leads to the existence of supersymmetric black brane solutions with horizon.

\subsection{Magnetic BPS black branes with scalars}
A class of BPS solutions with genuine horizons, corresponding to black branes and toroidal black holes in abelian gauged supergravity with FI terms, can be derived from the example in \cite{Cacciatori:2009iz} and following the steps in \cite{Hristov:2010ri}. The solutions are in complete analogy to the ones found in section \ref{6:Klemm} with the only exception that the flat horizon forces the magnetic charge carried by the graviphoton to vanish, $\xi_{\Lambda} p^{\Lambda} = 0$, as already mentioned above\footnote{The solution holds in gauged supergravity with FI parameters $P^a_{\Lambda} = \xi_{\Lambda} = const$. See section \ref{6:Klemm} for all technical details.}. In short, one can find a class of 1/4 BPS solutions, given by
 \begin{align}\label{our_solution}
 \begin{split}
 {\rm d} s^2 = e^{\mathcal{K}} \left(g r+\frac{c}{2 g r} \right)^2 {\rm d}t^2 - \frac{e^{-\mathcal{K}} {\rm d}r^2}{\left(g r+\frac{c}{2 g r} \right)^2} - e^{-\mathcal{K}} r^2 {\rm d} \sigma^2\ ,\\
 {\rm Re} X^{\Lambda} = H^{\Lambda} = \alpha^{\Lambda} + \frac{\beta^{\Lambda}}{r}, \qquad \qquad {\rm Re} F_{\Lambda} = 0\ ,\\
 \xi_{\Lambda} \alpha^{\Lambda} = - 1\ , \qquad \xi_{\Lambda}
  \beta^{\Lambda} = 0\ , \qquad F_{\Lambda} \left( -2 g^2 r \beta^{\Lambda} + c
  \alpha^{\Lambda}+g p^{\Lambda}\right)= 0\ ,
 \end{split}
 \end{align}
under the restriction $\xi_{\Lambda} p^{\Lambda} = 0$, with the toroidal area element given by \eqref{area-element}.

\section{Higher genus black holes}\label{6:higher}
The story of higher genus black holes is very similar in spirit to the cases with spherical and toroidal topology. We will therefore be very brief in this section, just marking some small differences. We again refer the interested reader to \cite{Caldarelli:1998hg} and \cite{Cacciatori:2009iz} for more details on higher genus black holes in minimal and non-minimal gauged supergravity.

In order to outline the main differences, we again consider minimal gauged supergravity and metrics of the form
\begin{equation}
{\rm d}s^2= U^2(r) \, {\rm d}t^2 -\frac{1}{U^2(r)} {\rm d}r^2 - r^2 {\rm d}\sigma^2\ , \qquad \qquad
U^2(r)= -1 + g^2 r^2 - \frac{2 \eta}{r} + \frac{q^2 + p^2}{r^2}\,,
\end{equation}
this time with a hyperbolic plane area element
\begin{equation}
{\rm d}\sigma^2 = {\rm d}\theta^2 + \sinh^2 \theta {\rm d}\varphi^2) \,.
\end{equation}
The electromagnetic 1-form (the graviphoton) and its corresponding field strength are:
\begin{equation}
A = \frac{q}{r} {\rm d}t + p \cosh \theta\ {\rm d}\varphi \,, \qquad F = \frac{q}{r^2}\ {\rm d}t \wedge {\rm d}r + p \sinh \theta\ {\rm d}\theta \wedge {\rm d}\varphi\ .
\end{equation}
In order to have a compact horizon, one again needs to periodically identify $\theta$ and $\varphi$ just as in the toroidal case. The main difference here is that the function $U(r)$ includes the term $-1$, which makes it much more easier for these solutions to develop event horizons. It turns out that the supersymmetry in this case requires \begin{equation}\eta = 0\ , \qquad p = \pm 1/(2 g)\ ,\end{equation}with an arbitrary electric charge $q$. For vanishing electric charge, this becomes a higher genus BPS black hole with a genuine event horizon and finite entropy. This is the only example of static supersymmetric black hole in minimal gauged supergravity. Again, these configurations have a generalization to non-constant scalars in gauged supergravity with vector multiplets, following precisely the same steps as in section \ref{6:Klemm}.


\part{Superalgebras and BPS Bounds}\label{part::3}

\chapter{Conserved supercharges} \label{chapter::BPS-general}
\section{Introduction}\label{sect7:intro}
In this short chapter we outline a novel procedure for determining the superalgebra of a given background solution in supergravity\footnote{To a certain extent this procedure was already used in \cite{Hollands:2006zu}, but we further extend it for the purpose of finding superalgebra commutation relations.}. This procedure relies on the Noether theorem and is completely general. There is no assumption about the supergravity theory or the background of interest. In the scope of this thesis we are going to apply our method to $D=4$ $N=2$ supergravity, focusing on the asymptotically Minkowski and AdS solutions of the previous chapters. We will see that our approach leads not only to unique answer for the superalgebra, but also to explicit expressions for the asymptotic charges in a given vacuum. These turn out to be automatically finite and well-defined, something previously not known to be obvious. In fact, there exists an explicit procedure for holographic renormalization \cite{Henningson:1998gx,Henningson:1998ey,deHaro:2000xn,Skenderis:2000in,Bianchi:2001de,Papadimitriou:2004rz,Papadimitriou:2005ii,Cheng:2005wk} of quantities on the boundary of AdS, developed for AdS/CFT applications. In our analysis such renormalization is shown to be superfluous.

Before going into details, it is worth explaining that a well-defined and algorithmic procedure for determining superalgebras was already developed in \cite{FigueroaO'Farrill:1999va} (see also \cite{Ortin}). It relies on finding the Killing vectors, $k^i$,  and spinors, $\epsilon_A$, of a given vacuum solution, which correspond to the spacetime and fermionic isometries. Each isometry has its corresponding abstract symmetry generator in the superalgebra, $B^i$ or $Q_A$. The commutator algebra of the Killing vectors and spinors directly gives the superalgebra structure constants,
\begin{align}
\begin{split}
           [B^i, B^j ] = f^{i j}{}_k B^k\ ,\\
            [Q_A, B^i ] = f_A{}^{i B} Q_B\ ,\\ \{Q_A,Q_B \} = f_{A B i} B^i\ .
        \end{split}
        \end{align}
This is similarly extended to include any other internal bosonic symmetries, such as gauge invariance.

While our proposed method is of course in full agreement with (and to a certain extent relying on) the procedure of \cite{FigueroaO'Farrill:1999va,Ortin}, we extend the known results and make them more explicit. The improvement is that we provide a more direct way of including possible central charges in the supercharge anticommutator. There is no longer need to use more general arguments and knowledge of the possible solutions (e.g.\ the historic path to finding the extension of the M-algebra via the M2 and M5 solutions can be substantially simplified). Furthermore, our method allows for very explicit evaluation of the anticommutator on given states, which to our best knowledge was not possible earlier.

In the next section we will therefore explain carefully our procedure. We keep discussion fully general and minimize notational details that will inadvertently depend on the explicit supergravity theory of interest.

\section{Supercurrents and charges from the Noether theorem}\label{sect:Noether procedure}
Given a lagrangian $\mathcal{L}(\phi,\partial_\mu\phi)$, depending on fields collectively denoted by $\phi$, we have that under general field
variations \begin{equation}\label{generic_field_variation} \delta \mathcal{L}= \sum_{\phi} \mathcal{E}_{\phi} \delta \phi + \partial_{\mu}
N^{\mu} \,\,\, , \end{equation} where $\mathcal{E}_{\phi}$ vanishes upon using the equation of motion of $\phi$, and \begin{equation}
N^\mu=\frac{\delta \mathcal{L}}{\delta (\partial_\mu \phi)}\delta\phi\ . \end{equation} Under a symmetry variation, parametrized by $\epsilon$,
the lagrangian must transform into a total derivative, such that the action is invariant for appropriate boundary conditions,
\begin{equation}\label{specific_field_variation} \delta_{\epsilon} \mathcal{L}= \partial_{\mu} K^{\mu}_{\epsilon} \,\, . \end{equation} Combining
this with  \eqref{generic_field_variation} for symmetry variations, we obtain \begin{equation} \sum_{\phi} \mathcal{E}_{\phi} \delta_{\epsilon}
\phi = \partial_{\mu} (K^{\mu}_{\epsilon}-N^{\mu}_{\epsilon}) \, . \end{equation} From the previous expression we see that the quantity
\begin{equation} J^{\mu}_{\epsilon} \equiv K^{\mu}_{\epsilon}-N^{\mu}_{\epsilon}\ , \end{equation} is the (on-shell) conserved current associated
with symmetry transformations. For the case of supersymmetry, we call $J^\mu_\epsilon$ the supercurrent. It depends on the (arbitrary) parameter
$\epsilon$ and is defined up to improvement terms of the form $\partial_{\nu} I^{\mu \nu}$ where $I$ is an antisymmetric tensor, as usual for
conserved currents. The associated conserved supercharge is then \begin{equation}\label{Q_from_J} 	\mathcal{Q} \equiv \int {\rm d}^3x
\,J^0_{\epsilon}(x)  \,. \end{equation} This supercharge should also generate the supersymmetry transformations of the fields,
\begin{equation}\label{7susy_variation_from_Q} 	\delta_{\epsilon} \phi = \{\mathcal{Q},\phi \}\ , \end{equation} via the classical Poisson (or
Dirac in case of constraints) brackets. Since the supercurrent and correspondingly the supercharge are defined up to improvement terms and
surface terms respectively, it is not directly obvious that the Noether procedure will lead to the correct supersymmetry variations using
\eqref{7susy_variation_from_Q}. In practice, one always has the information of the supersymmetry variations together with the supergravity
lagrangian. It is then possible to cross check the answers and thus derive uniquely the correct expression of the supercharge.

The supercharge $\mathcal{Q}$ as derived above is unique for each different theory, i.e.\ every supersymmetric lagrangian leads to a different $\mathcal{Q}$. For any background solution of a given theory, the supercharge will be a conserved quantity evaluated as a surface integral at the boundary of spacetime. However, at this stage $\mathcal{Q}$ is a bosonic quantity, containing in its definition the supersymmetry transformation parameter $\epsilon$. Thus, in order to evaluate $\mathcal{Q}$ on a given background solution (let $\phi_0$ denote the collection of fields), one also needs to know the parameter $\epsilon_{\phi_0}$ that corresponds to the field configuration $\phi_0$. Since $\mathcal{Q}$ is evaluated asymptotically, it turns out that we only need the asymptotic value of $\epsilon_{\phi_0}$, in fact the parameter $\epsilon_{\phi_0}$ needs not exist anywhere else except on the boundary of spacetime. This asymptotic spinor is the Killing spinor of the asymptotic background. It must satisfy the equation
\begin{equation}\label{7:susy_Killing_spinor_def} \delta_{\epsilon_{\phi_0}} \phi_0 \mid_{boundary} = 0\ .\end{equation}
The solution for the Killing spinor $\epsilon_{\phi_0}$ is always of the form
\begin{equation}\label{7:susy_Killing_spinor_const} \epsilon_{\phi_0} = \mathbf{M}(x) \epsilon_0\ ,\end{equation}
with $\mathbf{M}(x)$ a general spacetime dependent matrix, possibly carrying spinor and other types of indices, and $\epsilon_0$ an arbitrary constant spinor. Since we need only the asymptotic Killing spinor, it will turn out that there are large classes of different field configurations with the same $\epsilon_{\phi_0}$. To give a simple example, all asymptotically Minkowski solutions will have $\epsilon_{Minkowski}$. It is crucial therefore that the asymptotic vacuum preserves some supersymmetry such that a corresponding $\epsilon_{\phi_0}$ indeed exists. One can therefore define supercharges for every different vacuum configuration $\phi_{vac}$,
\begin{equation}\label{7:susy_charge_vac} \mathcal{Q}_{vac} (\phi_0) \equiv \mathcal{Q} (\phi_0|_{boundary} = \phi_{vac}, \epsilon_{\phi_{vac}})\ .\end{equation} This definition keeps the field configuration $\phi_0$ arbitrary, as long as it asymptotes to the given vacuum. Therefore at this stage $\mathcal{Q}_{vac}$ is a field dependent scalar quantity, whose dependence on the spinor $\epsilon$ is fixed by the asymptotic vacuum. We can also define its spinorial analog that typically appears in the superalgebra by stripping off the constant spinors $\epsilon_0$ in \eqref{7:susy_Killing_spinor_const},
\begin{equation}\label{7:susy_fermionic_charge_vac} \mathcal{Q}_{vac} (\phi_0) \equiv Q^T_{vac} (\phi_0) \epsilon_0 = \epsilon^T_0 Q_{vac} (\phi_0)\ .\end{equation}
The abstract supercharge $\tilde{Q}_{vac}$ in the superalgebra of the vacuum $\phi_{vac}$ can act explicitly on the state $| \phi_0 >$ whose fields asymptote to the vacuum via the eigenvalue equation,
\begin{equation}\label{7:eigenvalue} \tilde{Q}_{vac} | \phi_0 > = Q_{vac} (\phi_0) | \phi_0 >\ .\end{equation}
In explicit calculations one needs to know the Killing spinors of the asymptotic vacuum and the field configuration at hand to be able to evaluate $Q_{vac} (\phi_0)$.

It is then clear how the calculation of the supercharge anticommutators proceeds. The quantity $\{\mathcal{Q},\mathcal{Q} \}$ can be explicitly found via the Poisson (Dirac) brackets of the fundamental fields. It will result in a field dependent boundary integral that again includes the supersymmetry parameters $\epsilon$ and is fixed by the lagrangian. For a given asymptotic vacuum with known Killing spinor, one can then evaluate \begin{equation}\{\mathcal{Q}_{vac},\mathcal{Q}_{vac} \} (\phi_0) = \epsilon^T_0 \{Q_{vac}, Q_{vac} \} (\phi_0) \epsilon_0\ . \end{equation}
The (bosonic) matrix $\{Q_{vac}, Q_{vac} \} (\phi_0)$ now depends on the asymptotic vacuum and on the field configuration at hand. Every vacuum defines its own different superalgebra, where the anticommutator of the abstract quantities $\tilde{Q}_{vac}$ is given in terms of bosonic conserved charges, coming from the bosonic symmetries of the lagrangian. The eigenvalue equation
\begin{equation}\label{7:eigenvalue_new} \{ \tilde{Q}_{vac}, \tilde{Q}_{vac} \} | \phi_0 > = \{Q_{vac}, Q_{vac} \} (\phi_0) | \phi_0 > = \mathbf{B}_{vac} (\phi_0) | \phi_0 > \ ,\end{equation}
produces a matrix of bosonic symmetry eigenvalues $\mathbf{B}_{vac}$ for each field configuration $\phi_0$. Since it is a square of a hermitean operator, we further require $\mathbf{B}_{vac}$ to be positive definite for each field configuration that asymptotes to the given vacuum. The inequality
\begin{equation}\label{7:BPS_bound} \mathbf{B}_{vac} (\phi_0) \geq 0\ , \quad \forall \phi_0 |_{boundary} = \phi_{vac}\ ,\end{equation}
is the BPS bound for the corresponding vacuum. It projects out all negative norm states and keeps the states that are stably asymptotic to the vacuum state.

\chapter{BPS bounds in minimal gauged supergravity}\label{chapter::BPS-minimal}
\section{Introduction}
The procedure described in the previous chapter might seem somewhat abstract at this moment, but here and in the next chapter we will see very explicitly how to perform it both for minimal and non-minimal gauged supergravity for the Minkowski and AdS vacua. In this chapter we realize the general procedure outlined above in the case of minimal gauged supergravity. Due to some technical reasons, e.g.\ the form of the Killing spinors, we use slightly different conventions from the rest of the thesis. This simplifies substantially the discussion and allows us to concentrate on the physical meaning of our analysis.

The main motivation for our analysis comes from the following paradox. Similarly to BPS states in asymptotically flat spacetimes, the authors of
\cite{Kostelecky:1995ei} provided a BPS bound in asymptotically anti-de Sitter spacetime that in the static case reduces to:
\begin{equation}\label{bound_kost} M \geq \sqrt{Q_e^2+Q_m^2}\, . \end{equation} Supersymmetric configurations would have to saturate the bound
with $M^2=Q_e^2+Q_m^2$, for a given mass $M$, electric charge $Q_e$, and magnetic charge $Q_m$ in appropriate units. However, in  $N=2$ minimal
gauged supergravity, Romans \cite{Romans:1991nq} found two supersymmetric solutions, one of which does not saturate the BPS bound, and therefore
an apparent paradox arises. The main aim of this chapter is to resolve this conflict. The resolution of the paradox will lie in understanding the
BPS ground states of gauged supergravity, the associated superalgebras, and in a proper definition of the mass in asymptotically AdS$_4$
spacetimes, as we will explain in the coming sections.

Minimally gauged supergravity has only two bosonic fields, the metric and the graviphoton $A_\mu$. Shortly repeating our analysis from the previous part of the thesis to ease the reading, the most general static and spherically
symmetric solution of Einstein's equation with a negative cosmological constant and an electromagnetic field is given by, \begin{equation}\label{form_of_metric}
    {\rm d}s^2= U^2(r)\, {\rm d}t^2 - U^{-2}(r)\, {\rm d}r^2 - r^2 ({\rm d}\theta^2 + \sin^2 \theta {\rm d}\phi^2)\,,
\end{equation} with \begin{equation}\label{prefactor2}
 U^2(r) = 1- \frac{2M}{r} +\frac{Q_e^2+Q_m^2}{r^2} +g^2r^2 \ ,
\end{equation} and with nonvanishing components of the graviphoton \begin{equation}\label{ansatz_vectors}
    A_{t} = \frac{Q_e}{r} \, ,\,\,\,\,\,\,\,\,\,\,\,\,\, A_{\phi} = -Q_m \cos\theta \,\,.
\end{equation} In this class, there are two solutions that preserve some fraction of supersymmetry \cite{Romans:1991nq}. The first one is the
so-called AdS ``electric Reissner--Nordstr\"om (RN)''  solution, for which the magnetic charge $Q_m$ is set to zero and $M = Q_e$ so that the
factor $U$ has the form: \begin{equation}\label{prefactor} U^2 = \left(1- \frac{Q_e}{r}\right)^2  +g^2r^2 \, ,\,\,\,\,\,\,\,\,\,\,\,\,\, Q_m = 0
\,\,. \end{equation} This solution preserves one half of the supersymmetries (it is 1/2 BPS). Clearly, it saturates the BPS bound
\eqref{bound_kost}. Notice that the function $U(r)$ has no zeros. Therefore, there is no horizon and the point $r=0$ is a naked
singularity\footnote{In AdS spacetimes, supersymmetry does not seem to provide a cosmic censorship, contrary to most cases in asymptotically flat
spacetimes \cite{Kallosh:1992ii}. Whether cosmic censorship in AdS$_4$ can be violated is still an open problem, see e.g. \cite{Hertog:2004gz}.
This issue however has nothing to do with the paradox or contradiction mentioned above.}. Asymptotically, for $r\rightarrow \infty$, the solution
is that of pure AdS$_4$, with cosmological constant $\Lambda=-3g^2$ in standard conventions.

The second supersymmetric solution is the so-called  ``cosmic dyon'', having zero mass $M$ but nonzero fixed magnetic charge $Q_m = \pm 1/(2g)$.
Such a solution will never satisfy the BPS bound \eqref{bound_kost}. Moreover, the electric charge $Q_e$ can take an arbitrary value:
\begin{equation}\label{prefactor_magn} M=0\,,\,\,\,\,\,\,\, Q_m=\pm1/(2g)\,,\,\,\,\,\,\,\,\,   U^2  = \left(gr+\frac{1} {2gr}  \right)^2 +
\frac{Q_e^2}{r^2} \,\,. \end{equation} Again, there is a naked singularity at $r=0$. However, asymptotically, when $r\rightarrow \infty$, the
solution does not approach pure AdS$_4$, due to the presence of the magnetic charge. Instead, the solution defines another vacuum, since $M=0$,
but this vacuum is topologically distinct from AdS$_4$ in which $M=Q_m=0$. For this reason\footnote{One may argue that ground states should not
have naked singularities. Clearly, this discussion is related to cosmic censorship in AdS, which we mentioned in the previous footnote. It is
important to disentangle this discussion from the derivation of the BPS bounds. In fact, we already know that in matter coupled supergravity there is
a magnetic ground state without naked singularities, c.f.\ chapter \ref{6:Klemm}.}, we call this vacuum magnetic anti-de Sitter,
or mAdS$_4$.

The cosmic dyon solution is 1/4 BPS, i.e. it preserves two out of eight supercharges. For both the electric RN-AdS and the cosmic dyon, the
Killing spinors were explicitly constructed in \cite{Romans:1991nq}. The fact that the BPS bound \eqref{bound_kost} is not satisfied for the
cosmic dyon leads to a contradiction since states that admit a Killing spinor should saturate the BPS bound.

In this chapter we show that the cosmic dyon in fact satisfies a different BPS bound that follows from a superalgebra different from the usual
AdS$_4$ superalgebra. We will determine the new BPS bound starting from the explicit calculation of the supercharges and computing the
anticommutator\footnote{An alternative approach based on the Witten-Nester energy was proposed in an unpublished paper by Izquierdo, Meessen and Ort\'{\i}n, leading to similar conclusions.}. In summary, to state the main result of this chapter, for stationary configurations, the new BPS bounds are: \begin{itemize} \item For asymptotically AdS$_4$ solutions with vanishing magnetic charge, $Q_m=0$, the BPS bound is \begin{equation}\label{bound1} M \geq |Q_e|+g
|\vec{J}|\ , \end{equation} where $\vec{J}$ is the angular momentum. \item For asymptotically magnetic AdS$_4$ solutions with $Q_m=\pm 1/(2g)$,
the BPS bound is simply \begin{equation}\label{bound2} M \geq 0 \ , \end{equation} with unconstrained electric charge $Q_e$ and angular momentum
$\vec{J}$. \end{itemize}

Other values for the magnetic charges are not considered. The quantization condition requires it to be an integer multiple of the minimal unit,
$Q_m=n /(2g); n \in {\mathbb Z}$, but it is not known if any other supersymmetric vacua can exist with $n\neq 0,1$.

The meaning of the BPS bound is not that all solutions to the equations of motion must automatically satisfy \eqref{bound1} or \eqref{bound2}.
Rather, one constructs a physical configuration space consisting of solutions that satisfy a BPS bound like  \eqref{bound1} or \eqref{bound2}, as mentioned in the previous chapter.

Our procedure also provides  a new way of defining asymptotic charges in AdS$_4$ backgrounds with automatically built-in holographic
renormalization, somewhat different than the procedure developed in \cite{Henningson:1998gx,Henningson:1998ey,deHaro:2000xn,Skenderis:2000in,Bianchi:2001de,Papadimitriou:2004rz,Papadimitriou:2005ii,Cheng:2005wk}. The same technique can also be applied to non-minimal gauged
supergravity, which will be the topic of the next chapter.

Additional motivation to study more closely the magnetic AdS$_4$ case and its superalgebra is provided by the AdS/CFT correspondence. There are
suggestions in the literature \cite{Hartnoll:2007ai} that excitations of the dual theory are relevant for condensed matter physics in the presence of
external magnetic field, e.g. quantum Hall effect and Landau level splitting at strong coupling. A better understanding of the mAdS superalgebra
could then provide us with more insights about the dual field theory.

\subsubsection{Plan of this chapter}
First, in section \ref{sect8:minimal}, we explain our conventions about the minimal gauged supergravity we consider. We follow the procedure of chapter \ref{chapter::BPS-general} and derive the form of the asymptotic supercharges and their anticommutator. We then take special interest in deriving explicitly the form of the anticommutator for spherically symmetric AdS configurations, showing how the difference between AdS and magnetic AdS arises. We then extend the results to include the solutions with non-spherical symmetry in section \ref{sect8:non-spherical}. We see that for toroidal topology we find somewhat different results compared to both the AdS and the mAdS cases, while the hyperbolic solutions are in complete analogy to the spherical mAdS. In the last section of this chapter we take a more abstract approach and write down the full superalgebra for all the asymptotic vacua that are discussed. Some of the intermediate calculations that facilitate the discussion in this chapter are left for the appendices and referred to when needed.

\section{Minimal gauged supergravity}\label{sect8:minimal}
First we compute the supercurrent from the lagrangian of minimal $D=4$ $N=2$ gauged supergravity following the conventions of \cite{Ortin} (which
is written in 1.5-formalism): \begin{align} \begin{split} S = \int {\rm d}^4x& \,e \big[R(e,\omega) +6g^2 + \frac2e \epsilon^{\mu \nu \rho
\sigma} \overline{\psi}_{\mu} \gamma_5 \gamma_{\nu} (\hat{\mathcal{D}}_{\rho}+igA_{\rho}\sigma^2)\psi_{\sigma}- \mathcal{F}^2 \\
 &- \frac{1}{2e}\epsilon^{\mu\nu\rho\sigma}\overline{\psi}_{\rho} \gamma_5 \sigma^2 \psi_{\sigma}(i \overline{\psi}_{\mu} \sigma^2 \psi_{\nu} -
 \frac{1}{2e}\epsilon_{\mu\nu}{}^{\tau \lambda}  \overline{\psi}_{\tau} \gamma_5 \sigma^2 \psi_{\lambda}   ) \big]\,,
\end{split} \end{align} where $\overline{\psi}= i {\psi}^{\dagger} \gamma_0, e=\sqrt{{\rm det}g_{\mu \nu}},$  \begin{equation}\label{minimal_supercovariant_derivative}
\hat{\mathcal{D}}_{\rho} = \partial_{\rho} - \frac14 \omega_{\rho}^{ab} \gamma_{ab}-\frac{i}{2} g \gamma_{\rho}\ , \end{equation} and
\begin{align} \begin{split} \mathcal{F}_{\mu \nu} & = \partial_{\mu} A_{\nu} - \partial_{\nu} A_{\mu}+ i \overline{\psi}_{\mu} \sigma^2
\psi_{\nu} -\frac{1}{2e} \epsilon_{\mu\nu}{}^{\rho \sigma} \overline{\psi}_{\rho} \gamma_5 \sigma^2 \psi_{\sigma} = \\ &= F_{\mu \nu} + i
\overline{\psi}_{\mu} \sigma^2 \psi_{\nu} -\frac{1}{2e} \epsilon_{\mu\nu}{}^{\rho \sigma} \overline{\psi}_{\rho} \gamma_5 \sigma^2
\psi_{\sigma}\,. \end{split} \end{align} The spin connection satisfies \begin{equation} {\rm d}e^a - \omega^a{}_b \wedge e^b =0 \,\,
\end{equation} for a given vielbein $e^a = e_{\mu}^a {\rm d}x^{\mu}$. The $g^2$-term in the Lagrangian is related to the presence of a negative
cosmological constant $\Lambda= -3 g^2$.

In most of our calculations, such as in the supercurrents and supercharges, we only work to lowest order in fermions since higher order terms
vanish in the expression of the (on shell) supersymmetry algebra, where we set all fermion fields to zero. The supersymmetry variations are:
\begin{equation}\label{susy_gravitino} \delta_{\epsilon} \psi_{\mu} = \widetilde{\mathcal{D}}_\mu \epsilon= (\partial_{\mu} - \frac14
\omega_{\mu}^{ab} \gamma_{ab}-\frac{i}{2} g \gamma_{\mu} + ig A_{\mu} \sigma^2 +\frac14 F_{\lambda \tau} \gamma^{\lambda \tau} \gamma_{\mu}
\sigma^2) \epsilon \,\,\,, \end{equation} \begin{equation} \delta_{\epsilon} e_{\mu}^a = -i \overline{\epsilon} \gamma^a \psi_{\mu}\,\,\,,
\end{equation} \begin{equation}\label{susy_vector} \delta_{\epsilon} A_{\mu} = -i \overline{\epsilon} \sigma^2 \psi_{\mu}\,\,\,. \end{equation}
$U(1)$ gauge transformations act on the gauge potential and on the spinors in this way: \begin{equation} A_{\mu}'= A_{\mu} + \partial_{\mu}
\alpha\,\,, \end{equation} \begin{equation} \psi_{\mu}'= e^{-ig\alpha \sigma^2} \psi_{\mu}\,\,. \end{equation} We use the conventions in which
all the spinors are real Majorana ones\footnote{In conventions here, the two real gravitini in the gravity multiplet are packaged together in the
notation: \begin{equation} \psi_{\mu}=\left( \begin{array}{c}
 \psi_{\mu}{}^1 \\
\psi_{\mu}{}^2  \\ \end{array} \right)\, , \end{equation} where each gravitino is itself a 4-component Majorana spinor. Similar conventions are
used for the supersymmetry parameters. In other words, the $SU(2)_R$ indices are completely suppressed. }, and the gamma matrix
conventions and identities of appendix \ref{appendixA}.

The quantities $N^{\mu}$ and $K^{\mu}$ for this theory are: \begin{equation} 	N^{\mu} = \frac {\partial\mathcal{L}}{\partial_{\mu}\omega}
\delta\omega+ 2 \epsilon^{\mu \nu \rho \sigma} \overline{\psi}_{\nu} \gamma_5 \gamma_{\rho} \widetilde{\mathcal{D}}_{\sigma} \epsilon  +4 i e
F^{\mu\nu}\overline{\epsilon}\sigma^2 \psi_{\nu} \,\,. \end{equation} \begin{equation} 	K^{\mu} = \frac
{\partial\mathcal{L}}{\partial_{\mu}\omega} \delta\omega - 2 \epsilon^{\mu \nu \rho \sigma} \overline{\psi}_{\nu} \gamma_5 \gamma_{\rho}
\widetilde{\mathcal{D}}_{\sigma} \epsilon  +4 i e F^{\mu\nu}\overline{\epsilon}\sigma^2 \psi_{\nu} \,\,. \end{equation} Hence the supercurrent
has the form: \begin{equation}\label{scurrent} 	J^{\mu} = - 4 \epsilon^{\mu \nu \rho \sigma} \overline{\psi}_{\nu} \gamma_5 \gamma_{\rho}
\widetilde{\mathcal{D}}_{\sigma} \epsilon  \,\,. \end{equation} This expression is gauge invariant due to the cancelation between the variation
of the gravitino, the vector field and the supersymmetry parameter. Furthermore we can show that the supercurrent is conserved
\begin{equation}
        \partial_{\mu}J^{\mu} = \partial_{\mu} (-4 \epsilon^{\mu \nu \rho \sigma}
\overline{\psi}_{\nu} \gamma_5 \gamma_{\rho} \widetilde{\mathcal{D}}_{\sigma} \epsilon) = 0 \,
\end{equation}
if we enforce the equation of motion for $\overline{\psi}_{\nu}$ and use the antisymmetry of the Levi-Civita symbol.

The Dirac brackets defined for the given theory read (we only need those containing gravitinos): \begin{equation} 	\{
\psi_{\mu}(x),{\psi}_{\sigma}(x') \}_{t=t'}= 0\,\,, \end{equation} \begin{equation} 	\{ \overline{\psi}_{\mu}(x), \overline{\psi}_{\sigma}(x')
\}_{t=t'}= 0\,\,, \end{equation} \begin{equation}\label{conjugate} 	\{ \psi_{\mu}(x), 2 \epsilon^{0 \nu \rho \sigma} \overline{\psi}_{\rho}(x')
\gamma_5 \gamma_{\sigma}
 \}_{t=t'}= \delta_{\mu}{}^{\nu} \delta^3(\vec{x}-\vec{x'})\,\,.
\end{equation} We can now check if \eqref{7susy_variation_from_Q} holds with the above form of the supercurrent. It turns out that, up to overall
normalization, we indeed have the right expression without any ambiguity of improvement terms.  We only need to rescale,  since the factor of $4$
in \eqref{scurrent} does not appear in the supersymmetry variations \eqref{susy_gravitino}-\eqref{susy_vector}. The supercharge is then defined
as the volume integral\footnote{For volume and surface integrals in this chapter, we use the notation that
\begin{align}
{\rm d}\Sigma_\mu=\frac{1}{6 e^2}\epsilon_{\mu\nu\rho\sigma}\,{\rm d}x^\nu \wedge {\rm d}x^\rho\wedge {\rm d}x^\sigma\ ,\qquad
{\rm d}\Sigma_{\mu\nu}=\frac{1}{2 e^2}\epsilon_{\mu\nu\rho\sigma}\,{\rm d}x^\rho\wedge {\rm d}x^\sigma\ .
\end{align}}
\begin{align}\label{7susy_charge}
    \mathcal{Q} \equiv 2 \int_V {\rm d} \Sigma_{\mu} \epsilon^{\mu \nu \rho \sigma} \overline{\psi}_{\sigma} \gamma_5 \gamma_{\rho}
    \widetilde{\mathcal{D}}_{\nu} \epsilon  \,\, \stackrel{e.o.m.}{=} 2 \oint_{\partial V} {\rm d} \Sigma_{\mu \nu} \epsilon^{\mu \nu \rho
    \sigma} \overline{\psi}_{\sigma} \gamma_5 \gamma_{\rho} \epsilon\ ,
\end{align} where the second equality follows from the Gauss theorem via the equations of motion (in what follows we will always deal with
classical solutions of the theory). The Dirac bracket of two supersymmetry charges is then straightforwardly derived as the supersymmetry
variation of \eqref{7susy_charge}: \begin{equation}\label{basic_susy_anticommutator_minimal} \{\mathcal{Q},\mathcal{Q} \}  = 2 \oint_{\partial V}
{\rm d}\Sigma_{\mu \nu}(\epsilon^{\mu \nu \rho \sigma} \overline{\epsilon} \gamma_5 \gamma_{\rho} \widetilde{\mathcal{D}}_{\sigma} \epsilon)\ ,
\end{equation} which is again a boundary integral.

The above formula is reminiscent of the expression for the Witten-Nester energy \cite{Witten:1981mf,Nester:1982tr}, which has already been implicitly assumed
to generalize for supergravity applications \cite{Argurio:2008zt,Sezgin:2009dj} (see also \cite{Cheng:2005wk}). Thus, the correspondence between BPS bounds and positivity of
Witten-Nester energy is confirmed also in the case of minimal gauged $N=2$ supergravity by our explicit calculation of the supercharge
anticommutator.

\section{Two different BPS bounds with spherical symmetry}\label{sect8:twobounds}
In this section, we derive two BPS bounds based on the two BPS sectors that we consider. What is relevant for the BPS bound are the properties of
the asymptotic geometries and corresponding Killing spinors. The Killing spinors of AdS$_4$ and magnetic AdS$_4$ (``cosmic monopole'') are given
in App. \ref{app:E}, see also \cite{Romans:1991nq}. Since only the asymptotics are important, we can set $Q_e=0$ in the cosmic dyon
solution. The AdS$_4$ solution is characterized by $M=Q_e=Q_m=0$, while mAdS$_4$ by $M=0$, $Q_e=0$, $Q_m=\pm 1/(2g)$. The corresponding Killing
spinors take a very different form: \begin{equation}\label{Killing_spinors_AdS1}
    \epsilon_{AdS} = e^{\frac{i}{2} arcsinh (g r) \gamma_1} e^{\frac{i}{2} g t \gamma_0} e^{-\frac{1}{2} \theta \gamma_{1 2}}
e^{-\frac{1}{2} \varphi \gamma_{2 3}} \epsilon_0\ , \end{equation} \begin{equation}\label{Killing_spinors_magnAdS1}
    \epsilon_{mAdS} =  \frac{1}{4} \sqrt{g r + \frac{1}{2 g r}} (1 + i \gamma_1) (1\mp i \gamma_{2 3} \sigma^2) \epsilon_0\ ,
\end{equation} where $\epsilon_0$ is a doublet of constant Majorana spinors, carrying $8$ arbitrary parameters. From here we can see that AdS$_4$
is fully supersymmetric and its Killing spinors show dependence on all the four coordinates. mAdS$_4$ on the other hand is only $1/4$ BPS: its
Killing spinors satisfiy a double projection that reduces the independent components to $1/4$ and there is no angular or time dependence. We will
come back to this remarkable fact in section \ref{sect:Superalgebra}.

The form of the Killing spinors is important because the bracket of two supercharges is a surface integral at infinity
\eqref{basic_susy_anticommutator_minimal}. Writing out the covariant derivative in \eqref{basic_susy_anticommutator_minimal}, one obtains
\begin{equation}\label{basic_susy_anticommutator_expanded_deriv} \{ \mathcal{Q},\mathcal{Q}\}  = 2 \oint_{\partial V} {\rm d}\Sigma_{\mu \nu}
\left[ \epsilon^{\mu \nu \rho \sigma} \overline{\epsilon} \gamma_5 \gamma_{\rho}(\partial_{\sigma} - \frac14 \omega_{\sigma}^{ab}
\gamma_{ab}-\frac{i}{2} g \gamma_{\sigma} + ig A_{\sigma} \sigma^2 +\frac14 F_{\lambda \tau} \gamma^{\lambda \tau} \gamma_{\sigma} \sigma^2)
\epsilon \right]\ , \end{equation} and it depends on the asymptotic value of the Killing spinors of the solution taken into consideration.
Therefore the superalgebra will be different in the two cases and there will be two different BPS bounds.

The procedure to compute the BPS bound is the following. From \eqref{7susy_charge} we have a definition of the supercharges $\mathcal{Q}_{AdS}
(\epsilon_{AdS})$ and $\mathcal{Q}_{mAdS} (\epsilon_{mAdS})$. We will then make use of the following definition for the fermionic supercharges
$Q_{AdS}, Q_{mAdS}$: \begin{equation}
    \mathcal{Q}_{AdS} \equiv Q^{T}_{AdS} \epsilon_0 = \epsilon^T_0 Q_{AdS}\ , \qquad \mathcal{Q}_{mAdS} \equiv Q^{T}_{mAdS} \epsilon_0 =
    \epsilon^T_0 Q_{mAdS}\ ,
\end{equation} i.e. any spacetime and gamma matrix dependence of the bosonic supercharges $\mathcal{Q}$ is left into the corresponding fermionic
$Q$. We are thus able to strip off the arbitrary constant $\epsilon_0$ in any explicit calculations and convert the Dirac brackets for
$\mathcal{Q}$ into an anticommutator for the spinorial supercharges $Q$ that is standardly used to define the superalgebra. Therefore now we
compute the surface integrals \eqref{basic_susy_anticommutator_minimal} for the Killing spinors of AdS$_4$ and mAdS$_4$ respectively. After
stripping off the $\epsilon_0$'s, we find the anticommutator of fermionic supercharges given explicitly in terms of the other conserved charges
in the respective vacua. The BPS bound is then derived in the standard way by requiring the supersymmetry anticommutator to be positive definite,
see e.g. \cite{Gauntlett:2000ch} for details.

\subsection{Asymptotically AdS$_4$ states}

We now derive the resulting supersymmetry algebra from the asymptotic spinors of AdS$_4$. For this we use the general expression
\eqref{basic_susy_anticommutator_expanded_deriv} for the Dirac brackets of the supersymmetry charges, together with the asymptotic form of the
Killing spinors, \eqref{Killing_spinors_AdS1}. Inserting the Killing spinors $\epsilon_{AdS}$ of  \eqref{Killing_spinors_AdS1} in
\eqref{basic_susy_anticommutator_expanded_deriv}, we recover something that can be written in the following form:
\begin{equation}\label{AdS_susy_anticommutator_minimal} \{ \mathcal{Q}, \mathcal{Q} \}  = -i \, \overline{\epsilon_0} ( A + B_{a} \gamma^{a} + C
\gamma^5+ D_{i j} \gamma^{i j} + E_{i} \gamma^{0 i} + F_{a} \gamma^{a 5}){\epsilon_0}\ , \end{equation} where the charges $A,B,...$ can be
written down explicitly from the surface integral \eqref{basic_susy_anticommutator_expanded_deriv}. They will define the electric charge ($A$),
momentum ($B$), angular momentum ($D$, with $i,j=1,2,3$ spatial indices), and boost charges ($E$). The charge $C$ would correspond to a magnetic
charge, which we assumed to vanish by construction. Without the charge $F_a$, the above bracket will fit in the $OSp(2|4)$ superalgebra (see more
below). We will therefore take as definition of asymptotically AdS solutions  the ones for which $F_{a}$ vanishes. This choice of fall-off
conditions is similar to the case of $N=1$ supergravity, where the asymptotic charges are required to generate the $Osp(1|4)$ superalgebra
\cite{Henneaux:1985tv}. Extensions of the $N=2$ superalgebra where the charges $C$ and $F_a$ are non-zero have been discussed in \cite{Dibitetto:2010sp}.

From the previous expression we see that conserved charges like $Q_e$, $M$ et cetera will arise as surface integrals of the five terms (or their
combinations) appearing in the supercovariant derivative. We are going to see how this works analyzing each term in the supercovariant
derivative, explicitly in terms of the ansatz of the metric \eqref{form_of_metric} and vector fields \eqref{ansatz_vectors}. This will provide us
with a new definition of the asymptotic charges in AdS$_4$ with no need to use the holographic renormalization procedure anywhere. As an explicit
example one can directly read off the definition of mass $M \equiv B_0/(8\pi)$ from the explicit form of the asymptotic Killing spinors. In the
stationary case,
\begin{align} \label{mass_ads} \begin{split} M &= \frac{1}{8\pi}\lim_{r\rightarrow\infty} \oint e\ {\rm d}\Sigma_{t r} \{ e^t_{[ 0} e^r_1
e^{\theta}_{2]} + \sin \theta e^t_{[ 0} e^r_1 e^{\varphi}_{3]} \\ &+2 g^2 r e^t_{[0} e^r_{1]} - \sqrt{g^2 r^2 +1} (\omega_{\theta}^{a b} e^t_{[ 0}
e^r_a e^{\theta}_{b]} + \omega_{\varphi}^{a b} e^t_{[ 0} e^r_a e^{\varphi}_{b]} ) \}\ .\end{split} \end{align}

We are going to take into consideration both static and rotating solutions, but we will carry out our procedure and explain the calculation in
full detail only the case of the electric RN-AdS black hole and comment more briefly on the rotating generalizations.

\begin{itemize}
  \item Electric RN-AdS\\

Here we take into consideration solutions of the form \eqref{form_of_metric} - \eqref{ansatz_vectors} with zero magnetic charge, $Q_m=0$. We
now evaluate the various terms in \eqref{basic_susy_anticommutator_expanded_deriv}.

To begin, it is easy to determine the piece concerning the field strength, namely \begin{equation}\label{basic_susy_anticommu} 2\oint {\rm
d}\Sigma_{\mu \nu} \left[\epsilon^{\mu \nu \rho \sigma} \overline{\epsilon}(t,r,\theta,\varphi) \gamma_5 \gamma_{\rho} \frac14 F_{\lambda
\tau} \gamma^{\lambda \tau} \gamma_{\sigma} \sigma^2 \epsilon(t,r,\theta,\varphi) \right]\ . \end{equation} Inserting the Killing spinors for AdS$_4$ described in \eqref{Killing_spinors_AdS1}, and exploiting the Clifford algebra relations we get
\begin{equation}\label{AdS_strippedQ} 2\oint {\rm d}\Sigma_{tr} \, \overline{\epsilon}_{AdS}(t, r,\theta,\varphi) e \, F^{tr} \sigma^2 \epsilon_{AdS} (t,r,\theta,\varphi)\ = 8 i \pi \overline{\epsilon_0} Q_e \sigma^2 {\epsilon_0} \ ,
\end{equation} with the definition of the electric charge
\begin{equation}
Q_e= \frac{1}{4\pi}\oint_{S^2}F^{tr} r^2 \sin \theta\ {\rm d}\theta {\rm d}\phi \ .
\end{equation}

From here we can identify the term $A = - 8 \pi Q_e \sigma^2$ appearing in \eqref{AdS_susy_anticommutator_minimal}.

Next we consider the term containing the ``bare" gauge field $A_{\mu}$. This gives a possible contribution to the charge $F_a$ in
\eqref{AdS_susy_anticommutator_minimal}. As we mentioned above, we assumed this contribution to vanish for asymptotically AdS solutions. One
can explicitly check this for the class of electric RN-AdS solutions given in \eqref{form_of_metric}, since the only nonzero component of the
vector field is $A_t$ (see \eqref{ansatz_vectors}), hence \begin{equation}\label{A_term} \oint {\rm d}\Sigma_{tr} \left[ \epsilon^{tr \rho
\sigma} \overline{\epsilon}_{AdS}(t,r,\theta,\varphi) \gamma_5 \gamma_{\rho}  i g A_{\sigma} \sigma^2 \epsilon_{AdS}(t,r,\theta,\varphi)
\right] =0\,. \end{equation}

The term with the partial derivative $\partial_{\sigma} $ in \eqref{basic_susy_anticommutator_expanded_deriv} in the supercovariant
derivative gives nonvanishing contributions for $\sigma=\theta, \varphi$ and  it amounts to the integral:
\begin{equation}\label{AdS_strippedderivative} 2\oint {\rm d}\Sigma_{tr} \left[ \epsilon^{tr \rho \sigma}
\overline{\epsilon}_{AdS}(t,r,\theta,\varphi) \gamma_5 \gamma_{\rho}  \partial_{\sigma} \epsilon_{AdS}(t,r,\theta,\varphi) \right] = -2 i
\oint  r \overline{\epsilon_0}  \gamma_0 {\epsilon_0}  \sin \theta \, {\rm d}\theta {\rm d}\varphi \ . \end{equation} Clearly, this term will
contribute, together with other terms, to the mass.

The integral containing the spin connection is:
$$
-\frac24 \oint {\rm d}\Sigma_{tr} \epsilon^{t r \rho \sigma}   \overline{\epsilon}_{AdS} (t,r,\theta,\varphi) \gamma_5 \gamma_{\rho} \, \omega_{\sigma}^{ab} \, \gamma_{ab} \, \epsilon_{AdS}(t,r,\theta,\varphi) =
$$ \begin{equation}\label{AdS_strippedmass} = 2 i \oint \overline{\epsilon_0} \gamma^0  r
\sqrt{1+g^2r^2} \sqrt{1 +g^2r^2-\frac{2M}{r} +\frac{Q_e^2}{r^2}}\, {\epsilon_0} \sin \theta \, {\rm d}\theta {\rm d}\varphi \,,
\end{equation} where we have used \begin{equation}\label{sinh_relation} e^{ i\, arcsinh(gr) \gamma_{1}} = \sqrt{1+g^2r^2} +ig\gamma_1 r \,,
\end{equation} and the value of the spin connection: \begin{equation}\label{omega_AdS1} \omega_t^{0 1} = U \partial_r U, \quad
\omega_{\theta}^{1 2} = - U , \quad \omega_{\varphi}^{13} = - U \sin \theta, \quad \omega_{\varphi}^{23} = - \cos \theta\ . \end{equation}
Also \eqref{AdS_strippedmass} will contribute to $B^0$, and therefore to the mass.

The last contribution of the supercovariant derivative, the term proportional to $g\gamma_\sigma$, yields
\begin{equation}\label{contrib_gamma_explicit}
-2\oint {\rm d}\Sigma_{tr} \epsilon^{t r \rho \sigma} \overline{\epsilon}_{AdS}(t,r,\theta,\varphi) \frac{ig}{2} \gamma_5 \gamma_{\rho} \gamma_{\sigma} \, \epsilon_{AdS}(t,r,\theta,\varphi) =
-2 i \oint \overline{\epsilon_0} \gamma^0 r^3 g^2 {\epsilon_0}\,  \sin \theta \, {\rm d}\theta {\rm d}\varphi\ .
\end{equation} In deriving this we have used the formula $ \gamma^{tr \mu} \gamma_{\mu} = 2
\gamma^{tr}$. Again, this term contributes to the mass formula.

Collecting all the terms that contribute to the mass (the derivative term \eqref{AdS_strippedderivative}, the sum of the spin connection term
\eqref{AdS_strippedmass}, and the gamma term \eqref{contrib_gamma_explicit}) gives rise to: \begin{equation}\label{contrib_total_mass}
 2i \oint  \overline{\epsilon_0} \gamma^0 \left[ r \sqrt{1+g^2r^2} \sqrt{1 +g^2r^2-\frac{2M}{r}
+\frac{Q_e^2}{r^2}}- r^3 g^2- r \right] {\epsilon_0}\, \sin \theta \, {\rm d}\theta {\rm d}\varphi\,. \end{equation} The integral has to be
performed on a sphere with $r \rightarrow \infty$. Taking this limit one can  see that in this expression all the positive powers of $r$ are
canceled. Hence all possible divergences cancel out, and we are left with a finite contribution. In this sense, our method provides a
holographic renormalization of the mass. In the cases we can compare, our method agrees with previously known results.

Performing the integral on the remaining finite part we find: \begin{equation}\label{AdS_mass_cancellation} -8i \pi \overline{\epsilon_0} M
\gamma^{0} {\epsilon_0} = -i \overline{\epsilon_0}\gamma^0  B_0 \, {\epsilon_0} \,. \end{equation} To sum up, for the electric RN-AdS
solution, the brackets between supercharges read: \begin{align}\label{AdS_susy_anticommutator_minimal_AdSRN} \begin{split} \{ \mathcal{Q},
\mathcal{Q} \} = -8 \pi i\overline{\epsilon_0} (M \gamma^{0} - Q_e \sigma^2 ){\epsilon_0}\, \\ \Rightarrow \{\epsilon^T_0 Q, Q^T \epsilon_0
\} = 8 \pi \epsilon_0^{T} (M - Q_e \gamma^0 \sigma^2)\ \epsilon_0\ . \end{split} \end{align} Now we can strip off the constant linearly
independent doublet of spinors $\epsilon_0$ on both sides of the above formula to restore the original $SO(2)$ and spinor indices:
\begin{equation}\label{AdS_susy_anticommutator_minimal_AdSRN_indices} \{ Q^{A \alpha}, Q^{B \beta} \} = 8 \pi \left( M \delta^{AB}
\delta^{\alpha \beta} - i\,Q_e \epsilon^{AB} (\gamma^0)^{\alpha \beta}\right)\ . \end{equation} This expression coincides with the one
expected from the algebra $OSp(2|4)$ (see \eqref{osp24} in the next section) if we identify $M_{-1 0} = 8 \pi M$ and $T^{1 2} = 8 \pi Q_e$.

The BPS bound for the electric RN-AdS solution is then\footnote{See e.g. \cite{Gauntlett:2000ch} for details on the general procedure of
deriving of BPS bounds from the superalgebra.}: \begin{equation}\label{BPS_bound_AdSRN} M \geq |Q_e|\ . \end{equation} The state that
saturates this bound, for which $M=|Q_e|$, preserves half of the supersymmetries, i.e. it is half-BPS. It is the ground state allowed by
\eqref{BPS_bound_AdSRN} and represents a naked singularity. All the excited states have higher mass and are either naked singularities or
genuine black holes.

It is interesting to look at the case of extremal black holes, in which inner and outer horizon coincide. This yields a relation between the
mass and charge, which can be derived from the solution given in \eqref{form_of_metric}. Explicit calculation gives
the following result \cite{Caldarelli:1998hg}:
  \begin{equation}\label{extremality_bound_RNAdS}
    M_{extr} = \frac{1}{3 \sqrt{6} g} (\sqrt{1+12 g^2 Q_e^2} +2) (\sqrt{1+12 g^2 Q_e^2} - 1)^{1/2}\ .
  \end{equation}
This lies above the BPS bound unless $Q_e = 0$, in which case we recover the fully supersymmetric AdS$_4$ space. Thus,
  \begin{equation}\label{RNAdSBPSvsExtr}
    M_{extr}> M_{BPS} \ .
  \end{equation}

  \item Kerr-AdS\\
  The Kerr-AdS black hole is an example of a stationary spacetime without
charges but with non-vanishing angular momentum. It is most standardly written in Boyer-Lindquist-type coordinates and we refer to
\cite{Caldarelli:1998hg} for more details. More details on how to calculate the angular momenta from the anticommutator of the supercharges
can be found in App. \ref{app:F}. The BPS bound is straightforward to find also in this case, leading to
  \begin{equation}\label{kerr-AdS_BPS_bound}
    M \geq g |\vec{J}|\ ,
  \end{equation}
  where the BPS state satisfies $M = g |\vec{J}|$ and in fact corresponds to a
singular limit of the Kerr-AdS black hole because the AdS boundary needs to rotate as fast as the speed of light \cite{Hawking:1998kw}. Note
that in general for the Kerr black hole we have $|\vec{J}| = a M$, where $a$ is the rotation parameter appearing in the Kerr solution in
standard notation. Thus $M = g |\vec{J}|$ implies $a = 1/g$, which is exactly the singular case. All the excited states given by $a < 1/g$
are however proper physical states, corresponding to all the regular Kerr-AdS black holes, including the extremal one. Thus the BPS bound is
always satisfied but never saturated by any physical solution of the Kerr-AdS type,
  \begin{equation}\label{KerrAdSBPSvsExtr}
    M_{extr} > M_{BPS}\ ,
  \end{equation}
as is well-known.

  \item KN-AdS\\
  The BPS bound for Kerr-Newman-AdS (KN-AdS) black holes\footnote{See again \cite{Caldarelli:1998hg}
for more detailed description of the KN-AdS black holes.} is a bit more involved due to the presence of both electric charge and angular
momentum. We will not elaborate on the details of the calculation which is straightforward. The resulting
  BPS bound is
  \begin{equation}\label{KN-AdS_BPS_bound}
    M \geq |Q_e| + g |\vec{J}| = |Q_e| + a g M\ ,
  \end{equation}
  and the ground (BPS) state is in fact quarter-supersymmetric. The BPS bound in general does not
coincide with the extremality bound, which in the case of the KN-AdS black holes is a rather complicated expression that can be found in
\cite{Caldarelli:1998hg,Hawking:1998kw}. Interestingly, the BPS bound and the extremality bound coincide at a finite non-zero value for the
mass and charge (with $ag<1$),
  \begin{equation}\label{critical_charge_KNAdS}
    |Q_{e, crit}| \equiv \sqrt{\frac{a}{g}} \frac{1}{1 - a g}\ .
  \end{equation}
  Now we have two distinct possibilities for the relation between the BPS
state and the extremal KN-AdS black hole depending on the actual value for the electric charge (there is exactly one BPS state and exactly
one extremal black hole for any value of charge $Q_e$):
  \begin{align}\label{KNAdSBPSvsExtr}
   \begin{split}
    M_{extr} > M_{BPS}\ ,& \qquad  |Q_e| \neq |Q_{e, crit}| \ ,\\
      M_{extr} = M_{BPS}\ ,& \qquad  |Q_e| = |Q_{e, crit}|\ .
  \end{split}
  \end{align}
So for small or large enough electric charge the BPS solution will be a naked singularity and the extremal black hole will satisfy but not
saturate the BPS bound, while for the critical value of the charge the extremal black hole is supersymmetric and all non-extremal solutions
with regular horizon will satisfy the BPS bound.

\end{itemize}

\subsection{Magnetic AdS$_4$}\label{sect8:magnAdS} Unlike the standard AdS$_4$ case above, the Killing spinors of magnetic AdS$_4$ already break $3/4$ of the
supersymmetry, c.f. \eqref{Killing_spinors-magnAdS}. The projection that they obey is, \begin{equation}\label{projection_magneticAdS}
    \epsilon_{mAdS} = P \epsilon_{mAdS}\ , \qquad \qquad P \equiv \frac{1}{4}(1+ i \gamma_1) (1\mp i \gamma_{2 3} \sigma^2)\ ,
\end{equation} for either the upper or lower sign, depending on the sign of the magnetic charge. Furthermore, one has the following properties of
the projection operators, \begin{align}\label{proj} \begin{split}
    P^{\dag} P&= P^{\dag} i \gamma_1 P = \pm P^{\dag}
i \gamma_{2 3} \sigma^2  P = \pm P^{\dag} (-i \gamma_0 \gamma_5 \sigma^2)  P= P\ , \end{split} \end{align} and all remaining quantities of the
form $P^{\dag} \Gamma P$ vanish, where $\Gamma$ stands for any of the other twelve basis matrices generated by the Clifford algebra.

These identities allow us to derive, from \eqref{basic_susy_anticommutator_expanded_deriv}, the bracket
\begin{equation}\label{anticommut_magnetic_AdS}
  \{ \mathcal{Q} , \mathcal{Q} \} =  \overline{P\epsilon}_{0} \gamma_0 (-i \,8 \pi )  M P \epsilon_0 \qquad
  \Rightarrow \qquad\{\epsilon^T_0 P Q, (P Q)^T \epsilon_0 \} = \epsilon^T_0 (8 \pi M) P \epsilon_0\ .
\end{equation} provided that the mass is given by
\begin{align} \label{mass_mads}\begin{split} M &= \frac{1}{8\pi} \lim_{r\rightarrow\infty}  \oint e\ {\rm d}\Sigma_{t r} \left( g r + \frac{1}{2 g r} \right) \bigg(
2 g (A_{\theta} e^t_{[ 0} e^r_2 e^{\theta}_{3]} + A_{\varphi} e^t_{[ 0} e^r_2 e^{\varphi}_{3]}) \\ &+ \frac{\sin \theta}{g} e^t_{[ 0} e^r_1 e^{\theta}_{2}
e^{\varphi}_{3]} + 2 g e^t_{[0} e^r_{1]} - (\omega_{\theta}^{a b} e^t_{[ 0} e^r_a e^{\theta}_{b]} + \omega_{\varphi}^{a b} e^t_{[ 0} e^r_a
e^{\varphi}_{b]} ) \bigg)\ . \end{split} \end{align}
This expression simplifies further if we choose to put the vielbein matrix in an upper triangular form, such that we have nonvanishing
$e_t^{0,1,2,3}, e_r^{1,2,3}, e_\theta^{2,3}, e_\varphi^3,$ and the inverse vielbein has only components $e_0^{t,r,\theta,\varphi},
e_1^{r,\theta,\varphi}, e_2^{\theta,\varphi}, e_3^{\varphi}$. The mass is then \begin{align} M &= \frac{1}{8\pi} \lim_{r\rightarrow\infty}  \oint e\ {\rm d} \Sigma_{t r}\left( g r + \frac{1}{2 g r} \right)\left( \frac{\sin \theta}{g} e^t_{0} e^r_1 e^{\theta}_{2} e^{\varphi}_{3} +2 g e^t_{0} e^r_{1} -
(\omega_{\theta}^{1 2} e^t_{0} e^r_1 e^{\theta}_{2} + \omega_{\varphi}^{1 3} e^t_{0} e^r_1 e^{\varphi}_{3} ) \right)\ . \end{align} Notice that
this mass formula is different from the one for asymptotically AdS$_4$ spacetimes.

Stripping off the parameters $\epsilon_0$ in \eqref{anticommut_magnetic_AdS}, leaves us with a matrix equation in spinor space. Due to the
projection operators, one is effectively reducing the number of supercharges to two instead of eight. These two supercharges are scalars, since
the Killing spinors are invariant under rotation as they don't depend on the angular coordinates (see also in the next section). Denoting them by
$Q^1$ and $Q^2$,  the anticommutator then becomes \begin{equation}\label{QQ-mads} \{ Q^I , Q^J \} = 8 \pi M \delta^{IJ} \,\,,\qquad I,J=1,2\ .
\end{equation} Hence the BPS bound is just \begin{equation}\label{BPS_bound_magnetic_AdS}
  M \geq 0\, .
\end{equation} Saturating the bound leads to a quarter-BPS solution. None of the other conserved charges, i.e. the electric charge and (angular)
momentum, influences the BPS bound due to the projection relation \eqref{proj}. Thus $Q_e$ and $\vec{J}$ can be completely arbitrary.

In particular, for the case of of the Reissner-Nordstr\"om solution \eqref{form_of_metric} with fixed magnetic charge $Q_m= \pm 1/(2g)$ and $Q_e$
arbitrary, the mass integral \eqref{mass_mads} yields $$ \frac{1}{4\pi} \lim_{r\rightarrow\infty}  \oint  \left( gr+ \frac{1}{2gr} \right) \left(
r \sqrt{1 + g^2 r^2 - \frac{2M}{r}+ \frac{Q_e^2+1/(4g^2)}{r^2} }- g \,r^2- \frac{1}{2g} \right) \sin \theta \, {\rm d}\theta {\rm d}\varphi = $$
\begin{equation} = \frac{1}{4\pi} \oint \left(-g^2r^3 - \frac{r}{2} + M - \frac{r}{2} + g^2r^3 + r \right) \sin \theta \, {\rm d}\theta {\rm
d}\varphi = M \,. \end{equation} This is exactly the mass parameter $M$ appearing in \eqref{prefactor2}. The supersymmetric solutions found by
Romans (the so-called cosmic monopole/dyons) have vanishing mass parameter hence indeed saturate the BPS bound \eqref{BPS_bound_magnetic_AdS}.

Of course in the context of a rotating black hole vanishing mass results in vanishing angular momentum due to the proportionality between the
two, i.e. an asymptotically mAdS Kerr-Newman with non-zero angular momentum spacetime can never saturate the BPS bound
\eqref{BPS_bound_magnetic_AdS}. Nevertheless, excitations over the magnetic AdS$_4$ include all Reissner-Nordstr\"{o}m and Kerr-Newman AdS black
holes that have fixed magnetic charge $2 g Q_m = \pm 1$ and arbitrary (positive) mass, angular momentum and electric charge. All these solutions
satisfy the magnetic AdS$_4$ BPS bound.

\section{Non-spherical BPS bounds}\label{sect8:non-spherical}
As explained in the previous part of the thesis, the situation with supersymmetric asymptotically AdS solutions with toroidal and hyperbolic symmetry is slightly different. Magnetic ground states as mAdS$_4$, considered above for spherical symmetry, do not appear in the case of toroidal topology. Contrary to this, in the case of hyperbolic topology the only BPS ground state has non-vanishing magnetic charge, i.e.\ pure hyperbolic AdS does not admit Killing spinors. Thus, apart from the two cases with spherical symmetry discussed above, we have only two AdS-like ground states with non-spherical symmetry. Their corresponding Killing spinors can be found in appendix \ref{app:E}. We now turn to each of them separately, keeping the discussion short since it is very similar in spirit and outcome to the previous section.

\subsection{Toroidal AdS$_4$}
As explained in chapter \ref{chapter::blackholes_AdS}, the AdS ground state with flat topology is called RiAdS and has vanishing mass, electric and magnetic charges. Its Killing spinor breaks half of the supersymmetry due to the toroidal compactification of coordinates. It is derived in details in appendix \ref{app:E} and reads
\begin{equation}\label{spinortorus}
\epsilon_{RiAdS}  = \sqrt{r} \left( \frac{1+i\gamma_1}{2} \right) \epsilon_0 = \sqrt{r}P \epsilon_0 \,.
\end{equation}
This result agrees with the one found in \cite{Caldarelli:1998hg}. The toroidal black holes can be seen as excitations over the background characterized by these Killing spinors. To find the BPS bound, we plug the spinors \eqref{spinortorus} in the formula for the Dirac bracket of two supercharges \eqref{basic_susy_anticommutator_expanded_deriv}. We first note the relations involving the projector \eqref{projection-torus}:
\begin{align}\label{property1}
\begin{split}
 P i \gamma_1 P  =  P\ , \qquad    P \gamma_{02} P =  P ( -i\gamma_{012}) P = \gamma_{02} P\ ,\\  P \gamma_{03} P= P (-i\gamma_{013}) P = \gamma_{03} P \ , \qquad
P \gamma_{23} P= P i \gamma_{1 2 3} P =  \gamma_{23} P\ .
\end{split}
\end{align}
All the other gamma matrices between two projectors give zero: this strongly limits the number of terms present in the superalgebra. The anticommutator between two supercharges can now be computed. Due to the projection identities \eqref{property1} and the symmetries of the gamma matrices\footnote{$\gamma^{02},\gamma^{03}$ are symmetric in their spinor indices, while $\gamma^{23}$ is antisymmetric (see App.\ \ref{appendixA} for our gamma matrix conventions). The four terms in \eqref{susy_commutator} are therefore the only non-vanishing contributions from \eqref{basic_susy_anticommutator_expanded_deriv}.} only four terms appear, and the result is
\begin{equation}\label{susy_commutator}
\{Q,Q\}= 2 \overline{(P \epsilon_0)}(- i M \gamma^0 - i P_2 \gamma^{2} - i P_3 \gamma^{3} - Z \gamma^{5} \sigma^2) P \epsilon_0 \,.
\end{equation}
 The mass $M$ has the following expression:
\begin{equation}\label{mass_min}
M= \frac12 \lim_{r\to \infty}{\oint {\rm d} \Sigma_{tr} e_{0}^{t} e_{1}^{r} \left(\, 2 g r -r (\omega_{x}^{1 2} e_{2}^{x} + \omega_{y}^{13} e_{3}^{y} + \omega_{y}^{12} e_{2}^{y} ) \,  \right)}\ ,
\end{equation}
and the central charge $Z$ reads:
\begin{equation}\label{centralcharge}
Z=  \lim_{r\to \infty}\oint_{T^2} r F\ ,
\end{equation}
where $F$ is the vector field strength written as a two-form. The above formulas are valid after choosing an upper triangular vielbein, as in \eqref{vielbeins}. We omit the formulas for the momenta $P_2$ and $P_3$, which are straightforward to derive from \eqref{basic_susy_anticommutator_expanded_deriv} and \eqref{spinortorus}, but are not particularly insightful since they vanish for static solutions.

The main outcome of our analysis is therefore that for static black branes and toroidal black holes the BPS bound
\begin{align}\label{BPS-riads}
    M \geq |Z|\ ,
\end{align}
must be satisfied.

To give an explicit example how the BPS bound constrains the solutions space, we now restrict our attention to static solutions of the form \eqref{formbrane} with zero magnetic charge. The above mass formula can be explicitly evaluated:
\begin{align}
\begin{split}
M & =  \frac{1}{2} \lim_{r\to \infty} {\oint {\rm d}x {\rm d}y \mathcal{V} \,r^2 \,\left[ 2 g r +r \left(\omega_x^{12} \frac{|\tau|}{r \sqrt{ {\rm Im} \tau}} + \omega_y^{13} \frac{\sqrt{{\rm Im}\tau}}{r |\tau|} \right) \right]} = \\
& =  \lim_{r\to \infty} {\oint {\rm d}x {\rm d}y \mathcal{V} \, \left( gr^3 - r^2 \sqrt{r^2 g^2 -\frac{2 \eta}{r} + \frac{q_e^2}{r^2}} \right)=  \mathcal{V} \int_0^1 {\rm d}x \int_0^1{\rm d}y \, \, \eta} = \mathcal{V}\, \eta \,.
\end{split}
\end{align}
We see that the divergent part cancels (notice that the dependence on $\tau$ drops out, as a consequence of the modular symmetry) and we are left with the finite quantity $\eta$ for the mass density $M/\mathcal{V}$. Furthermore, the formula for the central charge \eqref{centralcharge} gives zero when computed on the ansatz \eqref{formbrane},
\begin{equation}\label{cencharge}
Z = p r \mathcal{V} =  0\ ,
\end{equation}
since $p$ is forced to vanish for asymptotically RiAdS solutions. We then have $M = \eta \mathcal{V}$, $P_2=P_3=Z =0$ for these solutions. Consequently, the BPS bound is just:
\begin{equation}
  \eta \geq 0 \,.
\end{equation}
 Note that the BPS bound does not involve the electric charge, similarly to the mAdS case. Moreover, this bound also holds in the decompactification limit for black branes, where the mass density $\eta$ is a finite number even if the mass $M$ is infinite. The BPS bound is saturated for $\eta=0$ with an arbitrary electric charge $q$. The resulting spacetime has a naked singularity whenever $q \neq 0$, as already explained in chapter \ref{chapter::blackholes_AdS}.

\subsection{Hyperbolic mAdS$_4$}
For completeness sake, this short subsection is devoted to the hyperbolic version of the magnetic AdS vacuum. From the explicit calculation of the Killing spinor of the vacuum solution, with vanishing mass and electric charge and magnetic charge $p = \pm 1/2 g$,
\begin{equation}
    \epsilon_{Hyp-mAdS} =  \frac{1}{4} \sqrt{g r - \frac{1}{2 g r}} (1 + i \gamma_1) (1\mp i \gamma_{2 3} \sigma^2) \epsilon_0\ ,
\end{equation}
one can see that the situation is very much analogous to the spherical mAdS case. The supersymmetry commutator and BPS bound trivially follow from the same steps taken in section \ref{sect8:magnAdS}. The bound is again
 \begin{equation}
  M \geq 0\, .
\end{equation}

\section{Superalgebras}\label{sect:Superalgebra}

\subsection{AdS$_4$}\label{sect:AdS_algebra}

The procedure we used to find the BPS bound determines also the superalgebras of AdS$_4$ and mAdS$_4$, which are found to be different. In fact,
given the Killing spinors and Killing vectors, there is a general algorithm to determine the superalgebra, see \cite{FigueroaO'Farrill:1999va} and chapter 13 in \cite{Ortin}.

For what concerns the pure AdS$_4$, in $N=2$ gauged supergravity the superalgebra is $OSp(2|4)$, which contains as bosonic subgroup $SO(2,3)
\times SO(2)$: the first group is the isometry group of AdS$_4$ and the second one corresponds to the gauged R-symmetry group that acts by
rotating the two gravitinos. The algebra contains the  generators of the $SO(2,3)$ group $M_{MN}$ ($M,N = -1, 0,1,2,3$),  and
$T^{AB}=-T^{BA}=T\epsilon^{AB}, A,B=1,2$, the generator of  $SO(2)$. Furthermore, we have supercharges $Q^{A \alpha}$ with $A=1,2$ that are
Majorana spinors. The non-vanishing (anti-)commutators of the $OSp(2|4)$ superalgebra are: \begin{align}\label{osp24} \begin{split} [Q^{A
\alpha},T] &= \epsilon^{AB} Q^{B \alpha} \\ [M_{MN},M_{PQ} ] &= - \eta_{MP} M_{NQ} - \eta_{NQ} M_{MP} + \eta_{MQ} M_{NP} +\eta_{NP} M_{MQ} \\
[Q^{A \alpha},M_{MN} ] &= \frac12 { (\hat{\gamma}_{MN})}^{\alpha}{}_{\beta} Q^{A \beta} \\ \{ Q^{A \alpha},Q^{B \beta} \} &=  \delta^{AB}
(\hat{\gamma}^{MN}C^{-1})^{\alpha \beta} M_{MN} - (C^{-1})^{\alpha\beta} T \epsilon^{AB}\ , \end{split} \end{align} where $\eta_{MN}=  {\rm
diag}(1,1,-1,-1,-1)$, the gamma matrices are $\hat{\gamma}_M\equiv \{\gamma_5,  i \gamma_{\mu} \gamma_5\}$, and ${\hat
\gamma}_{MN}=\frac{1}{2}[\hat\gamma_M,\hat\gamma_N]$. $T$ does not have the role of a central
charge, as it does not commute with the supercharges. Nevertheless it is associated to the electric charge\footnote{If we perform a
Wigner--In\"{o}n\"{u} contraction of the algebra, $T^{AB}$ gives rise to a central charge in the Poincar\'e superalgebra. See \cite{Ortin}
for further details.}. The isometry group of AdS$_4$ is $SO(2,3)$, isomorphic to the conformal group in three dimensions, whose generators are 3
translations, 3 rotations, 3 special conformal transformations (conformal boosts) and the dilatation.

\subsection{mAdS$_4$}\label{sect:mAdS_algebra} In the case of mAdS$_4$, the symmetry group is reduced.
Spatial translations and boosts are broken, because of the presence of a magnetic monopole. There are 4 Killing vectors related to the invariance
under time translations and rotations. The isometry group of this spacetime is then $ \mathbb{R} \times SO(3)$. Furthermore, we have also gauge
invariance. The projector \eqref{projection_magneticAdS} reduces the independent components of the Killing spinors to 1/4, consequently the
number of fermionic symmetries of the theory is also reduced. We have denoted the remaining two real supercharges with
$Q^I$ ($I=1,2$). To sum up, the symmetry generators of mAdS$_4$ are: \begin{itemize}
\item the angular momentum $ J_i $, $ i=1,2,3$, \item the Hamiltonian $H$, \item the gauge transformation generator $T$, \item the two
supercharges $ Q^I$ where $I=1,2$. \end{itemize} From \eqref{QQ-mads} the anticommutator between two supercharges is
\begin{equation}\label{magnetic_AdS_superalgebra} \{ Q^I , Q^J \} = H \delta^{IJ} \ . \end{equation} Since  mAdS$_4$ is static and spherically
symmetric, we have the commutation relations \begin{equation} [ H , J_i ] = 0\ ,\qquad [ J_i , J_j ] = \epsilon_{ijk} \,J_k \ . \end{equation}
The following commutators are then determined by imposing the Jacobi identities: \begin{equation}\label{scalar_charges} [ Q^I , J_i ] = [ Q^I , H
] = 0 \ . \end{equation} Next, we add the gauge generator $T$ to the algebra. Because of gauge invariance, we have the commutators
\begin{equation} [ T , J_i ] = [T , H ] = 0 \ . \end{equation} From the Jacobi identities one now derives that \begin{equation} [Q^I , T ] =
\epsilon^{IJ} Q^{J} \ , \end{equation} with a fixed normalization of $T$. This commutator also follows from the observation that gauge
transformations act on the supersymmetry parameters in gauged supergravity, together with the fact that $T$ commutes with the projection operator
$P$ defined in the previous section.

The first commutator in \eqref{scalar_charges} implies that the supercharges $Q^I$ are singlet under rotations. This is a consequence of the fact
that the mAdS$_4$ Killing spinors have no angular dependence \cite{Romans:1991nq}. Group theoretically, this follows from the fact that the group
of rotations entangles with the $SU(2)_R$ symmetry, as explained in \cite{deWit:2011gk}.

\subsection{RiAdS$_4$}
The full superalgebra of RiAdS can be most clearly presented as follows. After the projection we have only 4 real supercharges present, which we label  $Q_1,Q_2,Q_3,Q_4$. The non-vanishing supercharge anticommutators can then be read from \eqref{susy_commutator}:
\begin{align}
\begin{split}
\{Q_1, Q_1 \} &= \{Q_3, Q_3 \}= M + P_2\,, \qquad \{Q_2, Q_2 \} = \{Q_4, Q_4 \} = M - P_2\,,\\
 \{Q_1, Q_2 \} &=  \{Q_3, Q_4 \} = P_3\,, \qquad \{Q_1, Q_4 \} = - Z\ , \qquad \{Q_2, Q_3 \} = Z\ ,
\end{split}
\end{align}
 and the action of the gauged $U(1)_R$ symmetry leads to
\begin{equation}
 [Q_1,T]= Q_3\,, \quad [Q_2,T]= Q_4\ , \quad  [Q_3,T]= -Q_1\,, \quad [Q_4,T]= -Q_2\ .
\end{equation}
The other commutators vanish due to the form of the Killing vectors, the fact that gauge transformations commute with translations and compatibility with the super Jacobi identities. This shows that indeed $Z$ is a central charge, similar to the magnetic central charge in the Poincar\'{e} superalgebra.

Due to the toroidal compactification, the theory is endowed also with modular invariance. The metric and the Killing spinor are invariant under transformations that act on both the parameter $\tau$ and the coordinates $(x,y)$:
\begin{equation}\label{transfmodular}
\tau \rightarrow  \frac{a \tau +b}{ c \tau +d} \,, \qquad
\left( \begin{array}{c} x \\
y \\
\end{array} \right) \rightarrow
M
\left( \begin{array}{c} x \\
y \\
\end{array} \right) =
\left( \begin{array}{cc} a & b \\
c & d \\
\end{array} \right)
\left( \begin{array}{c} x \\
y \\
\end{array} \right)\,,
\end{equation}
with the condition $ad-bc=1$, i.e. $M \in SL(2, Z)$. For a finite volume of the torus, the superalgebra can be interpreted as corresponding to a modular invariant quantum mechanics theory in one dimension, since one can further reduce the $3d$ theory, dual to RiAdS, on the two compact spatial dimensions. This interpretation is no longer valid in the infinite volume limit where the boundary is an infinite flat plane.

\subsection{Hyperbolic mAdS$_4$}
As explained above, the situation for hyperbolic mAdS is fully analogous to the one of spherical mAdS. This also holds for their superalgebras, the only difference of course being that spherical symmetry encoded in the structure constants for the angular momenta commutator changes into hyperbolic symmetry. We will therefore not elaborate further on hyperbolic mAdS$_4$ solutions as their behavior is already captured in our analysis of spherical mAdS. This remains being true in the next chapter when we consider more complicated solutions allowing for running scalars.

\chapter{BPS bounds in the matter-coupled theory}\label{chapter::BPS-matter}
\section{Introduction and general results}\label{sect:BPS_bounds}
We continue our analysis of BPS bounds in the most general case of matter coupled electrically gauged supergravity, i.e.\ coming back to the lagrangian \eqref{lagr} and our original notation. After understanding more clearly the vacuum structure of gauged supergravities in the previous chapter, here we concentrate more on black hole solutions with non-trivial scalar profiles. We find an unexpected similarity between BPS solutions with very different asymptotics.

As we show in the following, the superalgebra structure does not change when considering more general matter
couplings in the theory. However, the explicit definition of the asymptotic charges ($M, Q_e,$ etc.) of a given solution depends directly on the field content. We first derive the form of the supersymmetry anticommutator for all possible solutions of gauged supergravity
with vectors and hypers. Then we focus on the special cases of Minkowski, AdS, mAdS, and RiAdS asymptotics where we
evaluate the anticommutator explicitly. These calculations show that the hypermultiplets do not produce
additional central charges in the superalgebra. We are also able to formulate renormalized expressions for the
mass in AdS and mAdS. Our results in AdS are in exact agreement with the techniques of holographic
renormalization \cite{Henningson:1998gx,Henningson:1998ey,deHaro:2000xn,Skenderis:2000in,Bianchi:2001de,Papadimitriou:2004rz,Papadimitriou:2005ii,Cheng:2005wk}. On the other hand, the mAdS mass takes a different form and in some examples leads to qualitatively different results that have no analog in previous literature.

From our knowledge of the minimal case and with the help of the susy variations we can derive explicitly the supercharge, as done in appendix \ref{app:G}. The original expression for the supercharge is somewhat lengthy and non-suggestive. However, using the equations of motion for the gravitinos we can cast the supercharge into a much simpler form as a surface integral (see the appendix for the technical details).

The important quantity for our purposes here is the Dirac bracket of two supercharges. It can be derived from the supercharge \eqref{susy_charge} and takes the remarkably simple form
\begin{align}\label{basic_susy_anticommutator_general}
\{\mathcal{Q},\mathcal{Q} \}  = \oint_{\partial V} {\rm d}\Sigma_{\mu \nu} \epsilon^{\mu \nu \rho \sigma}
\overline{\varepsilon}^A \gamma_{\rho} \widetilde{\mathcal{D}}_{\sigma} \varepsilon_A + h.c.
\end{align}
This is the main general result of this chapter. It can be explicitly evaluated on every spacetime that has an asymptotic Killing spinor.

Compared with the corresponding expression in the minimal case, \eqref{basic_susy_anticommutator_general} is just a straightforward generalization. A priori, one could expect some more radical changes due to the presence of vector and hypermultiplets, but this is not the case. We already see that the main conclusions of the previous chapter remain the same, with the difference that the definition of the asymptotic charges will generalize to accommodate for the possibility of non-constant scalars\footnote{Note that for a solution with constant scalars (both in the vector and in the hypermultiplet sector) \eqref{basic_susy_anticommutator_general} is equivalent with the result for the minimal case. Thus, the only difference between the asymptotic charges in minimal and non-minimal supergravity lies in the possibility for non-constant scalar profiles.}. In order to give more precise statements, we need to plug in the explicit Killing spinors of interest in the general Dirac bracket \eqref{basic_susy_anticommutator_general}.

 In the following sections we consider more carefully the cases of Minkowski, AdS$_4$, mAdS$_4$, and RiAdS$_4$ asymptotics, paying special attention to the asymptotic charges in stationary solutions. In each of the cases we give an explicit example from the study of black holes
as an application of our results. Somewhat surprisingly, we are able to find a very simple unified formula for
the mass of supersymmetric black hole spacetimes in all spherically symmetric cases. This also leads to a better conceptual
understanding of the difference in the mass in AdS and mAdS spacetimes. We conclude with some remarks on the
connection of our results to alternative approaches in literature.

 \section{Asymptotically flat solutions}\label{sect:flat}
 \subsection{General analysis}
 Here we will be interested in the superalgebra and asymptotic charges of Minkowski spacetime. In the context of electrically gauged supergravity
with vector and hypermultiplets the necessary conditions for a Minkowski vacuum were derived in chapter \ref{chapter::n2vac},
\begin{align}
k^i_\Lambda \bar{L}^\Lambda =0\ , \qquad \tilde k^u_\Lambda L^\Lambda &=0\ , \qquad
P^x_\Lambda =0\ ,
\end{align}
together with constant scalars, vanishing field strengths and flat $\mathbb{R}^{1,3}$ metric. These are now the conditions that asymptotically flat solutions will have to satisfy as $r\rightarrow \infty$ (we always work in spherical coordinates in this section).

The Majorana Killing spinors of Minkowski in spherical coordinates are
\begin{equation}\label{Killing_spinors-Minkowski}
    \tilde{\epsilon}^{1,2}_{M} = e^{-\frac{1}{2} \theta \gamma_{1 2}} e^{-\frac{1}{2}
    \varphi \gamma_{2 3}}\ \tilde{\epsilon}^{1,2}_0\ ,
\end{equation}
where $\tilde{\epsilon}^{1,2}_0$ are two arbitrary and linearly independent constant Majorana spinors. We will use the notation $\tilde{\epsilon}^A$ for Majorana spinors and $\varepsilon_A, \varepsilon^A$ for the positive/negative chirality Weyl spinors that are used in our notation. The chiral spinors are related to the Majorana ones through
\begin{align}\label{spinor_relation}
    \varepsilon_A \equiv \frac{1+\gamma_5}{2}\ \tilde{\epsilon}^A\ , \qquad \varepsilon^A \equiv \frac{1-\gamma_5}{2}\ \tilde{\epsilon}^A\ , \qquad (\varepsilon_A)^* = \varepsilon^A\ .
\end{align}

Having the Killing spinors we can now in principle plug \eqref{Killing_spinors-Minkowski} in \eqref{basic_susy_anticommutator_general} and derive the supercharge anticommutator directly. Of course, we already know the general answer from the Poincar\'{e} superalgebra,
\begin{align}\label{poincare_algebra}
\{ Q^{A \alpha},Q^{B \beta} \} =  \delta^{AB}
(i \gamma^{M}C^{-1})^{\alpha \beta} P_{M} - \epsilon^{AB} (({\rm Re}\mathcal{Z} + i \gamma^5 {\rm Im} \mathcal{Z})(C^{-1}))^{\alpha\beta}\ ,
\end{align}
where $C$ is the charge conjugation matrix, $P_M$ is the momentum operator, and $\mathcal{Z}$ is the complex central extension of the superalgebra. The explicit eigenvalues of the operators $P_M$ and $\mathcal{Z}$ for any asymptotically flat solution can be computed now from \eqref{basic_susy_anticommutator_general}. The additional $U(1)$ and $Sp(1)$ connections in \eqref{basic_susy_anticommutator_general} from the matter multiplets can potentially lead to contributions to the supersymmetry anticommutator that are not of the type \eqref{poincare_algebra}. Since we know that Minkowski asymptotics will necessarily lead to the Poincar\'{e} superalgebra it follows that these additional connections must fall off fast enough so that they do not contribute. \eqref{poincare_algebra} can in fact be taken as a definition for asymptotically flat spacetimes. In practice, the condition for the fall off of the connections will be equivalent with imposing the metric to approach Minkowski space. This will be illustrated more clearly with an explicit example.

In the next subsection we give the explicit expressions for $P_0, \mathcal{Z}$ in \eqref{poincare_algebra} for the stationary case, but one can straightforwardly derive the asymptotic charges in full generality if needed.

 \subsection{Stationary solutions}
 For stationary solutions we find that the supersymmetry anticommutator takes the following form\footnote{We rescale the central charges for convenience.}:
 \begin{align}\label{susy_commut_Mink}
    \{ Q^{A \alpha}, Q^{B \beta} \} = \delta^{A B} 8 \pi M (i \gamma^0 \mathcal{C}^{-1})^{\alpha \beta} - \epsilon^{A B} 8 \pi (({\rm Re}\mathcal{Z} + i \gamma^5 {\rm Im} \mathcal{Z}) (\mathcal{C}^{-1}))^{\alpha \beta}\ ,
 \end{align}
  where the complex central charge is given by
  \begin{align}\label{central_charge}
    \mathcal{Z} = \frac{1}{4\pi} \lim_{r \rightarrow \infty} \oint_{S^2} T^- = \lim_{r \rightarrow \infty}  \left( L^{\Lambda} q_{\Lambda} - M_{\Lambda} p^{\Lambda} \right)\ ,
  \end{align} as derived in detail in \cite{Ceresole:1995ca}\footnote{Note that the charges $q_{\Lambda}$ and $p^{\Lambda}$ in \eqref{central_charge} are the standard electric and magnetic charges as commonly defined in literature. The electric charges come from the dual field strengths $G_{\Lambda \mu \nu} \equiv i \epsilon_{\mu \nu \rho \sigma} \frac{\delta \mathcal{L}}{\delta F^{\Lambda}_{\rho \sigma}}$. See chapter \ref{chapter::supergravity} for more details.}. The derivation of the central charge from \eqref{basic_susy_anticommutator_general} is a bit subtle and uses the fact that $\widetilde{\mathcal{D}}_{\mu} \varepsilon_A$ contains a $T_{\mu \nu}^-$ term, while $\widetilde{\mathcal{D}}_{\mu} \varepsilon^A$ contains $T_{\mu \nu}^+$. This eventually leads to $\int \left(T^- (1+ \gamma_5) + T^+ (1-\gamma_5) \right) \sim {\rm Re}\mathcal{Z} + i \gamma^5 {\rm Im} \mathcal{Z}$. This calculation picks out the electric and magnetic charge carried by the graviphoton, which explicitly depend on the asymptotic values of the vector multiplet scalars.

 The mass, on the other hand, remains unaffected by scalars,
 \begin{align}\label{mass_Mink}
    M = \frac{1}{8\pi}\lim_{r\rightarrow\infty} \oint {\rm d}\Sigma_{t r} \bigg( e^t_{[ 0} e^r_1 e^{\theta}_{2]} + \sin \theta\ e^t_{[ 0} e^r_1 e^{\varphi}_{3]} - (\omega_{\theta}^{a b} e^t_{[ 0} e^r_a e^{\theta}_{b]} + \omega_{\varphi}^{a b} e^t_{[ 0} e^r_a e^{\varphi}_{b]} ) \bigg)\ ,
 \end{align} just as in the minimal case.

 The BPS bound, as always for stationary asymptotically flat solutions, is
 \begin{align}\label{bound_Mink}
    M \geq |\mathcal{Z}|\ .
 \end{align}
Note that the hypermultiplet sector seems to be completely decoupled from the above calculations since the hypers do not influence the asymptotic charges. This suggests that the stabilization of the hypers at a particular supersymmetric point in moduli space as described in chapter \ref{chapter::solution_generator} might be the generic situation in this case.

\subsection{Black hole example}
As a standard example we can just briefly glance through the single-centered supersymmetric black holes of chapter \ref{chapter::blackholes_flat}, c.f.\ \cite{Behrndt:1997ny}. First we take the most standard case of a static black hole as a warm up for the static examples in AdS and mAdS. We then also explain the case of a rotating BPS saturated Kerr-Newman metric, which provides a non-trivial test of the BPS bound \eqref{bound_Mink}.

The metric and symplectic sections in spherical coordinates are
 \begin{align}\label{bls_solution}
 \begin{split}
 {\rm d} s^2 = e^{\mathcal{K}} ({\rm d} t^2 + \omega {\rm d} \varphi^2) - e^{-\mathcal{K}} {\rm d} r^2 -e^{-\mathcal{K}} r^2 {\rm d} \Omega_2^2\ ,\\
2\ {\rm Im} (X^{\Lambda}) = H^{\Lambda} = h^{\Lambda} + \frac{p^{\Lambda}}{r}, \qquad 2\ {\rm Im} (F_{\Lambda}) = H_{\Lambda} = h_{\Lambda} + \frac{q_{\Lambda}}{r}\ ,
 \end{split}
 \end{align}
 where $h^{\Lambda}, h_{\Lambda}$ are arbitrary constants that decide the asymptotic value of the scalars, usually chosen such that $e^{-\mathcal{K}}$ asymptotes exactly to $1$\footnote{One does not really need to stick to a particular choice for $h^{\Lambda}, h_{\Lambda}$. We can always perform a coordinate transformation to make sure that we have the correct asymptotics at $r \rightarrow \infty$. This has exactly the same effect.}. The rotation $\omega$ is present only when the K\"{a}hler connection \eqref{gaugedU1} is non-vanishing.

Let us consider as a first simple example the prepotential $F = - \frac{(X^1)^3}{X^0}$ with non-vanishing magnetic charge $p^0$ and electric charge $q_1$ (also non-vanishing $h^0, h_1$). This implies that $X^0 = \frac{i}{2} H^0, X^1 = \frac{1}{2} \sqrt{\frac{H^0 H_1}{3}}$ and $e^{-\mathcal{K}} = \frac{2}{3 \sqrt{3}} \sqrt{H^0 (H_1)^3}$. The $U(1)$ connection vanishes and therefore the metric is static, $\omega = 0$. To normalize the K\"{a}hler potential we choose $h^0 (h_1)^3 = \frac{27}{4}$ and find for the central charge
\begin{align}
    \mathcal{Z} = \frac{1}{4} \left(\frac{p_0}{h_0}+3 \frac{q^1}{h^1}\right)\ .
\end{align}
The mass can be calculated from \eqref{mass_Mink} with the metric \eqref{bls_solution} and spin connection $$\omega_{\theta}^{12} = \frac{\omega_{\varphi}^{13}}{\sin \theta} = e^{\mathcal{K}/2} \partial_r (r e^{-\mathcal{K}/2})$$ and becomes
\begin{align}\label{simple_mass_Mink}
    M = \lim_{r\rightarrow\infty} (- r^2 \partial_r e^{-\mathcal{K}/2}) = \frac{1}{4} \left(\frac{p_0}{h_0}+3 \frac{q^1}{h^1}\right)\ .
\end{align}
This illustrates that the above spacetime is supersymmetric since $M = |\mathcal{Z}|$.

A slightly more challenging example is provided if we take the supersymmetric Kerr-Newman spacetime from section 4.2 of \cite{Behrndt:1997ny}. We will literally consider the same solution, taken in minimal supergravity with a prepotential $F = - \frac{i}{4} (X^0)^2$, such that $e^{-\mathcal{K}} = X^0 \bar{X}^0$. In oblate spheroidal coordinates (c.f.\ (59) of \cite{Behrndt:1997ny}), the harmonic functions that give the solution are $$H_0 = 1 + \frac{m r}{r^2+\alpha^2 \cos^2 \theta}\ , \qquad H^0 = \frac{2 \alpha \cos \theta}{r^2 + \alpha^2 \cos^2 \theta}\ .$$ Solving for the vector field strengths from this, we find that $q_0 = m, p^0 = 0$. This means that
\begin{align}
    \mathcal{Z} = e^{\mathcal{K}/2} X^0 m \quad \Rightarrow \quad |\mathcal{Z}| = m\ .
\end{align}
The K\"{a}hler connection (c.f.\ \eqref{gaugedU1}) in this example is in fact non-vanishing, $$A_{\theta} = \frac{1}{2} e^{\mathcal{K}/2} (H_0 \partial_{\theta} H^0 - H^0 \partial_{\theta} H_0)\ .$$ However, it goes as $r^{-2}$ as $r \rightarrow \infty$ and therefore does not contribute to the supercharge anticommutator and keeps the Minkowski asymptotics. If we further perform a redefinition $r \rightarrow r - m$, we obtain a stationary supersymmetric metric in the familiar form
\begin{align}\label{Kerr-newman}
\begin{split}
    {\rm d} s^2 &= \frac{(r-m)^2+\alpha^2 \cos^2 \theta}{r^2 + \alpha^2 \cos^2 \theta} ({\rm d} t^2 + \frac{(2 m r - m^2) \alpha \cos^2 \theta}{(r-m)^2 + \alpha^2 \cos^2 \theta} {\rm d} \varphi^2) - \frac{r^2 + \alpha^2 \cos^2 \theta}{(r-m)^2+\alpha^2} {\rm d} r^2 \\ &- (r^2 + \alpha^2 \cos^2 \theta) {\rm d} \theta^2 - (r^2 + \alpha^2 \cos^2 \theta) \frac{(r-m)^2+\alpha^2}{(r-m)^2 + \alpha^2 \cos^2 \theta} \sin^2 \theta {\rm d} \varphi^2 \ ,
\end{split}
\end{align}
which is the Kerr-Newman metric with equal mass and charge, leading to a nakedly singular rotating asymptotically flat spacetime. The mass can be again found by
\begin{align}\label{simple_mass_kerr}
    M = ... = \lim_{r\rightarrow\infty} (- r^2 \partial_r e^{-\mathcal{K}/2}) = m = |\mathcal{Z}|\ ,
\end{align}
after converting back to spherical coordinates\footnote{Eq.\ \eqref{simple_mass_kerr} holds also in the given set of Boyer-Lindquist coordinates, but in order to use \eqref{mass_Mink} one needs to first convert the relevant asymptotic quantities in spherical coordinates.}. This confirms that the Kerr-Newman metric \eqref{Kerr-newman} is supersymmetric and that the angular momentum, $J = \alpha m$, indeed does not enter in the BPS bound \eqref{bound_Mink} and remains unconstrained by supersymmetry.

 \section{AdS$_4$ asymptotics}\label{sect:AdS}
 \subsection{General analysis}
The necessary conditions for AdS$_4$ vacuum, derived in chapter \ref{chapter::n2vac}, are:
\begin{align}\label{AdS_asymptotic}
\begin{split}
k^i_\Lambda \bar{L}^\Lambda  =0\ ,& \qquad
\tilde k^u_\Lambda L^\Lambda =0 \\
P^x_\Lambda f_i^\Lambda =0\ ,& \qquad  \epsilon^{xyz} P^y_{\Lambda} P^z_{\Sigma} L^{\Lambda} \bar{L}^{\Sigma} = 0\ ,
\end{split}
\end{align}
with constant scalars, vanishing field strengths $F^\Lambda_{\mu \nu}=0$ and AdS$_4$ metric with cosmological constant\footnote{$\Lambda$ is the cosmological constant of pure AdS$_4$ with constant scalars. The curvature of all asymptotic AdS solutions will approach this value as $r \rightarrow \infty$. The reason for defining $g'$ is because the AdS Killing spinors explicitly contain this constant instead of the gauge coupling constant $g$.} $\Lambda \equiv -3 g'^2  = -3 g^2 P^x_{\Lambda}
P^x_{\Sigma} L^{\Lambda} \overline{L}^{\Sigma}$. \eqref{AdS_asymptotic} will have to hold at $r \rightarrow \infty$ for all asymptotically AdS spacetimes, together with the usual conditions on the metric. Note that we do not allow for asymptotic magnetic charge for the graviphoton, i.e.\ $P^x_{\Lambda} A^{\Lambda}_{\varphi} = 0$. Unlike in the minimal case, this does not rule out the existence of magnetic charges but only restricts them.

The last condition in \eqref{AdS_asymptotic} tells us that the $P^x_{\Lambda} L^{\Lambda}$'s are restricted in a certain way. We will assume that they are aligned in one particular direction asymptotically\footnote{$P^x \equiv P^x_{\Lambda} L^{\Lambda}$ rotates under $Sp(1) \simeq SU(2)$ and can always be put in a particular direction. This however does not mean that existing solutions in literature are automatically written in such a way.} (direction $a$), i.e. only $P^a \equiv P^a_{\Lambda} L^{\Lambda} \neq 0$. The Majorana Killing spinors for AdS were derived in App. \ref{app:E},
\begin{equation}\label{Killing_spinors-AdS}
    \tilde{\epsilon}^{1,2}_{AdS} = e^{\frac{i}{2} arcsinh (g' r) \gamma_1} e^{\frac{i}{2} g' t \gamma_0} e^{-\frac{1}{2} \theta \gamma_{1 2}} e^{-\frac{1}{2}
    \varphi \gamma_{2 3}}\ \tilde{\epsilon}^{1,2}_0\ ,
\end{equation}
where it was implicitly assumed that $a=2$ for the gauging in the minimal case. The end result for the supercharge anticommutator will of course not depend on which direction for the moment maps is chosen, but when $a=2$ the Killing spinors (the chiral ones can again be found using \eqref{spinor_relation}) take the simplest form. In the explicit formulas for the asymptotic charges it is clear how to leave the choice for the direction $a$ completely arbitrary. The basic anticommutator for asymptotically AdS solutions can be again derived directly using the chiral version of \eqref{Killing_spinors-AdS} in \eqref{basic_susy_anticommutator_general}. The result takes the expected form from the $OSp(2|4)$ superalgebra,
\begin{align}\label{osp_algebra}
\{ Q^{A \alpha},Q^{B \beta} \} &=  \delta^{AB}
(\hat{\gamma}^{MN}C^{-1})^{\alpha \beta} M_{MN} - \epsilon^{AB} T (C^{-1})^{\alpha\beta}\ ,
\end{align}
as discussed in detail in the case of minimal gauged supergravity. Here we also require that the $U(1)$ and $Sp(1)$ gauged conections in \eqref{basic_susy_anticommutator_general} fall off fast enough as $r \rightarrow \infty$ in order to precisely recover the above expression. \eqref{osp_algebra} can be taken as a definition of asymptotically AdS spacetimes. Any spacetime, whose Dirac bracket \eqref{basic_susy_anticommutator_general} does not simplify to \eqref{osp_algebra} is therefore not asymptotically AdS. In the explicit example that follows the fall off will already be of the correct type, but in principle one needs to always make sure that the spacetime in question really is asymptotically AdS in the sense of \eqref{AdS_asymptotic} and \eqref{osp_algebra}. Each of the asymptotic charges $M_{M N}$ and $T$ can be explicitly derived, but we will again concentrate on the mass and charge in the stationary case.

 \subsection{Stationary solutions}
 For asymptotically AdS solutions with vanishing magnetic charge $\lim_{r\rightarrow\infty} P^x_{\Lambda} p^{\Lambda} = 0$, the supersymmetry anticommutator is
 \begin{align}\label{susy_commut_AdS}
    \{ Q^{A \alpha}, Q^{B \beta} \} = \delta^{A B} 8 \pi  ((M \gamma^0 + g' J_{i j} \gamma^{i j}) \mathcal{C}^{-1})^{\alpha \beta} - \epsilon^{A B} 8 \pi T (\mathcal{C}^{-1})^{\alpha \beta}\ ,
 \end{align}
 with\footnote{Note that the following expression includes both the gauge coupling constant $g$ and the asymptotic cosmological constant $g'$.}
 \begin{align}\label{mass_AdS}
 \begin{split}
 M &= \frac{1}{8\pi}\lim_{r\rightarrow\infty} \oint {\rm d}\Sigma_{t r} \bigg( e^t_{[ 0} e^r_1 e^{\theta}_{2]} + \sin \theta\ e^t_{[ 0} e^r_1 e^{\varphi}_{3]} \\
&+2 g g' r |P^a_{\Lambda} L^{\Lambda}|\ e^t_{[0} e^r_{1]} - \sqrt{g'^2 r^2 +1} (\omega_{\theta}^{a b} e^t_{[ 0} e^r_a e^{\theta}_{b]} + \omega_{\varphi}^{a b} e^t_{[ 0} e^r_a e^{\varphi}_{b]} ) \bigg)\ ,
 \end{split}
 \end{align}
 and
 \begin{align}
    T = \frac{1}{4\pi} \lim_{r \rightarrow \infty} \oint_{S^2}{\rm Re} \left( T^- \right) = \lim_{r \rightarrow \infty}  {\rm Re}\left( L^{\Lambda} q_{\Lambda} - M_{\Lambda} p^{\Lambda} \right)\ .
  \end{align}
 The angular momenta $J_{i j}$ remain exactly as given in App. \ref{app:F}, unaffected directly by the scalars. The BPS bound is given by
 \begin{align}\label{bound_AdS}
    M \geq |T| + g' |\vec{J}|\ .
 \end{align}
 Note that the scalars enter explicitly in the definition of the mass \eqref{mass_AdS}, unlike for the asymptotically flat solutions.

 \subsection{Static example}
 Here we will explicitly consider the static supersymmetric spacetimes with non-constant scalars constructed by Sabra in
\cite{Sabra:1999ux}\footnote{As explained in chapter \ref{chapter::blackholes_AdS}, they do not correspond to black holes but rather to naked singularities due to the
absence of an event horizon.}. Unlike in the asymptotically flat case, one cannot easily find what the mass is
just from looking at the metric.

 Briefly summarized, the solution is in gauged supergravity with constant FI parameters $P^a_{\Lambda} = \xi_{\Lambda}$ and an arbitrary number of vector multiplets. The metric and symplectic sections are
 \begin{align}\label{sabra_solution}
 \begin{split}
 {\rm d} s^2 = e^{\mathcal{K}} \left(1 + g^2 r^2 e^{-2 \mathcal{K}}\right) {\rm d} t^2 - \frac{e^{-\mathcal{K}} {\rm d} r^2}{\left(1 + g^2 r^2 e^{-2 \mathcal{K}}\right)} -e^{-\mathcal{K}} r^2 {\rm d} \Omega_2^2\ ,\\
 {\rm Im} X^{\Lambda} = 0, \qquad  2\ {\rm Im} F_{\Lambda} = H_{\Lambda} = \xi_{\Lambda} + \frac{q_{\Lambda}}{r}\ .
 \end{split}
 \end{align}
 It is immediately clear that the charge $T$ of this configuration will be
 \begin{align}\label{T}
    T = \lim_{r \rightarrow \infty}  {\rm Re}\left( L^{\Lambda} q_{\Lambda} - M_{\Lambda} p^{\Lambda} \right) = \lim_{r \rightarrow \infty} L^{\Lambda} q_{\Lambda} = e^{\mathcal{K}(\xi)/2} X^{\Lambda} (\xi) q_{\Lambda}\ ,
 \end{align}
where $\mathcal{K}(\xi), X^{\Lambda} (\xi)$ denote the corresponding asymptotic values that will only depend on the gauge parameters via the second row of \eqref{sabra_solution}. Since the solutions are supersymmetric and static ($J_{i j} = 0$) it follows that the mass takes the exact same value as the charge $T$. We can show this explicitly for any given solution.

Let us for simplicity take the prepotential $F =-2 i \sqrt{X^0 (X^1)^3}$ with electric charges $q_0,q_1$ and FI parameters $\xi_0,\xi_1$. The sections are therefore $X^0 = \frac{1}{6 \sqrt{3}} \sqrt{\frac{(H_1)^3}{H_0}}, X^1 = \frac{1}{2 \sqrt{3}} \sqrt{H_0 H_1}$ with $e^{-\mathcal{K}} = \frac{2}{3 \sqrt{3}} \sqrt{H_0 (H_1)^3}$ and $g' = \frac{2^{1/2}}{3^{3/4}} g (\xi_0 (\xi_1)^3)^{1/4}$. The asymptotic charge $T$ from \eqref{T} becomes
\begin{align}
    T = \frac{(\xi_0 (\xi_1)^3)^{1/4}}{2^{3/2} 3^{3/4}} \left(\frac{q_0}{\xi_0} + 3 \frac{q_1}{\xi_1} \right)\ .
\end{align}

In order to find the mass of this configuration we first need to perform a simple coordinate rescaling to make sure that the metric asymptotes to AdS in spherical coordinates (equivalently we could insist that $e^{-\mathcal{K}}$ asymptotes to $1$). Transforming $r \rightarrow a r, t \rightarrow  t/a$, with $a = \lim_{r \rightarrow \infty} e^{-\mathcal{K}/2} = \frac{2^{1/2}}{3^{3/4}} (\xi_0 (\xi_1)^3)^{1/4}$ we achieve
 \begin{align}\label{sabra_solution-example}
 {\rm d} s^2 =  \left(a^2 e^{\mathcal{K}} + g^2 r^2 e^{-\mathcal{K}}\right) {\rm d} t^2 - \frac{{\rm d} r^2}{\left(a^2 e^{\mathcal{K}} + g^2 r^2 e^{- \mathcal{K}}\right)} -\frac{e^{-\mathcal{K}}}{a^2} r^2 {\rm d} \Omega_2^2\ ,
 \end{align}
which exactly asymptotes to AdS with cosmological constant $- 3 g'^2$ in spherical coordinates. The functions that further define the metric now take the form $$H_0 = \xi_0 + \frac{a q_0}{r}\ , \qquad H_1 = \xi_1 + \frac{a q_1}{r}\ .$$ The relevant spin connection components in this case are $$\omega_{\theta}^{12} = \frac{\omega_{\varphi}^{13}}{\sin \theta} = \sqrt{a^2 e^{\mathcal{K}} + g^2 r^2 e^{- \mathcal{K}}} \partial_r (\frac{r e^{-\mathcal{K}/2}}{a})\ $$ Now we can use \eqref{mass_AdS} to find the mass of this configuration:
$$ M = \lim_{r\rightarrow\infty} \frac{e^{-\mathcal{K}/2}}{a^2} r^2 \left( \frac{a}{r} + g g' r (\xi_0 X^0 + \xi_1 X^1) - \frac{1}{r} \sqrt{g'^2 r^2 + 1} \sqrt{a^2 e^{\mathcal{K}} + g^2 r^2 e^{-\mathcal{K}}} \partial_r (r e^{-\mathcal{K}/2}) \right) $$
 \begin{equation}= ... = \frac{(\xi_0 (\xi_1)^3)^{1/4}}{2^{3/2} 3^{3/4}} \left(\frac{q_0}{\xi_0} + 3 \frac{q_1}{\xi_1} \right) = T\ ,
\end{equation}
as expected. This is a rather non-trivial check that \eqref{mass_AdS} gives the correct expression for the AdS mass, and therefore reproduces correctly results from holographic renormalization \cite{Henningson:1998gx,Henningson:1998ey,deHaro:2000xn,Skenderis:2000in,Bianchi:2001de,Papadimitriou:2004rz,Papadimitriou:2005ii,Cheng:2005wk}. Interestingly, we note that in the process of simplifying the above formula, in ``$...$'' one finds the mass to be
\begin{align}\label{simple_mass_AdS}
    M = \lim_{r\rightarrow\infty} (- \frac{r^2}{a} \partial_r e^{-\mathcal{K}/2}) = \frac{(\xi_0 (\xi_1)^3)^{1/4}}{2^{3/2} 3^{3/4}} \left(\frac{q_0}{\xi_0} + 3 \frac{q_1}{\xi_1} \right)\ ,
\end{align}
i.e. picking the first subleading term of the K\"{a}hler potential after normalizing it to asymptote to $1$.
This simple formula turns out to give the mass for the static solutions both in Minkowski (c.f.\
\eqref{simple_mass_Mink} and \eqref{simple_mass_kerr}) and in AdS. We now turn to magnetic AdS asymptotics and
show that the same formula effectively gives the mass also for supersymmetric solutions in mAdS.

 \section{mAdS$_4$ asymptotics}\label{sect:mAdS}
 \subsection{General analysis}
 Here the asymptotic conditions on the spacetime remain as in \eqref{AdS_asymptotic} with constant scalars, only now the magnetic field strengths are $F^{\Lambda}_{\theta \varphi} = p^{\Lambda} \sin \theta$ under the restriction $g P^a_{\Lambda} p^{\Lambda} = \mp 1$ coming from Dirac quantization. As before, we have the redefinition of the cosmological constant to be $\Lambda \equiv -3 g'^2$ and assume the moment map in direction $P^a$ to be non-zero.

 For $a=2$, the Killing spinors of mAdS$_4$ were given App. \ref{app:E}. Here we can give the projections obeyed by the chiral Killing spinors as straightforward generalization of the analysis in chaper \ref{6:Klemm}:
 \begin{equation}\label{KS-ansatz}
\varepsilon_{mAdS, A} =  e^{i \alpha}\, \epsilon_{A B} \gamma^0
\varepsilon^B_{mAdS}\ , \qquad \varepsilon_{mAdS, A} =  \pm e^{i
\alpha}\,{\sigma}^a_{A B}\,\, {\gamma}^1\, \varepsilon^B_{mAdS}\ ,
\end{equation}
where $\alpha$ is an arbitrary constant phase, and the choice of sign of the second projection corresponds to the choice of sign for the charge quantization condition. We choose to set $\alpha = 0$, which can be done without any loss of generality. However, some explicit solutions in literature might implicitly use $\alpha = \pi/2$ or other choices, which results in rotation of the symplectic sections $\{F_{\Lambda}, X^{\Lambda} \}$ by $e^{i \alpha}$ in all the equations that follow. The functional dependence of the Killing spinors is only radial, $\sqrt{g' r + \frac{g'}{2 g^2 r}}$. This can be seen explicitly by analyzing the Killing spinor equation $\widetilde{\mathcal{D}}_{\mu} \varepsilon_A = 0$. Solving it also forces all asymptotically mAdS spacetimes to satisfy $P^a_{\Lambda} X^{\Lambda} = \pm 1, 2 g e^{\mathcal{K}} F_{\Lambda} p^{\Lambda} = \pm i$ as $r \rightarrow \infty$.

 For asymptotically mAdS solutions with non-vanishing magnetic charge, the supersymmetry anticommutator is just
 \begin{align}\label{susy_commut_mAdS}
    \{ Q^I, Q^J \} = \delta^{I J} 8 \pi M\ ,
 \end{align}
 with only two supercharge singlets as discussed above. The mass is given by explicitly plugging \eqref{KS-ansatz} in \eqref{basic_susy_anticommutator_general} for any asymptotically mAdS solution. It again turns out that the expression takes more convenient form if we choose an upper triangular vielbein\footnote{Note that the mass can only be defined upto an overall multiplicative constant, since one can always rescale the asymptotic Killing spinor by $k$, changing the mass by $k^2$. For Minkowski and AdS, there are already well-established conventions that fix $k$, but this is not the case for mAdS.}:
 \begin{align}\label{mass_mAdS}
 \begin{split}
M &= \frac{1}{8\pi} \lim_{r\rightarrow\infty}  \oint {\rm d}\Sigma_{t r} \left( g' r + \frac{g'}{2 g^2 r} \right)
\bigg( 2\ {\rm Im}\left( L^{\Lambda} q_{\Lambda} - M_{\Lambda} p^{\Lambda} \right) \sin \theta\ e^t_{0} e^r_1 e^{\theta}_{2} e^{\varphi}_{3}  \\
&+ 2 g |P^a_{\Lambda} L^{\Lambda}|\ e^t_{0} e^r_{1} - (\omega_{\theta}^{1 2} e^t_{0} e^r_1 e^{\theta}_{2} +
\omega_{\varphi}^{1 3} e^t_{0} e^r_1 e^{\varphi}_{3} ) \bigg)\ .
 \end{split}
 \end{align}
The BPS bound in this case is simply
 \begin{align}\label{bound_mAdS}
    M \geq 0\ .
 \end{align}
 Note that there is a crucial difference between the AdS and the mAdS masses since the scalars enter differently in the expressions, e.g.\ in the first term on the r.h.s.\ of \eqref{mass_mAdS}. We will see in the next subsection that this ultimately leads to a different notion of the mass in the two cases and that the standard holographic renormalization technique is equivalent to the mass definition \eqref{mass_AdS}, but does not reproduce correctly \eqref{mass_mAdS}. Note however that one can define another conserved charge for asymptotically mAdS spacetimes, in analogy to the central charge that appears in the Riemann AdS superalgebra (see section \ref{sect:RiAdS_scalar}),
\begin{equation}\label{z}
Z \equiv \lim_{r \rightarrow \infty} r  (\frac{1}{2} \pm g' Im (L^{\Lambda} q_{\Lambda} - M_{\Lambda} p^{\Lambda}))\ .
\end{equation}
This is always finite, since we have the identity $\lim_{r \rightarrow \infty} g' Im (L^{\Lambda} q_{\Lambda} - M_{\Lambda} p^{\Lambda}) = \mp 1/2$. In this case $Z$ does not play any role in the superalgebra, but seems to be relevant when one computes masses via the holographic renormalization procedure (see more later).  

 \subsection{Black hole example}
 Here we concentrate on the static supersymmetric black holes with magnetic charges of chapter \ref{6:Klemm}. The theory is again gauged supergravity with an arbitrary number of vector multiplets and FI gaugings $\xi_{\Lambda}$. The magnetic charges are restricted by the equation $g \xi_{\Lambda} p^{\Lambda} = 1$\footnote{We just choose the positive sign here without any loss of generality.}, and the metric and scalars are given by
 \begin{align}\label{our_solution}
 \begin{split}
 {\rm d} s^2 = e^{\mathcal{K}} \left(g r+\frac{c}{2 g r} \right)^2 {\rm d}t^2 - \frac{e^{-\mathcal{K}} {\rm d}r^2}{\left(g r+\frac{c}{2 g r} \right)^2} - e^{-\mathcal{K}} r^2 {\rm d} \Omega_2^2\ ,\\
 {\rm Re} X^{\Lambda} = H^{\Lambda} = \alpha^{\Lambda} + \frac{\beta^{\Lambda}}{r}, \qquad \qquad {\rm Re} F_{\Lambda} = 0\ ,\\
 \xi_{\Lambda} \alpha^{\Lambda} = - 1\ , \qquad \xi_{\Lambda}
  \beta^{\Lambda} = 0\ , \qquad F_{\Lambda} \left( -2 g^2 r \beta^{\Lambda} + c
  \alpha^{\Lambda}+g p^{\Lambda}\right)= 0\ .
 \end{split}
 \end{align}
 If we evaluate the mass of this solutions from \eqref{mass_mAdS} we get the supersymmetric value $M=0$.

 To see this in some detail, let us again consider the simplest case of prepotential $F =-2 i \sqrt{X^0 (X^1)^3}$ that was also discussed carefully in chapter \ref{6:Klemm}. We have $X^0 = H^0 =$ $\alpha^0+\frac{\beta^0}{r}$, $X^1 = H^1 = \alpha^1+\frac{\beta^1}{r}$ and $e^{-\mathcal{K}} = 8 \sqrt{H^0 (H^1)^3}$, with
 \begin{equation}\label{constants}
\beta^0 = -\frac{\xi_1 \beta^1}{\xi_0}, \qquad \alpha^0 = -\frac{1}{4 \xi_0}, \qquad \alpha^1 = -\frac{3}{4 \xi_1}, \qquad c = 1 - \frac{32}{3} (g \xi_1 \beta^1)^2\ ,
\end{equation}
and magnetic charges
\begin{equation}\label{magn-charges}
p^0 = \frac{2}{g \xi_0} \left(\frac{1}{8}+\frac{8 (g \xi_1 \beta^1)^2}{3} \right), \quad p^1 = \frac{2}{g \xi_1} \left(\frac{3}{8}-\frac{8 (g \xi_1 \beta^1)^2}{3} \right)\ .
\end{equation}
We again need to rescale $t$ and $r$ in order to have the metric asymptote to mAdS in spherical coordinates just as above: $r \rightarrow a r, t \rightarrow  t/a$, with $a = \lim_{r \rightarrow \infty} e^{-\mathcal{K}/2} = \frac{2^{1/2}}{3^{3/4}} (\xi_0 (\xi_1)^3)^{-1/4}$ and cosmological constant coming from $g' = \frac{3^{3/4}}{2^{1/2}} g  (\xi_0 (\xi_1)^3)^{1/4}$. The metric is then
\begin{align}\label{rescaled_solution}
 {\rm d} s^2 = e^{\mathcal{K}} \left(g r+\frac{a^2 c}{2 g r} \right)^2 {\rm d}t^2 - \frac{e^{-\mathcal{K}} {\rm d}r^2}{\left(g r+\frac{a^2 c}{2 g r} \right)^2} - \frac{e^{-\mathcal{K}}}{a^2} r^2 {\rm d} \Omega_2^2\ ,
 \end{align}
and $H^0 = \alpha^0+\frac{a \beta^0}{r}, H^1 = \alpha^1+\frac{a \beta^1}{r}$. Evaluating \eqref{mass_mAdS} now gives
\begin{align}
\begin{split}
M &= \lim_{r\rightarrow\infty} \frac{e^{-\mathcal{K}/2}}{a^2} r^2 \left(g' r +\frac{g'}{2 g^2 r} \right) \left(g - \frac{a^2 e^{\mathcal{K}}}{r^2} (F_0 p^0 + F_1 p^1) - \frac{e^{\mathcal{K}/2}}{r} \left( g r+\frac{a^2 c}{2 g r}  \right) \partial_r (r e^{-\mathcal{K}/2}) \right)\\
 &= 0\ .
\end{split}
\end{align}
 We are now in position to compare this result with the one obtained via the holographic renormalization techniques of \cite{Henningson:1998gx,Henningson:1998ey,deHaro:2000xn,Skenderis:2000in,Bianchi:2001de,Papadimitriou:2004rz,Papadimitriou:2005ii,Cheng:2005wk,Batrachenko:2004fd}. As found in section 9 of \cite{Hristov:2010ri}, the mass of the above black holes is non-vanishing if one uses the explicit formulas provided in \cite{Batrachenko:2004fd} based on the procedure of holographic renormalization \cite{Henningson:1998gx,Henningson:1998ey,deHaro:2000xn,Skenderis:2000in,Bianchi:2001de,Papadimitriou:2004rz,Papadimitriou:2005ii,Cheng:2005wk}. In fact these formulas give the same result as if \eqref{mass_AdS} were used, i.e.\ the holographic renormalization procedure does not consider the case of magnetic AdS asymptotics separately. More precisely, the holographically renormalized energy of asymptotically mAdS spacetimes is given by $\frac{g}{g'} M + Z$, i.e.\ one needs to combine \eqref{mass_mAdS} and \eqref{z} in a quantity that cannot be directly associated with the time-translation symmetry. 

 Remarkably, the effective formula that worked in the static cases for Minkowski and AdS (see \eqref{simple_mass_Mink} and \eqref{simple_mass_AdS}) turns out to give the correct result once again,
 \begin{align}\label{simple_mass_mAdS}
    M = \lim_{r\rightarrow\infty} (- \frac{r^2}{a} \partial_r e^{-\mathcal{K}/2}) = 0\ .
\end{align}
Although the fundamental mass formulas \eqref{mass_Mink},\eqref{mass_AdS} and \eqref{mass_mAdS} are a priori considerably different, it turns out that the corresponding supersymmetric solutions have such properties that in each case the mass reduces to exactly the same simple formula.

\section{RiAdS$_4$ asymptotics}\label{sect:RiAdS_scalar}
\subsection{General analysis and static solutions}
Here yet again the asymptotic conditions on the spacetime remain as in \eqref{AdS_asymptotic} with constant scalars and vanishing electromagnetic field strength. As before, we assume the moment map in direction $P^a$ to be non-zero. Analogously to the cases above, the superalgebra remains exactly the same as in the minimal case in the previous chapter so we directly focus on the definition of the asymptotic charges, which changes accordingly. They can be derived by realizing that the RiAdS Killing spinor, derived in appendix \ref{app:E}, in the standard conventions adopted in this thesis, obeys
 \begin{equation}\label{RiAdSKS-ansatz}
\varepsilon_{RiAdS, A} = e^{i
\alpha}\,{\sigma}^a_{A B}\,\, {\gamma}^1\, \varepsilon^B_{RiAdS}\ ,
\end{equation}
where $\alpha$ is an arbitrary constant phase.

We are mostly interested in describing objects with vanishing $P_2,P_3$ like static black holes and branes. The relevant asymptotic charges in this case are the mass $M$ and the central charge $Z$. In the general case with arbitrary vector and hypermultiplets, they are defined as:
\begin{equation}\label{mass}
M= \frac12 \lim_{r\to \infty}{\oint {\rm d} \Sigma_{tr} e_{0}^{t} e_{1}^{r} \left(\, 2 g r |P^a_{\Lambda} L^{\Lambda}| -r (\omega_{x}^{1 2} e_{2}^{x} + \omega_{y}^{13} e_{3}^{y} + \omega_{y}^{12} e_{2}^{y} ) \,  \right)}\ ,
\end{equation}
and
\begin{equation}\label{ccharge}
Z= \lim_{r \rightarrow \infty} \oint_{T^2} r\ {\rm Im} \left( T^- \right) = \lim_{r \rightarrow \infty} r \mathcal{V}\ {\rm Im}\left( L^{\Lambda} q_{\Lambda} - M_{\Lambda} p^{\Lambda} \right)\ ,
\end{equation}
where $T^-$ is the anti-selfdual part of the graviphoton field strength.

Compared to \eqref{mass_min}, the expression for the mass with vector and hypermultiplets is changed only slightly in order to accommodate for the cosmological constant, which is now dependent on the scalar fields via the expression $P^a_{\Lambda} L^{\Lambda}$.

The expression for the central charge is reminiscent of the expression for the magnetic charge caused by the graviphoton field strength\footnote{$q_{\Lambda}, p^{\Lambda}$ are the electric and magnetic charge densities of the vector field strengths appearing in the lagrangian.}, just as in \eqref{cencharge}. The magnetic charge, $\lim_{r \rightarrow \infty} {\rm Im}\left( L^{\Lambda} q_{\Lambda} - M_{\Lambda} p^{\Lambda} \right)$, is forced to vanish due to supersymmetry of the vacuum as proven in \cite{Cacciatori:2009iz}. $Z$ is in fact the first subleading term in the expression for the magnetic charge due to the extra $r$ factor and is automatically finite in the limit $r \rightarrow \infty$. For constant scalars, \eqref{ccharge} clearly reduces to \eqref{cencharge} and the central charge vanishes. For non-constant scalar profiles, however, it is now possible to generate a non-zero $Z$, which turns out to be crucial for generating massive BPS black brane with an event horizon of chapter \ref{6:branes}.

The BPS bound for static asymptotically RiAdS solutions when a central charge is allowed is therefore
\begin{align}\label{BPS-riads1}
    M \geq |Z|\ ,
\end{align}
as already predicted. In case when both $M$ and $Z$ vanish we recover a 1/2 BPS solution like the ones in the previous section, while in case $M = |Z| \neq 0$ we have a 1/4 BPS excitation. All other cases result in non-supersymmetric excitations over RiAdS.

\subsection{Black brane example}
As already explained in chapter \ref{6:branes}, one can find a class of 1/4 BPS solutions, given by
 \begin{align}\label{our_solution}
 \begin{split}
 {\rm d} s^2 = e^{\mathcal{K}} \left(g r+\frac{c}{2 g r} \right)^2 {\rm d}t^2 - \frac{e^{-\mathcal{K}} {\rm d}r^2}{\left(g r+\frac{c}{2 g r} \right)^2} - e^{-\mathcal{K}} r^2 {\rm d} \sigma^2\ ,\\
 {\rm Re} X^{\Lambda} = H^{\Lambda} = \alpha^{\Lambda} + \frac{\beta^{\Lambda}}{r}, \qquad \qquad {\rm Re} F_{\Lambda} = 0\ ,\\
 \xi_{\Lambda} \alpha^{\Lambda} = - 1\ , \qquad \xi_{\Lambda}
  \beta^{\Lambda} = 0\ , \qquad F_{\Lambda} \left( -2 g^2 r \beta^{\Lambda} + c
  \alpha^{\Lambda}+g p^{\Lambda}\right)= 0\ ,
 \end{split}
 \end{align}
under the restriction $\xi_{\Lambda} p^{\Lambda} = 0$, with the toroidal area element given by \eqref{area-element}. These solutions satisfy the BPS bound $M = |Z|$, where both asymptotic charges are non-vanishing. Thus, unlike their spherical analogs, the magnetic black branes have a non-vanishing mass.

To see this in some detail, consider the simple case of a prepotential $F =-2 i \sqrt{X^0 (X^1)^3}$. We have $X^0 = H^0 = \alpha^0+\frac{\beta^0}{r}, X^1 = H^1 = \alpha^1+\frac{\beta^1}{r}$ and $e^{-\mathcal{K}} = 8 \sqrt{H^0 (H^1)^3}$, with
 \begin{equation}\label{constants}
\beta^0 = -\frac{\xi_1 \beta^1}{\xi_0}, \qquad \alpha^0 = -\frac{1}{4 \xi_0}, \qquad \alpha^1 = -\frac{3}{4 \xi_1}, \qquad c = - \frac{32}{3} (g \xi_1 \beta^1)^2\ ,
\end{equation}
and magnetic charges
\begin{equation}\label{magn-charges}
p^0 = \frac{16 (g \xi_1 \beta^1)^2}{3 g \xi_0}, \quad p^1 = -\frac{16 (g \xi_1 \beta^1)^2}{3 g \xi_1}\ .
\end{equation}
Note that this solution is in almost complete analogy to the one discussed above in the spherical case. It has a double horizon at $r_h = \frac{4}{\sqrt{3}} \xi_1 \beta^1$, which shields the singularity for any positive value of the arbitrary parameter $\xi_1 \beta^1$. Just as in the spherical example, we have to rescale the radial coordinate $r$ with $a = \lim_{r \rightarrow \infty} e^{-\mathcal{K}/2}$ in order to have the proper asymptotics. Evaluating \eqref{mass} and \eqref{ccharge} eventually leads to:
\begin{align}
M/\mathcal{V} = \lim_{r\rightarrow\infty} \frac{e^{-\mathcal{K}/2}}{a^2} r^2 \left(g r- e^{\mathcal{K}/2} \left( g r+\frac{a^2 c}{2 g r}  \right) \partial_r (r e^{-\mathcal{K}/2}) \right)
 = \frac{128}{9} g (\xi_1 \beta^1)^3\ ,
\end{align}
\begin{align}
Z/\mathcal{V} = \lim_{r\rightarrow\infty} r e^{\mathcal{K}/2} \sqrt{\frac{H^1}{H^0}} (p^0 H^1 + 3 p^1 H^0) = \frac{128}{9} g (\xi_1 \beta^1)^3\ .
\end{align}
This proves that the mass is equal to the central charge. The solution is a 1/4 BPS toroidal black hole in RiAdS for any finite value of $\mathcal{V}$ and black brane in AdS as $\mathcal{V} \to \infty$. Note that, unlike the spherically symmetric examples, the mass of the BPS black branes is not given by \eqref{simple_mass_mAdS}.

From the form of the superalgebra it is clear that one should in principle be able to add arbitrary electric charges to these solutions and still keep them supersymmetric. To our best knowledge, such solutions have not been yet constructed (see however \cite{Charmousis:2010zz,Barisch:2011ui} for supersymmetric and extremal electric black branes that do not strictly asymptote to AdS).

 \section{Final remarks}
To summarize, the main results of this chapter are the general mass formulas \eqref{mass_Mink}, \eqref{mass_AdS}, \eqref{mass_mAdS}, and \eqref{mass} for asymptotically flat, AdS, mAdS, and RiAdS spacetimes, respectively. We confirmed the
well-known result \cite{Ceresole:1995ca} for the central charge in Minkowski, showing that the hypermultiplets do
not alter it. We also showed that supergravity does make a clear distinction between masses in AdS and mAdS. Our
analysis in AdS generalizes some previous works that did not allow for non-trivial scalars, e.g.\ \cite{Abbott:1981ff}. The
results for asymptotically AdS solutions are in fact equivalent to performing the procedure of holographic
renormalization \cite{Henningson:1998gx,Henningson:1998ey,deHaro:2000xn,Skenderis:2000in,Bianchi:2001de,Papadimitriou:2004rz,Papadimitriou:2005ii,Cheng:2005wk,Batrachenko:2004fd}, i.e.\ \eqref{mass_AdS} can be directly used in AdS/CFT applications.
In the asymptotically mAdS case, to our best understanding, \eqref{mass_mAdS} is the relevant mass formula that needs to be used. Physically, the mass formula in mAdS might seem a bit counter-intuitive as it allows for black hole solutions with vanishing mass. However, from the point of view of the superalgebra this is the only
possibility for BPS objects in mAdS. Therefore $M=0$ should not come as a surprise for the static magnetic black
holes.

It is important to observe that the scalar profiles as functions of the radial coordinate enter explicitly in the mass formulas \eqref{mass_AdS} and \eqref{mass_mAdS}. Thus, the AdS and mAdS masses not only depend on the asymptotic values of the scalars, but also on how the scalars approach these values. This feature provides a new point of view towards the attractor mechanism in AdS/mAdS. It shows that scalars are much more restricted to behave in a particular way in comparison with the Minkowski case. Nevertheless, for the spherically symmetric supersymmetric solutions it turned out that the mass can be described by the same formula in all three asymptotic vacua,
\begin{align}\label{simple_mass}
    M = \lim_{r\rightarrow\infty} (- \frac{r^2}{a} \partial_r e^{-\mathcal{K}/2})\ ,
\end{align}
where $a \equiv \lim_{r\rightarrow\infty} e^{-\mathcal{K}/2}$ is usually chosen to be $1$. This essentially means that the mass is the first subleading term of the K\"{a}hler potential expansion, no matter what the details of the solution and its asymptotics are. It is interesting to understand the physical reasons behind this.

\chapter{Black hole superalgebras} \label{chapter::no-go}
\section{Introduction}\label{sect10:intro}
In this chapter we are going to discuss a simple application of the analysis of superalgebras, which results in a no-go theorem for static supersymmetric black holes in AdS$_4$ in theories without hypermultiplets. The theorem is based purely on superalgebras and holds very generally for $D=4$ $N=2$ supergravity, including possible higher derivative theories. In the case of hypermultiplet gaugings, we give (very constraining) conditions on the gauging that must be met in order for a genuine BPS black hole in AdS$_4$ to exist. Additionally, we prove that the attractors for BPS black holes in Minkowski and mAdS are unique. There are no possibilities for near-horizon geometries, other than the ones already known from chapters \ref{chapter::blackholes_flat} and \ref{chapter::blackholes_AdS}. We also briefly comment on $5d$ black hole superalgebras as these turn out to be very closely related to their four-dimensional analogs and the main conclusions about non-rotating BPS black holes and rings in AdS$_5$ are the same. 

The proof is based on a simple observation: a supersymmetric black hole interpolates between two different supersymmetric vacua. One is the asymptotic spacetime which is typically at an infinite distance away from the black hole itself, while the other is the near-horizon geometry right outside the black hole horizon. These two vacua are solutions of the equations of motions by themselves (in the static case) and need to necessarily preserve some supersymmetry if the full solution is to be supersymmetric. The full black hole solution thus approaches these two vacua in the corresponding limits (usually in spherical coordinates $r \rightarrow \infty$ corresponds to the vacuum at infinity, e.g.\ Minkowski or AdS, and $r\rightarrow r_{horizon}$ to the near-horizon geometry, e.g.\ AdS$_2 \times$S$^2$). This leads to a very simple mathematical relation between the superalgebras of the asymptotic spaces - if the superalgebra of the vacuum at infinity is denoted by $\mathcal{A}_{\infty}$ and the near horizon superalgebra as $\mathcal{A}_{hor}$, then the full black hole symmetry algebra $\mathcal{A}_{BH}$ is a sub(super)algebra of both asymptotic superalgbras, $\mathcal{A}_{BH} \subseteq \mathcal{A}_{\infty}$ and $\mathcal{A}_{BH} \subseteq \mathcal{A}_{hor}$. Since we want the full solution to preserve some supersymmetry, $\mathcal{A}_{BH}$ must include some fermionic charges. This discussion can be repeated in a similar manner for all solutions in supergravity that interpolate between two distinct BPS vacua, e.g.\ different kinds of black objects and M/D-branes (see \cite{FigueroaO'Farrill:2008ka,Simon:2011rw}). Inverting the argument, it is clear that if two different superalgebras do not have a common subsuperalgebra, there can never be a supersymmetric solution that interpolates between the two corresponding vacua.

We use this observation in the context of gauged $D=4$ $N=2$ supergravity to show that there is no static near-horizon geometry (with spherical symmetry) that can match to the AdS superalgebra in absence of hypermultiplets, preserving a number of supercharges. There are two major steps in this proof that have already been accomplished in previous literature. The authors of \cite{Kunduri:2007vf} proved that in four dimensional supergravities the only allowed near horizon geometry for spherically symmetric static black holes in Minkowski or AdS is the product spacetime AdS$_2 \times$S$^2$. This result holds in full generality, including possible higher derivative terms. The other important step was achieved in \cite{deWit:2011gk}, where all possible supersymmetric AdS$_2 \times$S$^2$ solutions were analyzed in the context of general gauged $D=4$ $N=2$ supergravities with possible vector, hyper-, and tensor multiplets. It was shown that only two different classes exist, the fully BPS Bertotti-Robinson spacetime, and a half BPS solution where the radii of the AdS$_2$ and S$^2$ are different. These two solutions define their own distinct superalgebras. The important point is that, even though particular details of the solution might change depending on the theory, superalgebras remain the same in supergravity, i.e.\ no matter how many additional higher derivative corrections we consider they cannot influence the abstract superalgebra charges and structure constants. This ensures that there exist only two versions of $\mathcal{A}_{hor}$ which one can choose from in order to find a supersymmetric black hole with Minkowski or AdS asymptotics in four dimensions.

Our remaining job is therefore clear. We first need to consider carefully the two near horizon superalgebras, $\mathcal{A}_{hor}$, which is done in section \ref{sect10:horizon}. Then we list the possible asymptotic vacua at infinity, Minkowski, AdS$_4$, and magnetic AdS$_4$, and their corresponding superalgebras in section \ref{sect10:infinity}. In section \ref{sect10:black-holes} we analyze the possible black hole subalgebras, showing how the two near-horizon geometries match to the Minkowski and mAdS superalgebras. We then show that the AdS$_4$ superalgebra has no common (spherically symmetric) subsuperalgebra with the half-BPS near-horizon solution, while explicitly constructing the common $SU(1|2)$ algebra inside the AdS$_4$ and the Bertotti-Robinson superalgebras. We remind the reader of the conditions for existence of these two different vacua in the same theory, thus completing the proof that static supersymmetric solutions in AdS can never develop an event horizon and form black holes in the absence of hypermultiplets.

Note that we discuss only static spherically symmetric solutions. Thus all considered near-horizon and asymptotic geometries possess time translation and 3d rotation symmetries. Every superalgebra in consideration therefore includes the bosonic charges $M$ and $J_{i j}, i,j = 1, 2, 3$, obeying the commutation relations
\begin{equation}\label{bosonic_subalgebra}
    [M, J_{i j}] = 0\ , \qquad [J_{i j}, J_{k l}] = - \delta_{i k} J_{j l} - \delta_{j l} J_{i k} + \delta_{i l} J_{j k} +\delta_{j k} J_{i l}\ .
\end{equation}
We require these generators and commutation relations, corresponding to the group $\mathbb{R} \times SO(3)$ (or $U(1) \times SO(3)$ when time is compact as in AdS), to be present also in the black hole superalgebra, $\mathcal{A}_{BH}$. We are therefore looking for a common subalgebra, $\mathcal{A}_{hor} \supseteq \mathcal{A}_{BH} \subseteq \mathcal{A}_{\infty}$, such that $\mathbb{R} \times SO(3) \subset \mathcal{A}_{BH}$ and $\mathcal{A}_{BH}$ includes at least one fermionic symmetry.

In the end of this chapter, section \ref{5d} includes a short review of the known results about BPS black objects in $5d$ supergravity and we show that the existence of non-rotating solutions in AdS$_5$ is again related to the realization of some very particular hypermultiplet gaugings.

\section{Near-horizon superalgebras}\label{sect10:horizon}
In this section we consider more carefully the two available choices for supersymmetric near-horizon geometries. One is the fully BPS, i.e.\ $8$ conserved supercharges, Bertotti-Robinson solution, while the other one is a 1/2 BPS solution which we choose to call magnetic AdS$_2 \times$S$^2$. The spacetime in both cases is a direct product of AdS$_2$ and S$^2$, but their corresponding superalgebras $\mathcal{A}_{hor}$ are very distinct.

\subsection{Bertotti-Robinson superalgebra}
The superalgebra of the fully BPS AdS$_2 \times$S$^2$, $SU(1,1|2)$, was analyzed in detail in, e.g.\ \cite{Ortin,Zhou:1999sm}. Since this is the near-horizon geometry of static asymptotically flat black holes (c.f.\ \cite{Behrndt:1997ny}), the superalgebra is very well known. We therefore do not go into much depth here, just mentioning that the bosonic symmetry group is $SO(1,2) \times SO(3)$ which includes as a subgroup $U(1) \times SO(3)$. There are eight supercharges that are organized into two spinors $Q^A_{\alpha}, A=1,2$. The relevant commutation relations for our purposes here are\footnote{Note that we keep the gamma matrix conventions of appendix \ref{appendixA}. In particular, all gamma matrices are imaginary, and the charge conjugation matrix is $C = i \gamma^0$.}:
\begin{align}\label{10rob-bert}
\begin{split}
[Q^A_{\alpha},P_0 ] &= \frac{i}{2} \epsilon^{AB} { (\gamma^{1})}_{\alpha}{}^{\beta}  Q^B_{\beta}\ , \\
[Q^A_{\alpha},J_{13} ] &= \frac{i}{2} \epsilon^{AB} { (\gamma^{013})}_{\alpha}{}^{\beta}  Q^B_{\beta}\ , \\
[Q^A_{\alpha},J_{12} ] &= \frac{i}{2} \epsilon^{AB} { (\gamma^{012})}_{\alpha}{}^{\beta} Q^B_{\beta}\ , \\
[Q^A_{\alpha},J_{23} ] &= \frac{1}{2} { (\gamma^{23})}_{\alpha}{}^{\beta} Q^A_{\beta}\ , \\
\{ Q^A_{\alpha},Q^B_{\beta} \} &= (J_{12}\ \gamma^{0 2} + J_{13}\ \gamma^{0 3})_{\alpha \beta}\ \delta^{A B} + J_{23} (i \gamma^{1 2 3})_{\alpha \beta}\ \epsilon^{A B}\\ &+ P_0\ \delta_{\alpha \beta}\ \delta^{A B} + 2 X (\gamma^{01})_{\alpha \beta}\  \delta^{A B} + Y (i \gamma^0)_{\alpha \beta}\ \epsilon^{A B}\ .
\end{split}
\end{align}
The charges $X$ and $Y$ that appear in the supercharge anticommutator come from the $SO(1,2)$ symmetry of AdS$_2$. They are always broken in the full black hole solution. But the superalgebra (anti-)commutators can only produce symmetries on the right hand side. This means that the black hole superalgebra has to necessarily include only those supercharges whose anticommutator does not produce the broken symmetries $X$ and $Y$. We will see explicitly how this happens in section \ref{sect10:black-holes}.

\subsection{Magnetic AdS$_2 \times$S$^2$}
The magnetic AdS$_2 \times$S$^2$, analyzed carefully in \cite{deWit:2011gk}, is the near-horizon geometry of the static magnetic black holes in AdS$_4$ that were described in chapter \ref{6:Klemm}. These black holes do not asymptote to the usual fully supersymmetric AdS$_4$ solution, but rather to magnetic AdS - a topologically distinct vacuum preserving two of the original eight supercharges and distinguished by its non-vanishing magnetic charges (in the electrically gauged theory). This is the reason why we also choose to use the terminology magnetic AdS$_2 \times$S$^2$. Unlike its fully supersymmetric analog, the superalgebra of the magnetic AdS$_2 \times$S$^2$ is not generally known. It can be straightforwardly derived via the procedure of chapter \ref{chapter::BPS-general} (knowing that half of the Killing spinors components are projected out by the matrix $(1+\gamma^{2 3})$) and the use of super-Jacobi identities, but it will not be needed for our purposes here. It is enough to note the observation in \cite{deWit:2011gk} that the Killing spinors do not transform under rotations, i.e.\ the supercharges flip spin and become $SO(3)$ scalars. It is easiest to denote the four scalar supercharges by $Q^I, I = 1, 2, 3, 4$. The important commutator for us here is
\begin{equation}\label{10magnAdS2} [ Q^I , J_{i j} ] = 0\ . \end{equation}
Note that this makes the superalgebras of the Bertotti-Robinson and the magnetic AdS$_2 \times$S$^2$ very different. From \eqref{10rob-bert} we can see that spatial rotations do not leave any invariant spinor components, while \eqref{10magnAdS2} shows the opposite behavior.

\section{Asymptotic superalgebras}\label{sect10:infinity}
Here we concentrate on the asymptotic superalgebras, $\mathcal{A}_{\infty}$. We analyze the three BPS asymptotic vacua that are subject to the theorem of \cite{Kunduri:2007vf}, i.e.\ we know that in these spacetimes all static black holes with spherical symmetry become AdS$_2 \times$S$^2$ near the horizon. These are the fully supersymmetric Minkowski and AdS$_4$ solutions from chapter \ref{chapter::n2vac} and the quarter-BPS magnetic AdS$_4$, analyzed in appendix \ref{app:E}.

\subsection{Poincar\'{e} superalgebra}
The best known example of a superalgebra is of course the Poincar\'{e} superalgebra. In $D=4$ $N=2$, it consists of eight fermionic charges packed in two spinors $Q^A_{\alpha}, A=1,2$ and ten bosonic symmetries from the Poincar\'{e} group in four dimensions - one time translation $P_0$, space translations $P_i$, spatial rotations $J_{i j}$, and boosts $K_i$ ($i = 1,2,3$). The important relations of the superalgebra for us are:
\begin{align}\label{10poincare}
\begin{split}
[Q^A_{\alpha},J_{i j} ] &= \frac{1}{2} (\gamma_{i j})_{\alpha}{}^{\beta} Q^A_{\beta}\ , \quad [Q^A_{\alpha},P_0 ] = 0\ , \\
\{ Q^A_{\alpha},Q^B_{\beta} \} &= P_0\ \delta_{\alpha \beta}\ \delta^{A B} + P_i (\gamma^{0 i})_{\alpha \beta}\  \delta^{A B} + {\rm Re} Z (i \gamma^0)_{\alpha \beta}\ \epsilon^{A B} + {\rm Im} Z (i \gamma^{1 2 3})_{\alpha \beta}\ \epsilon^{A B}\ ,
\end{split}
\end{align}
where $Z$ is the complex central charge that accommodates for electric and magnetic charges in asymptotically flat solutions.

\subsection{AdS$_4$ superalgebra}
 A full account of the AdS$_4$ superalgebra, $OSp(2|4)$, was already presented in chapter \ref{chapter::BPS-minimal}. Here we just selectively repeat some of the facts needed for the purposes of the present chapter. The spacetime symmetries of AdS$_4$ have their exact counterparts in the Poincar\'{e} group, although translations no longer commute with each other. Unlike in the Poincar\'{e} case however, $OSp(2|4)$ does not allow for central charges. The gauge group generator, $T$, corresponds to the R-symmetry group and thus rotates the supercharges between each other. The relevant parts of the superalgebra here are:
\begin{align}\label{10osp24} \begin{split}
[Q^A_{\alpha},T] &= \frac{1}{2} \epsilon^{A B} Q^B_{\alpha}\ , \\
[Q^A_{\alpha},J_{i j} ] &= \frac{1}{2} (\gamma_{i j})_{\alpha}{}^{\beta} Q^A_{\beta}\ , \quad [Q^A_{\alpha},P_0 ] = - \frac{i}{2}(\gamma_0)_{\alpha}{}^{\beta} Q^A_{\beta}\ , \\ \{ Q^A_{\alpha},Q^B_{\beta} \} &= P_0\ \delta_{\alpha \beta}\ \delta^{A B} + P_i (\gamma^{0 i})_{\alpha \beta}\  \delta^{A B} + K_i (\gamma^i)_{\alpha \beta}\  \delta^{A B}\\&+ J_{i j} (i \gamma^{0 i j})_{\alpha \beta}\  \delta^{A B} +T (i \gamma^0)_{\alpha \beta}\ \epsilon^{A B}\ . \end{split} \end{align}

\subsection{mAdS$_4$ superalgebra}
The superalgebra of magnetic AdS$_4$ was also derived in chapter \ref{chapter::BPS-minimal}. It is very simple due to the fact that it consists of only two supercharges, time translations, rotations, and the gauge group generator $T$. The supercharges in this case are $SO(3)$ scalars, which we denote by $Q^I$, $I=1,2$. The important commutators are
\begin{align}\label{10magnetic_AdS_superalgebra}\begin{split}   [ Q^I , J_{i j} ] &= [ Q^I , H
] = 0 \ , \\ \{ Q^I , Q^J \} &= M\ \delta^{IJ}\ . \end{split}\end{align}

\section{Black hole superalgebras and a no-go theorem}\label{sect10:black-holes}
We finally turn into analysis of the superalgebras $\mathcal{A}_{BH}$, belonging to the full black hole solutions. We give the superalgebras of the known supersymmetric black holes in Minkowski (c.f.\ chapter \ref{chapter::blackholes_flat}) and magnetic AdS (c.f.\ chapter \ref{6:Klemm}). We then search for possible black hole superalgebras in AdS, proving a no-go theorem and showing how it can be potentially circumvented.

\subsection{Asymptotically flat black holes}

\subsubsection{Bertotti-Robinson horizon}
The most general static supersymmetric black holes in ungauged supergravity are half-BPS and interpolate between the maximally supersymmetric Bertotti-Robinson solution near the horizon and Minkowski at infinity. The superalgebra of these black holes can be found by projecting out half of the supersymmetries in \eqref{10rob-bert} and \eqref{10poincare}. At the horizon, it turns out the remaining supercharges obey $Q^A_{\alpha} = \frac{1}{2} (\delta_{\alpha}{}^{\beta} \delta^{A B}  - (i \gamma^0)_{\alpha}{}^{\beta} \epsilon^{A B}) Q^B_{\beta}$. Asymptotically, the four relevant supercharges from the black hole point of view are given by $Q^A_{\alpha} = \frac{1}{2} (\delta_{\alpha}{}^{\beta} \delta^{A B}  - a (i \gamma^0)_{\alpha}{}^{\beta} \epsilon^{A B} - b (i \gamma^{1 2 3})_{\alpha}{}^{\beta} \epsilon^{A B}) Q^B_{\beta}$, where $a = \frac{{\rm Re} Z}{\sqrt{({\rm Re} Z)^2 + ({\rm Im} Z)^2}}$ and $b = \frac{{\rm Im} Z}{\sqrt{({\rm Re} Z)^2 + ({\rm Im} Z)^2}}$. The black hole superalgebra is therefore
\begin{align}\label{10flat_BH}
\begin{split}
\quad [Q^A_{\alpha},M ] &= 0\ , \\
[Q^A_{\alpha},J_{i j} ] &=\frac{1}{2} (\gamma_{i j})_{\alpha}{}^{\beta} Q^A_{\beta}\ ,\\
\{ Q^A_{\alpha},Q^B_{\beta} \} &= M\ \delta_{\alpha \beta}\ \delta^{A B}\ ,
\end{split}
\end{align}
where only four of the spinor components $Q^A_{\alpha}$ are linearly independent. The time translations generator of the black hole superalgebra, $M$, has a different meaning when embedded in the two bigger superalgebras. Near the horizon, $M \equiv P_0 - Y$ from \eqref{10rob-bert}, while asymptotically $M \equiv P_0 - \sqrt{({\rm Re} Z)^2 + ({\rm Im} Z)^2}$ from \eqref{10poincare}. The rotation generators $J_{i j}$ remain exactly the same in \eqref{10flat_BH}, \eqref{10rob-bert}, and \eqref{10poincare}, while the remaining bosonic generators, $X$ in the Bertotti-Robinson superalgebra and $P_i, K_i$ in the Poincar\'{e} superalgebra, are broken\footnote{Of course one needs to make sure that the remaining commutation relations close under the super-Jacobi identities when breaking a given bosonic or fermionic symmetry. In other words, not every way of projecting out some symmetries from \eqref{10rob-bert} and \eqref{10poincare} leads to a consistent (sub)superalgebra. This is however the case at hand, due to the fact that $Y$ commutes with the rotations and the remaining supercharges in the Bertotti-Robinson superalgebra and $Z$ is a central charge in the Poincar\'{e} superalgebra.}. The black hole superalgebra can be rewritten in a more suggestive and clear form if we write the remaining four supercharges in two complex parameters, $Q^1$ and $Q^2$. The superalgebra then takes the form
\begin{align}\label{10flat_BHbetter}
\begin{split}
\quad [Q^A,M ] &= 0\ , \\
[Q^A,J_{i j} ] &= \frac{i}{2} \epsilon_{i j k}\ \sigma_k^{A B} Q^B\ ,\\
\{ Q^A,(Q^B)^* \} &= M\ \delta^{A B}\ .
\end{split}
\end{align}
We will soon discover that the $SU(1,1|2)$ superalgebra in fact admits another subalgebra that consists of the exact same fermionic and bosonic charges with different structure constants.

These facts are all well-understood and expected for the case of ungauged BPS black holes, yet they provide a clear and straightforward realization of the general idea in section \ref{sect10:intro}. This is an example where $\mathcal{A}_{hor} \supset \mathcal{A}_{BH} \subset \mathcal{A}_{\infty}$, i.e.\ the superalgebra of the full black hole solutions is smaller than the corresponding superalgebras of the two limiting spacetimes between which it interpolates. This need not be necessarily the case as we discuss next.

\subsubsection{Magnetic AdS$_2 \times$S$^2$ horizon}
One can show that Minkowski and magnetic AdS$_2 \times$S$^2$ do not share a common subalgebra that includes rotations and supercharges. The rigorous proof follows the same considerations as will be presented in detail for the case of $OSp(2|4)$ superalgebra. If one considers all possible projections and combinations of supercharges that exist in the Poincar\'{e} superalgebra, \eqref{10flat_BH} is the only consistent subalgebra that follows the requirements of $SO(3)$ symmetry and broken spatial translations and boost charges. It is clear that the supercharges in this case are not rotationally invariant and therefore magnetic AdS$_2 \times$S$^2$ cannot be a near-horizon geometry for any asymptotically flat BPS solution.

\subsection{Magnetic AdS black holes}
\subsubsection{Magnetic AdS$_2 \times$S$^2$ horizon}
The static BPS black holes in magnetic AdS, constructed originally in \cite{Cacciatori:2009iz}, are an example of supersymmetric solutions that interpolate between magnetic AdS$_2 \times$S$^2$ near the horizon and mAdS$_4$ at infinity, as shown in \cite{deWit:2011gk}. The black holes are quarter-BPS and their corresponding Killing spinors in fact obey the same projections as the ones of pure mAdS$_4$. This means that the black hole superalgebra is exactly the same as its asymptotic superalgebra, $\mathcal{A}_{BH} = \mathcal{A}_{\infty}$. On the other hand, there is still a supersymmetry enhancement near the horizon, where the solution preserves four instead of only two supercharges. The projection that relates \eqref{10magnAdS2} to \eqref{10magnetic_AdS_superalgebra} corresponds to $(1 + i \gamma^1)$ if we would have kept the spinor indices. Observe that the supercharges are scalars under rotations everywhere in spacetime, something that seems to distinguish magnetic solutions in gauged $D=4$ $N=2$ supergravity. We thus find that $\mathcal{A}_{hor} \supset \mathcal{A}_{BH} = \mathcal{A}_{\infty}$ with no need to repeat again the black hole superalgebra, \eqref{10magnetic_AdS_superalgebra}.

\subsubsection{Bertotti-Robinson horizon}
Magnetic AdS does not share a common superalgebra with the Bertotti-Robinson solution. This can be most easily seen when considering the possible subalgebras of $SU(1,1|2)$ that have a bosonic group $U(1) \times SO(3)$. It turns out there are two such superalgebras (one was given in \eqref{10flat_BH} and the other will be discussed in the coming subsection), but neither of them has rotationally invariant supercharges. Therefore, supersymmetric black holes in mAdS can never have a Bertotti-Robinson horizon.

\subsection{Black holes in AdS$_4$?}
Now we want to show that \eqref{10osp24} has no common subalgebra with \eqref{10magnAdS2} under the requirement that the common subalgebra must include some fermionic charges, as well as time translations and rotations. It is clear that $OSp(2|4)$ is a priori different from $SU(1,1|2)$ and the magnetic AdS$_2 \times$S$^2$ superalgebra, thus the only way of finding common subalgebras is to project away at least some of the supercharges. The AdS$_4$ superalgebra is written with two four-component spinors, $Q^A_{\alpha}$. Out of this eight supercharges, one can choose an arbitrary linear combination to be preserved. The remaining supercharges in any case will obey
\begin{equation}
Q^A_{\alpha} = P^{AB}{}_{\alpha}{}^{\beta} Q^B_{\beta}\ , \qquad P^{AC}{}_{\alpha}{}^{\gamma} P^{CB}{}_{\gamma}{}^{\beta} = P^{AB}{}_{\alpha}{}^{\beta}\ .
\end{equation}
The projection operator $P^{AB}{}_{\alpha}{}^{\beta}$ consists generally of combinations of the basis $2 \times 2$ matrices $\delta^{A B}$, $\sigma_1^{A B}, i \sigma_2^{A B} = \epsilon^{A B}, \sigma_3^{AB}$ and the $4 \times 4$ spinor space matrices spanned by $1, \gamma^0, \gamma^i, \gamma^5 \equiv i \gamma^0 \gamma^1 \gamma^2 \gamma^3, \gamma^{0 i}, \gamma^{i j}, \gamma^{0 5}, \gamma^{i 5}$. However, the requirements that $P^{AB}{}_{\alpha}{}^{\beta}$ is a projection operator and that we only consider rotation invariant subalgebras limits substantially the allowed choices for projection. An additional requirement for a potential black hole superalgebra in this case is that no space translations or boost charges are allowed, since these are certainly not present in the near-horizon superalgebras \eqref{10rob-bert} and  \eqref{10magnAdS2}. These have to be then taken out of the AdS superalgebra, and the choice of projection has to guarantee that the remaining supercharges do not produce $P_i$ and $K_i$ in the new supercharge anticommutator.

More technically speaking, the requirement of rotation invariant superalgebra means that the projection operator must include $\gamma^1, \gamma^2,$ and $\gamma^3$ in a symmetric way, as otherwise some of the angular momentum generators will be broken. A simple example is if one chooses a projection of the type $(1 + \gamma^{12})$, which results in a closed subalgebra only if $J_{13}$ and $J_{23}$ are absent. This means that the projection may only include combinations of matrices $\gamma^0, \gamma^5, \gamma^{05}, (\gamma^1 + \gamma^2 + \gamma^3), (\gamma^{12}+\gamma^{13}+\gamma^{23})$. However, matrices $(\gamma^1 + \gamma^2 + \gamma^3)$ and $\gamma^{05}$ will not be able to project out the charges $K_i$ in the supercharge anticommutator since they do not anticommute with $\gamma^i$. On the other hand, $\gamma^5$ and $(\gamma^{12}+\gamma^{13}+\gamma^{23})$ do not anticommute with $\gamma^{0i}$ and will not project the translations $P_i$ out of the supercharge anticommutator. We are therefore left only with $\gamma^0$ as a potential candidate for a projection that leads to a consistent rotationally invariant superalgebra with broken spatial translations and boosts. Thus we are left with four supercharges that obey the relation $Q^A_{\alpha} = \frac{1}{2} (\delta_{\alpha}{}^{\beta} \delta^{A B}  - (i \gamma^0)_{\alpha}{}^{\beta} \epsilon^{A B}) Q^B_{\beta}$. The resulting superalgebra commutators are
\begin{align}\label{10AdS_BH}
\begin{split}
[Q^A_{\alpha},J_{i j} ] &= \frac{1}{2} (\gamma_{i j})_{\alpha}{}^{\beta} Q^A_{\beta}\ , \quad [Q^A_{\alpha},M ] = \frac{1}{2} \epsilon^{A B}\ Q^B_{\alpha}, \\
\{ Q^A_{\alpha},Q^B_{\beta} \} &= M\ \delta_{\alpha \beta}\ \delta^{A B}\ + J_{i j} (\gamma^{ij})_{\alpha \beta} \epsilon^{A B},
\end{split}
\end{align}
where $M \equiv P_0 - T$ from \eqref{10osp24} and the $P_i$ and $K_i$ are indeed broken. This is the $SU(1|2)$ superalgebra of the electric Reissner-Nordstr\"{o}m-AdS (RN-AdS) solutions, described in \cite{Romans:1991nq}. These solutions, and their generalizations for arbitrary couplings with vector multiplets, \cite{Sabra:1999ux}, are static supersymmetric asymptotically AdS$_4$ solutions that represent nakedly singular spacetimes. This can be now more easily understood from the superalgebra point of view from the following argument.

Clearly, the $SU(1|2)$ superalgebra relations above do not correspond to a subalgebra of the magnetic near-horizon geometry \eqref{10magnAdS2}. More importantly, no additional rotationally symmetric projections of the supercharges in \eqref{10AdS_BH} can ever make the supercharges singlets under the angular momentum generators. Therefore, there does not exist a common rotationally invariant subsuperalgebra of $OSp(2|4)$ and the magnetic AdS$^2 \times$S$^2$ superalgebras. This could also be expected from the fact that AdS$_4$ and magnetic AdS$^2 \times$S$^2$ are topologically distinct for their magnetic charge. Interestingly, a common subalgebra of the AdS$_4$ and the Robonson-Bertotti superalgebras does exist, since $SU(1,1|2) \supset SU(1|2) \subset OSp(2|4)$. This can be seen by considering \eqref{10rob-bert} and imposing the breaking of the $X, Y$ symmetries, together with the projection $Q^A_{\alpha} = \frac{1}{2} (\delta_{\alpha}{}^{\beta}  - (\gamma^{0 1})_{\alpha}{}^{\beta}) Q^A_{\beta}$ for the supercharges. This again leads to the $SU(1|2)$ superalgebra, although written in a different basis for the fermionic supercharges and angular momentum parameters. This is therefore the superalgebra that static BPS black holes in AdS$_4$ must obey. The most intuitive form of the superalgebra is written with two complex supercharges, $Q^1$ and $Q^2$, similarly to the case of flat black holes:
\begin{align}\label{10su(1|2)}
\begin{split}
\quad [Q^A,M ] &= -\frac{i}{2} Q^A\ , \\
[Q^A,J_{i j} ] &= \frac{i}{2} \epsilon_{i j k}\ \sigma_k^{A B} Q^B\ ,\\
\{ Q^A,(Q^B)^* \} &= M\ \delta^{A B} + J_{i j}\ \epsilon_{i j k}\ \sigma_k^{A B}\ .
\end{split}
\end{align}
Now that we have established a possibility for a black hole in AdS$_4$ with a fully BPS near-horizon geometry, we need to remind ourselves the algebraic conditions for existence of the two fully BPS asymptotic vacua (AdS$_4$ and AdS$^2 \times$S$^2$). From chapter \ref{chapter::n2vac} we know that the AdS$^2 \times$S$^2$ solutions require
\begin{align}
k^i_\Lambda \overline L^\Lambda =0\,, \quad  \tilde k^u_\Lambda L^\Lambda =0\,, \quad P^x_\Lambda =0\,.
\end{align}
while the AdS$_4$ vacuum is realized when
\begin{align}
\begin{split}
 k^i_\Lambda \overline L^\Lambda &=0\,, \qquad
\tilde k^u_\Lambda L^\Lambda =0\,, \\
P^x_\Lambda f_i^\Lambda &=0\,, \qquad  \epsilon^{xyz} P^y \overline {P^z} = 0\ .
\end{split}
\end{align}
We therefore need to have a theory where both of these vacua are allowed, i.e.\ where $P^x_\Lambda =0$ and $P^x_\Lambda \neq 0$ can be realized in field space. Clearly, this cannot happen in a theory where the moment maps are constant. This is however conceivable for theories with hypermultiplets (although we are not aware of explicit examples), where the moment maps may vary in different points of spacetime via their dependence on the hypers. This concludes our proof that static supersymmetric black holes in AdS$_4$ cannot exist in theories without hypermultiplets, confirmed by the explicit solutions of \cite{Romans:1991nq} and \cite{Sabra:1999ux}. It remains to be seen whether explicit constructions of black holes with $SU(1|2)$ superalgebra in theories with gauged hypermultiplets can be found.

\section{A glance at black objects in $5d$}\label{5d}
Although slightly more complicated, the story of supersymmetric objects with event horizon in five-dimensional supergravity is not dissimilar. Due to the extra spatial dimension, in $5d$ event horizons can have both spherical (S$^3$) topology (black holes) and ring-shaped (S$^1 \times$S$^2$) topology (black rings). The previous discussion in this chapter about static BPS black holes in $4d$ has a natural extension to ``non-rotating'' black holes and rings in $5d$. A common term in $5d$ literature, non-rotating refers to solutions whose rotation vanishes at the event horizon. The near-horizon geometry is therefore again static and described by a product spacetime with AdS and spherical factors. For a good overview of the main results in this field one can read the introductory chapters of standard references, e.g.\ \cite{London:1995ib,Gauntlett:1998fz,LozanoTellechea:2002pn,Gauntlett:2002nw,Gutowski:2004ez,Kunduri:2009ud}.

In the asymptotically flat case, the non-rotating BPS black holes are given by the supersymmetric limit of the BMPV solutions \cite{Breckenridge:1996is} with near-horizon geometry AdS$_2 \times$S$^3_{sq}$ (here S$^3_{sq}$ is a squashed sphere for the generic cases with non-vanishing angular momentum, becoming the maximally symmetric sphere in the static case). On the other hand, the BPS black rings \cite{Elvang:2004rt} in Minkowski have a near-horizon geometry AdS$_3 \times$S$^2$. Similarly to the static solutions in AdS$_4$, at present there are no known non-rotating BPS black objects (i.e.\ solutions with event horizon) in AdS$_5$. 

Unlike in $4d$, the classification of near-horizon geometries in $5d$ (see e.g.\ \cite{Kunduri:2009ud}) is a much more involved and still ongoing research direction, so one cannot use the same arguments as above to formulate a no-go theorem for AdS solutions. However, we can again show that the flat near-horizon superalgebras can fit to the asymptotical superalgebra of AdS$_5$, $SU(2,2|1)$ in $5d$ $N=1$ gauged supergravity\footnote{In $5$ dimensions, $N=1$ supergravity has a total of $8$ supercharges and is the theory most closely related to $4d$ $N=2$. This is however not of special importance for us here, since it is trivial to extend the discussion to theories with more supersymmetry without changing the final outcome. The AdS$_5$ superalgebra in general $N$-extended $5d$ supergravity theories is $SU(2,2|N)$ (see, e.g.\ \cite{Aharony:1999ti}).}.

The flat near-horizon superalgebras were carefully considered in \cite{LozanoTellechea:2002pn}, where $4$, $5$, and $6$-dimensional near-horizon geometries were connected to each other in a very suggestive manner\footnote{This also implies that similar consideration will hold even in $6d$. We will however not pursue this subject further in this thesis.}. The relevant superalgebras turn out to be the following (table 1 of \cite{LozanoTellechea:2002pn}). The static AdS$_2 \times$S$^3$ exhibits $SU(1,1|2) \times SU(2)$ where one of the rotation groups ($SO(4) = SU(2) \times SU(2)$) is entagled with the supercharges and the other one is not. The rotating BMPV solution still preserves $SU(2) \times U(1)$ on the squashed sphere, which arrange themselves to preserve the same number of supercharges in the superalgebra $SU(1,1|2) \times U(1)$. The AdS$_3 \times$S$^2$ of the black ring curiously gives rise to $SU(1,1|2) \times SL(2,\mathbb{R})$, where the extra bosonic symmetries $SL(2,\mathbb{R})$ that decouple from the supercharges this time arise from the AdS$_3$ symmetry group.

Now it becomes clear that our discussion in section \ref{sect10:horizon} of the group $SU(1,1|2)$ becomes equally relevant in $5d$, where all of the possible near-horizon geometries share the same fermionic symmetry, together with some extra bosonic symmetries that are equally simple to understand and handle. It is also easy to see that the spherically symmetric way of breaking the AdS$_5$ symmetry $SU(2,2|1)$ leads to the group $SU(1|2)$$\times$$SU(2)$ (see e.g.\ \cite{London:1995ib} for more details), which can obviously be broken further to $SU(1|2)$$\times$ $U(1)$. From our previous discussion in $4d$ it is also clear that $SU(1|2)$$\times$$SU(2)$ and/or $SU(1|2)$$\times$$U(1)$ are subalgebras of the three different near-horizon superalgebras. Therefore we conclude that on the level of superalgebras it is allowed to have BPS black holes and rings and AdS$_5$ with flat horizons. One again needs to make sure that these vacua exist within the same theory, which once more leads to the requirement of some specific hypermultiplet gaugings in $D=5$ $N=1$ supergravity. To our best knowledge, such examples are not excluded from existence, but also not explicitly known at this moment.


\chapter{Discussion and Outlook}\label{chapter::conclusions}

\section{Lessons}
What are the lessons to be learned and the conclusions to be drawn from the case-study of $4$-dimensional $N=2$ supergravity? I will try to answer this broad question in the context of each separate direction in theoretical physics as outlined in chapter \ref{chapter::introduction}.

\subsubsection{General Relativity}
From the point of view of GR, $D=4$ $N=2$ supergravity is a particular supersymmetric way of coupling matter to gravity. However, supergravity does have to tell something about general GR solutions that are not necessarily supersymmetric. We saw that the concept of BPS bounds, which in the case of $N=1$ is equivalent with the Witten-Nester energy, provides a stability criterion for very large classes of solutions. This ensures that vacua such as Minkowski and AdS are stable and cannot decay into negative mass states such as (nakedly singular) Schwarzschild spacetimes. Our analysis in $N=2$ enriches this to include Einstein-Maxwell theories with or without cosmological constant. This is clearly related with the cosmological censorship conjecture. In fact, for static solutions in Minkowski, we see that the BPS bound projects all nakedly singular solutions. This is however no longer the case with rotations in Minkowski and with asymptotically AdS spacetimes, where a well-defined version of the cosmological censorship is still missing. Often, for AdS and other interesting nontrivial solutions in GR, physical intuition can be misleading and one needs to use all available mathematical tools to gain further insight. In this respect, I think that the study of classical solutions in supergravity can provide new ideas and better understanding of the vacuum structure in General Relativity.

\subsubsection{Quantum gravity}
We already connected the study of superalgebras and BPS bounds to the classical theory of gravity. However, superalgebras are even more interesting for their implications about the quantum aspects of gravity. Due to supersymmetry, superalgebras do not renormalize after considering quantum corrections. Thus, although explicitly derived from a particular classical solution, the abstract (super)symmetry algebras remain the same even if the original solution is deformed substantially in the quantum theory. This has been used to provide the microscopic description of entropy for the BPS black holes in chapter \ref{chapter::blackholes_flat}. The same principle should allow us to describe the black holes in chapter \ref{chapter::blackholes_AdS} on a quantum level (see more ideas in this direction in the following section). As seen in chapter \ref{chapter::no-go}, we are able to prove that certain classical solutions in AdS can never develop an event horizon even in the full quantum regime. This means that they remain fundamental objects (pure states with zero entropy) in the theory of quantum gravity, whatever it is. Such examples already show that one does not necessarily need to have the knowledge of a complete quantum theory in order to understand the nature of quantum gravity. In this sense, much remains to be learnt from classical (BPS) solutions of supergravity.

\subsubsection{String theory}
Although we did not directly touch the topic of string theory in this work, we provided some string motivated examples, e.g.\ in sections \ref{3.4}, \ref{5:trivial}, \ref{6:sect:M-theory reduction}. We saw that de Sitter space is not a supersymmetric background and therefore it does not enjoy any nice stability properties from BPS bounds. This is in a sense bad news for string theory and supergravity as candidates to describe the real world around us, but more research effort is needed before making any conclusive statements. On the other hand, string theory does improve some of the supergravity implications in this thesis. We saw that any effective $D=4$ $N=2$ supergravity action coming from string theory includes at least one hypermultiplet. This is enough to (potentially) evade the no-go theorem for black holes in AdS of chapter \ref{chapter::no-go}. This may be an indication that string theory selects the more physically relevant supergravities, since we expect the real world black holes to have non-zero entropy.

\subsubsection{AdS/CFT correspondence}
For the applications of the AdS/CFT correspondence, this work is providing some new gravitational backgrounds that might have interesting field theory duals. Although the usual version of AdS/CFT would only apply to the asymptotically AdS$_4$ solutions of previous chapters, it is not hard to imagine that magnetic AdS has its own distinct dual theory, waiting to be discovered (see the next section for more explicit ideas in this direction). Additionally, our explicit method for finding conserved charges in AdS from chapters \ref{chapter::BPS-general}, \ref{chapter::BPS-minimal}, \ref{chapter::BPS-matter} can be used for explicit AdS/CFT calculations instead of the procedure of holographic renormalization for quantities at the AdS boundary. It would also be interesting to find BPS black holes in AdS with nontrivial scalar profiles (e.g.\ the ones suggested in chapter \ref{chapter::no-go}), since those are relevant for the understanding of superconductors and quantum phase transitions in condensed matter physics.

\subsubsection{Supersymmetry and supergravity}
In the broad area of supersymmetry and supergravity, this work of course mainly concentrates on a particular version of supergravity and its BPS vacuum structure. However, the methods used in our analysis can be easily applied in all other types of supergravity theories. In particular, the reasoning in chapters \ref{chapter::n2vac}, \ref{chapter::solution_generator} can be used to classify BPS solutions in general, while some of the techniques in chapters \ref{chapter::blackholes_flat}, \ref{chapter::blackholes_AdS} have their analogs in searching for higher dimensional black holes in different supergravities. Furthermore, the method of chapter \ref{chapter::BPS-general} and the main idea of chapter \ref{chapter::no-go} are immediately relevant for every other theory with supersymmetry. As we will see in the next section, the study of superalgebras can be particularly insightful when applied to $11$-dimensional supergravity. This unique highest dimensional theory is in a way the meeting point of supergravity and string/M-theory. The complete understanding of its classical solutions, which can be facilitated in many ways by the contents of this thesis, might have a profound physical meaning.

\section{Future directions}
There is a host of open questions left for future exploration: some old ones that were partially understood in the various chapters of this work, and some new ones this thesis has uncovered. The following is just a small list of topics of interest for me and some ideas on how to approach them.

\subsubsection{Classification of (BPS) black hole solutions with arbitrary gaugings}
The classification of black holes is one of the main motifs in this thesis. Still, many things about black holes remain unexplored and presently a full understanding of solutions does not seem a realistic goal. On the other hand, the proper classification of BPS black objects seems well under way. With the technical results in chapters \ref{chapter::blackholes_flat}, \ref{chapter::blackholes_AdS} and the superalgebra perspective of chapters \ref{chapter::BPS-matter}, \ref{chapter::no-go}, we made a reasonable progress towards the understanding of static BPS black holes. We can briefly summarize a few important advances:
\begin{itemize}
  \item BPS black holes in Minkowski\\
    In the static case we proved our expectations from chapter \ref{chapter::blackholes_flat} that the attractor mechanism and general form of the solutions in ungauged supergravity remain unchanged in the gauged theories. This is due to the fact that the near-horizon geometry of the Bertotti-Robinson spacetime is the only one allowed for static asymptotically flat BPS black holes. We further saw that one cannot find rotating black holes in Minkowski, since the BPS requirement $M = |\mathcal{Z}|$ leads to naked singularities whenever $J \neq 0$.
  \item Spherical BPS black holes in AdS\\
    We proved that no static black holes can exist in absence of hypermultiplets. Furthermore, in case of an appropriate hypermultiplet gauging, we predicted the existence of a black hole with a fully BPS near-horizon geometry, resulting in the same attractor mechanism as for asymptotically flat black holes.
  \item Spherical BPS black holes in mAdS\\
    We explicitly constructed a general class of static solutions in mAdS in chapter \ref{chapter::blackholes_AdS} and showed that they have a vanishing mass and a half-BPS magnetic AdS$_2 \times$S$^2$ near-horizon geometry. We proved that no rotating BPS solutions can exist, since the BPS bound restricts $M$ to vanish and the angular momentum is proportional to the mass, $J = a M = 0$.
  \item Toroidal black holes/black brane in RiAdS\\
    We presented a general class of solutions in RiAdS and showed that they obey the BPS criterion, $M = |Z|$. We further showed that no BPS solutions can exist if the graviphoton carries a magnetic charge.
  \item Higher genus black holes in magnetic hyperbolic AdS\\
    In this case the BPS solutions share the properties of their spherical analogs. We thus showed that supersymmetric higher genus black holes always carry a non-vanishing magnetic charge of the graviphoton.
\end{itemize}

What remains to be done in order to fully understand BPS black holes in $D=4$ $N=2$ supergravity is the following:
\begin{itemize}
  \item Investigate more carefully multicentered black hole solutions in Minkowski and potential rotating solutions with non-constant vector and hypermultiplet scalars.
  \item Find explicitly static BPS black holes in AdS$_4$. It seems that the correct Killing spinors are already available \cite{Romans:1991nq,Sabra:1999ux} and one only needs to find a suitable hypermultiplet gauging to write down the complete solution. Alternatively, if an example of such a theory cannot be found, it would be desirable to extend the no-go theorem to any supergravity action. 
  \item Understand better rotating black holes in AdS and how rotating attractors differ from their static analogs. Construct examples of rotating black holes with non-constant scalars.
  \item Analyze in more detail the attractor mechanism in mAdS. One can try to follow the ideas of chapter \ref{chapter::n2vac} in order to derive the algebraic conditions on the scalars in magnetic AdS$_2 \times$S$^2$.
  \item Examine the difference between toroidal black holes and black branes, e.g.\ how one would construct rotating BPS black branes in AdS that do not exist in the toroidal case. Find the possible near-horizon geometries for static solutions with flat horizons.
\end{itemize}

\subsubsection{Lifshitz superalgebra and conserved charges}
Due to the increasing interest in condensed matter applications of the gauge/gravity dualities, it is important to carefully analyze the Lifshitz spacetime. Lifschitz was shown to be a BPS solution in $D=4$ $N=2$ supergravity with gauged hypermultiplets \cite{Cassani:2011sv,Halmagyi:2011xh} and it can therefore be potentially subjected to the procedure of chapter \ref{chapter::BPS-general}. The proper definition of conserved charges on the boundary of Lifshitz spacetime is of particular importance in explicit applications, as emphasized in e.g.\ \cite{Baggio:2011cp,Baggio:2011ha}.

\subsubsection{Microscopic entropy counting in AdS$_4$}
An interesting question from a quantum gravity point of view is whether one can correctly reproduce black hole entropy in AdS$_4$ by counting BPS state degeneracies. Unlike the case of asymptotically flat black holes, the answer seems to lie in the dual 3-dimensional field theory rather than in brane constructions. Brane constructions rely on the fact that black holes in Minkowski exist for any value of the string coupling constant. In AdS this is no longer the case - we have seen that the scalar fields in AdS get stabilized exactly at the minimum of their potential. However, due to the existence of the AdS/CFT dictionary, one can try to analyze BPS states in the dual field theory providing an independent entropy calculation. In practice it turns out that the quantity that can be properly counted on the dual side is a certain supersymmetric index. This research programme has been undertaken in the case of rotating BPS black holes in AdS$_5$ \cite{Kinney:2005ej,Kunduri:2006ek,Berkooz:2006wc,Berkooz:2008gc}.

One can imagine similar considerations can be helpful in the four-dimensional case as well, although the dual three-dimensional field theory is more poorly understood. Still, the steps towards microscopic counting in the dual theory are in principle clear. One first needs to identify properly the black hole of interest from M-theory point of view. This can be done purely at a superalgebra level, showing how the black hole superalgebra fits in the 11-dimensional supergroup. The same BPS states correspondingly exist in the dual Bagger-Lambert \cite{Bagger:2006sk,Gustavsson:2007vu,Bagger:2007jr} or ABJM theories \cite{Aharony:2008ug} since they have isomorphic algebras. Once these states are identified, the problem of counting becomes rather technical and depends on the particular details of the theory, as seen in the higher dimensional case \cite{Kinney:2005ej}. Nevertheless, the question of microscopic counting is clearly approachable after the analysis of AdS superalgebras in part \ref{part::3} of this thesis.

\subsubsection{mAdS$_4$/CFT$_3$ and microscopic entropy counting in mAdS}
The questions whether mAdS$_4$ has its own field theory dual and what such a theory might look like seem to be correlated with the question of how to describe the microscopic degrees of freedom of BPS black holes in mAdS. A known field theory dual would immediately give us a different point of view towards entropy in mAdS. To construct such a dual is however a much more complex issue. However, I believe that some of the analysis in this thesis provides a good starting point to solving these problems. The fact that mAdS black holes were embedded in M-theory in chapter \ref{6:sect:M-theory reduction} is suggesting that a microscopic picture does exist. This embedding still needs to be related to a proper brane solution in 11-dimensional supergravity and its superalgebra has to be analyzed by the methods of part \ref{part::3}. This will allow us to position the dual theory in a broader perspective, since we already know how theories of multiple M2 branes look like \cite{Bagger:2006sk,Gustavsson:2007vu,Bagger:2007jr}. A particular possibility is that the theory dual to mAdS$_4$ is just a supersymmetry preserving deformation of ABJM theory. This is however just a speculation that will be confirmed only if the superalgebra of the conjectured bound state of M2's and Kaluza-Klein monopoles is a subalgebra of the M2 superalgebra \cite{Simon:2011rw}. Fortunately, these issues can be discussed on a purely algebraic level in the spirit of chapter \ref{chapter::no-go}. This means that we do not explicitly need to construct the conjectured bound state that describes mAdS in M-theory. Rather, it is good start to find its superalgebra, which is the common subalgebra of the 11-dimensional super-Poincar\'{e} group and the mAdS$_4 \ times$S$^7$ superalgebra.

We have therefore outlined the first few steps on the way of constructing the mAdS dual. However, this is in no way a proof that a dual theory needs to exist in the usual sense in which the AdS/CFT correspondence holds. It is important to stress again that mAdS is a different vacuum and its M-theory interpretation does not necessarily lead to a brane construction and a well-defined field theory description. At present this remains an interesting new possibility that deserves further attention.

\renewcommand{\publ}{}

\appendix

\chapter{Notation, conventions and spacetimes}\label{appendixA}

\section{Notation and conventions}\label{app:notations}
We mainly follow the notation and conventions
from~\cite{Andrianopoli:1996cm}. In particular, our spacetime has a $\{+,-,-,-\}$ signature. Self-dual and anti-self-dual tensors
are defined as
\begin{align}
  F^\pm_{\mu\nu} = \frac 12 \left(F_{\mu\nu} \pm \frac i 2
    \epsilon_{\mu\nu\rho\sigma} F^{\rho\sigma}\right),
\end{align}
where $\epsilon_{0123} = 1$.

The gamma matrices satisfy
\begin{align}
\begin{split}
  \{\gamma_a,\gamma_b\} &= 2 \eta_{ab}\ ,\\
  [\gamma_a,\gamma_b] &\equiv 2 \gamma_{ab}\ ,\\
\gamma_5 &\equiv - i \gamma_0 \gamma_1 \gamma_2 \gamma_3 = i
\gamma^0\gamma^1 \gamma^2 \gamma^3\ .
\end{split}
\end{align}
In addition, they can be chosen such that
\begin{align}
  \gamma_0^\dagger = \gamma_0, \quad \gamma_0 \gamma_i^\dagger
  \gamma_0 = \gamma_i,\quad \gamma_5^\dagger = \gamma_5,\quad   \gamma_\mu^* = -\gamma_\mu\ .
\end{align}
An explicit realization of such gamma matrices is the Majorana
basis, given by
\begin{align}
  \gamma^0 &= \begin{pmatrix}0 & \sigma^2\\ \sigma^2 &
    0 \end{pmatrix},&
  \gamma^1 &= \begin{pmatrix}i \sigma^3 & 0\\ 0&i\sigma^3 \end{pmatrix},&
  \gamma^2 &= \begin{pmatrix}0 & -\sigma^2\\ \sigma^2 &
    0 \end{pmatrix},\nonumber\\
  \gamma^3 &= \begin{pmatrix}-i \sigma^1 & 0\\
    0&-i\sigma^1 \end{pmatrix},&
  \gamma_5 &= \begin{pmatrix}\sigma^2 & 0\\ 0&-\sigma^2 \end{pmatrix}\ ,
\end{align}
where the $\sigma^i; i=1,2,3$ are the Pauli matrices. Their
$SU(2)$ matrix indices $A,B$ can be lowered or raised with the
antisymmetric tensor. We then obtain the following set of
matrices:
\begin{equation}
\sigma^1 = \left(
\begin{array}{cc}
0 & 1  \\
1 & 0  \end{array} \right), \quad \sigma^2 = \left(
\begin{array}{cc}
0 & -i  \\
i & 0  \end{array} \right), \quad \sigma^3 = \left(
\begin{array}{cc}
1 & 0  \\
0 & -1  \end{array} \right), \quad indices_A{}^B.
\end{equation}

\begin{equation}
\sigma^1 = \left(
\begin{array}{cc}
1 & 0  \\
0 & -1  \end{array} \right), \quad \sigma^2 = \left(
\begin{array}{cc}
-i & 0  \\
0 & -i  \end{array} \right), \quad \sigma^3 = \left(
\begin{array}{cc}
0 & -1  \\
-1 & 0  \end{array} \right), \quad \epsilon = \left(
\begin{array}{cc}
0 & 1  \\
-1 & 0  \end{array} \right), \quad indices_{A B}.
\end{equation}

\begin{equation}
\sigma^1 = \left(
\begin{array}{cc}
-1 & 0  \\
0 & 1  \end{array} \right), \quad \sigma^2 = \left(
\begin{array}{cc}
-i & 0  \\
0 & -i  \end{array} \right), \quad \sigma^3 = \left(
\begin{array}{cc}
0 & 1  \\
1 & 0  \end{array} \right), \quad \epsilon = \left(
\begin{array}{cc}
0 & 1  \\
-1 & 0  \end{array} \right), \quad indices^{A B}.
\end{equation}
Our conventions for the sigma matrices follow \cite{Andrianopoli:1996cm}; in
particular they are symmetric and satisfy
$\left(\sigma^{xAB}\right)^* = -{\sigma^x}_{AB}$, and we have the
relation
\begin{align}
  \sigma^x_{AB} \sigma^{yBC} = -{\delta_A^C} \delta^{xy} + i \epsilon_{AB}
  \epsilon^{xyz} \sigma^{zBC}\ .
\end{align}
Indices on bosonic quantities are raised and lowered as
\begin{equation}
\epsilon_{AB}V^B=V_A\ ,\qquad \epsilon^{AB}V_B=-V^A\ ,
\end{equation}
and similarly for quaternionic indices $\alpha$, raised and lowered with the antisymmetric symplectic metric $\mathbb{C}_{\alpha \beta}$.
As mentioned in the main text, all fermions with upper $SU(2)_R$
index have negative chirality and all fermions with lower index
have positive chirality. Since $\gamma_5$ was chosen to be purely imaginary, the complex conjugation interchanges chirality.

For the charge conjugation matrix, we choose \begin{equation} C=i \gamma^0\ ,
\end{equation} hence Majorana spinors have real components.

We also make use of the following identities, with curved indices: \begin{equation}
    \epsilon^{\mu \nu \rho \sigma} \gamma_5 \gamma_{\rho} = i e \gamma^{\mu \nu \sigma}  \,,
\end{equation} \begin{equation}
    \gamma_{\mu} \gamma_{\rho \sigma} = -\gamma_{\rho} g_{\mu \sigma}+ \gamma_{\sigma}
g_{\mu \rho}+ \frac{i}{e} \epsilon_{\mu \nu \rho \sigma} \gamma_5 \gamma^{\nu} \,, \end{equation} \begin{equation} \gamma_{\mu} \gamma_{\nu
\rho}\gamma_{\sigma} - \gamma_{\sigma} \gamma_{\nu \rho}\gamma_{\mu}= 2 g_{\mu \nu} g_{ \rho \sigma}- 2g_{\mu \rho} g_{\nu \sigma} +2 \frac{i}{e}
\epsilon_{\mu \nu \rho \sigma} \gamma_5 \,. \end{equation} Another important property that ensures the super-Jacobi identities of $OSp(2|4)$ hold is \begin{equation}
    (\hat{\gamma}^{MN} C^{-1})^{\alpha \beta} (\hat{\gamma}_{MN} C^{-1})^{\gamma \delta} = (C^{-1})^{\alpha \gamma} (C^{-1})^{\beta \delta} +
    (C^{-1})^{\alpha \delta} (C^{-1})^{\beta \gamma}\ ,
\end{equation} where $\hat{\gamma}_{MN}$ are defined in section \ref{sect:AdS_algebra}.

Antisymmetrizations are taken with weight one half, and the totally antisymmetric Levi-Civita symbol is defined by \begin{equation} \epsilon^{0123} = -1 = - \epsilon_{0123} \,. \end{equation} With curved indices, \begin{equation}
\epsilon^{\mu\nu\rho\sigma}\equiv e^\mu_ae^\nu_be^\rho_ce^\sigma_d\,\epsilon^{abcd}\ , \end{equation}
 is a tensor.

The action is defined by $S = \int \sqrt{|g|} \mathcal L$. We
consider the ungauged lagrangian, whose Einstein-Hilbert and
scalar derivative terms read
\begin{align}\label{lagrangian}
\mathcal L = \frac 1 2 R + g_{i\bar \jmath} \partial_{\mu} z^i
\partial^{\mu} z^{\bar \jmath} + h_{uv} \partial_{\mu} q^u \partial^{\mu} q^v\ .
\end{align}

We set the Newton constant $\kappa^2=1$. As we
use a $\{+,-,-,-\}$ metric signature, we have to choose $g_{i\bar \jmath}$ and
$h_{uv}$ positive definite to get positive kinetic
terms for the scalars.

We compute the Riemann curvature as follows\footnote{Note
that this definition, when applied to the Riemann curvature of the
quaternionic manifold, differs with a
factor of $2$ compared with~\cite{Andrianopoli:1996cm,D'Auria:2001kv}. As a consequence,
there one has $R(h_{uv}) = - 4 n (n+2)$.}
\begin{align}
\begin{split}
{R^\rho}_{\sigma\mu\nu} &= \epsilon \left[ \partial_\mu\Gamma^\rho_{\nu\sigma}
    - \partial_\nu\Gamma^\rho_{\mu\sigma}
    + \Gamma^\rho_{\mu\lambda}\Gamma^\lambda_{\nu\sigma}
    - \Gamma^\rho_{\nu\lambda}\Gamma^\lambda_{\mu\sigma} \right]\ ,\\
R_{\mu\nu}&={R^\rho}_{\mu\rho\nu}\ , \quad R=g^{\mu\nu}R_{\mu\nu}\
,
\end{split}
\end{align}
where $\epsilon = 1$ for Riemann spaces (the quaternionic and
special K\"ahler target spaces) and $\epsilon = -1$ for Lorentzian
spaces (space-time). The overall minus sign in the latter case is
needed to give AdS spaces a negative scalar curvature. This
gives a sphere in Euclidean space (with signature $\{+,+,+,+\}$) a
positive scalar curvature.

The spin connection
enters in the covariant derivative
\begin{align}\label{eq:app-covariant}
\begin{split}
  D_\mu &= \partial_\mu - \frac 14 \omega_\mu^{ab} \gamma_{ab}\ ,\\
  \omega_\mu^{ab}  &= \frac 12 e_{\mu c}\left(\Omega^{cab} -
    \Omega^{abc} - \Omega^{bca} \right)\ ,\\
\Omega^{cab} &= \left(e^{\mu a} e^{\nu b} - e^{\mu b} e^{\nu
    a}\right) \partial_\mu e^c{}_\nu\ .
\end{split}
\end{align}
The Lagrangian \eqref{lagrangian} is only supersymmetric if the
Riemann curvature of the hypermultiplet moduli space satisfies $
R(h_{uv}) = -8 n (n+2)\ , $ where $n$ is the number of
hypermultiplets, so the dimension of the quaternionic manifold is $4n$
(in applications to the universal hypermultiplet, we have $n=1$
and hence $R= - 24$).

\section{Metrics and field strengths}\label{app:metrics}
\begin{itemize}
\item AdS$_2 \times $S$^2$\\ The line element, in local coordinates
$\{t,x,\theta,\phi\}$, is
\begin{align}
  {\rm ds}^2 = q_0^2 \left( {\rm d}t^2 - \sin^2 (t) {\rm d}x^2 - {\rm
      d}\theta^2 - \sin^2 (\theta) {\rm d}\phi^2 \right)\ ,
\end{align}
where $q_0$ is a real, overall constant which determines the size
of both AdS$_2$ and S$^2$. From~\eqref{spacetime-Riemann} we find the
only non--vanishing components
\begin{align}
\begin{split}
  T^+_{tx} &= \frac 12 q_0 \sin(t) {\rm e}^{i\alpha}\ ,\\
  T^+_{\theta\phi} &= -\frac i 2 q_0 \sin(\theta) {\rm e}^{i\alpha}\ .
\end{split}
\end{align}

\item The pp-wave\\ The line element of a four--dimensional
Cahen-Wallach space~\cite{CahenWallach}, in local coordinates
$\{x^-,x^+,x^1,x^2\}$, is given by
\begin{align}
  {\rm ds}^2 = -2 {\rm d}x^+ {\rm d}x^- - A_{ij} x^i x^j ({\rm d}x^-)^2
  - ({\rm d}x^i)^2\ ,
\end{align}
where $A_{ij}$ is a symmetric matrix. Conformal flatness requires
$A_{11} = A_{22}$ and $A_{12} = 0$. We denote $A_{11} = -\mu^2$ as
$A_{11}$ should be negative. This space is known as the pp-wave. From~\eqref{spacetime-Riemann} we find the
only non--vanishing components
\begin{align}\begin{split}
  T^+_{x^- x^1} &= \frac \mu 2 {\rm e}^{i\alpha}\ ,\\
T^+_{x^- x^2} &= -i \frac \mu 2 {\rm e}^{i\alpha}\ .
\end{split}
\end{align}

\end{itemize}


\chapter{Integrability conditions}\label{app:integrability}

\section{Commutators of supersymmetry tranformations}
A killing spinor $\varepsilon_A$ satisfies
\begin{equation}
\delta_\varepsilon \psi_{\mu A}=\nabla_\mu\varepsilon_A +
T^-_{\mu\nu}\gamma^\nu
\epsilon_{AB}\varepsilon^B+igS_{AB}\gamma_\mu\varepsilon^B = \tilde{\mathcal{D}}_\mu\varepsilon_A =0\,,
\end{equation}
whence the commutator is
\begin{align}\label{eq:comm-killing}
\begin{split}
[ \nabla_\mu, \nabla_\nu] \varepsilon_A
=& -\epsilon_{AB} D_\mu T_{\nu\rho}^-
  \gamma^\rho \epsilon^B +  \frac g 2 \sigma^x_{AB} \nabla_\mu P^x \gamma_\nu
\varepsilon^B -  (\mu\nu)\\
&+T^-_{\nu\rho} \gamma^\rho T^+_{\mu\sigma} \gamma^\sigma
\varepsilon_A - (\mu\nu) \\
&-\frac g2  T^-_{\nu \rho} \gamma^\rho \gamma_\mu
 P^x_\Lambda \bar L^\Lambda\sigma^x{}_A{}^C\varepsilon_C + \frac g 2 T^+_{\mu
  \rho} \gamma_\nu  \gamma^\rho P^x_\Lambda L^\Lambda
 \sigma^x{}_A{}^C \varepsilon_C - (\mu\nu)\\
&+\frac {g^2} 2 \left(\delta_A{}^C P^x \overline {P^x} -i\epsilon^{xyz}
  \sigma^x{}_A{}^C P^y \overline {P^z}\right) \gamma_{\mu\nu}
\varepsilon_C\ .
\end{split}
\end{align}
From \eqref{scov-eps} we obtain
\begin{align}\label{eq:comm-definitie}
\begin{split}
  [ \nabla_\mu, \nabla_\nu] \varepsilon_A =& -\frac 14 R_{\mu\nu}{}^{ab}
  \gamma_{ab} \varepsilon_A - g_{i \bar \jmath} \nabla_{[\mu} z^i
  \nabla_{\nu]} z^{\bar \jmath} \varepsilon_A - i g F^\Lambda_{\mu\nu}
  P_\Lambda\\&+ 2i \Omega_{uvA}{}^B \nabla_{[\mu} q^u
  \nabla_{\nu]} q^v \varepsilon_B + ig \sigma^x{}_A{}^B
  F^\Lambda_{\mu\nu} P^x_\Lambda \varepsilon_B\ .
\end{split}
\end{align}

\section{Fully BPS vacua}
In the fully BPS case, all terms with a covariant derivative
in~\eqref{eq:comm-killing} and~\eqref{eq:comm-definitie}
vanish. We
furthermore see that~\eqref{eq:comm-definitie} does not contain a term
proportional to $\epsilon_{AB}$, so $D_\mu T^-_{\nu\rho} = 0$.

Some algebra now yields the necessary and sufficient conditions to
match the terms proportional to $\sigma^x{}_A{}^B$:
\begin{align}
\begin{split}
  T^-_{\mu \nu} \overline{P^x} &= 0\,,\\
  \epsilon^{xyz} P^y \overline {P^z} &= 0\,,
\end{split}
\end{align}
which give the first conditions of section \ref{3.2.3}. The other
conditions are obtained by comparing the parts proportional to
$\mathbf \delta_A{}^B$.

\section{Half BPS vacua}\label{sect:integrability}
We use~\eqref{eq:ansatz} to eliminate
$\varepsilon^A$ in terms of $\varepsilon_A$ and for convenience define
$b \equiv -i e^{i\alpha}$. The remaining equation
should hold for any choice of $\varepsilon_A$. We can then use the
independence of the gamma matrices and the SU(2) matrices
$\epsilon_{AB}, \sigma^x_{AB}$ to find the conditions
\begin{enumerate}
\item
Terms proportional to $\epsilon_{AB}$, no gamma.
\begin{align}
 b  D_\mu T^-_{\nu 0} - (\mu\nu)  = - g_{i \bar \jmath} \partial_{[\mu} z^i
  \partial_{\nu]} z^{\bar \jmath}\ .
\end{align}

\item
Terms proportional to $\epsilon_{AB}$, two gamma
\begin{align}
b D_\mu T^-_{\nu\rho} \gamma^{\rho 0} + T^-_{\nu\rho} T^+_{\mu\sigma}
  \gamma^{\rho\sigma} - (\mu\nu)
 + \frac {g^2} 2 P^x \overline{P^x} \gamma_{\mu\nu}=-\frac 14
 R_{\mu\nu}{}^{ab} \gamma_{ab}\ .
\end{align}

\item
Terms proportional to $\sigma^x_{AB}$, no gamma
\begin{align}
\begin{split}
& \frac g 2 b  \nabla_\mu P^x g_{\nu 0}
  - (\mu\nu) + g T^-_{\mu\nu} \overline{ P^x} + g T^+_{\mu\nu} P^x\\
&\hspace{40pt}=   g \left(L^\Lambda T^+_{\mu\nu} - \bar L^\Lambda T^-_{\mu\nu} - 2 i f_{\bar
  \imath}^\Lambda G^{i+}_{\mu\nu} + 2 i f_i^\Lambda
G^{i-}_{\mu\nu}\right) P^x_\Lambda\ ,
\end{split}
\end{align}
where we used that $ - \Omega_{uv}^x \nabla_{[\mu} q^u \nabla_{\nu]} q^v =
0$, which follows from~\eqref{eq:hyper-conditions}. Using $f_i^\Lambda P^x_\Lambda = 0$ from~\eqref{eq:Pfiszero} we therefore find
\begin{align}
& \frac g 2 b  \nabla_\mu P^x g_{\nu 0}
  - (\mu\nu)  =  -2g  T^-_{\mu\nu}  P^x_\Lambda \bar L^\Lambda\ .
\end{align}
We now take components $\mu=\theta$ and use $\nabla_\theta P^x = 0$
and $g_{\theta 0} = 0$. We then find $T^-_{\theta\nu} P^x = 0$,
whence $P^x = 0$ or $T^-_{\theta\nu} =0$. In the latter case also $T^-_{\mu\nu} = 0$,
because of the anti-self-duality property, and then
$T_{\mu\nu} = 0$. We conclude
\begin{align}\label{eq:pxl=0}
  T^-_{\mu\nu} P^x_\Lambda L^\Lambda = 0\ .
\end{align}

\item
Terms proportional to $\sigma^x_{AB}$, two gamma. Using~\eqref{eq:pxl=0}
we find
\begin{align}
\epsilon^{xyz} P^y \overline {P^z} \gamma_{\mu\nu}
= 0\ .
\end{align}
\end{enumerate}

To summarize: we found two cases, one with $T^-_{\mu\nu} = 0$, the other
with $P^x=0$. We now list the remaining conditions for each case.
\subsection{Case A: $F = 0$}\label{sect:tzero}
The remaining conditions are
\begin{align}
\begin{split}
 \frac {g^2} 2 P^x \overline{P^x} \gamma_{\mu\nu}&=-\frac 14
 R_{\mu\nu}{}^{ab} \gamma_{ab}\ ,\\
g_{i\bar \jmath} \partial_{[\mu} z^i \partial_{\nu]} z^{\bar \jmath}&=
0\ ,\\
\epsilon^{xyz} P^y \overline {P^z}&=0\ .
\end{split}
\end{align}
The first condition implies that the spacetime is maximally symmetric, with constant
curvature $\propto P^x \overline {P^x}$, and is then solved.

\subsection{Case B: $P^x = 0$}
The remaining conditions are
\begin{align}
\begin{split}
 b D_\mu T^-_{\nu 0} - (\mu\nu) &= - g_{i \bar \jmath} \partial_{[\mu} z^i
  \partial_{\nu]} z^{\bar \jmath}\ ,\\
b D_\mu T^-_{\nu\rho} \gamma^{\rho 0} + T^-_{\nu\rho} T^+_{\mu\sigma}
  \gamma^{\rho\sigma} - (\mu\nu) &=-\frac 14 R_{\mu\nu}{}^{ab}
  \gamma_{ab}\ .
\end{split}
\end{align}

From the second condition we find the Riemann tensor
\begin{align}
\begin{split}
R_{\mu\nu\rho\sigma} &= R^-_{\mu\nu\rho\sigma} +
R^+_{\mu\nu\rho\sigma}\ ,\\
 R^-_{\mu\nu\rho\sigma} =&  - b D_\mu T^-_{\nu \rho} e^0_\sigma +T^-_{\nu \rho} T^+_{\mu \sigma} - (\mu\nu)\\
&- b D_\nu T^-_{\mu \sigma} e^0_\rho +T^-_{\mu \sigma} T^+_{\nu \rho} - (\mu\nu)\\
&+ b i \epsilon_{\rho\sigma}{}^{\lambda\kappa} D_\mu T_{\nu \lambda}^-
e_\kappa^0 + i
\epsilon_{\rho\sigma}{}^{\lambda\kappa} T^-_{\nu \lambda} T^+_{\mu
  \kappa} - (\mu\nu)\ .
\end{split}
\end{align}


\chapter{Isometries of special K\"ahler manifolds}\label{app:specialkahler}

In this appendix, we present some further relevant identities that
are used in the main body of the thesis. First, we have defined the
moment maps on the special K\"ahler manifold as follows. Given an
isometry, with a symplectic embedding \eqref{Gtransf-sections}, we
can define the functions
\begin{equation}
P_\Lambda\equiv i(k^i_\Lambda \partial_i{\cal K}+r_\Lambda)\ .
\end{equation}
Since the K\"ahler potential satisfies \eqref{Gtransf-K}, it follows that $P_\Lambda$ is real. From this definition, we can verify that
\begin{equation}
k^i_\Lambda=-ig^{i\bar \jmath}\partial_{\bar \jmath}P_\Lambda\ .
\end{equation}
Hence the $P_\Lambda$ can be called moment maps, but they are {\it
not} subject to arbitrary additive constants. Using \eqref{equiv2}
and \eqref{mom-maps}, we find
\begin{equation}
k^i_\Lambda g_{i\bar\jmath}k^{\bar\jmath}_\Sigma-k^i_\Sigma
g_{i\bar\jmath}k^{\bar\jmath}_\Lambda=if_{\Lambda\Sigma}{}^\Pi
P_\Pi\ ,
\end{equation}
also called the equivariance condition.

We can obtain formulas for the moment maps in terms of the
holomorphic sections. For this, one needs the identities
\begin{equation}\label{kdX}
k^i_\Lambda\partial_iX^\Sigma=-f_{\Lambda\Pi}{}^\Sigma
X^\Pi+r_\Lambda X^\Sigma\ ,\qquad
k_\Lambda^i\partial_iF_\Sigma=c_{\Lambda,\Sigma\Pi}X^\Pi+f_{\Lambda\Sigma}{}^\Pi
F_\Pi+r_\Lambda F_\Sigma\ ,
\end{equation}
which follow from the gauge transformations of the sections, see
\eqref{Gtransf-sections}. Using the chain rule in
\eqref{mom-maps}, it is now easy to derive
\begin{equation}\label{mom-maps2}
P_\Lambda={\rm e}^{\cal K}\Big[f_{\Lambda\Pi}{}^\Sigma(X^\Pi{\bar
F}_\Sigma+F_\Sigma{\bar X}^\Pi)+c_{\Lambda,\Pi\Sigma}X^\Pi{\bar
X}^\Sigma\Big]\ ,
\end{equation}
and similarly
\begin{equation}\label{Kill-vect}
k^i_\Lambda =-ig^{i\bar \jmath}\Big[f_{\Lambda\Pi}{}^\Sigma(f^\Pi_{\bar \jmath} M_\Sigma+{
h}_{\Sigma|\bar\jmath}L^\Pi)+c_{\Lambda,\Sigma\Pi}{\bar
f}^\Pi_{\bar\jmath} L^\Sigma\Big]\ .
\end{equation}
The Killing vectors \eqref{Kill-vect} are not manifestly holomorphic. This needs not
be the case because otherwise we would have constructed isometries
for arbitrary special K\"ahler manifolds, since holomorphic vector
fields obtained from a (real) moment map solve the Killing
equation.

We now show that $P_\Lambda L^\Lambda = 0$, following the discussion
in~\cite{Hubscher:2008yz}. We start from the consistency conditions on
the symplectic embedding of the gauge transformations,
equations \eqref{kdX}. We eliminate $r_\Lambda$ using
\eqref{mom-maps}, and rewrite them as
\begin{align}\label{deriv-transformations}
-f_{\Lambda \Pi}{}^\Sigma L^\Pi &= k^i_\Lambda f_i^\Sigma+i
P_\Lambda L^\Sigma\ ,\\
f_{\Lambda \Gamma}{}^\Sigma M_\Sigma + c_{\Lambda,\Gamma\Sigma}
L^\Sigma &= k^i_\Lambda h_{\Gamma|i} + i P_\Lambda M_\Gamma\ ,
\end{align}
with $h_{\Gamma|i} = {\rm e}^{K/2} D_i F_\Gamma$. Multiplication of the first equation with $M_\Sigma$ and the second
with $L^\Gamma$ and subtracting leads to
\begin{align}\label{eq:comp}
2   f_{\Lambda\Gamma}{}^\Sigma L^\Gamma M_\Sigma + c_{\Lambda, \Gamma\Sigma}
 L^\Gamma L^\Sigma = 0\ ,
\end{align}
where we have used the identity $f_i^\Sigma M_\Sigma - h_{\Gamma|i}
  L^\Gamma = 0$. Contracting equation~\eqref{mom-maps2} with $L^\Lambda$ and
using~\eqref{eq:comp} and~\eqref{cyclicity} one finds
\begin{align}
  P_\Lambda L^\Lambda = 0\ ,
\end{align}
as announced below equation~\eqref{int-relation}. Contrating the first equation of~\eqref{deriv-transformations} with $L^\Lambda$ gives
$  L^\Lambda k^i_\Lambda f_i^\Sigma = 0$. It follows from contracting with $\text{Im}\, \mathcal N_{\Gamma \Sigma}
f^\Sigma_{\bar \jmath}$ that
\begin{align}
  L^\Lambda k^i_\Lambda = 0\ .
\end{align}
Here we have used the special geometry identities on the period
matrix~\eqref{eq:period-identities}.


\chapter{The universal hypermultiplet}\label{app:UHM}

The metric for the universal hypermultiplet is
\begin{align}\label{UHM-metric}
  {\rm d}s^2 
&= \frac 1 {r^2} \left( {\rm d}r^2 + r \, ({\rm d}\chi^2 + {\rm d}\varphi^2) +
  \bigl({\rm d}\sigma +  \chi
    {\rm d}\varphi\bigr)^2  \right)\ .
\end{align}
It describes the coset space $SU(2,1)/U(2)$. There are eight Killing vectors spanning the
isometry group $SU(2,1)$. In the coordinates of \eqref{UHM-metric}, they can be written as
\begin{align}
  \tilde{k}_{a=1} &= \partial_\sigma\ ,\nonumber\\
  \tilde{k}_{a=2} &= \partial_\chi - \varphi \partial_\sigma\ ,\nonumber\\
  \tilde{k}_{a=3} &= \partial_\varphi ,\nonumber\\
  \tilde{k}_{a=4} &= -\varphi \partial_\chi + \chi \partial_\varphi + \frac 12
  (\varphi^2 - \chi^2) \partial_\sigma\ ,\nonumber\\
  \tilde{k}_{a=5} &= 2 r \partial_r + \chi \partial_\chi +
  \varphi \partial_\varphi + 2\sigma \partial_\sigma\ ,\\
  \tilde{k}_{a=6} &= 2 r \varphi \partial_r + (-2 \sigma + \varphi
  \chi) \partial_\chi + \frac 12 (-3 r + \varphi^2 -3
  \chi^2) \partial_\varphi + (2 \sigma \varphi + 2 r \chi + \chi^3) \partial_\sigma\,,\nonumber\\
  \tilde{k}_{a=7} &= 2 r \chi \partial_r + \frac 12 (-4 r - 3 \varphi^2+
  \chi^2) \partial_\chi + (2 \sigma + 3 \varphi \chi) \partial_\varphi +
  \frac \varphi 2 ( \varphi^2 - 3 \chi^2) \partial_\sigma\,,\nonumber\\
  \tilde{k}_{a=8} &= r (2 \sigma + \varphi \chi) \partial_r + \frac 14(-4r
  \varphi - \varphi^3 +4 \sigma \chi + \varphi \chi^2) \partial_\chi +
  \frac 14 (4r \chi + 4 \sigma \varphi + 3 \varphi^2 \chi +
  \chi^3)\partial_\varphi \nonumber\\&\quad + \frac 1 {16} (-16 r^2 + 16 \sigma^2 +
  \varphi^4 - 2(8r+ 3 \varphi^2)\chi^2 - 3\chi^4) \partial_\sigma\,.\nonumber
\end{align}

The moment maps $P^x$ are computed from
\begin{align}
P^x = \Omega^x_{uv} D^u \tilde{k}^v\ .
\end{align}
The quaternionic two-forms $\Omega^x$  satisfy $\Omega^x \Omega^y = - \frac 14 \delta^{xy} + \frac 12 \epsilon^{xyz}
  \Omega^z$, and can be written as
\renewcommand{\d}{{\rm d}}
\begin{align}
\begin{split}
  \Omega^1 &= \frac 1 {2r^{3/2}} \left( \d r \wedge \d \chi + \d \varphi \wedge
    \d \sigma \right)\ ,\\
  \Omega^2 &= \frac 1 {2r^{3/2}} \left(- \d r \wedge \d \varphi + \d \chi \wedge
    \d \sigma - \chi \d \varphi \wedge  \d \chi \right)\ ,\\
\Omega^3 &= \frac 1 {2r^2} \left( \d r \wedge \d \sigma  + \chi \d r
  \wedge \d \varphi - r \d \varphi \wedge \d \chi
\right)\ .
\end{split}
\end{align}
We then find the moment maps
\begin{align}
\begin{split}
  P_{a=1} &= \left\{ 0, 0, -\frac 1 {2r}\right\}, \\
  P_{a=2} &= \left\{-\frac 1 {\sqrt r},0,\frac \varphi  {2r}\right\}, \\
  P_{a=3} &= \left\{0,\frac 1 {\sqrt r},-\frac \chi {2r}\right\}, \\
  P_{a=4} &= \left\{\frac \varphi {\sqrt r},\frac \chi {\sqrt r},1 -
   \frac{\chi^2+\varphi^2}{4r}\right\},\\
  P_{a=5} &= \left\{ -\frac \chi {\sqrt r}, \frac \varphi {\sqrt r},
    -\frac {\sigma + \frac 12 \varphi \chi} r
    \right\},\\
  P_{a=6} &= \left\{ \frac{2 \sigma - \varphi \chi}{\sqrt r}, \frac{4
      r + \varphi^2 - 3 \chi^2}{
     2 \sqrt r},\frac{-4 \sigma \varphi - (12r + \varphi^2)\chi + \chi^3}{4r} \right\},\\
  P_{a=7} &= \left\{ -\frac{4r - 3 \varphi^2+\chi^2}{
     2 \sqrt r}, \frac{2 \sigma + 3 \varphi \chi}{\sqrt r},-\frac{-12r
     \varphi + \varphi^3 + 4 \sigma \chi + 3 \varphi \chi^2}{4r} \right \},\\
  P_{a=8} &= \bigg\{\frac{-4 r \varphi + \varphi^3 - 4 \sigma \chi -
      \varphi \chi^2}{4
      \sqrt{r}},\frac{4 \sigma \varphi - 4 r \chi + 3 \varphi^2 \chi +
    \chi^3}{4 \sqrt r},\\
&\qquad\qquad-\frac{16 r^2 + 16 \sigma^2 + \varphi^4 + 16
    \sigma \chi \varphi + 6 \varphi^2 \chi^2 + \chi^4 - 24
    r(\varphi^2+\chi^2)}{32r}\bigg\}.
\end{split}
\end{align}
These formulas are needed for some of the examples that we consider in the main text of this thesis.


\chapter{Asymptotic Killing spinors}\label{app:E}

In this appendix we follow the spinor conventions of chapter \ref{chapter::BPS-minimal} for convenience. We therefore use a doublet of real Majorana spinors instead of the two complex chiral spinors used elsewhere in this work. See chapter \ref{chapter::BPS-matter} for more explanation of how to change between real and chiral spinors.

\section{AdS$_4$}\label{app:AdS} Here we give details about the Killing spinors for AdS$_4$. We consider the metric in spherical coordinates
\begin{equation}\label{metric-AdS} {\rm d} s^2 = (1+ g^2 r^2)\, {\rm d}t^2 - (1+g^2 r^2)^{-1}\, {\rm d}r^2 - r^2\, ({\rm d} \theta^2 + \sin^2
\theta {\rm d} \varphi^2)\ , \end{equation} and corresponding vielbein \begin{equation}\label{vierbein_AdS} e_{\mu}^a = {\rm
diag}\Big(\sqrt{1+g^2 r^2}, \sqrt{1+g^2 r^2}^{-1}, r, r \sin \theta\Big)\ . \end{equation} The non-vanishing components of the spin connection
turn out to be: \begin{equation}\label{omega-AdS} \omega_t^{0 1} = g^2 r, \quad \omega_{\theta}^{1 2} = -\sqrt{1+g^2 r^2}, \quad
\omega_{\varphi}^{13} = - \sqrt{1+g^2 r^2} \sin \theta, \quad \omega_{\varphi}^{23} = - \cos \theta\ . \end{equation} For the AdS$_4$ solution,
the field strength vanishes, \begin{equation}\label{field_strengths-AdS} F_{\mu \nu} = 0\ . \end{equation} To find the Killing spinors
corresponding to this spacetime we need to solve $\widetilde{\mathcal{D}}_{\mu} \epsilon = 0$. This equation has already been solved
in an unpublished paper by Izquierdo, Meessen and Ort\'{\i}n and one can explicitly check that the resulting Killing spinors are given by \begin{equation}\label{Killing_spinors-AdS}
    \epsilon_{AdS} = e^{\frac{i}{2} arcsinh (g r) \gamma_1} e^{\frac{i}{2} g t \gamma_0} e^{-\frac{1}{2} \theta \gamma_{1 2}} e^{-\frac{1}{2}
    \varphi \gamma_{2 3}} \epsilon_0\ ,
\end{equation} where $\epsilon_0$ is a doublet of arbitrary constant Majorana spinors, representing the eight preserved supersymmetries of the
configuration.

It is important to note that the asymptotic solution of the Killing spinor equations as $r\rightarrow\infty$ (given the same asymptotic metric)
cannot change unless $A_{\varphi} \neq 0$. This is easy to see from the form of the supercovariant derivative
\eqref{minimal_supercovariant_derivative} since any other term would necessarily vanish in the asymptotic limit. More precisely, any gauge field
carrying an electric charge that appears in the derivative vanishes asymptotically, the only constant contribution can come when a magnetic
charge is present. In other words, any spacetime with vanishing magnetic charge and asymptotic metric \eqref{metric-AdS} has asymptotic Killing
spinors given by \eqref{Killing_spinors-AdS}.

\section{Magnetic AdS$_4$}\label{app:magnetic AdS} Now we will show that the asymptotic Killing spinors take a very different form when
magnetic charge is present. In this case the metric is \begin{equation}\label{metric-magnAdS} {\rm d} s^2 = (1+ g^2 r^2+\frac{Q_m^2}{r^2})\, {\rm
d}t^2 - (1+g^2 r^2+\frac{Q_m^2}{r^2})^{-1}\, {\rm d}r^2 - r^2\, ({\rm d} \theta^2 + \sin^2 \theta {\rm d} \varphi^2)\ , \end{equation} with
corresponding vielbein: \begin{equation}\label{vierbein_magnAdS} e_{\mu}^a = {\rm diag}\Big(\sqrt{1+g^2 r^2+\frac{Q_m^2}{r^2}}, \sqrt{1+g^2
r^2+\frac{Q_m^2}{r^2}}^{-1}, r, r \sin \theta\Big)\ . \end{equation} The non-vanishing components of the spin connection turn out to be:
\begin{eqnarray}\label{omega-magnAdS} &&\omega_t^{0 1} = g^2 r-\frac{Q_m^2}{r^3}\ , \qquad \omega_{\theta}^{1 2} = -\sqrt{1+g^2
r^2+\frac{Q_m^2}{r^2}}\ ,\nonumber\\ &&\omega_{\varphi}^{1 3} = - \sqrt{1+g^2 r^2+\frac{Q_m^2}{r^2}} \sin \theta\ , \qquad \omega_{\varphi}^{2 3}
= - \cos \theta\ . \end{eqnarray} As opposed to the previous section, now we have a non-vanishing gauge field component $A_{\varphi} = - Q_m \cos
\theta$, resulting in $F_{\theta \varphi} = Q_m \sin \theta$. If we require $\widetilde{\mathcal{D}}_{\mu} \epsilon = 0$ and insist that $Q_m
\neq 0$, we get a solution described by Romans in \cite{Romans:1991nq} as a ``cosmic monopole'' (which we call magnetic AdS$_4$). The magnetic
charge satisfies $2 g Q_m = \pm 1$, such that the metric function is an exact square $(g r + \frac{1}{2 g r})^2$. The Killing spinors
corresponding to solutions with $Q_m=\pm 1/(2g)$ in our conventions are given by \begin{equation}\label{Killing_spinors-magnAdS}
    \epsilon_{mAdS} =  \frac{1}{4} \sqrt{g r + \frac{1}{2 g r}} (1+ i \gamma_1) (1\mp i \gamma_{2 3} \sigma^2) \, \epsilon_0\ ,
\end{equation} preserving two of the original eight supersymmetries. Note that in the limit $r\rightarrow\infty$ the Killing spinor projections
continue to hold. Furthermore, the functional dependence is manifestly different in the expressions \eqref{Killing_spinors-AdS} and
\eqref{Killing_spinors-magnAdS} for the Killing spinors of ordinary AdS$_4$ and its magnetic version. This leads to the conclusion that these two
vacua and their corresponding excited states belong to two separate classes, i.e. they lead to two independent superalgebras and BPS bounds. Note
that one can also add an arbitrary electric charge $Q_e$ to the above solution, preserving the same amount of supersymmetry (the ``cosmic dyon''
of \cite{Romans:1991nq}). The corresponding Killing spinors \cite{Romans:1991nq} have the asymptotic form of \eqref{Killing_spinors-magnAdS},
i.e. the cosmic dyons are asymptotically magnetic AdS$_4$.

\section{Riemann AdS$_4$}\label{app:RiAdS}
We now determine the Killing spinors of the ground state with toroidal topology and vanishing mass and charges, $\eta=q=p=0$. The metric is
\begin{equation}\label{RiAdS} {\rm d} s^2 = (1+ g^2 r^2)\, {\rm d}t^2 - (1+g^2 r^2)^{-1}\, {\rm d}r^2 - r^2\, \frac{\mathcal{V}}{{\rm Im}\tau} ({\rm d}x^2 + 2 {\rm Re}\tau \, {\rm d}x {\rm d}y + |\tau|^2 {\rm d}y^2)\ , \end{equation}
Choosing upper triangular vielbein $e_{\mu}^a$
\begin{equation}\label{vielbeins}
e_t^0= g r\,, \quad e_r^1= \frac{1}{g r}\,, \quad e_x^2= \frac{r \sqrt{ {\rm Im} \tau } \sqrt{\mathcal{V}}}{|\tau|} \,, \quad e_x^3 = \frac{r  \sqrt{\mathcal{V}}}{|\tau|} \frac{{\rm Re}\tau}{\sqrt{{\rm Im} \tau}}\,, \quad e_y^3= \frac{r |\tau|  \sqrt{\mathcal{V}}}{\sqrt{{\rm Im}\tau} }\ ,
\end{equation}
one can straightforwardly derive the (non-vanishing) components of the spin connection,
\begin{equation}\label{spinconn}
\omega_{t}^{01}= g^2 r, \quad \omega_{x}^{12}= - \frac{\sqrt{ {\rm Im} \tau} \sqrt{\mathcal{V}}}{|\tau|} g r\,, \quad \omega_{x}^{13}= - \frac{{\rm Re}\tau  \sqrt{\mathcal{V}}}{|\tau|\sqrt{ {\rm Im} \tau}} g r\,, \quad \omega_{y}^{13}= -\frac{|\tau| \sqrt{\mathcal{V}}}{\sqrt{{\rm Im} \tau}} g r\,.
\end{equation}
The Killing spinors can now be computed from $\widetilde{\mathcal{D}}_{\mu} \epsilon = 0$, and the solution, with arbitrary constant spinors $\epsilon_0$, is
\begin{equation}\label{prefactor2}
 \epsilon = e^{\frac{i}{2} log(r)\gamma_1}\left (1+ \frac{ig}{2} \left[ x \frac{\sqrt{{\rm Im} \tau}  \sqrt{\mathcal{V}}}{|\tau|} \left(\gamma_2 +\frac{{\rm Re} \tau}{ {\rm Im} \tau} \gamma_3 \right) +y \frac{|\tau| \sqrt{\mathcal{V}}}{\sqrt{ {\rm Im} \tau}}\gamma_3 + gt \gamma_0 \right] \left(1-i \gamma_1 \right) \right) \epsilon_0\,\,.
\end{equation}
Without restriction on $\epsilon_0$, all eight supercharges are preserved, but \eqref{prefactor2} does not respect the identification of the coordinates $x,y$ on the torus\footnote{Without the toroidal compactification, the spacetime is of course normal AdS written in Poincar\'{e} coordinates. It is therefore no surprise that \eqref{prefactor2} is similar to \eqref{Killing_spinors-AdS}.}. One can make the Killing spinors well-defined on the torus by imposing a projection on $\epsilon_0$, namely
\begin{equation}\label{projection-torus}
\epsilon_0= P\epsilon_0\ ,\qquad P\equiv \frac{1+i\gamma_1}{2}\ .
\end{equation}
Such a projection breaks half of the supersymmetries, and the Killing spinors of RiAdS are therefore
\begin{equation}
\epsilon_{RiAdS}  = \sqrt{r} \left( \frac{1+i\gamma_1}{2} \right) \epsilon_0 = \sqrt{r}P \epsilon_0 \,.
\end{equation}

\chapter{Rotations in AdS$_4$}\label{app:F}

Here we focus on stationary spacetimes with rotations. From the supersymmetry Dirac brackets in
asymptotic AdS$_4$ spaces, \begin{equation} \{ \mathcal{Q}, \mathcal{Q} \} =- 8 \pi i \overline{\epsilon}_{0}\left( ...+ g J_{i j} \gamma^{i j}+
... \right)  \epsilon_{0}\ , \end{equation} we can derive a definition of the conserved angular momenta. The explicit expressions are somewhat
lengthy and assume a much simpler form once we choose the vielbein matrix $e_{\mu}^a$ in an upper triangular form, such that its inverse
$e_a^{\mu}$ is also upper triangular. More explicitly, in spherical coordinates we choose nonvanishing $e_t^{0,1,2,3}, e_r^{1,2,3},
e_\theta^{2,3}, e_\varphi^3,$ such that the inverse vielbein has only non-vanishing components $e_0^{t,r,\theta,\varphi}, e_1^{r,\theta,\varphi},
e_2^{\theta,\varphi}, e_3^{\varphi}$. The resulting expressions for the angular momenta in this case become: \begin{align}\label{ang_mom}
\nonumber    J_{12} = \frac{1}{8 \pi} \lim_{r\rightarrow\infty} \int\limits_{0}^{2 \pi} {\rm d} \varphi \int\limits_{0}^{\pi} {\rm d} \theta &
\left( (e_0^t e_1^r e_{\varphi}^3 \omega_{\theta}^{01} + e_0^t e_1^r e_{\theta}^2 e_{\varphi}^3 e_2^{\varphi} \omega_{\varphi}^{01}) r \cos
\varphi + (e_0^t e_1^r e_{\theta}^2 \omega_{\varphi}^{01}) r \cos \theta \sin \varphi \right)\ , \\ \nonumber    J_{13} = \frac{1}{8 \pi}
\lim_{r\rightarrow\infty} \int\limits_{0}^{2 \pi} {\rm d} \varphi \int\limits_{0}^{\pi} {\rm d} \theta & \left( (e_0^t e_1^r e_{\varphi}^3
\omega_{\theta}^{01} + e_0^t e_1^r e_{\theta}^2 e_{\varphi}^3 e_2^{\varphi} \omega_{\varphi}^{01}) r \sin \varphi + (e_0^t e_1^r e_{\theta}^2
\omega_{\varphi}^{01}) r \cos \theta \cos \varphi \right)\ , \\
    J_{23} &= \frac{1}{8 \pi} \lim_{r\rightarrow\infty} \int\limits_{0}^{2 \pi} {\rm d} \varphi \int\limits_{0}^{\pi} {\rm d} \theta \left( e_0^t
    e_1^r e_{\theta}^2 \omega_{\varphi}^{01} r \sin \theta \right) \ .
\end{align} It is easy to see that in case of axisymmetric solutions around $\varphi$, such as the Kerr and Kerr-Newman metrics in AdS, the
angular momenta $J_{12}$ and $J_{13}$ automatically vanish due to $\int_0^{2 \pi} {\rm d} \varphi \sin \varphi = \int_0^{2 \pi} {\rm d} \varphi
\cos \varphi = 0$.

One can then use the formula for $J_{23}$ to derive the value of the angular momentum for the Kerr black hole. This is still somewhat non-trivial
because one needs to change the coordinates from Boyer-Lindquist-type to spherical. The leading terms at large $r$ were found in appendix B of
\cite{Henneaux:1985tv} and are enough for the calculation of the angular momentum since subleading terms vanish when the limit is taken in
\eqref{ang_mom}. The calculation of the relevant component of the spin connection leads to
 \begin{equation}
    \omega_{\varphi}^{01} = - \frac{3 a m \sin^2 \theta (1-g^2 a^2 \sin^2 \theta)^{-5/2}}{r^2} + \mathcal{O} (r^{-3})
 \end{equation}
 and gives the exact same result as in (B.8) of \cite{Henneaux:1985tv},
 \begin{equation}
   J_{23} = \frac{a m}{(1-g^2 a^2)^2}\ .
 \end{equation}
This expression has also been derived from different considerations in \cite{Caldarelli:1999xj}, thus confirming the consistency of our results.

One can also verify the result for the asymptotic mass of the Kerr and Kerr-Newman spacetimes using \eqref{mass_ads} and the metric in appendix B
of \cite{Henneaux:1985tv}. After a somewhat lengthy but straightforward calculation one finds \begin{align}
    M =  \frac{m}{(1-g^2 a^2)^2}\ ,
\end{align} as expected from previous studies (see, e.g., \cite{Caldarelli:1998hg,Caldarelli:1999xj}).

\chapter{Supercharge of the general gauged theory}\label{app:G}

\section{Additional details on $D=4$ $N=2$ gauged supergravity}\label{app:G1}
Here we will give more details on the theory in consideration. Alternatively, see \cite{Andrianopoli:1996cm} for a very detailed description. The bosonic part of the supergravity lagrangian was given in \eqref{lagr}-\eqref{eq:scalar-potential}. The supersymmetry variations under which the full action is invariant (upto higher order terms in fermions) are as follows. The fermionic variations were already given in chapter \ref{chapter::supergravity}: \eqref{susy-gravi},\eqref{susygluino},\eqref{susyhyperino}. The bosonic susy variations are as follows. The vielbein variation reads
\begin{equation}\label{susy-vielbein}
\delta_\varepsilon e_{\mu}^a=- i \overline{\psi}_{\mu A} \gamma^a \varepsilon^A - i \overline{\psi}_{\mu}^A \gamma^a \varepsilon_A\ .
\end{equation}
In the vector multiplet sector we have
\begin{equation}\label{susy-vecscalar}
\delta_\varepsilon z^i= \overline{\lambda}^{iA} \varepsilon_A\ ,
\end{equation}
and
\begin{equation}\label{susy-vectors}
\delta_\varepsilon A_{\mu}^{\Lambda}= 2 \bar{L}^{\Lambda} \overline{\psi}_{\mu A} \varepsilon_B \epsilon^{A B} + i f^\Lambda_i \overline{\lambda}^{iA} \gamma_{\mu} \varepsilon^B \epsilon_{A B} + h.c.\ .
\end{equation}
The susy variation of the hypermultiplet scalars (hypers) is
\begin{equation}\label{susy-hypers}
\delta_\varepsilon q_u = \mathcal{U}^{A\alpha}_u \left(\overline{\zeta}_\alpha \varepsilon^A + \mathbb{C}^{\alpha \beta} \epsilon^{A B} \overline{\zeta}^{\beta} \varepsilon_B \right) .
\end{equation}

In order to derive the supercharge of the theory from the procedure described in chapter \ref{chapter::BPS-general}, we additionally need the Poisson/Dirac brackets of the fundamental fields. It will suffice to list the non-vanishing fermionic Dirac brackets that follow from the full lagrangian\footnote{The brackets for the bosonic fields can be derived directly from \eqref{lagr} if needed.} (see e.g. \cite{Andrianopoli:1996cm}):
\begin{align}\label{Dirac_brackets}
\begin{split}
\{ \psi_{\mu A}(x), \epsilon^{0 \nu \rho \sigma} \overline{\psi}^B_{\rho}(x') \gamma_{\sigma}
 \}_{t=t'} &= \delta_{\mu}{}^{\nu} \delta_A{}^B \delta^3(\vec{x}-\vec{x'})\ ,\\
 \{ \lambda^{i}_A (x), - \frac{i}{2} g_{k \bar{\jmath}} \overline{\lambda}^{B \bar{\jmath}}(x')
\gamma_0
 \}_{t=t'} &= \delta_{A}{}^{B} \delta_k{}^i \delta^3(\vec{x}-\vec{x'})\ ,\\
 \{ \zeta_{\alpha}(x), - i \overline{\zeta}^{\beta}(x')
\gamma_0
 \}_{t=t'} &= \delta_{\alpha}{}^{\beta} \delta^3(\vec{x}-\vec{x'})\ .
\end{split}
\end{align}
Note in particular that we follow the original conventions of appendix \ref{appendixA} for $\epsilon^{\mu \nu \rho \sigma}$. Consequently, we define as a measure for the volume/surface integrals \begin{align}
{\rm d}\Sigma_\mu=\frac{1}{6}\epsilon_{\mu\nu\rho\sigma}\,{\rm d}x^\nu \wedge {\rm d}x^\rho\wedge {\rm d}x^\sigma\ ,\qquad
{\rm d}\Sigma_{\mu\nu}=\frac{1}{2}\epsilon_{\mu\nu\rho\sigma}\,{\rm d}x^\rho\wedge {\rm d}x^\sigma\ ,
\end{align}
which are defined differently in chapter \ref{chapter::BPS-minimal}.

\section{Supersymmetry charge}\label{app:G2}
From the susy variations one can fix uniquely the supersymmetry charge $\mathcal{Q}$ by the requirement that
\begin{equation}\label{susy_variation_from_Q}   \delta_{\epsilon} \phi = \{\mathcal{Q},\phi \}, \end{equation}
for all fundamental fields (here denoted by $\phi$) in the theory. From the supersymmetry variations \eqref{susy-gravi},\eqref{susygluino},\eqref{susyhyperino}, together with the Dirac brackets \eqref{Dirac_brackets}, one finds
\begin{align}\label{susy_charge_beg}
\begin{split}
    \mathcal{Q} = \int_V {\rm d} \Sigma_{\mu} &[ \epsilon^{\mu \nu \rho \sigma} \overline{\psi}_{\nu}^A \gamma_{\rho} \tilde{\mathcal{D}}_{\sigma} \epsilon_A + h.c. \\
    &- i g_{i \bar{\jmath}} \overline{\lambda}^{\bar{\jmath}}_A  \gamma^{\mu} (i\nabla_\nu z^i
\gamma^\nu\varepsilon^A + G_{\nu\rho}^{-i}
\gamma^{\nu\rho}\epsilon^{AB}\varepsilon_B+gW^{iAB}\varepsilon_B\ ) + h.c. \\
&- i \overline{\zeta}^{\alpha} \gamma^{\mu} (i\,
\mathcal{U}^{B\beta}_u\nabla_\nu q^u \gamma^\nu \varepsilon^A
\epsilon_{AB}\mathbb{C}_{\alpha\beta} + g N_\alpha^A\varepsilon_A\ ) + h.c. ]\ ,
\end{split}
\end{align}
up to higher order in fermions. The expression for the supercharge simplifies considerably when evaluated on shell, due to the very suggestive form of the equations of motion of the gravitinos:
\begin{align}\label{eom_gravitino}
\begin{split}
\epsilon^{\mu \nu \rho \sigma} \gamma_{\nu} \tilde{\mathcal{D}}_{\rho} \psi_{\sigma A} &= g_{i \bar{\jmath}} (\nabla^\mu \bar{z}^{\bar{\jmath}} \lambda^{i}_A-\nabla_{\nu} z^i \gamma^{\mu \nu} \lambda^{\bar{\jmath}}_A) - i g_{i \bar{\jmath}} ( G_{\mu \nu}^{+ \bar{\jmath}}
\gamma^{\nu}\epsilon_{AB} \lambda^{i B} + gW^{i}_{AB} \gamma^{\mu} \lambda^{\bar{\jmath} B} ) \\
& - (\mathcal{U}^{B\beta}_u\nabla^{\mu}q^u \epsilon_{AB}\mathbb{C}_{\alpha\beta} - \mathcal{U}^{B\beta}_u\nabla_{\nu}q^u \gamma^{\mu \nu} \epsilon_{AB}\mathbb{C}_{\alpha\beta} + i g N_{\alpha A} \gamma^{\mu}) \zeta^{\alpha}\ .
\end{split}
\end{align}
After performing a partial integration of the first term on the r.h.s.\ of \eqref{susy_charge_beg} and using \eqref{eom_gravitino}, the supercharge becomes a surface integral:
\begin{align}\label{susy_charge}
\mathcal{Q} \stackrel{e.o.m.}{=} \oint_{\partial V} {\rm d} \Sigma_{\mu \nu} \epsilon^{\mu \nu \rho \sigma} \left( \overline{\psi}_{\sigma}^A \gamma_{\rho} \varepsilon_A - \overline{\psi}_{\sigma A} \gamma_{\rho} \varepsilon^A \right)\ ,
\end{align}
similarly to \eqref{7susy_charge} in the minimal case.

\selectlanguage{english}
\renewcommand{\leftmark}{\it Acknowledgements}
\renewcommand{\rightmark}{\it Acknowledgements}

\chapter*{Acknowledgements}
\addcontentsline{toc}{chapter}{Acknowledgements}
\setlength{\parindent}{0pt}\setlength{\parskip}{1em}

There are hundreds of people I would like to thank to for their involvement with my life during my PhD years. The top of the list is reserved for my parents and sister, who have always been the most important and influential people for me. Unfortunately, I am bound to forget to mention some of the other names in the list and therefore I will only write about the people that have most directly contributed to the successful completion of this work. To all my friends in Bulgaria, Netherlands and the other parts of the world I would just like to say that you are all a-mazing!

Academically, I am most grateful to my supervisor, Stefan Vandoren, for allowing me to work with him and his scientific enthusiasm and insight, patience and guidance in the last five years. I hope I have become at least a bit sharper and more rigorous in my physics understanding due to our daily interactions. Another important person in my academic life that I need to mention is Bernard de Wit, who introduced me to the subject of quantum field theory in first place. The two of them, together with Biene Meijerman and the people in NUFFIC and the university of Sofia (especially Hristo Dimov and Radoslav Rashkov), backed me up crucially in the organizationally messy start of my PhD. I hope my thesis was worth their effort.

My academic career has been additionally boosted by many other colleagues through the years. I am very thankful to Stefanos Katmadas for the hundreds of hours spent for answering my questions and random physics discussions. I also learnt a lot during my collaboration with Hugo Looijestijn and Chiara Toldo, who always kept my motivation and enthusiasm at high levels. I further enjoyed the numerous discussions and journal clubs with Anne Franzen, Chris Wever and especially Philipp H\"{o}hn on non-string theory related physics. Other colleagues at the ITF that have suffered from my never-ending questions are Stijn van Tongeren, Alessandro Sfondrini, Ivano Lodato, Vivian Jacobs (special thanks to Vivian and Chris for their help with the Dutch version of my summary), Timothy Budd, Maaike van Zalk, Nabamita Banerjee, Erik Plauschinn, and Jeroen Diederix. Outside Utrecht, I have profited from various discussions and correspondence with Glenn Barnich, Gianguido Dall'Agata, Riccardo D'Auria, Jan de Boer, Giuseppe Dibitetto, Roberto Emparan, Stefano Giusto, Alessandra Gnecchi, Gary Horowitz, Dietmar Klemm, Andrei Micu, Petya Nedkova, Tomas Ort\'{i}n, Ashoke Sen, Ergin Sezgin, and Kostas Skenderis. I am particularly grateful to Tomas Ort\'{\i}n for passing us a set of unpublished notes written by him and two collaborators (Patrick Meessen and Jose Izquierdo). In the end, I would like to express my general gratitude to the people at the ITF, the DRSTP schools and physics communities elsewhere for creating a friendly work environment that welcomes young and inexperienced students.

\begin{flushright} Kiril Hristov,\\18 April 2012. \end{flushright}


\addcontentsline{toc}{chapter}{Bibliography}
\renewcommand{\leftmark}{\it Bibliography}
\renewcommand{\rightmark}{\it Bibliography}
\bibliographystyle{utphys}
\providecommand{\href}[2]{#2}\begingroup\raggedright\endgroup

\end{document}